\documentclass[12pt,a4paper]{book}
\usepackage{UniSa-phd}
\usepackage{amssymb}
\usepackage{graphicx}
\usepackage{amsfonts}
\usepackage{amsmath}
\usepackage[bf, small]{caption}

\begin{document}

\frontmatter

\dept{Universit\`{a} degli Studi di Salerno\\Dipartimento di
Fisica "E.R. Caianiello"}

\title{\textbf{The Weak Field Limit\\of\\Higher Order Gravity}}

\author{Arturo Stabile}

\submitdate{March 20, 2008}

\supervisors{Prof. Salvatore Capozziello\\Prof. Gaetano Scarpetta}

\coord{\\Prof. Gaetano Vilasi}

\phdy{VI Ciclo II Serie\\2004\,-\,2007}

\dedication{\emph{\textbf{To my lovely family}}:\\\emph{My
grandmothers Angela and Giuseppina}\\\emph{My grandfathers Antonio
and Arturo}\\\emph{My parents Giuseppina and Pasquale}\\\emph{My
brother Antonio}\\\emph{and ... She ... Georgia}}

\maketitle

\newpage\thispagestyle{empty}

\tableofcontents

\chapter{The Weak Field Limit of Higher Order Gravity}

The Higher Order Theories of Gravity -
$f(R, R_{\alpha\beta}R^{\alpha\beta})$ - theory, where $R$
is the Ricci scalar, $R_{\alpha\beta}$ is the Ricci tensor
and $f$ is any analytic function - have recently attracted
a lot of interest as alternative candidates to explain the
observed cosmic acceleration, the flatness of the rotation
curves of spiral galaxies and other relevant astrophysical
phenomena. It is a crucial point testing these alternative
theories in the so called weak field and newtonian limit
of a $f(R, R_{\alpha\beta}R^{\alpha\beta})$ - theory.
With this "perturbation technique" it is possible to find
spherically symmetric solutions and compare them with  the
ones of General Relativity. On both approaches we found a
modification of General Relativity: the behaviour of
gravitational potential presents a modification Yukawa - like
in the newtonian case and a massive propagation in the weak
field case. When the modification of the theory is removed
(i.e. $f(R, R_{\alpha\beta}R^{\alpha\beta})\,=\,R$, Hilbert -
Einstein lagrangian) we find the usual outcomes of General
Relativity. Also the Noether symmetries technique has been
investigated to find some time independent spherically
symmetric solutions.\\\\\\
Arturo Stabile\footnote{e -
mail address: arturo.stabile@gmail.com}

\newpage\thispagestyle{empty}

\chapter{Introduction: History and Motivations}
\markboth{Introduction: Historical motivations}{Introduction:
Historical motivations}

\section{General considerations on Gravitational theories}

General Relativity (GR) is a comprehensive theory of spacetime,
gravity and matter. Its formulation implies that space and time
are not "absolute" entities, as in Classical Mechanics, but
dynamical quantities strictly related to the distribution of
matter and energy. As a consequence, this approach gave rise to a
new conception of the Universe itself which, for the first time,
was considered as a dynamical system. In other words, Cosmology
has been enclosed in the realm of Science and not only of
Philosophy, as before the Einstein work. On the other hand, the
possibility of a scientific investigation of the Universe has led
to the formulation of the \emph{Standard Cosmological Model}
\cite{weinberg} which, quite nicely, has matched with
observations.

Despite of these results, the study of possible modifications of
Einstein's theory of gravitation has a long history which reaches
back to the early 1920s \cite{weyl, weyl1, pauli, bach, eddington,
lanczos}. While the proposed early amendments of Einstein's theory
were aimed toward the unification of gravity with the other
interactions of physics, like electromagnetism and whether GR is
the only fundamental theory capable of explaining the
gravitational interaction, the recent interest in such
modifications comes from cosmological observations (for a
comprehensive review, see \cite{schmidt}). Such issues come,
essentially, from Cosmology and Quantum Field Theory. In the first
case, the presence of the Big Bang singularity, the flatness and
horizon problems \cite{guth} led to the statement that
Cosmological Standard Model, based on GR and Standard Model of
Particle Physics, is inadequate to describe the Universe at
extreme regimes. These observations usually lead to the
introduction of additional \emph{ad-hoc} concepts like dark
energy/matter if interpreted within Einstein's theory. On the
other hand, the emergence of such stopgaps could be interpreted as
a first signal of a breakdown of GR at astrophysical and
cosmological scales \cite{cap-card-tro2, cap-card-tro3}, and led
to the proposal of several alternative modifications of the
underlying gravity theory (see \cite{noj-odi6} for a review).
Besides from Quantun Field Theory point view, GR is a
\emph{classical} theory which does not work as a fundamental
theory, when one wants to achieve a full quantum description of
spacetime (and then of gravity).

While it is very natural to extend Einstein's gravity to theories
with additional geometric degrees of freedom, (see for example
\cite{heh-von-ker-nes, heh-mcc-mie-nee, trautman} for general
surveys on this subject as well as \cite{puetzfeld} for a list of
works in a cosmological context), recent attempts focused on the
old idea of modifying the gravitational Lagrangian in a purely
metric framework, leading to higher order field equations. Such an
approach is the so-called \emph{Extended Theories of Gravity}
(ETG) which have become a sort of paradigm in the study of
gravitational interaction. They are based on corrections and
enlargements of the  Einstein theory. The paradigm consists,
essentially, \emph{in adding higher order curvature invariants and
minimally or non-minimally coupled scalar fields into dynamics
which come out from  the effective action of quantum gravity}
\cite{buc-odi-sha}.

The idea to extend Einstein's theory of gravitation is fruitful
and economic also with respect to several attempts which try to
solve problems by adding new and, most of times, unjustified
ingredients in order to give a self-consistent picture of
dynamics. The today observed accelerated expansion of the Hubble
flow and the missing matter of astrophysical large scale
structures, are primarily enclosed in these considerations. Both
the issues could be solved changing the gravitational sector,
\emph{i.e.} the \emph{l.h.s.} of field equations. The philosophy
is alternative to add new cosmic fluids (new components in the
\emph{r.h.s.} of field equations) which should give rise to
clustered structures (dark matter) or to accelerated dynamics
(dark energy) thanks to exotic equations of state. In particular,
relaxing the hypothesis that gravitational Lagrangian has to be a
linear function of the Ricci curvature scalar $R$, like in the
Hilbert-Einstein formulation, one can take into account an
effective action where the gravitational Lagrangian includes other
scalar invariants.

In summary, the general features of ETGs are that the Einstein
field equations result to be modified in two senses: $i)$ geometry
can be non-minimally coupled to some scalar field, and / or $ii)$
higher than second order derivative terms in the metric come out.
In the former case, we generically deal with scalar-tensor
theories of gravity; in the latter, we deal with higher order
theories. However combinations of non-minimally coupled and
higher-order terms can emerge as contributions in effective
Lagrangians. In this case, we deal with higher-order-scalar-tensor
theories of gravity.

Due to the increased complexity of the field equations in this
framework, the main amount of works dealt with some formally
equivalent theories, in which a reduction of the order of the
field equations was achieved by considering the metric and the
connection as independent fields \cite{ama-elg-mot-mul,
mag-fer-fra, all-bor-fra, sot1, sot-lib}. In addition, many
authors exploited the formal relationship to scalar-tensor
theories to make some statements about the weak field regime,
which was already worked out for scalar-tensor theories more than
ten years ago \cite{dam-esp}.

Other motivations to modify GR come from the issue of a full
recovering of the Mach principle which leads to assume a varying
gravitational coupling. The principle states that the local
inertial frame is determined by some average of the motion of
distant astronomical objects \cite{bondi}. This fact implies that
the gravitational coupling can be scale-dependent and related to
some scalar field. As a consequence, the concept of ``inertia''
and the Equivalence Principle have to be revised. For example, the
Brans-Dicke theory \cite{bra-dic} is a serious attempt to define
an alternative theory to the Einstein gravity: it takes into
account a variable Newton gravitational coupling, whose dynamics
is governed by a scalar field non-minimally coupled to the
geometry. In such a way, Mach's principle is better implemented
\cite{bra-dic, cap-der-rub-scu, sciama}.

As already mentioned, corrections to the gravitational Lagrangian,
leading to higher order field equations, were already studied by
several authors \cite{weyl1, eddington, lanczos} soon after the GR
was proposed. Developments in the 1960s and 1970s \cite{buchdahl1,
dewitt, bicknell, havas, stel}, partially motivated by the
quantization schemes proposed at that time, made clear that
theories containing {\it only} a $R^2$ term in the Lagrangian were
not viable with respect to their weak field behavior. Buchdahl, in
1962 \cite{buchdahl1} rejected pure $R^2$ theories because of the
non-existence of asymptotically flat solutions.

Another concern which comes with generic \emph{higher order
gravity} (HOG) theories is linked to the initial value problem. It
is unclear if the prolongation of standard methods can be used in
order to tackle this problem in every theory. Hence it is doubtful
that the Cauchy problem could be properly addressed in the near
future, for example within $1/R$ theories, if one takes into
account the results already obtained in fourth order theories
stemming from a quadratic Lagrangian \cite{tey-tou, dur-ker}.

Starting from the Hilbert-Einstein Lagrangian

\begin{eqnarray}
\mathcal{L}_{GR}\,=\,\sqrt{-g}R
\end{eqnarray}
the following terms

\begin{eqnarray}
\left\{\begin{array}{ll}
\mathcal{L}_1\,=\,\sqrt{-g}R^2\\\\
\mathcal{L}_2\,=\,\sqrt{-g}R_{\alpha\beta}R^{\alpha\beta}\\\\
\mathcal{L}_3\,=\,\sqrt{-g}R_{\alpha\beta\gamma\delta}R^{\alpha\beta\alpha\delta}
\end{array}\right.
\end{eqnarray}
and combinations of them, represent a first obvious choices for an
extended gravity theory with improved dynamics with respect to GR.
Since the variational derivative of $\mathcal{L}_3$ can be
linearly expressed \cite{bach, lanczos2} via the variational
derivatives of $\mathcal{L}_1$ and $\mathcal{L}_2$, one can omit
$\mathcal{L}_3$ in the final Lagrangian of a HOG without loss of
generality.

Besides, every unification scheme as Superstrings, Supergravity or
Grand Unified Theories, takes into account effective actions where
non-minimal couplings to the geometry or higher order terms in the
curvature invariants are present. Such contributions are due to
one-loop or higher loop corrections in the high curvature regimes
near the full (not yet available) quantum gravity regime
\cite{buc-odi-sha}. Specifically, this scheme was adopted in order
to deal with the quantization on curved spacetimes and the result
was that the interactions among quantum scalar fields and
background geometry or the gravitational self-interactions yield
corrective terms in the Hilbert-Einstein Lagrangian
\cite{bir-dav}. Moreover, it has been realized that such
corrective terms are inescapable in order to obtain the effective
action of quantum gravity at scales closed to the Planck one
\cite{vilkovisky}. All these approaches are not the ``\emph{full
quantum gravity}" but are needed as working schemes toward it.

In summary, higher order terms in curvature invariants (such as
$R^{2}$, $R_{\alpha\beta}R^{\alpha\beta}$,
$R_{\alpha\beta\gamma\delta}R^{\alpha\beta\gamma\delta}$, $R
\,\Box R$, or $R \,\Box^{k}R$) or non-minimally coupled terms
between scalar fields and geometry (such as $\phi^{2}R$) have to
be added to the effective Lagrangian of gravitational field when
quantum corrections are considered. For instance, one can notice
that such terms occur in the effective Lagrangian of strings or in
Kaluza-Klein theories, when the mechanism of dimensional reduction
is used \cite{gas-ven}.

On the other hand, from a conceptual viewpoint, there are no
\emph{a priori} reason to restrict the gravitational Lagrangian to
a linear function of the Ricci scalar $R$, minimally coupled with
matter \cite{mag-fer-fra}. More precisely, higher order terms
appear always as contributions of order two in the field
equations. For example, a term like $R^{2}$ gives fourth order
equations \cite{ruz-ruz}, $R\,\Box R$ gives sixth order equations
\cite{got-sch-sta, ame-bat-cap-got-mul-occ-sch}, $R\,\Box^{2}R$
gives eighth order equations \cite{bat-sch} and so on. By a
conformal transformation, any 2nd order derivative term
corresponds to a scalar field\footnote{The dynamics of such scalar
fields is usually given by the corresponding Klein-Gordon
equation, which is  second order.}: for example, fourth order
gravity gives Einstein plus one scalar field, sixth order gravity
gives Einstein plus two scalar fields and so on \cite{got-sch-sta,
schmidt1}. Furthermore, the idea that there are no ``exact'' laws
of physics could be taken into serious account: in such a case,
the effective Lagrangians of physical interactions are
``stochastic'' functions. This feature means that the local gauge
invariances (\emph{i.e.} conservation laws) are well approximated
only in the low energy limit and the fundamental physical
constants can vary \cite{bar-ott}.

\section{Issues from Cosmology}

Beside fundamental physics motivations, these theories have
acquired a huge interest in Cosmology due to the fact that they
``naturally" exhibit inflationary behaviors able to overcome the
shortcomings of  Cosmological Standard Model (based on GR). The
related cosmological models seem  realistic and capable of
matching with the Cosmic Microwave Background Radiation (CMBR)
observations \cite{ dur-ker, starobinsky, la-ste}. Furthermore, it
is possible to show that, via conformal transformations, the
higher order and non-minimally coupled terms always correspond to
the Einstein gravity plus one or more than one minimally coupled
scalar fields \cite{tey-tou, got-sch-sta, maeda, wands,
cap-der-mar}.

Furthermore, it is possible to show that the $f(R)$-gravity
($f$-gravity) is equivalent not only to a scalar-tensor one but
also to the Einstein theory plus an ideal fluid
\cite{cap-noj-odi}. This feature results very interesting if we
want to obtain multiple inflationary events since an early stage
could select ``very'' large-scale structures (clusters of galaxies
today), while a late stage could select ``small'' large-scale
structures (galaxies today) \cite{ame-bat-cap-got-mul-occ-sch}.
The philosophy is that each inflationary era is related to the
dynamics of a scalar field. Finally, these extended schemes  could
naturally solve the problem of ``graceful exit" bypassing the
shortcomings of former inflationary models \cite{la-ste,
am-cap-lit-occ}.

In recent years, the efforts to give a  physical explanation to
the today observed cosmic acceleration \cite{rie, rie1, per, per1}
have attracted a good amount of interest in $f$-gravity,
considered as a viable mechanism to explain the cosmic
acceleration by extending the geometric sector of field equations
\cite{cap, cap-car-tro, cap-car-car-tro, car-duv-tro-tur,
noj-odi1, noj-odi3, cap-card-tro1, car-dun-cap-tro, vollick,
men-wan2, men-wan3, fla, fla1,kre-alv, all-bor-fra1, cap-card-fra,
col, spe, spe1}. There are several physical and mathematical
motivations to enlarge GR by these theories. For comprehensive
reviews, see \cite{cap-fra, noj-odi, farhoudi}.

Specifically, cosmological models coming from $f$-gravity were
firstly introduced by Staro-binsky \cite{starobinsky} in the early
80'ies to build up a feasible inflationary model where geometric
degrees of freedom had the role of the scalar field ruling the
inflation and the structure formation.

In addition to the  revision of Standard Cosmology at early epochs
(leading to the Inflation),  a new approach is necessary also at
late epochs. ETGs could play a fundamental role also in this
context. In fact,  the increasing bulk of data that have been
accumulated in the last few years have paved the way to the
emergence of a new cosmological model usually referred to as the
\emph{Cosmological Concordance Model} ($\Lambda$ \emph{Cold Dark
Matter}: $\Lambda$CDM).

The Hubble diagram of Type Ia Supernovae (hereafter SNeIa),
measured by both the Supernova Cosmology Project \cite{per, kno}
and the High\,-\,\emph{z} Team \cite{rie, ton} up to redshift $z
\sim 1$, the luminosity distance of Ia Type Supernovae \cite{rie,
rie1, per, per1}, the Large Scale Structure \cite{col} and the
anisotropy of CMBR \cite{spe, spe1} are the evidence that the
Universe is undergoing a phase of accelerated expansion. On the
other hand, balloon born experiments, such as BOOMERanG \cite{deb}
and MAXIMA \cite{sto}, determined the location of the first and
second peak in the anisotropy spectrum of the CMBR strongly
pointing out that the geometry of the Universe is spatially flat.
If combined with constraints coming from galaxy clusters on the
matter density parameter $\Omega_M$, these data indicate that the
Universe is dominated by a non-clustered fluid with negative
pressure, generically dubbed \emph{dark energy}, which is able to
drive the accelerated expansion. This picture has been further
strengthened by the recent precise measurements of the CMBR
spectrum, due to the WMAP experiment \cite{spe, spe1, hin}, and by
the extension of the SNeIa Hubble diagram to redshifts higher than
1 \cite{rie}. From this amount of data, the widely accepted
$\Lambda$CDM is a spatially flat Universe, dominated by cold dark
matter (CDM $(\sim 0.25\div 0.3\%)$ which should explain the
clustered structures) and dark energy $(\Lambda$ $\sim 0.65\div
0.7\%)$, in the form of an ``effective" cosmological constant,
giving rise to the accelerated behavior.

After these observational evidences, an  overwhelming flood of
papers has appeared: they present a great variety of models trying
to explain this phenomenon. In any case,  the simplest explanation
is claiming for the well known cosmological constant $\Lambda$
\cite{sah-sta}. Although it is the best fit to most of the
available astrophysical data \cite{spe, spe1, hin}, the
$\Lambda$CDM model fails in explaining why the inferred value of
$\Lambda$ is so tiny (120 orders of magnitude lower than the value
of quantum gravity vacuum state!) if compared with the typical
vacuum energy values predicted by particle physics and why its
energy density is today comparable to the matter density  (the so
called \emph{coincidence problem}).

Although the cosmological constant \cite{pad, pee-rat,
cop-sam-tsu} remains the most relevant candidate to interpret the
accelerated behavior, several proposals have been suggested in the
last few years: quintessence models, where the cosmic acceleration
is generated by means of a scalar field, in a way similar to the
early time inflation \cite{starobinsky}, acting at large scales
and recent epochs \cite{pee-rat1, cal-dav-ste}; models based on
exotic fluids like the Chaplygin-gas \cite{kam-mos-pas,
bil-tup-vio, ben-ber-sen}, or non-perfect fluids
\cite{car-tor-tro-cap}); phantom fields, based on scalar fields
with anomalous signature in the kinetic term \cite{cal,
sin-sam-dad, noj-odi-tsu, faraoni2}, higher dimensional scenarios
(braneworld) \cite{dva-gab-por, dva-gab-kol-nit, dva-gab-kol-nit1,
maa}. These results can be achieved in metric and Palatini
approaches \cite{all-bor-fra, cap-car-tro, vollick, men-wan2, fla,
fla1, men-wan3, kre-alv, all-bor-fra1, mul-vil}. In addition,
reversing the problem, one can reconstruct the form of the gravity
Lagrangian by observational data of cosmological relevance through
a "back scattering" procedure. All these facts suggest that the
function $f$ should be more general than the linear
Hilbert-Einstein one implying that HOG could be a suitable
approach to solve GR shortcomings without introducing mysterious
ingredients as dark energy and dark matter (see e.g.
\cite{cap-car-tro1, cap-car-car-tro1}).

Actually, all of these models, are based on the peculiar
characteristic of introducing new sources into the cosmological
dynamics, while it would be preferable to develop scenarios
consistent with observations without invoking further parameters
or components non-testable (up to now) at a fundamental level.

Moreover, it is not clear where this scalar field originates from,
thus leaving a great uncertainty on the choice of the scalar field
potential. The subtle and elusive nature of  dark energy has led
many authors to look for completely different scenarios able to
give a quintessential behavior without the need of exotic
components. To this aim, it is worth stressing that the
acceleration of the Universe only claims for a negative pressure
dominant component, but does not tell anything about the nature
and the number of cosmic fluids filling the Universe.

Actually, there is still a different way to face the problem of
cosmic acceleration. As stressed in \cite{lue-sco-sta}, it is
possible that the observed acceleration is not the manifestation
of another ingredient in the cosmic pie, but rather the first
signal of a breakdown of our understanding of the laws of
gravitation (in the infra-red limit).

From this point of view, it is thus tempting to modify the
Friedmann equations to see whether it is possible to fit the
astrophysical data with  models comprising only the standard
matter. Interesting examples of this kind are the Cardassian
expansion \cite{fre-lew} and the Dvali-Gabadaze-Porrati gravity
\cite{dva-gab-por}. Moving in this same framework, it is possible
to  find alternative schemes where a quintessential behavior is
obtained by taking into account effective models coming from
fundamental physics giving rise to generalized or HOG actions
\cite{cap, noj-odi1, car-duv-tro-tur, all-bor-fra} (for a
comprehensive review see \cite{noj-odi}).

For instance, a cosmological constant term may be recovered as a
consequence of a non\,-\,vanishing torsion field thus leading to a
model which is consistent with both SNeIa Hubble diagram and
Sunyaev\,-\,Zel'dovich data coming from clusters of galaxies
\cite{cap-car-pie-ser-tro}. SNeIa data could also be efficiently
fitted including higher-order curvature invariants in the gravity
Lagrangian \cite{cap-car-car-tro, li-bar, noj-odi2, li-bar-mot}.
It is worth noticing that these alternative models provide
naturally a cosmological component with negative pressure whose
origin is related to the geometry of the Universe thus overcoming
the problems linked to the physical significance of the scalar
field.

It is evident, from this short overview, the large number of
cosmological models which are viable candidates to explain the
observed accelerated expansion. This abundance of models is, from
one hand, the signal of the fact that we have a limited number of
cosmological tests to discriminate among rival theories, and, from
the other hand, that a urgent degeneracy problem has to be faced.
To this aim, it is useful to remark that both the SNeIa Hubble
diagram and the angular size\,-\,redshift relation of compact
radio sources \cite{che-rat, pod-dal-mor-rat} are distance based
methods to probe cosmological models so then systematic errors and
biases could be iterated. From this point of view, it is
interesting to search for tests based on time-dependent
observables.

For example, one can take into account the \emph{lookback time} to
distant objects since this quantity can discriminate among
different cosmological models. The lookback time is
observationally estimated as the difference between the present
day age of the Universe and the age of a given object at redshift
$z$. Such an estimate is possible if the object is a galaxy
observed in more than one photometric band since its color is
determined by its age as a consequence of stellar evolution. It is
thus possible to get an estimate of the galaxy age by measuring
its magnitude in different bands and then using stellar
evolutionary codes to choose the model that reproduces the
observed colors at best.

The resort to modified gravity theories, which extend in some way
the GR, allows to pursue this different approach (no further
unknown sources) giving rise to suitable cosmological models where
a late time accelerated expansion naturally arises.

The idea that the Einstein gravity should be extended or corrected
at large scales (infrared limit) or at high energies (ultraviolet
limit) is suggested by several theoretical and observational
issues. Quantum field theories in curved spacetimes, as well as
the low energy limit of string theory, both imply
semi\,-\,classical effective Lagrangians containing  higher-order
curvature invariants or scalar-tensor terms. In addition, GR has
been tested only at  solar system scales while it shows several
shortcomings if checked at higher energies or larger scales.

Of course modifying the gravitational action asks for several
fundamental challenges. These models can exhibit instabilities
\cite{faraoni3, cog-zer, cog-gas-zer} or ghost\,-\,like behaviors
\cite{stel}, while, on the other side, they should be matched with
the low energy limit observations and experiments (solar system
tests). Despite of all these issues, in the last years, several
interesting results have been achieved in the framework of $f$-
gravity  at cosmological, galactic and solar system scales.

For example, models based on generic functions of the Ricci scalar
$R$ show cosmological solution with late time accelerating
dynamics \cite{ all-bor-fra, car-duv-tro-tur, noj-odi1, noj-odi3,
cap-card-tro1, car-dun-cap-tro}, in addition, it has been shown
that some of them could agree with CMBR observational
prescriptions \cite{li-bar1, car-bun-tro}, nevertheless this
matter is still argument of debate \cite{bea-ber-pog-sil-tro,
son-hu-saw}. For a review of the models and their cosmological
applications see, e.g.,\cite{noj-odi, son-hu-saw}.

Moreover, considering $f$-gravity in the low energy limit, it is
possible to obtain corrected gravitational potentials capable of
explaining the flat rotation curves of spiral galaxies without
considering huge amounts of dark matter \cite{cap-card-tro2,
cap-card-tro, fri-sal, sobouti, men-ros} and, furthermore, this
seems the only self-consistent way to reproduce the universal
rotation curve of spiral galaxies \cite{sal-lap-ton-gen-yeg-kle}.
On the other hand, several anomalies in Solar System experiments
could be framed and addressed in this picture
\cite{ber-boe-har-lob, noj-odi5}.

Summarizing, almost $95\%$ of matter-energy content of the
universe is unknown in the framework of Standard Cosmological
Model while we can experimentally probe only gravity and ordinary
(baryonic and radiation) matter. Considering another point of
view, anomalous acceleration (Solar System), dark matter (galaxies
and galaxy clusters), dark energy (cosmology) could be nothing
else but the indications that shortcomings are present in GR and
gravity is an interaction depending on the scale. The assumption
of a linear Lagrangian density in the Ricci scalar $R$ for the
Hilbert-Einstein action could be too simple to describe gravity at
any scale and more general approaches should be pursued to match
observations. Among these schemes, several motivations suggest to
generalize GR by considering gravitational actions where generic
functions of curvature invariants are present.

\section{The Weak Field Limit of Higher Order Gravity}

It is well known that GR is the cornerstone theory among several
attempts proposed to describe gravity. It represents an elegant
approach giving several phenomenological predictions, but its
validity, in the Newtonian limit regime, is experimentally probed
only at Solar System scales. However, also at these scales, some
conundrums come out as an apparent, anomalous, long-range
acceleration revealed from the data analysis of Pioneer 10/11,
Galileo, and Ulysses spacecrafts. Such a feature is difficult to
be framed in the standard scheme of GR and its low energy limit
\cite{anderson, anderson1}, while it could be framed in a general
theoretical scheme by taking corrections to the Newtonian
potential into account \cite{ber-boh-lob}. Furthermore, at
galactic distances, huge bulks of dark matter are needed to
provide realistic models matching with observations. In this case,
retaining GR and its low energy limit, implies the introduction of
an actually unknown ingredient. We face a similar situation even
at larger scales: clusters of galaxies are gravitationally stable
and bounded only if large amounts of dark matter are supposed in
their potential wells.

Taking into account the weak field limit approximation, ETGs are
expected to reproduce GR \cite{will}. This fact is matter of
debate since several relativistic theories \emph{do not} reproduce
exactly the Einstein results in the Newtonian approximation but,
in some sense, generalize them. As it was firstly noticed by
Stelle \cite{stel}, a $R^2$-theory gives rise to Yukawa-like
corrections in the Newtonian potential. Such a feature could have
interesting physical consequences. For example, some authors claim
to explain the flat rotation curves of galaxies by using such
terms \cite{san}. Others \cite{man-kaz} have shown that a
conformal theory of gravity is nothing else but a HOG model
containing such terms in the Newtonian limit.

In general, any relativistic theory of gravitation  yields
corrections to the Newton potential (see for example
\cite{qua-sch}) which, in the post-Newtonian formalism, could be a
test for the same theory \cite{will}. Furthermore the newborn
\emph{gravitational lensing astronomy} \cite{sch-ehl-fal} is
giving rise to additional tests of gravity over small, large, and
very large scales which soon will provide direct measurements for
the variation of the Newton coupling \cite{kra-whi}, the potential
of galaxies, clusters of galaxies and several other features of
self-gravitating systems. Such data could be, very likely, capable
of confirming or ruling out the physical consistency of GR or of
any ETG.

In recent papers, some authors  have confronted this kind of
theories even with the Post Parameterized Newtonian (PPN)
prescriptions in metric and Palatini approaches. The results seem
controversial since in some cases \cite{olm, olm2} it is argued
that GR is always valid at Solar System scales and there is no
room for other theories; nevertheless, some other studies
\cite{cap-tro, all-fra-rug-tar} find that recent experiments as
Cassini and Lunar Laser Ranging allow the possibility that ETGs
could be seriously taken into account. In particular, it is
possible to define PPN-parameters in term of $f$-gravity functions
and several classes of fourth order theories result compatible
with experiments in Solar System \cite{cap-tro}.

In a recent paper \cite{mul-vil1}, spherically symmetric solutions
for $f$-gravity in vacuum have been found considering relations
among functions defining the spherical metric or imposing a
constant Ricci curvature scalar. The authors have been able to
reconstruct, at the end, the form of some $f$-theories, discussing
their physical relevance. In \cite{mul-vil2}, the same authors
have discussed static spherically symmetric perfect fluid
solutions for $f$-gravity in metric formalism. They showed that a
given matter distribution is not capable of determining the
functional form of $f$.

The discussion about the short scale behavior of HOG has been
quite vivacious in the last years since  GR shows is best
predictions just at the Solar System level. As matter of fact,
measurements coming from weak field limit tests like the bending
of light, the perihelion shift of planets, frame dragging
experiments represent inescapable tests for whatever theory of
gravity. Actually, in our opinion, there are sufficient
theoretical predictions to state that HOG can be compatible with
Newtonian and post-Newtonian prescriptions [\textbf{A}]. In other
papers \cite{cap-tro}, it has been that this result can be
achieved by means of the analogy of $f$\,-\,models with
scalar\,-\,tensor gravity.

Nevertheless, up to now, the discussion on the weak field limit of
$f$\,-\,theories is far to be definitive and there are several
papers claiming for opposite results \cite{noj-odi5,
all-fra-rug-tar, sot, chi, eri-smi-kam, cli-bar, faraoni1,
jin-liu-li,chi-eri-smi}, or stating that no progress has been
reached in the last years due to the several common misconceptions
in the various theories of gravity \cite{sot-far-lib}.

In the last few years, several authors have dealt with this matter
with contrasting conclusions, in particular with respect to the
PPN limit \cite{olm, cap-tro, all-fra-rug-tar, chi, faraoni4}. On
the other hand, the investigation of spherically symmetric
solutions for such kind of models has been developed in several
papers \cite{mul-vil1, mul-vil2, sot, kai-pii-rei-sun}. Such an
analysis deserves particular attention since it can allow to draw
interesting conclusions on the effective modification of the
gravitational potential induced by HOG at low energies and, in
addition, it could shed new light on the PPN limit of such
theories at least in a preliminary way. For example, theories like
${\displaystyle f\,=\,R+\frac{\mu}{R}}$, which fairly address the
cosmic acceleration issue \cite{cap, cap-car-tro,
cap-car-car-tro}, suffer a ill-defined PPN-limit since a theory
containing terms like $R^{-1}$ is singular in $R\,=\,0$ and does
not admit any Minkowski limit and then  any other background
solution which is Ricci flat. On the other hand, affirming that
the unique $f$-gravity spherically symmetric "static" solution,
corresponding to a realistic mass source and matching the present
cosmic background at infinity has the PPN parameter $\gamma=1/2$,
in conflict with experiments which give $\gamma\simeq 1$ could be
misleading  since by assuming, for example, $f\,=\,R^{1+\epsilon}$
with $\epsilon\rightarrow 0$ has to give results compatible with
GR (\emph{i.e.} $\gamma\sim 1$).

As a matter of fact, defining a PPN-limit in such a case is a
quite delicate issue, since in order to develop an analytical
study of the deviation from the Newtonian approximation requires
that the spacetime should be, at least, asymptotically Minkowski.
Finally, approaching the problem considering a curvature constant
metric, as in the case of the Schwarzschild\,-\,de Sitter
solution, could induce to flawed conclusions. As a consequence,
understanding the properties of HOG with respect to spherically
symmetric solutions in the weak field limit turns out to be  a key
issue from several points of view.

This is only one example about the debate on the weak limit field
limit: authors approached the weak limit issue following different
schemes and developing different parameterizations which, in some
cases, turn out to be not necessarily correct.

The purpose of this thesis (also referring to the published papers
[\textbf{C}], [\textbf{D}]) is to take part to the debate,
building up a rigorous formalism which deals with the formal
definition of weak field and small velocities limit applied to HOG
gravity. In a series of papers [\textbf{C}], [\textbf{D}],
[\textbf{E}], the aim is to pursue a systematic discussion
involving: $i)$ the Newtonian limit of $f$-gravity, $ii)$
spherically symmetric solutions toward the weak filed limit of
$f$-gravity; and, finally, $iii)$ considering general HOG theories
where also invariants as $R_{\mu\nu}R^{\mu\nu}$ or
$R_{\alpha\beta\mu\nu}R^{\alpha\beta\mu\nu}$ are considered.
Besides the Birkhoff theorem is not a general result in HOG models
[\textbf{C}] also if it holds for several interesting classes of
these theories as discussed, for example, in \cite{whitt,
bar-ham1}.

The analysis is based on the metric approach, developed in the
Jordan frame, assuming that the observations are performed in it,
without resorting to any conformal transformation as done in
several cases \cite{olm}. This point of view is adopted in order
to avoid dangerous variable changes which could compromise the
correct physical interpretation of the results.

A relevant aspect of HOG thoeries, in the post-Minkowskian limit
is the propagation of the gravitational fields. It turns out that
wave signals can be characterized with both tensorial and scalar
mode [\textbf{F}]. This issue represents a quite striking
difference between GR-like models and extended gravity models
since in the standard Einstein scheme only tensorial degrees of
freedom are allowed. As matter of facts, the gravitational wave
limit of these models can represent an interesting framework to
study, in order to discuss the physical observable footprints
which discriminate between GR and HOG experimental predictions.

\emph{In this PhD thesis, we are going to analyze and discuss, in
a general way and without specifying a priori the form of the
Lagrangian, the relation between the spherical symmetry and the
weak field limit, pointing out the differences and the relations
with respect to the post-Newtonian and the post-Minkowskian limits
of $f$-gravity. Our aim is to develop a systematic approach
considering the theoretical prescriptions to obtain a correct weak
field limit in order to point out  the analogies and the
differences with respect to GR. A fundamental issue is to recover
the asymptotically flat solution in absence of gravity and the
well-known results related to the specific case $f\,=\,R$, i.e.
GR.} Only in this situation a correct comparison between GR and
any ETG is possible from an experimental and a theoretical
viewpoint. For example,

In literature, there are several definitions and several claims in
this direction but clear statements and discussion on these
approaches urge in order to find out definite results to be tested
by experiments \cite{sot-far-lib}.

\section{Plan of thesis}

The layout of the PhD thesis is organized as follows. In the first
chapter we report a general review of ETGs and the fundamental
aspects of GR. In particular we display all fundamental tools:
Einstein Equation, Bianchi Identity, Conformal transformations,
Metric and Palatini formalism, ETG theories (Scalar-tensor, HOG
theories and so on), Coordinates system transformations and the
relations between them (for example standard, isotropic
coordinates etc).

In the second chapter some "exact" spherically symmetric solution
of GR is shown (\emph{Schwar-zschild, Schwarzschild-de Sitter,
Reissner-Nordstrom solution}). On the other hand, we show the
technicality of development of field equation [\textbf{C}] with
respect to Newtonian and Post-Newtonian approach: in such case we
also introduce the Eddington parameters. Finally we perform the
post-Minkowskian limit: the gravitational waves. The developments
are computed in generic coordinates systems and in the gauge
harmonic.

The third chapter is devoted to some general remarks on spherical
symmetry in $f$-gravity [\textbf{D}]. In particular, the
expression of the Ricci scalar and the general form of metric
components are derived in spherical symmetry discussing how
recovering the correct Minkowski flat limit. We discuss the
spherically symmetric background solutions with constant scalar
curvature considering, in particular Schwarzschild-like and
Schwarzschild-de Sitter-like solutions with constant curvature; we
discuss also the cases in which the spherical symmetry is present
also for the Ricci scalar depending on the radial coordinate $r$.
This is an interesting situation, not present in GR. In fact, as
it is well known, in the Einstein theory, the Birkhoff theorem
states that a spherically symmetric solution is always stationary
and static \cite{haw-ell} and the Ricci scalar is constant. In $f$
the situation is more general and then the Ricci scalar, in
principle, can evolve with radial and time coordinates. Finally,
the last part of chapter is devoted to the study of a perturbation
approach starting from a spherically symmetric background
considering the general case in which the Ricci scalar is
$R=R(r)$. The motivation is due to the fact that, in GR, the
Schwarzschild solution and the weak field limit coincide under
suitable conditions.

In the fourth one we want to seek for a general method to find out
spherically symmetric solutions in $f$-gravity and, eventually, in
generic ETG [\textbf{B}]. Asking for a certain symmetry of the
metric, we would like to investigate if such a symmetry holds for
a generic theory of gravity. In particular for the $f$-theories.
Specifically, we want to apply the Noether Symmetry Approach
\cite{cap-der-rub-scu} in order to search for spherically
symmetric solutions in generic $f$-theories of gravity. This means
that we consider the spherical symmetry for the metric as a
Noether symmetry and search for $f$ Lagrangians compatible with
it.

In the fifth chapter, we follow a different approach. Starting
from the definitions of PPN-parameters in term of a generic
analytic function $f$ and its derivatives, we deduce a class HOG
theories, compatible with data, by means of an inverse procedure
which allows to compare PPN-conditions with data [\textbf{A}]. As
a matter of fact, it is possible to show that a third order
polynomial, in the Ricci scalar, is compatible with observational
constraints on PPN-parameters. The degree of deviation from GR
depends on the experimental estimate of PPN-parameters. The second
part of chapter is dedicated to very strong debate about the
analogy or not between $f$- and Scalar-tensor gravity
[\textbf{H}]. In fact for some authors the Newtonian limit of
$f$-gravity is equivalent to the one of Brans-Dicke gravity with
$\omega_{BD}\,=\,0$, so that the PPN parameters of these models
turn out to be ill defined. We don't agree with this claim. We
show that this is indeed not true. We discuss that HOG models are
dynamically equivalent to a O'Hanlon Lagrangian which is a special
case of Brans-Dicke theory characterized by a self-interaction
potential and that, in the low energy and small velocity limit,
this will imply a non-standard behaviour. This result turns out to
be completely different from the one of a pure Brans-Dicke model
and in particular suggests that it is completely misleading to
consider the PPN parameters of this theory with
$\omega_{BD}\,=\,0$ in order to characterize the homologous
quantities of $f$-gravity.

In the sixth one we analyze the Newtonian limit of HOG theory. We
are going to focus exclusively on the weak field limit within the
metric approach [\textbf{C}], [\textbf{E}], [\textbf{G}]. At this
point we remind the readers that it was already shown in
\cite{buchdahl} that different variational procedures do not lead
to equivalent results in the case of quadratic order Lagrangians,
casting a shadow on several newer works in which this equivalence
was implicitly assumed. In the first part we use the development
shown in the second chapter for a generic analytic function $f$
and find the solution in the vacuum. For the sake of completeness
we will treat the problem also by imposing the harmonic gauge on
the field equations. Besides, we show that the Birkhoff theorem is
not a general result for $f$-gravity since time-dependent
evolution for spherically symmetric solutions can be achieved
depending on the order of perturbations. In the second part we
find again the Newtonian limit but for a so-called \emph{quadratic
gravity lagrangian} with the Green function method. We find the
internal and external potential generated by an extended
spherically symmetric matter source. In the last part we outline
the general approach to find the expression of metric tensor at
fourth order perturbative for a generic $f$-theory.

In the seventh chapter we develop a formal description of the
gravitational waves propagation in HOG models focusing on the
scalar degrees of freedom and the characteristic of such scalar
candidate in the gravity sector of gauge bosons [\textbf{F}]. As
in the previously chapter we performed the Newtonian limit in
vacuum with a spherically symmetric solution in standard
coordinates, now we repeat the development but in the
post-Minkowskian limit. In addition we discuss, in such a
framework, the definition of the energy-momentum tensor of gravity
which is a fundamental quantity in order to calculate the
gravitational time delay in Pulsar timing. Some considerations on
the differences between GR and HOG in the post-Minkowskian limit
are addressed.

Finally in the last chapter we report the discussions and
conclusions.

\clearpage{\pagestyle{empty}\cleardoublepage}

\mainmatter

\chapter{Extended theories of gravity: a review}

\section{What a good theory of Gravity has to do: General Relativity and its extensions}

From a phenomenological point of view, there are some minimal
requirements that any relativistic theory of gravity has to match.
First of all, it has to explain the astrophysical observations
(e.g. the orbits of planets, the potential of self-gravitating
structures).

This means that it has to reproduce the Newtonian dynamics in the
weak-energy limit. Besides, it has to pass the classical Solar
System tests which are all experimentally well founded
\cite{will}.

As second step, it should reproduce galactic dynamics considering
the observed baryonic constituents (e.g. luminous components as
stars, sub-luminous components as planets, dust and gas),
radiation and Newtonian potential which is, by assumption,
extrapolated to galactic scales.

Thirdly, it should address the problem of large scale structure
(e.g. clustering of galaxies) and finally cosmological dynamics,
which means to reproduce, in a self-consistent way, the
cosmological parameters as the expansion rate, the Hubble
constant, the density parameter and so on. Observations and
experiments, essentially, probe the standard baryonic matter, the
radiation and an attractive overall interaction, acting at all
scales and depending on distance: the gravity.

The simplest theory which try to satisfies the above requirements
was formulated by Albert Einstein in the years 1915 - 1916
\cite{einstein1} and it is known as the Theory of General
Relativity. It is firstly based on the assumption that space and
time have to be entangled into a single spacetime structure,
which, in the limit of no gravitational forces, has to reproduce
the Minkowski spacetime structure. Einstein profitted also of
ideas earlier put forward by Riemann, who stated that the Universe
should be a curved manifold and that its curvature should be
established on the basis of astronomical observations
\cite{eisenhart}.

In other words, the distribution of matter has to influence point
by point the local curvature of the spacetime structure. The
theory, eventually formulated by Einstein in 1915, was strongly
based on three assumptions that the physics of gravitation has to
satisfy.

The "\emph{Principle of Relativity}", which states that amounts to
require all frames to be good frames for Physics, so that no
preferred inertial frame should be chosen a priori (if any exist).

The "\emph{Principle of Equivalence}", that amounts to require
inertial effects to be locally indistinguishable from
gravitational effects (in a sense, the equivalence between the
inertial and the gravitational mass).

The "\emph{Principle of General Covariance}", that requires field
equations to be "generally covariant" (today, we would better say
to be invariant under the action of the group of all spacetime
diffeomorphisms) \cite{schroedinger}.

And - on the top of these three principles - the requirement that
causality has to be preserved (the "\emph{Principle of
Causality}", i.e. that each point of spacetime should admit a
universally valid notion of past, present and future).

Let us also recall that the older Newtonian theory of spacetime
and gravitation - that Einstein wanted to reproduce at least in
the limit of small gravitational forces (what is called today the
"post-Newtonian approximation") - required space and time to be
absolute entities, particles moving in a preferred inertial frame
following curved trajectories, the curvature of which (i.e., the
acceleration) had to be determined as a function of the sources
(i.e., the "forces").

On these bases, Einstein was  led to postulate that the
gravitational forces have to be expressed by the curvature of a
metric tensor field
$ds^2\,=\,g_{\alpha\beta}dx^{\alpha}dx^{\beta}$ on a
four-dimensional spacetime manifold, having the same signature of
Minkowski metric, i.e., the so-called "Loren-tzian signature",
herewith assumed to be $(+,-,-,-)$. He also postulated that
spacetime is curved in itself and that its curvature is locally
determined by the distribution of the sources, i.e. - being
spacetime a continuum - by the four-dimensional generalization of
what in Continuum Mechanics is called the "matter stress-energy
tensor", i.e. a rank-two (symmetric) tensor $T_{\mu\nu}$.

Once a metric $g_{\mu\nu}$ is given, the inverse $g^{\mu\nu}$
satisfies the condition

\begin{eqnarray}\label{controvariant-metric-condition}
g^{\mu\alpha}g_{\alpha\beta}\,=\,\delta_\nu^\mu\,.\end{eqnarray}
Its curvature is expressed by the \emph{Riemann tensor}
(curvature)

\begin{eqnarray}\label{rienmanntensor}
R^{\alpha}_{\,\,\,\,\,\mu\beta\nu}\,=\,\Gamma^{\alpha}_{\mu\nu,\beta}-\Gamma^{\alpha}_{\mu\beta,\nu}+
\Gamma^{\sigma}_{\mu\nu}\Gamma^{\alpha}_{\sigma\beta}-\Gamma^{\alpha}_{\sigma\nu}\Gamma^{\sigma}_{\mu\beta}
\end{eqnarray}
where the comas are partial derivatives. The
$\Gamma^{\alpha}_{\mu\nu}$ are the Christoffel symbols given by

\begin{eqnarray}\label{christoffel}\Gamma^{\alpha}_{\mu\nu}\,=\,\frac{1}{2}g^{\alpha\sigma}(g_{\mu\sigma,\nu}+g_{\nu\sigma,
\mu}-g_{\mu\nu,\sigma})\,,\end{eqnarray}if the Levi-Civita
connection is assumed. The contraction of the Riemann tensor
(\ref{rienmanntensor})

\begin{eqnarray}\label{riccitensor}
R_{\mu\nu}\,=\,R^{\alpha}_{\,\,\,\,\,\mu\alpha\nu}=\,\Gamma^{\sigma}_{\mu\nu,\sigma}-\Gamma^{\sigma}_{\mu\sigma,\nu}+
\Gamma^{\sigma}_{\mu\nu}\Gamma^{\rho}_{\sigma\rho}-\Gamma^{\rho}_{\sigma\nu}\Gamma^{\sigma}_{\mu\rho},
\end{eqnarray}
is the \emph{Ricci tensor} and the scalar

\begin{eqnarray}\label{ricciscalar}
R=g^{\sigma\tau}R_{\sigma\tau}\,=\,R^{\sigma}_{\,\,\,\,\,\sigma}\,=\,g^{\tau\xi}\Gamma^{\sigma}_{\tau\xi,\sigma}-g^{\tau\xi}
\Gamma^{\sigma}_{\tau\sigma,\xi}+g^{\tau\xi}\Gamma^{\sigma}_{\tau\xi}\Gamma^{\rho}_{\sigma\rho}-g^{\tau\xi}\Gamma^{\rho}_
{\tau\sigma}\Gamma^{\sigma}_{\xi\rho}
\end{eqnarray}
is called the \emph{scalar curvature} of $g_{\mu\nu}$. The Riemann
tensor (\ref{rienmanntensor}) satisfies the so-called
\emph{Bianchi identities}:

\begin{eqnarray}\label{bianchi-identity}
\left\{\begin{array}{ll}
R_{\alpha\mu\beta\nu;\delta}+R_{\alpha\mu\delta\beta;\nu}+R_{\alpha\mu\nu\delta;\beta}=0
\\\\
R_{\alpha\mu\beta\nu}^{\,\,\,\,\,\,\,\,\,\,\,\,\,\,\,;\alpha}+R_{\mu\beta;\nu}-R_{\mu\nu;\beta}=0
\\\\
2R_{\alpha\beta}^{\,\,\,\,\,\,\,\,;\alpha}-R_{;\beta}=0
\\\\
2R_{\alpha\beta}^{\,\,\,\,\,\,\,\,;\alpha\beta}-\Box R=0
\end{array} \right.
\end{eqnarray}
where the covariant derivative is
$A^{\alpha\beta\dots\delta}_{\,\,\,\,\,\,\,\,\,\,\,\,\,\,\,\,;\mu}\,=\,\nabla_\mu
A^{\alpha\beta\dots\delta}\,=\,A^{\alpha\beta\dots\delta}_{\,\,\,\,\,\,\,\,\,\,\,\,\,\,\,\,,\mu}+\Gamma^\alpha_{\sigma\mu}
A^{\sigma\beta\dots\delta}+\Gamma^\beta_{\sigma\mu}
A^{\alpha\sigma\dots\delta}+\dots+\Gamma^\delta_{\sigma\mu}
A^{\alpha\beta\dots\sigma}$ and
$\nabla_\alpha\nabla^\alpha\,=\,\Box\,=\,\frac{\partial_\alpha(\sqrt{-g}g^{\alpha\beta}\partial_\beta)}{\sqrt{-g}}$
is  the d'Alembert operator with respect to the metric
$g_{\mu\nu}$.

Einstein was led to postulate the following equations for the
dynamics of gravitational forces

\begin{eqnarray}\label{wrong}
R_{\mu\nu}\,=\,\mathcal{X}\,T_{\mu\nu}
\end{eqnarray}
where $\mathcal{X}\,=\,8\pi G$ is a coupling constant (we will use
the convention $c\,=\,1$). These equations turned out to be
physically and mathematically unsatisfactory.

As Hilbert pointed out \cite{schroedinger}, they have not a
variational origin, \emph{i.e.} there was no Lagrangian able to
reproduce them exactly (this is  slightly wrong, but this remark
is unessential here). Einstein replied that he knew that the
equations were physically unsatisfactory, since they were
contrasting with the continuity equation of any reasonable kind of
matter. Assuming that matter is given as a perfect fluid, that is

\begin{eqnarray}\label{perfectfluid}
T_{\mu\nu}\,=\,(p+\rho)u_\mu u_\nu-pg_{\mu\nu}
\end{eqnarray}
where $u_\mu u_\nu$ define a comoving observer, $p$ is the
pressure and $\rho$ the density of the fluid, then the continuity
equation requires $T_{\mu\nu}$ to be covariantly constant, i.e. to
satisfy the conservation law

\begin{eqnarray}\label{conservationlaw}T^{\mu\sigma}_{\,\,\,\,\,\,\,\,\,\,;\sigma}\,=\,0\,.\end{eqnarray}

In fact, it is not true that
$R^{\mu\sigma}_{\,\,\,\,\,\,\,\,\,\,;\sigma}$ vanishes (unless $R
= 0$). Einstein and Hilbert reached independently the conclusion
that the wrong field equations (\ref{wrong}) had to be replaced by
the correct ones

\begin{eqnarray}\label{fieldequationGR}
G_{\mu\nu}\,=\,\mathcal{X}\,T_{\mu\nu}
\end{eqnarray}
where

\begin{eqnarray}\label{einstein-tensor}
G_{\mu\nu}\,=\,R_{\mu\nu}-\frac{R}{2}g_{\mu\nu}
\end{eqnarray}
that is currently called the "\emph{Einstein tensor}" of
$g_{\mu\nu}$. These equations are both variational and satisfy the
conservation laws (\ref{conservationlaw}) since the following
relation holds

\begin{eqnarray}G^{\mu\sigma}_{\,\,\,\,\,\,\,\,\,\,;\sigma}\,=\,0\,,\end{eqnarray} as a byproduct of the
so-called \emph{Bianchi identities} that the curvature tensor of
$g_{\mu\nu}$ has to satisfy \cite{weinberg, landau}.

The Lagrangian that allows to obtain the field equations
(\ref{fieldequationGR}) is the sum of a \emph{matter Lagrangian}
$\mathcal{L}_m$ and of the Ricci scalar:

\begin{eqnarray}\label{HElagrangian}\mathcal{L}_{HE}=\sqrt{-g}(R+\mathcal{X}\mathcal{L}_m)\,,\end{eqnarray}
where $\sqrt{-g}$ denotes the square root of the value of the
determinant of the metric $g_{\mu\nu}$. The action of GR is

\begin{eqnarray}\label{HEaction}\mathcal{A}\,=\,\int d^4x\sqrt{-g}(R+\mathcal{X}\mathcal{L}_m)\,.\end{eqnarray}
From the action principle, we get the field equations
(\ref{fieldequationGR}) by the variation:

\begin{eqnarray}\label{variationprincipleGR}\delta\mathcal{A}=&&\delta\int d^4x\sqrt{-g}(R+\mathcal{X}\mathcal{L}_m)=
\int
d^4x\sqrt{-g}\biggr[R_{\mu\nu}-\frac{R}{2}g_{\mu\nu}+\nonumber\\\nonumber\\&&\,\,\,\,\,\,\,\,\,\,\,\,\,\,\,\,\,\,\,\,\,\,\,\,
\,\,\,\,\,\,\,\,\,\,\,\,\,\,\,\,\,\,\,\,\,\,\,\,-\mathcal{X}\,T_{\mu\nu}\biggr]\delta
g^{\mu\nu}+\int d^4x\sqrt{-g}g^{\mu\nu}\delta
R_{\mu\nu}=0\,,\end{eqnarray} where $T_{\mu\nu}$ is energy
momentum tensor of matter:

\begin{eqnarray}\label{definitiontensormatter}T_{\mu\nu}\,=\,-\frac{1}{\sqrt{-g}}\frac{\delta (\sqrt{-g}\mathcal{L}_m)}
{\delta g^{\mu\nu}}\,.\end{eqnarray} The last term in
(\ref{variationprincipleGR}) is a 4-divergence

\begin{eqnarray}\int d^4x\sqrt{-g}g^{\mu\nu}\delta
R_{\mu\nu}\,=\,\int d^4x\sqrt{-g}[(-\delta
g^{\mu\nu})_{;\mu\nu}-\Box(g^{\mu\nu}\delta
g_{\mu\nu})]\end{eqnarray} then we can neglect it and we get the
field equation (\ref{fieldequationGR}). For the variational calculus (\ref{variationprincipleGR}) we used the following relations

\begin{eqnarray}\label{variationalcalculus}
\left\{\begin{array}{ll}
\delta\sqrt{-g}\,=\,-\frac{1}{2}\,\sqrt{-g}\,g_{\alpha\beta}\,\delta g^{\alpha\beta}
\\\\
\delta\,R\,=\,R_{\alpha\beta}\,\delta g^{\alpha\beta}+g^{\alpha\beta}\,\delta R_{\alpha\beta}
\\\\
\delta\,R_{\alpha\beta}\,=\,\frac{1}{2}(\delta g^\rho_{\,\,\,\,\alpha;\beta\rho}+\delta g^\rho_{\,\,\,\,\beta;\alpha\rho}-\Box\,\delta g_{\alpha\beta}-g^{\rho\sigma}\delta\,g_{\rho\sigma;\alpha\beta})
\end{array} \right.
\end{eqnarray}

The choice of Hilbert and Einstein was completely arbitrary (as
it became clear a few years later), but it was certainly the
simplest one both from the mathematical and the physical
viewpoint. As it was later clarified by Levi-Civita in 1919,
curvature is  not a "purely metric notion" but, rather, a notion
related to the "linear connection"  to which "parallel transport"
and "covariant derivation" refer \cite{levicivita}.

In a sense, this is the precursor idea of what, in the sequel,
would be called a "gauge theoretical framework" \cite{kle}, after
the pioneering work by Cartan in 1925 \cite{cartan}. But at the
time of Einstein, only metric concepts were at hands and his
solution was the only viable.

It was later clarified that the three principles of relativity,
equivalence and covariance, together with causality, just require
that the spacetime structure has to be determined by either one or
both of two fields, a Lorentzian metric  $g$  and a linear
connection $\Gamma$, assumed to be torsionless for the sake of
simplicity.

The metric $g$ fixes the causal structure of spacetime (the light
cones) as well as its metric relations (clocks and rods); the
connection  $\Gamma$  fixes the free-fall, i.e. the locally
inertial observers. They have, of course, to satisfy a number of
compatibility relations which amount to require that photons
follow the null geodesics of $\Gamma$, so that $\Gamma$ and $g$
can be independent, \emph{a priori}, but constrained, \emph{a
posteriori}, by some physical restrictions. These, however, do not
impose that $\Gamma$ has necessarily to be the Levi Civita
connection of  $g$ \cite{palatini}.

This justifies - at least on a purely theoretical basis - the fact
that one can envisage the so-called "alternative theories of
gravitation", that we prefer  to call "\emph{Extended Theories of
Gravitation}" since their starting points are exactly those
considered by Einstein and Hilbert: theories in which gravitation
is described by either a metric (the so-called "purely metric
theories"), or by a linear connection (the so-called "purely
affine theories") or by both fields (the so-called "metric-affine
theories", also known as "first order formalism theories"). In
these theories,  the Lagrangian is a scalar density  of the
curvature invariants constructed out of both $g$ and $\Gamma$.

The choice (\ref{HElagrangian}) is by no means unique and it turns
out that the Hilbert-Einstein Lagrangian is in fact the only
choice that produces an invariant that is linear in second
derivatives of the metric (or first derivatives of the
connection). A Lagrangian that, unfortunately, is rather singular
from the Hamiltonian viewpoint, in much than same way as
Lagrangians, linear in canonical momenta, are rather singular in
Classical Mechanics (see e.g. \cite{arnold}).

A number of attempts to generalize GR (and unify it to
Electromagnetism) along these lines were  followed by Einstein
himself and many others (Eddington, Weyl, Schrodinger, just to
quote the main contributors; see, e.g., \cite{app-fre}) but they
were eventually given up in the fifties of XX Century, mainly
because of a number of difficulties related to the definitely more
complicated structure of a non-linear theory (where by
"non-linear" we mean here a theory that is based on non-linear
invariants of the curvature tensor), and also because of the new
understanding of physics that is currently based on four
fundamental forces and requires the more general "gauge framework"
to be adopted (see \cite{kaku}).

Still a number of sporadic investigations about "alternative
theories" continued even after 1960 (see \cite{will} and refs.
quoted therein for a short history). The search of a coherent
quantum theory of gravitation or the belief that gravity has to be
considered as a sort of low-energy limit of string theories (see,
e.g., \cite{gre-sch-wit}) - something that we are not willing to
enter here in detail - has more or less recently revitalized the
idea that there is no reason to follow the simple prescription of
Einstein and Hilbert and to assume that gravity should be
classically governed by a Lagrangian linear in the curvature.

Further curvature invariants or non-linear functions of them
should be also considered, especially in view of the fact that
they have to be included in both the semi-classical expansion of a
quantum Lagrangian or in the low-energy limit of a string
Lagrangian.

Moreover, it is clear from the recent astrophysical observations
and from the current cosmological hypotheses that Einstein
equations are no longer a good test for gravitation  at Solar
System, galactic, extra-galactic and cosmic scale, unless one does
not admit that the matter side of Eqs.(\ref{fieldequationGR})
contains some kind of exotic matter-energy which is the "dark
matter" and "dark energy" side of the Universe.

The idea which we propose here is much simpler. Instead of
changing the matter side of Einstein Equations
(\ref{fieldequationGR}) in order to fit the "missing
matter-energy" content of the currently observed Universe (up to
the $95\%$ of the total amount!), by adding any sort of
inexplicable and strangely behaving matter and energy, we claim
that it is simpler and more convenient to change the gravitational
side of the equations, admitting corrections coming from
non-linearities in the Lagrangian. However, this is nothing else
but a matter of taste and, since it is possible, such an approach
should be explored. Of course, provided that the Lagrangian can be
conveniently tuned up (i.e., chosen in a huge family of allowed
Lagrangians) on the basis of its best fit with all possible
observational tests, at all scales (solar, galactic, extragalactic
and cosmic).

Something that - in spite of some commonly accepted but disguised
opinion - can and should be done before rejecting a priori a
non-linear theory of gravitation (based on a non-singular
Lagrangian) and insisting that the Universe has to be necessarily
described by a rather singular gravitational Lagrangian (one that
does not allow a coherent perturbation  theory from a good
Hamiltonian viewpoint) accompanied by matter that does not follow
the behavior that standard baryonic matter, probed in our
laboratories, usually satisfies.

\section{The Extended Theories of Gravity: $\digamma(R,\Box R,...,\Box^k R,\phi)$}

With the above considerations in mind, let us start with a general
class of higher-order-scalar-tensor theories in four dimensions
given by the action

\begin{eqnarray}\label{HOGaction}\mathcal{A}=\int d^{4}x\sqrt{-g}[\digamma(R,\Box R,\Box^{2}R, ..., \Box^k R,\phi)
+\omega(\phi) \phi_{;\alpha}\phi^{;\alpha}
+\mathcal{X}\mathcal{L}_m]\,,
\end{eqnarray} where $\digamma$ is an unspecified function of curvature invariants and of a scalar
field $\phi$. The term $\mathcal{L}_m$, as above, is the minimally
coupled ordinary matter contribution and $\omega(\phi)$ is a
generic function of the scalar field $\phi$. For example its
values could be $\omega(\phi)=\pm 1, 0$ fixing the nature and the
dynamics of the scalar field which can be a standard scalar field,
a phantom field or a field without dynamics (see \cite{faraoni3,
rub-scu} for details).

In the metric approach, the field equations are obtained by
varying (\ref{HOGaction}) with respect to  $g_{\mu\nu}$. We get

\begin{eqnarray}\label{fieldequationHOG}
&&\hat{\digamma}G_{\mu\nu}-\frac{1}{2}g_{\mu\nu}(\digamma-\hat{\digamma}R)-\hat{\digamma}_{;\mu\nu}+g_{\mu\nu}\Box\hat{
\digamma}+g_{\mu\nu}\biggl[(\Box^{j-1}R)^{;\alpha}\Box^{i-j}\frac{\partial\digamma}{\partial\Box^iR}\biggr]_{;\alpha}+
\nonumber\\\nonumber\\\nonumber\\&&-\frac{1}{2}\sum_{i=1}^{k}\sum_{j=1}^{i}\biggl[g_{\mu\nu}(\Box^{j-i})^{;\alpha}\biggl(
\Box^{i-j}\frac{
\partial\digamma}{\partial\Box^{i}R}\biggr)_{;\alpha}+(\Box^{j-i})_{;\nu}\biggl(\Box^{i-j}\frac{\partial\digamma}{\partial
\Box^{i}R}\biggr)_{;\mu}\biggr]+\nonumber\\\nonumber\\&&\,\,\,\,\,\,\,\,\,\,\,\,\,\,\,\,\,\,\,\,\,\,\,\,\,\,\,\,\,\,\,\,\,\,\,
\,\,\,\,\,\,\,\,\,\,\,\,\,\,\,\,\,\,\,\,\,\,\,\,\,\,\,\,\,\,\,\,\,\,\,\,\,\,\,\,\,-\frac{\omega(\phi)}{2}\phi_{;\alpha}\phi^
{;\alpha}g_{\mu\nu}+\omega(\phi)\phi_{;\mu}\phi_{;\nu}=\mathcal{X}\,T_{\mu\nu}
\end{eqnarray}
where $G_{\mu\nu}$ is the above Einstein tensor (\ref{einstein-tensor}) and

\begin{eqnarray}\hat{\digamma}=\sum_{j=0}^{n}\Box^{j}\frac{\partial\digamma}{\partial\Box^jR}\,.\end{eqnarray}
The differential Equations (\ref{fieldequationHOG}) are of order
$(2k+4)$. The (eventual) contribution of a potential $V(\phi)$ is
contained in the definition of $\mathcal{F}$. By varying with
respect to the scalar field $\phi$, we obtain the Klein - Gordon
equation

\begin{eqnarray}\Box\phi=\frac{1}{2}\frac{\delta\ln\omega(\phi)}{\delta\phi}\phi_{;\alpha}\phi^{;\alpha}+\frac{1}{2\omega
(\phi)}\frac{\delta\digamma(R,\Box R,\Box^{2}R, ..., \Box^k
R,\phi)}{\delta\phi}\,.\end{eqnarray}

Several approaches can be used to deal with such equations. For
example, as we said, by a conformal transformation, it is possible
to reduce an ETG to a (multi) scalar - tensor theory of gravity
\cite{dam-esp, got-sch-sta, wands, qua-sch, schmidt2}.

The simplest extension of GR is achieved assuming

\begin{eqnarray}\digamma\,=\,f(R)\,,\qquad \omega(\phi)=0\,,\end{eqnarray}
in the action (\ref{HOGaction}); $f$ is an arbitrary (analytic)
function of the Ricci curvature scalar $R$. Then

\begin{eqnarray}\label{actionfR}
\mathcal{A}^{f}\,=\,\int
d^4x\sqrt{-g}\biggl[f+\mathcal{X}\mathcal{L}_m\biggr]\end{eqnarray}
where the standard Hilbert-Einstein action is, of course,
recovered for $f\,=\,R$. Varying the (\ref{actionfR}) with respect
to $g_{\mu\nu}$, we get the field equations

\begin{eqnarray}\label{fe}
H_{\mu\nu}\,\doteq\,f'R_{\mu\nu}-\frac{1}{2}fg_{\mu\nu}-f'_{;\mu\nu}+g_{\mu\nu}\Box
f'=\mathcal{X}\,T_{\mu\nu}\,,
\end{eqnarray} which are fourth-order equations due to the terms $f'_{;\mu\nu}$ and $\Box f'$; the
prime indicates the derivative with respect to $R$. The trace of
(\ref{fe}) is

\begin{eqnarray}\label{fetr}
H\,=\,g^{\alpha\beta}H_{\alpha\beta}\,=\,3\Box
f'+f'R-2f=\mathcal{X}\,T\,.\end{eqnarray}

The peculiar behavior of $f\,=\,R$ is  due to the particular form
of the Lagrangian itself which, even though it is a second order
Lagrangian, can be non-covariantly rewritten as the sum of a first
order Lagrangian plus a pure divergence term. The Hilbert-Einstein
Lagrangian can be in fact recast as follows:

\begin{eqnarray}
\mathcal{L}_{HE}=\sqrt{-g}R=\sqrt{-g}g^{\alpha\beta}
(\Gamma^{\rho}_{\alpha \sigma} \Gamma^{\sigma}_{\rho
\beta}-\Gamma^{\rho}_{\rho \sigma} \Gamma^{\sigma}_{\alpha
\beta})+ \nabla_\sigma
(\sqrt{-g}g^{\alpha\beta}{u^{\sigma}}_{\alpha \beta})\,;
\end{eqnarray}
$\Gamma$ is the Levi - Civita connection of $g$ and
$u^{\sigma}_{\alpha \beta}$ is a quantity constructed out with the
variation of $\Gamma$ \cite{weinberg, landau}. Since
$u^{\sigma}_{\alpha \beta}$ is not a tensor, the above expression
is not covariant; however a standard procedure has been studied to
recast covariance in the first order theories \cite{fer-fra}. This
clearly shows that the field equations should consequently be
second order and the Hilbert-Einstein Lagrangian is thus
degenerate.

From the action (\ref{HOGaction}), it is possible to obtain
another interesting case by choosing

\begin{eqnarray}\digamma\,=\,F(\phi)R+V(\phi)\,,\end{eqnarray}
where $V(\phi)$ and $F(\phi)$ are generic functions describing
respectively the potential and the coupling of a scalar field
$\phi$. In this case, we get

\begin{eqnarray}\label{TSaction} \mathcal{A}^{ST}\,=\,\int d^4x \sqrt{-g}[F(\phi)R+\omega(\phi)\phi_{;\alpha}\phi^{;\alpha}+
V(\phi)+\mathcal{X}\mathcal{L}_m]\,.\end{eqnarray} The Brans-Dicke
theory of gravity is a particular case of the action
(\ref{TSaction}) in which we have $V(\phi)=0$ and
$\omega(\phi)=-\frac{\omega_{BD}}{\phi}$. In fact we have

\begin{eqnarray}\label{BD-action}
\mathcal{A}^{BD}=\int d^4x\sqrt{-g}\left[\phi
R-\omega_{BD}\frac{\phi_{;\alpha}\phi^{;\alpha}}{\phi}+\mathcal{X}
\mathcal{L}_m\right]\,.
\end{eqnarray}
The variation of (\ref{TSaction}) with respect to $g_{\mu\nu}$ and
$\phi$ gives the second-order field equations

\begin{eqnarray}\label{TSfieldequation}
\left\{\begin{array}{ll}
F(\phi)G_{\mu\nu}-\frac{1}{2}V(\phi)g_{\mu\nu}+\omega(\phi)\biggl[\phi_{;\mu}\phi_{;\nu}-\frac{1}{2}\phi_{;\alpha}\phi^
{;\alpha}g_{\mu\nu}\biggr]-F(\phi)_{;\mu\nu}+\\\\
\,\,\,\,\,\,\,\,\,\,\,\,\,\,\,\,\,\,\,\,\,\,\,\,\,\,\,\,\,\,\,\,\,\,\,\,\,\,\,\,\,\,\,\,\,\,\,\,\,\,\,\,\,\,\,\,\,\,\,\,\,\,
\,\,\,\,\,\,\,\,\,\,\,\,\,\,\,\,\,\,\,\,\,\,\,\,\,\,\,\,\,\,\,\,\,\,\,\,\,\,\,\,\,\,\,\,\,\,\,\,+g_{\mu\nu}\Box
F(\phi)=\mathcal{X}\,T_{\mu\nu}
\\\\
2\omega(\phi)\Box\phi-\omega_{,\phi}(\phi)\phi_{;\alpha}\phi^{;\alpha}-[F(\phi)R+V(\phi)]_{,\phi}=0
\\\\
3\Box
F(\phi)-F(\phi)R-2V(\phi)-\omega(\phi)\phi_{;\alpha}\phi^{;\alpha}=\mathcal{X}\,T
\\\\
2\omega(\phi)\Box\phi+3\Box
F(\phi)-[\omega_{,\phi}(\phi)+\omega(\phi)]\phi_{;\alpha}\phi^{;\alpha}-[F(\phi)R+V(\phi)]_{,\phi}+\\\\
\,\,\,\,\,\,\,\,\,\,\,\,\,\,\,\,\,\,\,\,\,\,\,\,\,\,\,\,\,\,\,\,\,\,\,\,\,\,\,\,\,\,\,\,\,\,\,\,\,\,\,\,\,\,\,\,\,\,\,\,\,\,
\,\,\,\,\,\,\,\,\,\,\,\,\,\,\,\,\,\,\,\,\,\,\,\,\,\,\,\,\,\,\,\,\,\,\,\,\,\,\,\,\,\,-F(\phi)R-2V(\phi)=\mathcal{X}\,T
\end{array} \right.
\end{eqnarray}
The third equation in (\ref{TSfieldequation}) is the trace of
field equation for $g_{\mu\nu}$ and the last one is a combination
of the trace and of the one for $\phi$. This last equation is
equivalent to the Bianchi contracted identity \cite{cap-der}.
Standard fluid matter can be treated as above.

\section{Conformal transformations}\label{conformal-transformation-general-approach}

Let us now introduce conformal transformations to show that any
higher-order or scalar-tensor theory, in absence of ordinary
matter, e.g. a perfect fluid, is conformally equivalent to an
Einstein theory plus minimally coupled scalar fields. If standard
matter is present, conformal transformations allow to transfer
non-minimal coupling to the matter component \cite{mag-sok}. The
conformal transformation on the metric $g_{\mu\nu}$ is

\begin{eqnarray}\label{transconf}\tilde{g}_{\mu\nu}\,=\,A(x^\lambda)g_{\mu\nu}\end{eqnarray}
with $A(x^\lambda)>0$. $A$ is the conformal factor. Obviously the
transformation rule for the contravariant metric tensor is
$\tilde{g}^{\mu\nu}=A^{-1}g^{\mu\nu}$. The various mathematical
quantities in the so-called Einstein frame (EF) (quantities
referred to $\tilde{g}_{\mu\nu}$) are linked to the ones in the
so-called Jordan Frame (JF) (quantities referred to $g_{\mu\nu}$)
as follows

\begin{eqnarray}
\left\{\begin{array}{ll}
\tilde{\Gamma}^\alpha_{\mu\nu}=\Gamma^\alpha_{\mu\nu}+\phi_{,\mu}\delta^\alpha_\nu+\phi_{,\nu}\delta^\alpha_\mu
-\phi^{,\alpha}g_{\mu\nu}
\\\\
\tilde{R}^{\alpha}_{\,\,\,\,\,\mu\beta\nu}\,=\,R^{\alpha}_{\,\,\,\,\,\mu\beta\nu}-\delta^\alpha_\beta(\phi_{;\mu\nu}-\phi_{;\mu}
\phi_{;\nu}+g_{\mu\nu}\phi^{;\sigma}\phi_{;\sigma})+\delta^\alpha_\nu(\phi_{;\mu\beta}-\phi_{;\mu}\phi_{;\beta}+\\\\\,\,\,\,\,\,
\,\,\,\,\,\,\,\,\,\,\,\,\,\,\,\,\,\,\,\,+g_{\mu\beta}\phi^{;\sigma}\phi_{;\sigma})-g_{\mu\nu}(\phi^{;\alpha}_{\,\,\,\,\,\beta}
-\phi^{;\alpha}\phi_{;\beta})+g_{\mu\beta}(\phi^{;\alpha}_{\,\,\,\,\,\nu}
-\phi^{;\alpha}\phi_{;\nu})
\\\\
\tilde{R}_{\mu\nu}=R_{\mu\nu}-2\phi_{;\mu\nu}+2\phi_{;\mu}\phi_
{;\nu}-g_{\mu\nu}\Box\phi-2g_{\mu\nu}\phi_{;\sigma}\phi^{;\sigma}\\\\
\tilde{R}=e^{-2\phi}(R-6\Box\phi-6\phi_{;\sigma}\phi^{\sigma})
\\\\
\tilde{\phi_{;\mu\nu}}\,=\,\phi_{,\mu\nu}-\tilde{\Gamma}^\sigma_{\alpha\beta}\phi_{,\sigma}\,=\,\phi_{;\mu\nu}-2\phi_{;\mu}
\phi_{;\nu}+g_{\mu\nu}\phi^{;\sigma}\phi_{;\sigma}
\\\\
\tilde{G}_{\mu\nu}\,=\,G_{\mu\nu}-2\phi_{;\mu\nu}+2\phi_{;\mu}\phi_{;\nu}+2g_{\mu\nu}\,\Box\phi+
g_{\mu\nu}\phi^{;\sigma}\phi_{;\sigma}
\end{array} \right.
\end{eqnarray}
where $\phi\,\doteq\,\ln A^{1/2}$. But we can have also the inverse
relations

\begin{eqnarray}
\left\{\begin{array}{ll}
\Gamma^\alpha_{\mu\nu}=\tilde{\Gamma}^\alpha_{\mu\nu}-\phi_{,\mu}\delta^\alpha_\nu-\phi_{,\nu}\delta^\alpha_\mu
+\tilde{\phi^{,\alpha}}\tilde{g}_{\mu\nu}
\\\\
R^{\alpha}_{\,\,\,\,\,\mu\beta\nu}\,=\,\tilde{R}^{\alpha}_{\,\,\,\,\,\mu\beta\nu}+\delta^\alpha_\beta(\tilde{\phi_{;\mu\nu}}+
\phi_{;\mu}\phi_{;\nu})-\delta^\alpha_\nu(\tilde{\phi_{;\mu\beta}}+\phi_{;\mu}\phi_{;\beta})+\\\\\,\,\,\,\,\,\,\,\,\,\,\,\,\,\,
\,\,\,\,\,\,\,\,\,\,\,+\tilde{g}_{\mu\nu}(\tilde{\phi^{;\alpha}_{\,\,\,\,\,\,\,\beta}}-\tilde{\phi^{;\alpha}}\phi_{;\beta})-
\tilde{g}_{\mu\beta}(\tilde{\phi^{;\alpha}_{\,\,\,\,\,\,\,\nu}}-\tilde{\phi^{;\alpha}}\phi_{;\nu})
\\\\
R_{\mu\nu}=\tilde{R}_{\mu\nu}+2\tilde{\phi_{;\mu\nu}}+2\phi_{;\mu}\phi_{;\nu}+\tilde{g}_{\mu\nu}\,\tilde{\Box}\phi-2\tilde{g}_
{\mu\nu}\tilde{\phi^{;\sigma}\phi_{;\sigma}}\\\\
R=e^{2\phi}(\tilde{R}+6\tilde{\Box}\phi-6\tilde
{\phi^{;\sigma}\phi_{;\sigma}})
\\\\
\phi_{;\mu\nu}\,=\,\phi_{,\mu\nu}-\Gamma^\sigma_{\alpha\beta}\phi_{,\sigma}\,=\,\tilde{\phi_{;\mu\nu}}+2\phi_{;\mu}\phi_{;\nu}
-\tilde{g}_{\mu\nu}\tilde{\phi^{;\sigma}\phi_{;\sigma}}
\\\\
G_{\mu\nu}\,=\,\tilde{G}_{\mu\nu}+2\tilde{\phi_{;\mu\nu}}+2\phi_{;\mu}\phi_{;\nu}-2\tilde{g}_{\mu\nu}\,\tilde{\Box}\phi\,
+\tilde{g}_{\mu\nu}\tilde
{\phi^{;\sigma}\phi_{;\sigma}}
\end{array} \right.
\end{eqnarray}
where $\Box$ and $\tilde{\Box}$ are the d'Alembert operators with respect to the metric $g_{\mu\nu}$ and $\tilde{g}_{\mu\nu}$. The transformation between the operators is $\Box\,=\,e^{2\phi}\tilde{\Box}-2\phi^{;\nu}\partial_{\nu}$.

Under these transformations, the action in (\ref{TSaction}) can be
reformulated as follows

\begin{eqnarray}\label{TS-EF-action}
\mathcal{A}_{EF}^{ST}\,=\,\int d^4x \sqrt{-\tilde{g}}
\biggl[\Lambda\,\tilde{R}
+\Omega(\varphi)\varphi_{;\alpha}\varphi^{;\alpha}+W(\varphi)+\mathcal{X}
\tilde{\mathcal{L}}_{m}\biggr]\,.
\end{eqnarray}
in which $\tilde{R}$ is the Ricci scalar relative to the metric
$\tilde{g}$ and $\Lambda$ is a generic constant. The relations
between the quantities in two frames are

\begin{eqnarray}\label{transconfTS}
\left\{\begin{array}{ll}\Omega(\varphi){d\varphi}^2\,=\,\Lambda\biggl[\frac{\omega(\phi)}{F(\phi)}-\frac{3}{2}\biggl(\frac{
d\ln F(\phi)}{d\phi}\biggr)^2\biggr]{d\phi}^2\\\\W(\varphi)=\frac{\Lambda^2}{F(\phi(\varphi))^2}V(\phi(\varphi))\\\\
\tilde{\mathcal{L}}_m=\frac{\Lambda^2}{F(\phi(\varphi))^2}\mathcal{L}_m\biggl(\frac{\Lambda\,\tilde{g}_{\rho\sigma}}{F(\phi(
\varphi))}\biggr)\\\\F(\phi){A(x^\lambda)}^{-1}=\Lambda\end{array}\right.
\end{eqnarray}
The field equations for the new fields $\tilde{g}_{\mu\nu}$ and
$\varphi$ are

\begin{eqnarray}\label{fe-EF-conformal-transformed}
\left\{\begin{array}{ll}
\Lambda\tilde{G}_{\mu\nu}-\frac{1}{2}W(\varphi)\tilde{g}_{\mu\nu}+\Omega(\varphi)\biggl[\varphi_{;\mu}\varphi_{;\nu}-
\frac{1}{2}\varphi_{;\alpha}\varphi^{;\alpha}\tilde{g}_{\mu\nu}\biggr]=\mathcal{X}\,\tilde{T}^\varphi_{\mu\nu}
\\\\
2\Omega(\varphi)\tilde{\Box}\varphi-\Omega_{,\varphi}(\varphi)\varphi_{;\alpha}\varphi^{;\alpha}-W_{,\varphi}(\varphi)=
\mathcal{X}\tilde{\mathcal{L}}_{m,\varphi}
\\\\
\tilde{R}=-\frac{1}{\Lambda}\biggl(\mathcal{X}\tilde{T}^\varphi+2W(\varphi)+\Omega(\varphi)\tilde{g}^{\sigma\tau}\varphi_
{;\sigma}\varphi_{;\tau}\biggr)
\end{array} \right.
\end{eqnarray}

Therefore, every non-minimally coupled scalar-tensor theory, in
absence of ordinary matter, e.g. perfect fluid, is conformally
equivalent to an Einstein theory, being the conformal
transformation and the potential suitably defined by
(\ref{transconfTS}). The converse is also true: for a given
$F(\phi)$, such that is valid the relations (\ref{transconfTS}),
we can transform a standard Einstein theory into a non-minimally
coupled scalar-tensor theory. This means that, in principle, if we
are able to solve the field equations in the framework of the
Einstein theory in presence of a scalar field with a given
potential, we should be able to get the solutions for the
scalar-tensor theories, assigned by the coupling $F(\phi)$, via
the conformal transformation (\ref{transconf}) with the
constraints given by (\ref{transconfTS}). Following the standard
terminology, the ``Einstein frame'' is the framework of the
Einstein theory with the minimal coupling and the ``Jordan frame''
is the framework of the non-minimally coupled theory
\cite{mag-sok}.

This procedure can be extended to more general theories. If the
theory is assumed to be higher than fourth order, we may have
Lagrangian densities of the form \cite{got-sch-sta, buchdahl},

\begin{eqnarray}\mathcal{L}\,=\,\mathcal{L}(R,\Box R, ..., \Box^{k} R)\,. \end{eqnarray}
Every $\Box$ operator introduces two further terms of derivation
into the field equations. For example, a theory like

\begin{eqnarray}\mathcal{L}\,=\,\sqrt{-g}\,R\Box R\,,\end{eqnarray}
is a sixth-order theory and the above approach can be pursued by
considering a conformal factor of the form

\begin{eqnarray}
A\,=\,\biggl|\frac{\partial(R\,\Box R)}{\partial
R}+\Box\frac{\partial(R\,\Box R)}{\partial\,\Box R}\biggr|\,.
\end{eqnarray}

In general, increasing two orders of derivation in the field
equations (\emph{i.e.} for every term $\Box R$), corresponds to
adding a scalar field in the conformally transformed frame
\cite{got-sch-sta}. A sixth-order theory can be reduced to an
Einstein theory with two minimally coupled scalar fields; a
$2n$-order theory can be, in principle, reduced to an Einstein
theory plus $(n-1)$ scalar fields. On the other hand, these
considerations can be directly generalized to higher - order -
scalar - tensor theories in any number of dimensions as shown in
\cite{maeda}.

The analogy between scalar-tensor gravity and HOG, although
mathematically straightforward, requires a careful physical
analysis. Recasting fourth - order gravity as a scalar - tensor
theory, often the following  steps, in terms of a generic scalar
field $\phi$, are considered

\begin{eqnarray}\label{f(R)-analogy-TS}
f+\mathcal{L}_m \,\rightarrow\,
f'(\phi)R+f(\phi)-f'(\phi)\phi+\mathcal{L}_m \,\rightarrow\,
f'(\phi)R+V(\phi)+\mathcal{L}_m\,,
\end{eqnarray}
where, by analogy, $\phi \rightarrow R$ and the "potential" is
$V(\phi)\,=\,f(\phi)-f'(\phi)\phi$. Clearly the kinetic term is
not present so that (\ref{f(R)-analogy-TS}) is usually referred as
a scalar-tensor description of $f$ - gravity where
$\omega(\phi)\,=\,0$. This is the so-called O'Hanlon Lagrangian
\cite{tey-tou}:

\begin{eqnarray}\label{ohanlon}
\mathcal{A}^{OH}\,=\,\int d^4x\sqrt{-g}\biggl[\phi
R+V(\phi)+\mathcal{L}_m\biggr]\,.\end{eqnarray}

As concluding remarks, we can say that conformal transformations
work at three levels: $i)$ on the Lagrangian of the given theory;
$ii)$ on the field equations; $iii)$ on the solutions. The table
\ref{picture-f(R)-TS-GR-approach} summarizes the situation for
HOG, non-minimally coupled scalar-tensor theories (ST) and
standard Hilbert-Einstein (HE) theory. Clearly, direct and inverse
transformations correlate all the steps of the table but no
absolute criterion, at this point of the discussion, is able to
select which is the ``physical" framework since, at least from a
mathematical point of view, all the frames are equivalent
\cite{mag-sok}.

However, the typical Brans-Dicke action is the (\ref{BD-action})
where no scalar field potential is present and $\omega_{BD}$ is a
constant, while  the O'Hanlon Lagrangian (\ref{ohanlon}) has a
potential but has no kinetic term. The most general situation is
in (\ref{TSaction}) where we have non-minimal coupling, kinetic
term, and scalar field potential. This means that fourth-order
gravity and scalar tensor gravity can be "compared" only by means
of conformal transformations where kinetic and potential terms are
preserved. In particular, it is misleading to state that PPN -
limit of HOG is not working since these models provide
$\omega_{BD}\,=\,0$ and this is in contrast with observations
\cite{olm, olm2}.

\begin{table}
\begin{center}
\begin{tabular}{ccccc}
  \hline\hline\hline
  $\mathcal{L}_{HOG}$ & $\longleftrightarrow$ & $\mathcal{L}_{ST}$ & $\longleftrightarrow$ & $\mathcal{L}_{HE}+\varphi$ \\
  $\updownarrow$ & & $\updownarrow$ & & $\updownarrow$ \\
  HOG Eqs. & $\longleftrightarrow$ & ST Eqs. & $\longleftrightarrow$ & Einstein Eqs. $+\varphi$ \\
  $\updownarrow$ & & $\updownarrow$ & & $\updownarrow$ \\
  HOG Solutions & $\longleftrightarrow$ & ST
  Solutions & $\longleftrightarrow$ & Einstein Solutions \\
  \hline\hline\hline
\end{tabular}
\end{center}
\caption{Summary of the three approaches: Scalar-Tensor $(ST)$,
Einstein $+\varphi$, and $f$  and their relations at Lagrangians,
field equations and solutions levels. The solutions are in the
Einstein frame for the minimally coupled case while they are in
Jordan frame for $f$ and $ST$ - gravity. Clearly, $f$ and $ST$
theories can be rigorously compared only recasting them in the
Einstein frame.\label{picture-f(R)-TS-GR-approach}}
\end{table}

Scalar-tensor theories and $f$-theories can be rigorously
compared, after conformal transformations, in the Einstein frame
where both kinetic and potential terms are present.

\section{The Palatini Approach and the Intrinsic Conformal Structure}

As we said, the Palatini approach, considering $g$ and $\Gamma$ as
independent fields, is ``intrinsically" bi-metric and capable of
disentangling the geodesic structure from the chronological
structure of a given manifold. Starting from these considerations,
conformal transformations assume a fundamental role in defining
the affine connection which is merely ``Levi - Civita" only for
the Hilbert-Einstein theory.

In this section, we work out examples showing how conformal
transformations assume a fundamental physical role in relation to
the Palatini approach in ETGs \cite{all-cap-cap-fra}.

Let us start from the case of fourth-order gravity where Palatini
variational principle is straightforward in showing the
differences with Hilbert-Einstein variational principle, involving
only metric. Besides, cosmological applications of $f$-gravity
have shown the relevance of Palatini formalism, giving physically
interesting results with singularity - free solutions
\cite{all-bor-fra, vollick, men-wan2, men-wan3, fla, fla1,
kre-alv}. This last nice feature is not present in the standard
metric approach.

An important remark is in order at this point. The Ricci scalar
entering in $f$ is $R\equiv
R(g\,,\Gamma)\,\,=g^{\alpha\beta}R_{\alpha \beta}(\Gamma)$ that is
a \emph{generalized Ricci scalar} and $R_{\mu\nu}(\Gamma)$ is the
Ricci tensor of a torsion-less connection $\Gamma$, which, \emph{a
priori}, has no relations with the metric $g$ of spacetime. The
gravitational part of the Lagrangian is controlled by a given real
analytical function of one real variable $f$, while $\sqrt{-g}$
denotes a related scalar density of weight $1$. Field equations,
deriving from the Palatini variational principle are:

\begin{eqnarray}\label{fieldequationpalatini}
\left\{\begin{array}{ll}
f'R_{(\mu\nu)}(\Gamma)-\frac{1}{2}fg_{\mu\nu}\,=\,\mathcal{X}\,T_{\mu\nu}\\\\
\nabla _{\alpha }^{\Gamma }(\sqrt{-g}f'g^{\mu\nu})=0
\end{array} \right.
\end{eqnarray}
where $\nabla^{\Gamma}$ is the covariant derivative with respect
to $\Gamma$. We shall use the standard notation denoting by
$R_{(\mu\nu)}$ the symmetric part of $R_{\mu\nu}$, \emph{i.e.}
$R_{(\mu\nu)}\equiv \frac{1}{2}(R_{\mu\nu}+R_{\nu\mu})$.

In order to get the first one of (\ref{fieldequationpalatini}),
one has to additionally assume that $\mathcal{L}_m$ is
functionally independent of $\Gamma$; however it may contain
metric covariant derivatives $\stackrel{g}{\nabla}$ of fields.
This means that the matter stress-energy tensor
$T_{\mu\nu}=T_{\mu\nu}(g,\Psi)$ depends on the metric $g$ and some
matter fields denoted here by $\Psi$, together with their
derivatives (covariant derivatives with respect to the Levi -
Civita connection of $g$). From the second one of
(\ref{fieldequationpalatini}) one sees that
$\sqrt{-g}f'g^{\mu\nu}$ is a symmetric twice contravariant tensor
density of weight $1$. As previously discussed in
\cite{all-cap-cap-fra, fer-fra-vol}, this naturally leads to
define a new metric $h_{\mu\nu}$, such that the following relation
holds:

\begin{eqnarray}\label{newmetricpalatini1}
\sqrt{-g}f'g^{\mu\nu}=\sqrt{-h}h^{\mu\nu}\,.
\end{eqnarray}
This \emph{ansatz} is suitably made in order to impose $\Gamma$ to
be the Levi - Civita connection of $h$ and the only restriction is
that $\sqrt{-g}f'g^{\mu\nu}$ should be non-degenerate. In the case
of Hilbert-Einstein Lagrangian, it is $f'=1$ and the statement is
trivial.

Eq.(\ref{newmetricpalatini1}) imposes that the two metrics $h$ and
$g$ are conformally equivalent. The corresponding conformal factor
can be easily found to be $f'$ (in dim $\mathcal{M}=4$) and the
conformal transformation results to be ruled by:

\begin{eqnarray}\label{newmetricpalatini2}
h_{\mu\nu }=f'g_{\mu\nu}
\end{eqnarray}
Therefore, as it is well known, Eq.(\ref{fieldequationpalatini})
implies that $\Gamma =\Gamma _{LC}(h)$ and
$R_{(\mu\nu)}(\Gamma)=R_{\mu \nu }(h)\equiv R_{\mu\nu}$. Field
equations can be supplemented by the scalar-valued equation
obtained by taking the trace of (\ref{fieldequationpalatini})

\begin{eqnarray} \label{structR}
f'R-2f=\mathcal{X}T\end{eqnarray} which controls solutions of
(\ref{fieldequationpalatini}).

We shall refer to this scalar-valued equation as \emph{the
structural equation} of the spacetime. In the vacuum case (and
spacetimes filled with radiation, such that $T=0$) this
scalar-valued equation admits constant solutions, which are
different from zero only if one add a cosmological constant. This
means that the universality of Einstein field equations holds
\cite{fer-fra-vol}, corresponding to a theory with cosmological
constant \cite{sah-sta}.

In the case of interaction with matter fields, the structural
equation (\ref{newmetricpalatini2}), if explicitly solvable,
provides an expression of $R=R(T)$ and consequently both $f$ and
$f'$ can be expressed in terms of $T$. The matter content of
spacetime thus rules the bi-metric structure of spacetime and,
consequently, both the geodesic and metric structures which are
intrinsically different. This behavior generalizes the vacuum case
and corresponds to the case of a time-varying cosmological
constant. In other words, due to these features, conformal
transformations, which allow to pass from a metric structure to
another one, acquire an intrinsic physical meaning since ``select"
metric and geodesic structures which, for a given ETG, in
principle, \emph{do not} coincide.

Let us now try to extend the above formalism to  the case of
non-minimally coupled scalar-tensor theories. The effort is to
understand if and how the bi-metric structure of spacetime behaves
in this cases and which could be its geometric and physical
interpretation.

We start by considering scalar-tensor theories in the Palatini
formalism, calling $\mathcal{A}_1$ the action functional. After,
we take into account the case of decoupled non-minimal interaction
between a scalar-tensor theory and a $f$-theory, calling
$\mathcal{A}_2$ this action functional. We finally consider the
case of non-minimal-coupled interaction between the scalar field
$\phi$ and the gravitational fields $(g, \Gamma)$, calling
$\mathcal{A}_3$ the corresponding action functional. Particularly
significant is, in this case, the limit of low curvature $R$. This
resembles the physical relevant case of present values of
curvatures of the Universe and it is important for cosmological
applications.

The action (\ref{TSaction}) for scalar-tensor gravity  can be
generalized, in order to better develop the Palatini approach, as:

\begin{eqnarray} \label{lagrfR1}
\mathcal{A}_1=\int
d^4x\sqrt{-g}[F(\phi)R+\omega(\phi)\stackrel{g}\nabla_\mu
\phi\stackrel{g}\nabla^\mu\phi+V(\phi)+\mathcal{X}\mathcal{L}_{m}(\Psi,
\stackrel{g}{\nabla}\Psi)]
\end{eqnarray}

As above, the values of $\omega(\phi)=\pm 1$ selects between
standard scalar field theories and quintessence (phantom) field
theories. The relative ``signature" can be selected by conformal
transformations. Field equations for the gravitational part of the
action are, respectively for the metric $g$ and the connection
$\Gamma$:

\begin{eqnarray}\label{fieldequationpalatiniTS}
\left\{\begin{array}{ll}F(\phi)[R_{(\mu\nu)}-\frac{R}{2}g_{\mu
\nu}]\,=\,\mathcal{X}T_{\mu\nu}+\frac{1}{2}\omega(\phi)\stackrel{g}\nabla_\mu
\phi\stackrel{g}\nabla^\mu\phi g_{\mu\nu}+\frac{1}{2}V(\phi)g_{\mu\nu}\\\\
\nabla _\alpha^\Gamma(\sqrt{-g}F(\phi)g^{\mu\nu})=0
\end{array} \right.
\end{eqnarray}
$R_{(\mu\nu)}$ is the same defined in
(\ref{fieldequationpalatini}). For matter fields we have the
following field equations:

\begin{eqnarray}
\left\{\begin{array}{ll}2\omega(\phi)\stackrel{g}\Box\phi+\omega_{,\phi}(\phi)\stackrel{g}\nabla_\alpha
\phi\stackrel{g}\nabla^\alpha\phi+V_{,\phi}(\phi)+F_{,\phi}(\phi)R=0\\\\
\frac{\delta{\mathcal{L}}_{m}}{\delta \Psi}=0
\end{array} \right.
\end{eqnarray}
In this case, the structural equation of spacetime implies that:

\begin{eqnarray}\label{fieldequationpalatiniTStrace}
R=-\frac{\mathcal{X}T+2\omega(\phi)\stackrel{g}\nabla_\alpha
\phi\stackrel{g}\nabla^\alpha\phi+2V(\phi)}{F(\phi)}
\end{eqnarray}
which expresses the value of the Ricci scalar curvature in terms
of the traces of the stress-energy tensors of standard matter and
scalar field (we have to require $F(\phi)\neq 0$). The bi-metric
structure of spacetime is thus defined by the ansatz:

\begin{eqnarray}
\sqrt{- g}F(\phi)g^{\mu\nu}=\sqrt{- h}h^{\mu\nu}
\end{eqnarray}
such that $g$ and $h$ result to be conformally related

\begin{eqnarray}\label{conformalmetric}
h_{\mu\nu}=F(\phi)g_{\mu\nu }
\end{eqnarray}

The conformal factor is exactly the interaction factor. From
(\ref{fieldequationpalatiniTStrace}), it follows that in the
vacuum case $T=0$ and $\omega(\phi)\stackrel{g}\nabla_\alpha
\phi\stackrel{g}\nabla^\alpha\phi+V(\phi)=0$: this theory is
equivalent to the standard Einstein one without matter. On the
other hand, for $F(\phi)=F_0$ we recover the Einstein theory plus
a minimally coupled scalar field: this means that the Palatini
approach intrinsically gives rise to the conformal structure
(\ref{conformalmetric}) of the theory which is trivial in the
Einstein, minimally coupled case.

As a further step, let us generalize the previous results
considering the case of a non-minimal coupling in the framework of
$f$-theories. The action functional can be written as:

\begin{eqnarray}
\mathcal{A}_2=\int
d^4x\sqrt{-g}[F(\phi)f(R)+\omega(\phi)\stackrel{g}\nabla_\alpha
\phi\stackrel{g}\nabla^\alpha\phi+V(\phi)+2\mathcal{X}\mathcal{L}_m(\Psi,
\stackrel{g}\nabla\Psi)]
\end{eqnarray}
where $f$ is, as usual, any analytical function of the  Ricci
scalar $R$. Field equations (in the Palatini formalism) for the
gravitational part of the action are:

\begin{eqnarray}
\left\{\begin{array}{ll}
F(\phi)[f'R_{(\mu\nu)}-\frac{f}{2}g_{\mu\nu}]=\mathcal{X}T_{\mu\nu}+\frac{1}{2}\omega(\phi)\stackrel{g}\nabla_\alpha
\phi\stackrel{g}\nabla^\alpha\phi\,\,g_{\mu\nu}+\frac{1}{2}V(\phi)g_{\mu\nu}\\\\
\nabla _{\alpha }^{\Gamma }(\sqrt{ -g} F(\phi)f'g^{\mu \nu })=0
\end{array} \right.
\end{eqnarray}
For scalar and matter fields we have, otherwise, the following
field equations:

\begin{eqnarray}
\left\{\begin{array}{ll}
2\omega(\phi)\stackrel{g}\Box\phi+\omega_{,\phi}(\phi)\stackrel{g}\nabla_\alpha
\phi\stackrel{g}\nabla^\alpha\phi+V_{,\phi}(\phi)+F_{,\phi}(\phi)f(R)=0\\\\
\frac{\delta{\mathcal{L}}_{m}}{\delta \Psi}=0
\end{array} \right.
\end{eqnarray}
where the non-minimal interaction term enters into the modified
Klein-Gordon equations. In this case the structural equation of
spacetime implies that:

\begin{eqnarray}\label{fieldequationpalatiniTStrace2}
f'R-2f=\frac{\mathcal{X}T+2\omega(\phi)\stackrel{g}\nabla_\alpha
\phi\stackrel{g}\nabla^\alpha\phi+2V(\phi)}{F(\phi)}
\end{eqnarray}

We remark again that this equation, if solved,  expresses the
value of the Ricci scalar curvature in terms of traces of the
stress-energy tensors of standard matter and scalar field (we have
to require again that $F(\phi)\neq 0$). The bi-metric structure of
spacetime is thus defined by the ansatz:

\begin{eqnarray}
\sqrt{- g}F(\phi)f'g^{\mu\nu}=\sqrt{- h}h^{\mu\nu}
\end{eqnarray}
such that $g$ and $h$ result to be conformally related by:

\begin{eqnarray}
h_{\mu\nu}=F(\phi)f'g_{\mu\nu}
\end{eqnarray}

Once the structural equation is solved, the conformal factor
depends  on the values of the matter fields ($\phi$, $\Psi$) or,
more precisely, on the traces of the stress-energy tensors and the
value of $\phi$. From equation
(\ref{fieldequationpalatiniTStrace2}), it follows that in the
vacuum case, \emph{i.e.}  both $T=0$ and
$\omega(\phi)\stackrel{g}\nabla_\alpha
\phi\stackrel{g}\nabla^\alpha\phi+V(\phi)=0$, the universality of
Einstein field equations still holds as in the case of minimally
interacting $f$-theories \cite{fer-fra-vol}. The validity of this
property is related to the decoupling of the scalar field and the
gravitational field.

Let us finally consider the case where the gravitational
Lagrangian is a general function of $\phi$ and $R$. The action
functional can thus be written as:

\begin{eqnarray}
\mathcal{A}_3=\int
d^4x\sqrt{-g}[K(\phi\,,R)+\omega(\phi)\stackrel{g}{\nabla}_\alpha\phi\stackrel{g}{\nabla}^\alpha
\phi+V(\phi)+\mathcal{X}\mathcal{L}_{m}(\Psi,\stackrel{g}{\nabla}\Psi)]
\end{eqnarray}
Field equations for the gravitational part of the action are:

\begin{eqnarray}
\left\{\begin{array}{ll} \frac{\partial K(\phi,R)}{\partial R}
R_{(\mu\nu)}-\frac{K(\phi, R)}{2}g_{\mu\nu}
=\mathcal{X}T_{\mu\nu}+\frac{1}{2}\omega(\phi)\stackrel{g}\nabla_\alpha
\phi\stackrel{g}\nabla^\alpha\phi\,\,g_{\mu\nu}+\frac{1}{2}V(\phi)g_{\mu\nu}\\\\
\nabla _\alpha^\Gamma\biggl[\sqrt{ -g}\biggl(\frac{\partial
K(\phi,R)}{\partial R}\biggr)g^{\mu\nu}\biggr]=0
\end{array} \right.
\end{eqnarray}
For matter fields, we have:

\begin{eqnarray}
\left\{\begin{array}{ll}
2\omega(\phi)\stackrel{g}\Box\phi+\omega_{,\phi}(\phi)\stackrel{g}\nabla_\alpha
\phi\stackrel{g}\nabla^\alpha\phi+V_{,\phi}(\phi)+\frac{\partial
K(\phi,R)}{\partial\phi}=0\\\\
\frac{\delta{\mathcal{L}}_{m}}{\delta \Psi}=0
\end{array} \right.
\end{eqnarray}

The structural equation of spacetime can be expressed as:

\begin{eqnarray}\label{fieldequationpalatiniTStrace3}
\frac{\partial K(\phi, R)}{\partial R}R-2K(\phi,
R)=\mathcal{X}T+2\omega(\phi)\stackrel{g}\nabla_\alpha
\phi\stackrel{g}\nabla^\alpha\phi+2V(\phi)
\end{eqnarray}
This equation, if  solved,  expresses again the form of the Ricci
scalar curvature in terms of traces of the stress-energy tensors
of matter and scalar field (we have to impose regularity
conditions and, for example, $K(\phi,R) \ne 0$). The bi-metric
structure of spacetime is thus defined by the ansatz:

\begin{eqnarray}
\sqrt{-g}\frac{\partial K(\phi, R)}{\partial
R}g^{\mu\nu}=\sqrt{-h}h^{\mu\nu}
\end{eqnarray}
such that $g$ and $h$ result to be conformally related by

\begin{eqnarray}\label{conformalmetric1}
h_{\mu\nu }=\frac{\partial K(\phi, R)}{\partial R}g_{\mu\nu}\,.
\end{eqnarray}
Again, once the structural equation is solved, the conformal
factor depends just on the values of the matter fields and (the
trace of) their stress energy tensors. In other words, the
evolution, the definition of the conformal factor and the
bi-metric structure is ruled by the values of traces of the
stress-energy tensors and by the value of the scalar field $\phi$.
In this case, the universality of Einstein field equations does
not hold anymore in general. This is evident from
(\ref{fieldequationpalatiniTStrace3}) where the strong coupling
between $R$ and $\phi$ avoids the possibility, also in the vacuum
case, to achieve  simple constant solutions.

We consider, furthermore, the case of small values of $R$,
corresponding to  small curvature spacetimes. This limit
represents, as a good approximation, the present epoch of the
observed Universe under suitably regularity conditions. A Taylor
expansion of the analytical function $K(\phi, R)$ can be
performed:

\begin{eqnarray}
K(\phi, R)=K_0(\phi)+K_1(\phi)R+o(R^2)
\end{eqnarray}
where only the first leading term in $R$ is considered and we have
defined:

\begin{eqnarray}
\left\{\begin{array}{ll}
K_0(\phi)=K(\phi,R)_{R=0}\\\\
K_1(\phi)=\biggl(\frac{\partial K(\phi, R)}{\partial R}
\biggr)_{R=0}
\end{array} \right.
\end{eqnarray}
Substituting this expression in
(\ref{fieldequationpalatiniTStrace3}) and (\ref{conformalmetric1})
we get (neglecting higher order approximations in $R$) the
structural equation and the bi-metric structure in this particular
case. From the structural equation, we get:

\begin{eqnarray} \label{Rgenapr}
R=-\frac{T+2\omega(\phi)\stackrel{g}\nabla_\alpha
\phi\stackrel{g}\nabla^\alpha\phi+2V(\phi)+2K_0(\phi)}{K_1(\phi)}
\end{eqnarray}
such that the value of the Ricci scalar is always determined, in
this first order approximation, in terms of
$\omega(\phi)\stackrel{g}\nabla_\alpha
\phi\stackrel{g}\nabla^\alpha\phi+V(\phi)$, $T$ e $\phi$. The
bi-metric structure is, otherwise, simply defined by means of the
first term of the Taylor expansion, which is

\begin{eqnarray}
h_{\mu\nu}=K_1(\phi)g_{\mu\nu}\,.
\end{eqnarray}

It reproduces, as expected, the scalar-tensor case
(\ref{conformalmetric}). In other words, scalar-tensor theories
can be recovered in a first order approximation of a general
theory where gravity and non-minimal couplings are any (compare
(\ref{Rgenapr}) with (\ref{fieldequationpalatiniTStrace})). This
fact agrees with the above considerations where Lagrangians of
physical interactions can be considered as stochastic functions
with local gauge invariance properties \cite{bar-ott}.

Finally we have to say that there are also bi-metric theories
which cannot be conformally related (see for example the summary
of alternative theories given in \cite{will}) and torsion field
should be taken into account, if one wants to consider the most
general viewpoint \cite{heh-von-ker-nes, cap-lam-sto}. We will not
take into account these general theories in this review.

After this short review of ETGs in metric and Palatini approach,
we are going to face some  remarkable applications to cosmology
and astrophysics. In particular, we deal with the straightforward
generalization of GR, the $f$-gravity, showing that, in principle,
no further ingredient, a part a generalized gravity, could be
necessary to address issues as missing matter (dark matter) and
cosmic acceleration (dark energy).  However what we are going to
consider here are nothing else but toy models which are not able
to fit the whole expansion history, the structure growth law and
the CMB anisotropy  and polarization. These issues require more
detailed  theories which, up to now, are not available but what we
are discussing could be a useful working paradigm as soon as
refined experimental tests to probe such theories will be proposed
and pursued. In particular, we will outline an independent test,
based on the stochastic background of gravitational waves, which
could be extremely useful to discriminate between ETGs and GR or
among the ETGs themselves. In this latter case, the data delivered
from  ground-based interferometers, like VIRGO and LIGO, or the
forthcoming space interferometer LISA, could be of extreme
relevance in such a discrimination.

Finally, we do not take into account the well known inflationary
models based on ETGs (e.g. \cite{starobinsky}) since we want to
show that also the last cosmological epochs, directly related to
the so called \emph{Precision Cosmology}, can be framed in such a
new "economic" scheme.

\section{The general $f$ $-$ theory}

Let $f$ be an analytic function of Ricci scalar $R$. We can
formulate a HOG starting from the action principle
(\ref{actionfR}). By varying the action (\ref{actionfR}) and by using the properties (\ref{variationalcalculus}) we get
the field equations:

\begin{eqnarray}
\delta\mathcal{A}\,=&&\delta\int d^4x\sqrt{-g}[f+\mathcal{X}\,\mathcal{L}_m]=\nonumber\\=&&\int d^4x\sqrt{-g}
\biggr[\biggl(f'R_{\mu\nu}-\frac{f}{2}g_{\mu\nu}-\mathcal{X}\,T_{\mu\nu}\biggr)\delta
g^{\mu\nu}+g_{\mu\nu}f'\delta R^{\mu\nu}\biggr]=\nonumber\\=&&\int
d^4x\sqrt{-g}\biggr\{\biggl(f'R_{\mu\nu}-\frac{f}{2}g_{\mu\nu}-\mathcal{X}\,T_{\mu\nu}\biggr)\delta
g^{\mu\nu}+\nonumber\\&&\,\,\,\,\,\,\,\,\,\,\,\,\,\,\,\,\,\,\,\,\,\,\,\,\,\,\,\,\,\,\,\,\,\,\,\,\,\,\,\,\,\,\,\,\,\,\,\,\,\,
\,\,\,\,\,\,\,\,\,\,\,\,\,\,\,\,\,+f'[-(\delta
g^{\mu\nu})_{;\mu\nu}-\Box(g^{\mu\nu}\delta
g_{\mu\nu})]\biggr\}\sim\nonumber\\\sim&&\int
d^4x\sqrt{-g}\biggr\{f'R_{\mu\nu}-\frac{f}{2}g_{\mu\nu}-\mathcal{X}\,T_{\mu\nu}-f'_{;\mu\nu}+g_{\mu\nu}\Box
f'\biggr\}\delta g^{\mu\nu}=\nonumber\\=&&\int
d^4x\sqrt{-g}(H_{\mu\nu}-\mathcal{X}\,T_{\mu\nu})\delta
g^{\mu\nu}=0
\end{eqnarray}
where the symbol $\sim$ means that we neglected a pure divergence;
then we obtain the field equation (\ref{fe}). Eq. (\ref{fe})
satisfies the condition
$H^{\alpha\mu}_{\,\,\,\,\,\,\,\,;\alpha}\,=\,\mathcal{X}\,T^{\alpha\mu}_{\,\,\,\,\,\,\,\,;\alpha}\,=\,0$.
In fact it is easy to check that

\begin{eqnarray}H^{\alpha\mu}_{\,\,\,\,\,\,\,\,;\alpha}\,=&&f'_{;\alpha}R^{\alpha\mu}+f'R^{\alpha\mu}_{\,\,\,\,\,\,\,\,;
\alpha}-\frac{1}{2}f'^{;\mu}-f'^{;\alpha\mu}_{\,\,\,\,\,\,\,\,\,\,\,\,\alpha}+f'^{;\alpha\,\,\,\,\,\,\mu}_{\,\,\,\,\,\,\,\,
\alpha}\,=\nonumber\\&&f''R^{\alpha\mu}R_{;\alpha}-f'^{;\alpha\mu}_{\,\,\,\,\,\,\,\,\,\,\,\,\alpha}+f'^{;\alpha\,\,\,\,\,\,
\mu}_{\,\,\,\,\,\,\,\,\alpha}\,=\nonumber\\&&f''R^{\alpha\mu}R_{;\alpha}-f'^{;\alpha}R^{\,\,\,\,\,\mu}_{\alpha}\,=\nonumber
\\&&f''R^{\alpha\mu}R_{;\alpha}-f''R^{;\alpha}R^{\,\,\,\,\,\mu}_{\alpha}\,=\,0\,;
\end{eqnarray}
where we used the properties
$G^{\alpha\mu}_{\,\,\,\,\,\,\,\,\,;\alpha}\,=\,0$ and
$[\nabla^\mu,\nabla_\alpha]f'^{;\alpha}\,=\,-f'^{;\alpha}R_{\alpha}^{\,\,\,\,\mu}$.
If we develop the covariant derivatives in (\ref{fe}) and in
(\ref{fetr}) we obtain the complete expression for a generic
$f$-theory

\begin{eqnarray}\label{fe4}
\left\{\begin{array}{ll}
H_{\mu\nu}\,=\,f'R_{\mu\nu}-\frac{1}{2}fg_{\mu\nu}+\mathcal{H}_{\mu\nu}\,=\,\mathcal{X}\,T_{\mu\nu}\\\\
H\,=\,f'R-2f+\mathcal{H}\,=\,\mathcal{X}\,T
\end{array} \right.
\end{eqnarray}
where the two quantities $\mathcal{H}_{\mu\nu}$ and $\mathcal{H}$
read\,:

\begin{eqnarray}\label{highterms1}
\left\{\begin{array}{ll}
\mathcal{H}_{\mu\nu}\,=\,-f''\biggl\{R_{,\mu\nu}-\Gamma^\sigma_{\mu\nu}R_{,\sigma}-g_{\mu\nu}\biggl[\biggl({g^{\sigma\tau}}
_{,\sigma}+g^{\sigma\tau}\ln\sqrt{-g}_{,\sigma}\biggr)R_{,\tau}+\\\,\,\,\,\,\,\,\,\,\,\,\,\,\,\,\,\,\,\,\,\,\,\,\,\,\,\,\,\,\,
\,\,\,\,\,\,\,\,\,\,\,\,\,\,\,\,\,\,\,\,+g^{\sigma\tau}R_{,\sigma\tau}\biggr]\biggr\}-f'''\biggl(R_{,\mu}R_{,\nu}-g_{\mu\nu}
g^{\sigma\tau}R_{,\sigma}R_{,\tau}\biggr)\\\\\mathcal{H}\,=\,3f''\biggl[\biggl({g^{\sigma\tau}}_{,\sigma}+g^{\sigma\tau}
\ln\sqrt{-g}_{,\sigma}\biggr)R_{,\tau}+g^{\sigma\tau}R_{,\sigma\tau}
\biggr]+3f'''g^{\sigma\tau}R_{,\sigma}R_{,\tau}
\end{array} \right.
\end{eqnarray}
$\Gamma^\alpha_{\mu\nu}$ are the standard Christoffel's symbols
defined by (\ref{christoffel}). We conclude, then, this paragraph
having shown the most general expression of field equations of
$f$-gravity in metric formalism.

\section{The field equations for the $R_{\alpha\beta}R^{\alpha\beta}$ and $R_{\alpha\beta\gamma\delta}R^{\alpha\beta\gamma
\delta}$ $-$ invariants}

The technicality is ever the same one. We start from the action
principle for a Lagrangian densities
$\sqrt{-g}\,R_{\alpha\beta}R^{\alpha\beta}$ and
$\sqrt{-g}\,R_{\alpha\beta\gamma\delta}R^{\alpha\beta\gamma\delta}$
and we get their field equations:

\begin{eqnarray}\delta\mathcal{A}\,=&&\delta\int d^4x\sqrt{-g}[R_{\alpha\beta}R^{\alpha\beta}+\mathcal{X}\,\mathcal{L}_m]=
\nonumber\\=&&\delta\int
d^4x\sqrt{-g}[R_{\alpha\beta}g^{\alpha\rho}g^{\beta\sigma}R_{\rho\sigma}+\mathcal{X}\mathcal{L}_m]=\nonumber\\&&\nonumber\\=
&&\int d^4x\sqrt{-g}\biggr[\biggl(2R_\mu^{\,\,\,\,\alpha}R_{\alpha\nu}-\frac{R_{\alpha\beta}R^{\alpha\beta}}{2}g_{\mu\nu}-
\mathcal{X}\,T_{\mu\nu}\biggr)\delta
g^{\mu\nu}+\nonumber\\&&\,\,\,\,\,\,\,\,\,\,\,\,\,\,\,\,\,\,\,\,\,\,\,\,\,\,\,\,\,\,\,\,\,\,\,\,\,\,\,\,\,\,\,\,\,\,\,\,\,\,
\,\,\,\,\,\,\,\,\,\,\,\,\,\,\,\,\,\,\,\,\,\,\,\,\,\,\,\,\,\,\,\,\,\,\,\,\,\,\,\,\,\,\,\,\,\,\,\,\,\,\,\,\,\,\,\,\,
+2R^{\mu\nu}\delta R_{\mu\nu}\biggr]=\nonumber\\=&&\int
d^4x\sqrt{-g}\biggr[\biggl(2R_\mu^{\,\,\,\,\alpha}R_{\alpha\nu}-\frac{R_{\alpha\beta}R^{\alpha\beta}}{2}g_{\mu\nu}-
\mathcal{X}\,T_{\mu\nu}\biggr)\delta
g^{\mu\nu}+\nonumber\\&&\,\,\,\,\,\,\,\,\,\,\,\,\,\,\,\,\,\,\,\,\,\,\,
\,\,\,\,\,\,\,\,+R^{\mu\nu}(2g^{\rho\sigma}\delta g_{\rho\,(\mu;\nu)\sigma}-
\Box\,\delta g_{\mu\nu}-g^{\rho\sigma}\delta g_{\rho\sigma;\mu\nu})\biggr]\sim\nonumber\\\sim &&\int
d^4x\sqrt{-g}\biggr[\biggl(2R_\mu^{\,\,\,\,\alpha}R_{\alpha\nu}-\frac{R_{\alpha\beta}R^{\alpha\beta}}{2}g_{\mu\nu}-
\mathcal{X}\,T_{\mu\nu}\biggr)\delta
g^{\mu\nu}+\nonumber\\&&\,\,\,\,\,\,\,\,\,\,\,\,\,\,\,\,\,\,\,\,
\,\,\,-2R^\sigma_{\,\,\,\,\,(\mu;\nu)\sigma}\delta g^{\mu\nu}+\Box
R_{\mu\nu}\delta g^{\mu\nu}+R^{\sigma\tau}_{\,\,\,\,\,\,\,\,\,\,;\sigma\tau}g_{\mu\nu}\delta g^{\mu\nu}\biggr]=\nonumber\\
=&&\,\int
d^4x\sqrt{-g}\biggr[2R_\mu^{\,\,\,\,\alpha}R_{\alpha\nu}-\frac{R_{\alpha\beta}R^{\alpha\beta}}{2}g_{\mu\nu}-2R^\sigma_{\,\,\,\,
\,(\mu;\nu)\sigma}+\nonumber\\&&\,\,\,\,\,\,\,\,\,\,\,\,\,\,\,\,\,\,\,\,\,\,\,\,\,\,\,\,\,\,\,\,\,\,\,\,\,\,\,\,\,\,\,\,\,\,\,\,
\,\,\,\,\,\,+\Box R_{\mu\nu}+g_{\mu\nu}R^{\sigma\tau}_{\,\,\,\,\,\,\,\,\,\,;\sigma\tau}-\mathcal{X}\,T_{\mu\nu}\biggr]\delta
g^{\mu\nu}=0\,.
\end{eqnarray}
Then, the field equations are

\begin{eqnarray}\label{fericcisquare}
H^{Ric}_{\mu\nu}=2R_\mu^{\,\,\,\,\alpha}R_{\alpha\nu}-\frac{R_{\alpha\beta}R^{\alpha\beta}}{2}g_{\mu\nu}-2R^\sigma_{\,\,\,\,\,
(\mu;\nu)\sigma}+\Box
R_{\mu\nu}+g_{\mu\nu}R^{\sigma\tau}_{\,\,\,\,\,\,\,\,\,\,;\sigma\tau}=\mathcal{X}\,T_{\mu\nu}
\end{eqnarray}
and the trace is

\begin{eqnarray}\label{fetrriccisquare}
H^{Ric}=2\Box R=\mathcal{X}\,T\,,\end{eqnarray} where we used the
Bianchi identity contracted (\ref{bianchi-identity}).

Let us calculate the field equations for the
$R_{\alpha\beta\gamma\delta}R^{\alpha\beta\gamma\delta}$ -
invariant:

\begin{eqnarray}\label{variationalRienmancalculus}\delta\mathcal{A}=&&\delta\int d^4x\sqrt{-g}[R_{\alpha\beta\gamma\delta}R^{\alpha\beta\gamma\delta}+
\mathcal{X}\mathcal{L}_m]=\nonumber\\=&&\delta\int d^4x\sqrt{-g}[R_{\alpha\beta\gamma\delta}g^{\alpha\rho}g^{\beta\sigma}g^{\gamma\tau}g^{\delta\xi}R_{\rho\sigma\tau\xi}+
\mathcal{X}\mathcal{L}_m]=\nonumber\\=&&\int
d^4x\sqrt{-g}\biggr[\biggl(4R_{\mu\alpha\beta\gamma}R_{\nu}^{\,\,\,\,\,\alpha\beta\gamma}-\frac{R_{\alpha\beta\gamma\delta}
R^{\alpha\beta\gamma\delta}}{2}g_{\mu\nu}-\mathcal{X}T_{\mu\nu}\biggr)\delta
g^{\mu\nu}+\nonumber\\&&\,\,\,\,\,\,\,\,\,\,\,\,\,\,\,\,\,\,\,\,\,\,\,\,\,\,\,\,\,\,\,\,\,\,\,\,\,\,\,\,\,\,\,\,\,\,\,\,\,
\,\,\,\,\,\,\,\,\,\,\,\,\,\,\,\,\,\,\,\,\,\,\,\,\,\,\,\,\,\,\,\,\,\,\,\,\,\,\,\,\,\,\,\,\,\,\,\,\,\,\,\,\,\,\,\,\,\,\,\,\,
\,\,\,\,\,\,\,\,\,\,\,\,+2R^{\alpha\beta\gamma\delta}\delta
R_{\alpha\beta\gamma\delta}\biggr]=\nonumber\\=&&\int
d^4x\sqrt{-g}\biggr[\biggl(4R_{\mu\alpha\beta\gamma}R_{\nu}^{\,\,\,\,\,\alpha\beta\gamma}-\frac{R_{\alpha\beta\gamma\delta}
R^{\alpha\beta\gamma\delta}}{2}g_{\mu\nu}-\mathcal{X}T_{\mu\nu}\biggr)\delta
g^{\mu\nu}+\nonumber\\&&+R^{\alpha\beta\gamma\delta}(\delta g_{\alpha\beta;\delta\gamma}+\delta g_{\alpha\delta;\beta\gamma}
-\delta g_{\beta\delta;\alpha\gamma}-\delta g_{\alpha\beta;\gamma\delta}-\delta g_{\alpha\gamma;\beta\delta}+\delta g_{\beta\gamma;\alpha\delta})\biggr]
\sim\nonumber\\\sim&&\int
d^4x\sqrt{-g}\biggr[2R_{\mu\alpha\beta\gamma}R_{\nu}^{\,\,\,\,\,\alpha\beta\gamma}-\frac{R_{\alpha\beta\gamma\delta}
R^{\alpha\beta\gamma\delta}}{2}g_{\mu\nu}-4R^{\,\,\,\,\,\alpha\beta}_{\mu\,\,\,\,\,\,\,\,\,\,\nu;\alpha\beta}+\nonumber\\&&
\,\,\,\,\,\,\,\,\,\,\,\,\,\,\,\,\,\,\,\,\,\,\,\,\,\,\,\,\,\,\,\,\,\,\,\,\,\,\,\,\,\,\,\,\,\,\,\,\,\,\,\,\,\,\,\,\,\,\,\,\,\,
\,\,\,\,\,\,\,\,\,\,\,\,\,\,\,\,\,\,\,\,\,\,\,\,\,\,\,\,\,\,\,\,\,\,\,\,\,\,\,\,\,\,\,\,\,\,\,\,\,\,\,\,\,\,\,\,\,\,\,\,\,\,\,
\,\,\,\,-\mathcal{X}T_{\mu\nu}\biggr]\delta g^{\mu\nu}=0\end{eqnarray}
We used the expressions

\begin{eqnarray}\label{variationalRienman}
\left\{\begin{array}{ll}
\delta\,R_{\alpha\beta\gamma\delta}\,=\,\delta(g_{\alpha\sigma}R^\sigma_{\,\,\,\,\,\beta\gamma
\delta})\,=\,R^\sigma_{\,\,\,\,\,\beta\gamma
\delta}\delta\,g_{\alpha\sigma}+g_{\alpha\sigma}\delta\,R^\sigma_{\,\,\,\,\,\beta\gamma
\delta}\\\\\delta R^\sigma_{\,\,\,\,\,\beta\gamma\delta}\,=\,\frac{1}{2}(\delta g^\alpha_{\,\,\,\beta;\delta\gamma}+\delta g^\alpha_{\,\,\,\delta;\beta\gamma}-\delta g_{\beta\delta\,\,\,\,\,\,\,\,\gamma}^{\,\,\,\,\,\,\,;\alpha}-\delta g^\alpha_{\,\,\,\beta;\gamma\delta}-\delta g^\alpha_{\,\,\,\gamma;\beta\delta}+{\delta g_{\beta\gamma}}^{;\alpha}_{\,\,\,\,\,\delta})
\end{array} \right.
\end{eqnarray}
Then, the field equations, from (\ref{variationalRienmancalculus}), are

\begin{eqnarray}\label{ferienmansquare}
H^{Rie}_{\mu\nu}\,=\,2R_{\mu\alpha\beta\gamma}R_{\nu}^{\,\,\,\,\,\alpha\beta\gamma}-\frac{R_{\alpha\beta\gamma\delta}
R^{\alpha\beta\gamma\delta}}{2}g_{\mu\nu}-4R^{\,\,\,\,\,\alpha\beta}_{\mu\,\,\,\,\,\,\,\,\,\,\nu;\alpha\beta}\,=\,
\mathcal{X}T_{\mu\nu}
\end{eqnarray}
and the trace is

\begin{eqnarray}\label{fetrrienmansquare}
H^{Rie}\,=\,-4R^{\,\,\,\alpha\beta\gamma}_{\gamma\,\,\,\,\,\,\,\,\,\,\,;\alpha\beta}\,=\,\mathcal{X}T\,.\end{eqnarray}

\section{Generalities on spherical symmetry}

Since we are interesting to understand the modifications of
predictions of GR when one considers a concentration of matter in
the space, it is fundamental requiring particular properties of
metric $g_{\mu\nu}$. The first step, and also the easiest, we will
study, in the next chapters, the gravitational potential generated
by spherically symmetric matter distribution (point-like and not)
and the choice of mathematical form of metric becomes very
important. Starting from the matter spherically symmetric we
expect also the metric has the same symmetries.

We conclude this chapter showing the principal relations between
some coordinates systems we will use in this PhD thesis.

The most general spherically symmetric metric\footnote{The metric
is spherically symmetric if it depends only on $\textbf{x}$ and
$d\textbf{x}$ only through the rotational invariants
$d\textbf{x}^2$, $\textbf{x}\cdot d\textbf{x}$ and
$\textbf{x}^2$.} can be written as follows\,:

\begin{eqnarray}\label{me0}
{ds}^2\,=\,g_1(t,|\textbf{x}|)\,dt^2+g_2(t,|\textbf{x}|)\,dt\,\textbf{x}\cdot
d\textbf{x}+g_3(t,|\textbf{x}|) (\textbf{x}\cdot
d\textbf{x})^2+g_4(t,|\textbf{x}|){d|\textbf{x}|}^2
\end{eqnarray}
where $g_i$ are functions of the distance $|\textbf{x}|$ and of
the time $t$. The set of coordinates is
$x^\mu\,=\,(t,x^1,x^2,x^3)$. The scalar product is defined as
usual form: $\textbf{x}\cdot d\textbf{x}=x^1dx^1+x^2dx^2+x^3dx^3$.
By spherically symmetric form of (\ref{me0}) it is convenient to
replace $\textbf{x}$ with spherical polar coordinates $r, \theta,
\phi$ defined as usual by

\begin{eqnarray}x^1\,=\,r \sin\theta\cos\phi\,,\,\,\,\,x^2\,=\,r \sin\theta\sin\phi\,,\,\,\,\,x^3\,=\,r\cos\phi\,.
\end{eqnarray}
The proper time interval (\ref{me0}) then becomes

\begin{eqnarray}\label{me1}
{ds}^2\,=\,g_1(t,r)\,dt^2+rg_2(t,r)\,dtdr+r^2g_3(t,r)
dr^2+g_4(t,r)(dr^2+r^2d\Omega)\,,
\end{eqnarray}
where $d\Omega=d\theta^2+\sin^2\theta d\phi^2$ is the solid angle.
We are free to reset our clocks by defining the time coordinate

\begin{eqnarray}t=t'+\zeta(t',r)\,,
\end{eqnarray}
with $\zeta(t',r)$ an arbitrary function of $t'$ and $r$. This
allows us to eliminate the off-diagonal element $g_{tr}$ in the
metric (\ref{me1}) by setting

\begin{eqnarray}\frac{d\zeta(t',r)}{dr}=-\frac{rg_2(t',r)}{2g_1(t',r)}\,,\end{eqnarray}
the metric (\ref{me1}) becomes

\begin{eqnarray}\label{me2}
{ds}^2\,=\,g_1(t',r)\biggl[1+\frac{d\zeta(t',r)}{dt'}\biggr]^2\,dt'^2+\biggl[r^2g_3(t',r)-\frac{r^2g_2(t',r)^2}{4g_1(t'
,r)}+g_4(t',r)\biggr]{dr}^2+\nonumber\\\nonumber\\+g_4(t',r)r^2d\Omega\,,
\end{eqnarray}
where if we introduce a new metric coefficients $g_{tt}(t',r)$,
$g_{rr}(t',r)$ and $g_{\Omega\Omega}(t',r)$ we can recast the
(\ref{me2}) as follows

\begin{eqnarray}\label{me3}
{ds}^2\,=\,g_{tt}(t',r)\,dt'^2-g_{rr}(t',r){dr}^2-g_{\Omega\Omega}(t',r)d\Omega\,;
\end{eqnarray}
if we introduce a new radial coordinate ($r'$) by considering a
further transformation

\begin{eqnarray}r'=(const)e^{\int dr\sqrt{\frac{g_{rr}(t',r)}{g_{\Omega\Omega}(t',r)}}}\end{eqnarray}
it is possible to recast Eq. (\ref{me3}) into the isotropic form
(isotropic coordinates)

\begin{eqnarray}\label{me4}
{ds}^2\,=\,g_{tt}(t',r')\,dt^2-g_{ij}(t',r')dx^i
dx^j\,;\end{eqnarray} and then it is possible also to choose
$g_{\Omega\Omega}(t',r)=r''^2$ (this condition allows us to obtain
the standard definition of the circumference with radius $r''$)
and to have the metric (\ref{me3}) in the standard form (standard
coordinates)

\begin{eqnarray}\label{me5}
{ds}^2\,=\,g_{tt}(t',r'')\,dt^2-g_{rr}(t',r''){dr''}^2-r''^2d\Omega\,.
\end{eqnarray} Obviously the functions
$g_{tt}(t',r'')$ and $g_{rr}(t',r'')$ are not the same of
(\ref{me3}). If we suppose $g_{ij}(t',r')=Y(t',r')\delta_{ij}$ we
note that it is possible pass from (\ref{me4}) to (\ref{me5}) by
the coordinate transformations:

\begin{eqnarray}\label{transf-stan-isotr-coord}r'=r'(r'')=(const)e^{\int dr'' \frac{\sqrt{\tilde{Y}(r'')}}{r''}}\,.
\end{eqnarray}
We can, then, affirm that the expressions (\ref{me3}), (\ref{me4})
and (\ref{me5}) are equivalent to the metric (\ref{me0}) and we
can consider them without loss of generality as the most general
definitions of a spherically symmetric metric compatible with a
pseudo\,-\,Riemannian manifold without torsion. The choice of the
form of the metric is only a practical issue. With this
hypothesis, by inserting these metrics into the field equations
(\ref{fe}), one obtains:

\begin{eqnarray}\label{asd}
\left\{\begin{array}{ll}
H_{\mu\nu}=f'R_{\mu\nu}-\frac{1}{2}fg_{\mu\nu}-f''\biggl\{R_{,\mu\nu}-\Gamma^t_{\mu\nu}R_{,t}-\Gamma^r_{\mu\nu}R_{,r}-
g_{\mu\nu}\biggl[\biggl({g^{tt}}_{,t}+\\\,\,\,\,\,\,\,\,\,\,\,\,\,\,\,\,\,\,+g^{tt}\ln\sqrt{-g}_{,t}\biggr)R_{,t}
+\biggl({g^{rr}}_{,r}+g^{rr}\ln\sqrt{-g}_{,r}\biggr)R_{,r}+g^{tt}R_{,tt}+\\\,\,\,\,\,\,\,\,\,\,\,\,\,\,\,\,\,\,
+g^{rr}R_{,rr}\biggr]\biggr\}-f'''\biggl[R_{,\mu}R_{,\nu}-g_{\mu\nu}\biggl(g^{tt}{R_{,t}}^2+g^{rr}{R_{,r}}^2\biggr)\biggr]\\\\
H=f'R-2f+3f''\biggl[\biggl({g^{tt}}_{,t}+g^{tt}\ln\sqrt{-g}_{,t}\biggr)R_{,t}+\biggl({g^{rr}}_{,r}+g^{rr}\ln\sqrt{-g}_{,r}
\biggr)R_{,r}\\\,\,\,\,\,\,\,\,\,\,\,\,\,\,+g^{tt}R_{,tt}+g^{rr}R_{,rr}\biggr]+3f'''\biggl[g^{tt}{R_{,t}}^2
+g^{rr}{R_{,r}}^2\biggr]
\end{array} \right.
\end{eqnarray}

Eqs. (\ref{asd}) are the starting-point for the next chapter. All
our studies in the next chapters are referred ever to field
equations (\ref{asd}), except the second part of sixth chapter
where we have to insert in the field equations also the
contribution of $R_{\alpha\beta}R^{\alpha\beta}$-invariant
(\ref{fericcisquare}).

We conclude having shown the most general spherically symmetric
metric tensor for our aim and before starting from third chapter
with a systematic study of $f$-gravity we want to stop, in the
second chapter, to consider the principal spherically symmetric
solutions in GR. Some of these solutions will be the
starting-point to find, with perturbative methods, the corrections
induced by $f$-theory.

\clearpage{\pagestyle{empty}\cleardoublepage}

\chapter{Exact and perturbative solutions in General Relativity}

In this chapter we show, starting from knowledge of outcomes of
GR, the mathematical tools needed for aims of present thesis.
First of all we present the particular spherically symmetric
solutions in GR and consequent Birkhoff theorem (\S\,
\ref{sch-desit-rei-nor-bir}). In \S\,\ref{post-new-new-formalism}
we show the technicality of development of field equations with
respect to Newtonian and Post-Newtonian approach [\textbf{C}].
Finally, in \S\,\ref{PM-limit-GR} we perform the post-Minkowskian
limit: the gravitational waves. The developments are computed in
generic coordinates systems and in the gauge harmonic.

The $f$-gravity theory, from mathematical point-view, is more
complicated than GR. Giving the exact solutions of Eqs. (\ref{fe})
is vary hard challenge. Nevertheless, known the basic solutions of
GR, we can try to find new solutions by requiring the Newtonian
and post-Newtonian limit approach. This approach is very useful
when we consider the astrophysical problems or the study of planet
motion in the Solar System. An another field of comparison between
GR and $f$-gravity is possible in the post-Minkowskian regime. In
this case we can study the propagation of gravitational field
induced by $f$-gravity.

We dedicate, then, this chapter to understanding the outcomes of
GR and to showing all mathematical tools needed for the next
chapters.

\section{The Schwarzschild, Schwarzschild $-$ de Sitter and Reissner $-$ Nordstrom solutions: the Birkhoff theorem in General
Relativity}\label{sch-desit-rei-nor-bir}

We can rewrite the metric (\ref{me5}) as follows

\begin{eqnarray}ds^2\,=\,e^{\nu(t,r)}\,dt^2-e^{\mu(t,r)}{dr}^2-r^2d\Omega\,,\end{eqnarray}
where we recalled the radial coordinate. The only nonvanishing
components of metric tensor $g_{\mu\nu}$ are

\begin{eqnarray}g_{tt}\,=\,e^{\nu(t,r)}\,,\,\,\,\,\,\,\,g_{rr}\,=\,-e^{\mu(t,r)}\,,\,\,\,\,\,\,\,g_{\theta\theta}\,=\,-r^2
\,,\,\,\,\,\,\,\,g_{\phi\phi}\,=\,-r^2\sin^2\theta\end{eqnarray}
with functions $\mu(t,r)$ and $\nu(t,r)$ that are to be determined
by solving the field equations in GR (\ref{fieldequationGR}).
Since $g_{\mu\nu}$ is diagonal, it is easy to write down all the
nonvanishing components of its inverse:

\begin{eqnarray}g^{tt}\,=\,e^{-\nu(t,r)}\,,\,\,\,\,\,\,\,g^{rr}\,=\,-e^{-\mu(t,r)}\,,\,\,\,\,\,\,\,g^{\theta\theta}\,=\,
-r^{-2}\,,\,\,\,\,\,\,\,g^{\phi\phi}\,=\,-r^{-2}\sin^{-2}\theta\,.\end{eqnarray}
Furthermore, the determinant of the metric tensor is

\begin{eqnarray}g\,=\,-e^{\mu(t,r)+\nu(t,r)}r^4\sin^2\theta\end{eqnarray}
so the invariant volume element is

\begin{eqnarray}\sqrt{-g}\,dr\,d\theta\,d\phi\,=\,r^2e^{-\frac{\mu(t,r)+\nu(t,r)}{2}}\,\sin\theta\,dr\,d\theta\,d\phi\,.
\end{eqnarray}
The only nonvanishing components of symbols Christoffel
(\ref{christoffel}) are

\begin{eqnarray}
\left\{\begin{array}{ll} \begin{array}{ccccc}
  \Gamma^{t}_{tt}\,=\,\frac{\dot{\nu}(t,r)}{2}\,, & \,\,\, & \Gamma^r_{rr}\,=\,\frac{\mu'(t,r)}{2}\,, & \,\,\, & \Gamma^r_{tt}
  \,=\,\frac{\nu'(t,r)}{2}e^{\nu(t,r)-\mu(t,r)}\,, \\
  & & & \\
  \Gamma^t_{rr}\,=\,\frac{\mu'(t,r)}{2}e^{\mu(t,r)-\nu(t,r)}\,,& \,\,\, & \Gamma^t_{tr}\,=\,\frac{\nu'(t,r)}{2}\,, & \,\,\, &
  \Gamma^r_{tr}\,=\,\frac{\nu(t,r)}{2}\,, \\
  & & & \\
  \Gamma^r_{\theta\theta}\,=\,-r\,e^{-\mu(t,r)}\,,& \,\,\, & \Gamma^\theta_{r\theta}\,=\,\Gamma^\phi_{r\phi}\,=\,\frac{1}{r}
  \,, & \,\,\, &
  \Gamma^r_{\phi\phi}\,=\,-r\,e^{-\mu(t,r)}\,\sin^2\theta\,, \\
  & & & \\
  \Gamma^\theta_{\phi\phi}\,=\,-\sin\theta\,\cos\theta\,,& \,\,\, & \Gamma^\phi_{\theta\phi}\,=\cot\theta\,, & \,\,\, & \\
  \end{array}
\end{array}\right.
\end{eqnarray}
and the field equations (\ref{fieldequationGR}) become

\begin{eqnarray}\label{fe-schwarz-sol}
\left\{\begin{array}{ll}
  \frac{1}{r^2}-e^{-\mu(t,r)}\biggl[\frac{1}{r^2}-\frac{\mu'(t,r)}{r}\biggr]\,=\,\mathcal{X}\,T^t_{\,\,\,t} \\
  \\
  \frac{\dot{\mu}(t,r)}{r}e^{-\mu(t,r)}\,=\,\mathcal{X}\,T^r_{\,\,\,\,t} \\
  \\
  \frac{1}{r^2}-e^{-\mu(t,r)}\biggl[\frac{\nu'(t,r)}{r}+\frac{1}{r^2}\biggr]\,=\,\mathcal{X}\,T^r_{\,\,\,\,r} \\
  \\
  \frac{e^{-\nu(t,r)}}{2}\biggl[\ddot{\mu}(t,r)+\frac{\dot{\mu}^2(t,r)}{2}-\frac{\dot{\mu}(t,r)\,\dot{\nu}(t,r)}{2}\biggr]+
  \\
  \,\,\,\,\,\,\,\,\,\,\,\,\,\,\,\,-\frac{e^{-\mu(t,r)}}{2}\biggl[\nu''(t,r)+\frac{\nu'^2(t,r)}{2}+\frac{\nu'(t,r)-\mu'(t,r)}
  {r}-\frac{\nu'(t,r)\,\mu'(t,r)}{2}\biggr]\,=\,\mathcal{X}\,T^\theta_{\,\,\,\,\theta}\,=\,\mathcal{X}\,T^\phi_{\,\,\,\,
  \phi}
\end{array}\right.
\end{eqnarray}
and if we suppose a tensor of matter like $T_{\mu\nu}\,=\,\rho
u_\mu u_\nu$ with $\rho\,=\,M\,\delta(\textbf{x})$ the density of
matter time-independent we obtain the socalled \emph{Schwarzschild
solution} in standard coordinates:

\begin{eqnarray}\label{schwarz-solution-stand-coord}ds^2\,=\,\biggl[1-\frac{r_g}{r''}\biggr]dt^2-\frac{dr''^2}{1-\frac{r_g}
{r''}}-r''^2d\Omega\end{eqnarray} where $r_g\,=\,2GM$ is the so
called \emph{Schwarzschild radius}.

Metric (\ref{schwarz-solution-stand-coord}) determines completely
the gravitational field in the vacuum generated by a spherically
matter density distribution. Furthermore the Schwarzschild
solution is valid also when we consider a moving source with a
spherical distribution. The spatial metric is determined by
expression of spatial distance element

\begin{eqnarray}dl^2\,=\,\frac{dr''^2}{1-\frac{r_g}
{r''}}+r''^2d\Omega\,.\end{eqnarray} We have to note that, while
the length of circumference with "radius" $r''$ is the usual one
$2\pi r''$, the distance between two points on the same radius is
given by the integral

\begin{eqnarray}\label{dist-curved-space}\int_{r''_1}^{r''_2}\frac{dr''}{\sqrt{1-\frac{r_g}{r''}}}\,>\,r''_2-r''_1\,;
\end{eqnarray}
then the space is curved. Besides we note that $g_{tt}\,\leq\,1$,
then, by the relation between the time coordinate $t$ and the
proper time $\tau$ ($d\tau\,=\,\sqrt{g_{tt}}\,dt$), we get the
condition

\begin{eqnarray}\label{proper-time}d\tau\leq dt\,.\end{eqnarray}
At infinity, the time coordinate coincides with physical time. We
can state that when we are at a finite distance from the the mass,
there is a slowdown of the time with respect to the time measured
at infinity.

In presence of matter the situation is the following. In fact from
the first equation in the (\ref{fe-schwarz-sol}), when
$r\rightarrow 0$, $\mu(t,r)$ has to vanish as $r^2$; otherwise
$T^t_{\,\,\,\,t}$ could have a singular point in the origin. By
integrating formally the equation with the condition
$\mu(t,r)|_{r\,=\,0}\,=\,0$, one get

\begin{eqnarray}\mu(t,r)\,=\,-\ln\biggl[1-\frac{\mathcal{X}}{r}\int_0^rT^t_{\,\,\,\,t}\,\hat{r}^2\,d\hat{r}\biggr]\,.
\end{eqnarray}
It is easy to demonstrate also in the matter with spherical
symmetry that the proprieties (\ref{dist-curved-space}),
(\ref{proper-time}) and $\mu(t,r)+\nu(t,r)\,\leq\,0$ are verified
\cite{landau}. If the gravitational field is created by spherical
body with "radius" $\xi$, we have $T^t_{\,\,\,\,t}\,=\,0$ outside
the body ($r\,>\,\xi$) and we can write

\begin{eqnarray}\mu_\xi(t,r)\,=\,-\ln\biggl[1-\frac{\mathcal{X}}{r}\int_0^\xi T^t_{\,\,\,\,t}\,\hat{r}^2\,d\hat{r}\biggr]
\end{eqnarray}
and obtain the analogous expression of
(\ref{schwarz-solution-stand-coord}) in the matter:

\begin{eqnarray}\label{schwarz-solution-stand-coord-matter}ds^2=\biggl[1-\frac{r_g(r'')}{r''}\biggr]dt^2-\frac{dr''^2}{1-
\frac{r_g(r'')}{r''}}-r''^2d\Omega\,,\end{eqnarray} where we
introduced the Schwarzschild radius linked to the quantity of
matter included in the sphere with radius $r''$:

\begin{eqnarray}r_g(r'')=\mathcal{X}\int_0^{r''}T^t_{\,\,\,\,t}\,\hat{r}^2\,d\hat{r}\,;\end{eqnarray}
obviously when the distance is bigger than the radius of the body,
the metric (\ref{schwarz-solution-stand-coord-matter}) is equal to
(\ref{schwarz-solution-stand-coord}).

If we consider the transformation (\ref{transf-stan-isotr-coord}),
which in the case of Schwarzschild solution is

\begin{eqnarray}r'\,=\,\frac{2r''-r_g+2\sqrt{r''^2-r_g\,r''}}{4}\,,\end{eqnarray}
it is possible to obtain the Schwarzschild solution
(\ref{schwarz-solution-stand-coord}) in isotropic coordinates:

\begin{eqnarray}\label{schwarz-solution-isot-coord}ds^2\,=\,\biggl[\frac{1-\frac{r_g}{4r'}}{1+\frac{r_g}{4r'}}
\biggr]^2dt^2-\biggl[1+\frac{r_g}{4r'}\biggr]^{4}(dr'^2+r'^2d\Omega)\,.\end{eqnarray}

In both cases, the solutions (\ref{schwarz-solution-stand-coord})
and (\ref{schwarz-solution-stand-coord-matter}) agree with the
trace equation of Einstein equation: $R\,=\,-\mathcal{X}\,T$.
Since in the vacuum the trace of matter tensor is vanishing
(except the origin, in which the trace is proportional to
$\delta(\textbf{x})$) we can state that the Schwarzschild solution
is "Ricci flat": $R\,=\,0$.

If we add in the Hilbert - Einstein lagrangian
(\ref{HElagrangian}) a term like ($-2\sqrt{-g}\,\Lambda$) with
$\Lambda$ a generic constant the field equations
(\ref{fieldequationGR}) are modified as follows

\begin{eqnarray}\label{fe-equationGR-lambda}G_{\mu\nu}+\Lambda g_{\mu\nu}\,=\,\mathcal{X}\,T_{\mu\nu}\,,\end{eqnarray}
and if we consider a point-like source, we find the
\emph{Schwarzschild - de Sitter solution}

\begin{eqnarray}\label{schwarz-desitt-solution}ds^2\,=\,\biggl[1-\frac{r_g}{r''}+\frac{\Lambda}{3}r''^2\biggr]dt^2-\frac{
dr''^2}{1-\frac{r_g}{r''}+\frac{\Lambda}{3}r''^2}-r''^2d\Omega\,.\end{eqnarray}
In this case the trace of (\ref{fe-equationGR-lambda}) is

\begin{eqnarray}R\,=\,4\Lambda-\mathcal{X}\,T\,\end{eqnarray}
from which we note that this solution does not admit solution in
the vacuum, since also in absence of ordinary matter
($T_{\mu\nu}\,=\,0$) we have a nonvanishing scalar curvature. The
contribution is given by cosmological constant $\Lambda$. It is
also possible in this case to find the analogous of
(\ref{schwarz-solution-stand-coord-matter}).

Finally let us consider as source a radial and static electric
field $\textbf{E}\,=\,Q\,\textbf{x}\,/|\textbf{x}|^3$. We know
that the Lagrangian of electromagnetic field is
$-\frac{1}{4\pi}F_{\alpha\beta}F^{\alpha\beta}$ where
$F_{\alpha\beta}$ is the electromagnetic tensor. Then, the Hilbert
- Einstein lagrangian is

\begin{eqnarray}\mathcal{L}_{HE}\,=\,\sqrt{-g}(R-\frac{1}{4\pi}\,F_{\alpha\beta}F^{\alpha\beta})\,,\end{eqnarray}
and the Einstein equation (\ref{fieldequationGR}) becomes

\begin{eqnarray}G_{\mu\nu}\,=\,-\frac{1}{8\pi}\,(g_{\mu\nu}F_{\alpha\beta}F^{\alpha\beta}-4F_{\mu\alpha}F^\alpha_{\,\,\,
\,\nu})\,.\end{eqnarray} The solution for a spherically symmetric
system is the \emph{Reissner - Nordstrom solution}:

\begin{eqnarray}\label{reis-nord-solution}ds^2\,=\,\biggl[1-\frac{r_g}{r''}+\frac{Q^2}{r''^2}\biggr]dt^2-\frac{
dr''^2}{1-\frac{r_g}{r''}+\frac{Q^2}{r''^2}}-r''^2d\Omega\,.\end{eqnarray}

In all the above cases shown the Birkhoff theorem holds: \emph{The
metric tensor generated in vacuum by a matter density distribution
with a spherical symmetry is time-independent}. Also a
time-dependent source with a spherical symmetry produces a static
metric. The curvature of spacetime in the matter, a distance $r$
from the origin, is proportional only to the matter inside the
sphere of radius $r$. This conclusion is compatible with the Gauss
theorem of classical mechanics.

One of the goal of the present thesis is to develop similar
considerations in the case of $f$-gravity.

\section{Perturbations of the Schwarzschild solution: The Eddington parameters $\beta$ and $\gamma$}

The Schwarzschild solution (\ref{schwarz-solution-isot-coord}) is
a mathematically exact solution and is true everywhere. But in
same cases the physical conditions could permit a "reduction" of
them. In fact the Schwarzschild radius $r_g$ is a scale-length
induced by theory. Then we could be at radial distance $r'$ for
the which we have $r_g/r'\ll1$ and the
(\ref{schwarz-solution-isot-coord}) becomes

\begin{eqnarray}\label{schwarz-isotropic-pertur}
ds^2\simeq\biggl[1-\frac{r_g}{r'}+\frac{1}{2}\biggl(\frac{r_g}{r'}\biggr)^2+\dots\biggr]dt^2-\biggl[1+
\frac{r_g}{r'}+\dots\biggr]\biggl[dr'^2+r'^2d\Omega\biggr]\,.\end{eqnarray}

Since we are interesting to investigate the deviations, induced by
$f$-gravity, from behavior (\ref{schwarz-isotropic-pertur}) it is
useful to introduce the method taking into account such deviations
with respect to GR. A standard approach is the
Parameterized-Post-Newtonian (PPN) expansion of the Schwarzschild
metric (\ref{schwarz-solution-isot-coord}). Eddington
parameterized deviations with respect to GR, considering a Taylor
series in term of $r_g/r'$ assuming that in Solar System, the
limit $r_g/r'\ll 1$ holds \cite{will}. The resulting metric is

\begin{eqnarray}\label{schwarz-isotropic-PPN}
ds^2\simeq\biggl[1-\alpha\frac{r_g}{r'}+\frac{\beta}{2}\biggl(\frac{r_g}{r'}\biggr)^2+\dots\biggr]dt^2-\biggl[1+\gamma
\frac{r_g}{r'}+\dots\biggr]\biggl[dr'^2+r'^2d\Omega\biggr]\,,\end{eqnarray}
where $\alpha$, $\beta$ and $\gamma$ are unknown dimensionless
parameters (Eddington parameters) which parameterize deviations
with respect to GR. The reason to carry out this expansion up to
the order $(r_g /r')^2$ in $g_{tt}$ and only to the order
$(r_g/r')$ in $g_{ij}$ is that, in applications to celestial
mechanics, $g_{ij}$  always appears multiplied by an extra factor
$v^2\backsim M/r')$. It is evident that the standard GR solution
for a spherically symmetric gravitational system in vacuum, is
obtained for $\alpha\,=\,\beta\,=\,\gamma\,=\,1$ giving again the
"perturbed" Schwarzschild solution
(\ref{schwarz-isotropic-pertur}). Actually, the parameter $\alpha$
can be settled to the unity due to the mass definition of the
system itself \cite{will}. As a consequence, the expanded metric
(\ref{schwarz-isotropic-PPN}) can be recast in the form\,:

\begin{eqnarray}
ds^2\simeq\biggl[1-\frac{r_g}{r''}+\frac{\beta-\gamma}{2}\biggl(\frac{r_g}{r''}\biggr)^2+\dots\biggr]dt^2-\biggl[1+\gamma
\frac{r_g}{r''}+\dots\biggr]dr''^2-r''^2d\Omega\,,
\end{eqnarray}
where we have restored the standard spherical coordinates by means
of the transformation
$r''\,=\,r'\biggl[1+\frac{r_g}{4r'}\biggr]^2$. The two parameters
$\beta,\,\gamma$ have a physical interpretation. The parameter
$\gamma$ measures the amount of curvature of space generated by a
body of mass $M$ at radius $r'$. In fact, the spatial components
of the Riemann curvature tensor are, at post-Newtonian order,

\begin{eqnarray}
R_{ijkl}=\frac{3}{2}\gamma\frac{r_g}{r'^3}N_{ijkl}
\end{eqnarray}
independently of the gauge choice, where $N_{ijkl}$ represents the
geometric tensor properties (e.g. symmetries of the Riemann tensor
and so on). On the other side, the parameter $\beta$ measures the
amount of non-linearity ($\sim (r_g/r')^2 $) in the $g_{tt}$
component of the metric. However, this statement is valid only in
the standard post-Newtonian gauge.

\section{General remarks on the Newtonian and the post $-$ Newtonian approximation of Einstein
equation}\label{post-new-new-formalism}

At this point, it is worth discussing  some general issues on the
Newtonian and post-Newtonian limits. Basically there are some
general features one has to take into account when approaching
these limits, whatever the underlying theory of gravitation is. In
fact here we are not interested in entering the theoretical
discussion on how to formulate a mathematically well founded
Newtonian limit (and post-Newtonian) of general relativistic field
theories, nevertheless we point the interested reader to
\cite{trautman, friedrichs, kilmister, dautcourt, kuenzle, ehlers,
ehlers1}. In this section, we provide the explicit form of the
various quantities needed to compute the approximations in the
field equations in GR theory and any metric theory of gravity. We
only mention that there is also been a discussion on alternative
ways to define the Newtonian and Post - Newtonian limit in
higher-order theories in the recent literature, see for example
\cite{dick}. In this work, the Newtonian and Post - Newtonian
limit is identified with the maximally symmetric solution, which
is not necessarily Minkowski spacetime in $f$ - theories which
could be singular.

If one consider a system of gravitationally interacting particles
of  mass $\bar{M}$, the kinetic energy
$\frac{1}{2}\bar{M}\bar{v}^2$ will be, roughly, of the same order
of magnitude as the typical potential energy
$U=G\bar{M}^2/\bar{r}$, with $\bar{M}$, $\bar{r}$, and $\bar{v}$
the typical average values of  masses, separations, and velocities
of these particles. As a consequence:

\begin{eqnarray}\bar{v^2}\sim \frac{G\bar{M}}{\bar{r}}\,,\end{eqnarray}
(for instance, a test particle in a circular orbit of radius $r$
about a central mass $M$ will have velocity $v$ given in Newtonian
mechanics by the exact formula $v^2=GM/r$.)

The post-Newtonian approximation can be described as a method for
obtaining the motion of the system to higher approximations than
the first order (approximation which coincides with the Newtonian
mechanics) with respect to the quantities $G\bar{M}/\bar{r}$ and
$\bar{v}^2$ assumed small with respect to the squared light speed.
This approximation is sometimes referred to as an expansion in
inverse powers of the light speed.

The typical values of the Newtonian gravitational potential $\Phi$
are nowhere larger (in modulus) than $10^{-5}$ in the Solar System
(in geometrized units, $\Phi$ is dimensionless). On the other
hand, planetary velocities satisfy the condition
$\bar{v}^2\lesssim-\Phi$, while the matter pressure $p$
experienced inside the Sun and the planets is generally smaller
than the matter gravitational energy density $-\rho\Phi$, in other
words \footnote{Typical values of $p/\rho$ are $\sim 10^{-5}$ in
the Sun and  $\sim 10^{-10}$ in the Earth \cite{will}.}
$p/\rho\lesssim -\Phi$. Furthermore one must consider that even
other forms of energy in the Solar System (compressional energy,
radiation, thermal energy, etc.) have small intensities and the
specific energy density $\Pi$ (the ratio of the energy density to
the rest-mass density) is related to $U$ by $\Pi\lesssim U$ ($\Pi$
is $\sim 10^{-5}$ in the Sun and $\sim 10^{-9}$ in the Earth
\cite{will}). As matter of fact, one can consider that these
quantities, as function of the velocity, give second order
contributions\,:

\begin{eqnarray}
-\Phi\sim v^2\sim p/\rho\sim \Pi\sim\text{O(2)}\,.
\end{eqnarray}
Therefore, the velocity $v$ gives O(1) terms in the velocity
expansions, $U^2$ is of order O(4), $Uv$ of O(3), $U\Pi$ is of
O(4), and so on. Considering these approximations, one has

\begin{eqnarray}
\frac{\partial}{\partial t}\sim\textbf{v}\cdot\nabla\,,
\end{eqnarray}
and

\begin{eqnarray}
\frac{|\partial/\partial t|}{|\nabla|}\sim\text{O(1)}\,.
\end{eqnarray}
Now, particles move along geodesics\,:

\begin{eqnarray}
\frac{d^2x^\mu}{ds^2}+\Gamma^\mu_{\sigma\tau}\frac{dx^\sigma}{ds}\frac{dx^\tau}{ds}=0\,,
\end{eqnarray}
which can be written in details as

\begin{eqnarray}
\frac{d^2x^i}{dt^2}=-\Gamma^i_{tt}-2\Gamma^i_{tm}\frac{dx^m}{dt}-
\Gamma^i_{mn}\frac{dx^m}{dt}\frac{dx^n}{dt}+\biggl[\Gamma^t_{tt}+
2\Gamma^t_{tm}\frac{dx^m}{dt}+2\Gamma^t_{mn}\frac{dx^m}{dt}\frac{dx^n}{dt}\biggr]\frac{dx^i}{dt}\,.
\end{eqnarray}
In the Newtonian approximation, that is vanishingly small
velocities and only  first-order terms in the difference between
$g_{\mu\nu}$ and the Minkowski metric $\eta_{\mu\nu}$, one obtains
that the particle motion equations reduce to the standard
result\,:

\begin{eqnarray}
\frac{d^2x^i}{dt^2}\simeq-\Gamma^i_{tt}\simeq-\frac{1}{2}\frac{\partial
g_{tt}}{\partial x^i}\,.
\end{eqnarray}
The quantity $1-g_{tt}$ is of order $G\bar{M}/\bar{r}$, so that
the Newtonian approximation gives
$\displaystyle\frac{d^2x^i}{dt^2}$ to the order
$G\bar{M}/\bar{r}^2$, that is, to the order $\bar{v}^2/r$. As a
consequence if we would like to search for the post-Newtonian
approximation, we need to compute
$\displaystyle\frac{d^2x^i}{dt^2}$ to the order
$\bar{v}^4/\bar{r}$. Due to the Equivalence Principle and the
differentiability of spacetime manifold, we expect that it should
be possible to find out a coordinate system in which the metric
tensor is nearly equal to the Minkowski one $\eta_{\mu\nu}$, the
correction being expandable in powers of
$G\bar{M}/\bar{r}\sim\bar{v}^2$. In other words one has to
consider the metric developed as follows\,:

\begin{eqnarray}\label{PPN-metric}
\left\{\begin{array}{ll}g_{tt}(t,\textbf{x})\simeq1+g^{(2)}_{tt}(t,\textbf{x})+g^{(4)}_{tt}(t,\textbf{x})+\text{O(6)}
\\\\g_{ti}(t,\textbf{x})\simeq g^{(3)}_{ti}(t,\textbf{x})+\text{O(5)}\\\\
g_{ij}(t, \textbf{x})\simeq-\delta_{ij}+g^{(2)}_{ij}(t,
\textbf{x})+\text{O(4)}
\end{array}\right.
\end{eqnarray}
where $\delta_{ij}$ is the Kronecker delta, and for the
controvariant form of $g_{\mu\nu}$, one has

\begin{eqnarray}\label{PPN-metric-contro}
\left\{\begin{array}{ll}g^{tt}(t,\textbf{x})\simeq 1+g^{(2)tt}(t,
\textbf{x})+g^{(4)tt}(t, \textbf{x})+\text{O(6)}
\\\\
g^{ti}(t,\textbf{x})\simeq g^{(3)ti}(t,\textbf{x})+\text{O(5)}\\\\
g^{ij}(t,\textbf{x})\simeq-\delta_{ij}+g^{(2)ij}(t,{\textbf{x}})+{\text{O(4)}}
\end{array}\right.
\end{eqnarray}
The inverse of the metric tensor (\ref{PPN-metric}) is defined by
(\ref{controvariant-metric-condition}). The relations among the
higher than first order terms turn out to be

\begin{eqnarray}\label{PPN-metric-contro-cov}
\left\{\begin{array}{ll}g^{(2)tt}(t,\textbf{x})=-g^{(2)}_{tt}(t,\textbf{x})\\\\g^{(4)tt}(t,\textbf{x})={g^{(2)}_{tt}
(t,\textbf{x})}^2-g^{(4)}_{tt}(t,\textbf{x})\\\\g^{(3)ti}=g^{(3)}_{ti}\\\\g^{(2)ij}(t,\textbf{x})=-g^{(2)}_{ij}(t,\textbf{x})
\end{array}\right.
\end{eqnarray}
In evaluating $\Gamma^\mu_{\alpha\beta}$ we must take into account
that the scale of distance and time, in our systems, are
respectively set by $\bar{r}$ and $\bar{r}/\bar{v}$, thus the
space and time derivatives should be regarded as being of order

\begin{eqnarray}
\frac{\partial}{\partial x^i}\sim\frac{1}{\bar{r}}\,, \ \ \ \ \ \
\ \frac{\partial}{\partial t}\sim\frac{\bar{v}}{\bar{r}}\,.
\end{eqnarray}
Using the above approximations (\ref{PPN-metric}),
(\ref{PPN-metric-contro}) and (\ref{PPN-metric-contro-cov}) we
have, from the definition (\ref{christoffel}),

\begin{eqnarray}\label{PPN-christoffel}
\left\{\begin{array}{ll} \begin{array}{ccc}
  {\Gamma^{(3)}}^t_{tt}=\frac{1}{2}g^{(2)}_{tt,t}\, & \,\, & {\Gamma^{(2)}}^{i}_{tt}=\frac{1}{2}g^{(2)}_{tt,i} \\
  & & \\
  {\Gamma^{(2)}}^{i}_{jk}=\frac{1}{2}\biggl(g^{(2)}_{jk,i}-g^{(2)}_{ij,k}-g^{(2)}_{ik,j}\biggr)\,
  & \,\, & {\Gamma^{(3)}}^{t}_{ij}=\frac{1}{2}\biggl
  (g^{(3)}_{ti,j}+g^{(3)}_{jt,i}-g^{(2)}_{ij,t}\biggr) \\
  & & \\
  {\Gamma^{(3)}}^{i}_{tj}=\frac{1}{2}\biggl(g^{(3)}_{tj,i}-g^{(3)}_{it,j}-g^{(2)}_{ij,t}\biggr)\,
  & \,\, & {\Gamma^{(4)}}^{t}_{ti}=\frac{1}{2}\biggl
  (g^{(4)}_{tt,i}-g^{(2)}_{tt}g^{(2)}_{tt,i}\biggr) \\
  & & \\
  {\Gamma^{(4)}}^{i}_{tt}=\frac{1}{2}\biggl(g^{(4)}_{tt,i}+g^{(2)}_{im}g^{(2)}_{tt,m}-2g^{(3)}_{it,t}\biggr)\, & \,\, &
  {\Gamma^{(2)}}^{t}_{ti}= \frac{1}{2}g^{(2)}_{tt,i}
  \end{array}
\end{array}\right.
\end{eqnarray}
The Ricci tensor components (\ref{riccitensor}) are

\begin{eqnarray}\label{PPN-ricci-tensor}
\left\{\begin{array}{ll}R^{(2)}_{tt}=\frac{1}{2}g^{(2)}_{tt,mm}\\\\R^{(4)}_{tt}=\frac{1}{2}g^{(4)}_{tt,mm}+\frac{1}{2}g^{(2)}
_{mn,m}g^{(2)}_{tt,n}+\frac{1}{2}g^{(2)}_{mn}g^{(2)}_{tt,mn}+\frac{1}{2}g^{(2)}_{mm,tt}-
\frac{1}{4}g^{(2)}_{tt,m}g^{(2)}_{tt,m}+\\\\\,\,\,\,\,\,\,\,\,\,\,\,\,\,\,\,\,\,\,\,\,\,\,\,\,\,\,\,\,\,\,\,\,\,\,\,\,\,\,\,
\,\,\,\,\,\,\,\,\,\,\,\,\,\,\,\,\,\,\,\,\,\,\,\,\,\,\,\,\,\,\,\,\,\,\,\,\,\,\,\,\,\,\,\,\,\,\,\,\,\,\,\,\,\,\,\,\,\,\,\,\,\,
\,\,\,\,\,\,\,\,\,-\frac{1}{4}g^{(2)}_{mm,n}g^{(2)}_{tt,n}-g^{(3)}_{tm,tm}\\\\
R^{(3)}_{ti}=\frac{1}{2}g^{(3)}_{ti,mm}-\frac{1}{2}g^{(2)}_{im,mt}-\frac{1}{2}g^{(3)}_{mt,mi}+\frac{1}{2}g^{(2)}_{mm,ti}\\\\
R^{(2)}_{ij}=\frac{1}{2}g^{(2)}_{ij,mm}-\frac{1}{2}g^{(2)}_{im,mj}-\frac{1}{2}g^{(2)}_{jm,mi}-\frac{1}{2}g^{(2)}_{tt,ij}+
\frac{1}{2}g^{(2)}_{mm,ij}
\end{array}\right.
\end{eqnarray}
and the Ricci scalar (\ref{ricciscalar}) is

\begin{eqnarray}\label{PPN-ricci-scalar}
\left\{\begin{array}{ll}R^{(2)}=R^{(2)}_{tt}-R^{(2)}_{mm}=g^{(2)}_{tt,mm}-g^{(2)}_{nn,mm}+g^{(2)}_{mn,mn}\\\\
R^{(4)}=R^{(4)}_{tt}-g^{(2)}_{tt}R^{(2)}_{tt}-g^{(2)}_{mn}R^{(2)}_{mn}=\\\\\,\,\,\,\,\,\,\,\,\,
\,\,\,=\,\frac{1}{2}g^{(4)}_{tt,mm}+\frac{1}{2}g^{(2)}_{mn,m}g^{(2)}_{tt,n}+\frac{1}{2}g^{(2)}_{mn}g^{(2)}_{tt,mn}+\frac{1}{2}
g^{(2)}_{mm,tt}-\frac{1}{4}g^{(2)}_{tt,m}g^{(2)}_{tt,m}+\\\\\,\,\,\,\,\,\,\,\,\,\,\,\,\,\,\,\,\,\,-\frac{1}{4}g^{(2)}_{mm,n}
g^{(2)}_{tt,n}-g^{(3)}_{tm,tm}-\frac{1}{2}g^{(2)}_{tt}g^{(2)}_{tt,mm}-\frac{1}{2}g^{(2)}_{mn}\biggl(g^{(2)}_{mn,ll}-g^{(2)}_
{ml,ln}+\\\\\,\,\,\,\,\,\,\,\,\,\,\,\,\,\,\,\,\,\,\,\,\,\,\,\,\,\,\,\,\,\,\,\,\,\,\,\,\,\,\,\,\,\,\,\,\,\,\,\,\,\,\,\,\,\,\,\,
\,\,\,\,\,\,\,\,\,\,\,\,\,\,\,\,\,\,\,\,\,\,\,\,\,\,\,\,\,\,\,\,\,\,\,\,\,\,\,\,\,\,\,\,\,\,\,\,\,\,\,\,\,\,-g^{(2)}_{nl,lm}
-g^{(2)}_{tt,mn}+g^{(2)}_{ll,mn}\biggr)
\end{array}\right.
\end{eqnarray}
The Einstein tensor components (\ref{einstein-tensor}) are

\begin{eqnarray}\label{PPN-einstein-tensor}
\left\{\begin{array}{ll}G^{(2)}_{tt}=R^{(2)}_{tt}-\frac{1}{2}R^{(2)}=\frac{1}{2}g^{(2)}_{mm,nn}+\frac{1}{2}g^{(2)}_{mn,mn}\\\\
G^{(4)}_{tt}=R^{(4)}_{tt}-\frac{1}{2}R^{(4)}-\frac{1}{2}g^{(2)}_{tt}R^{(2)}=...\\\\
G^{(3)}_{ti}=R^{(3)}_{ti}=\frac{1}{2}g^{(3)}_{ti,mm}-\frac{1}{2}g^{(2)}_{im,mt}-\frac{1}{2}g^{(3)}_{mt,mi}+\frac{1}{2}g^{(2)}
_{mm,ti}\\\\G^{(2)}_{ij}=R^{(2)}_{ij}+\frac{\delta_{ij}}{2}R^{(2)}=\frac{1}{2}g^{(2)}_{ij,mm}-\frac{1}{2}g^{(2)}_{im,mj}-
\frac{1}{2}g^{(2)}_{jm,mi}-\frac{1}{2}g^{(2)}_{tt,ij}+\frac{1}{2}g^{(2)}_{mm,ij}+\\\\\,\,\,\,\,\,\,\,\,\,\,\,\,\,\,\,\,\,\,\,\,
\,\,\,\,\,\,\,\,\,\,\,\,\,\,\,\,\,\,\,\,\,\,\,\,\,\,\,\,\,\,\,\,\,\,\,\,\,\,\,\,\,\,\,\,\,\,\,\,\,\,\,\,\,\,\,\,\,\,\,\,\,\,\,\,
\,\,\,\,\,\,\,\,\,\,\,\,\,\,\,\,\,+\frac{\delta_{ij}}{2}\biggl[g^{(2)}_{tt,mm}-g^{(2)}_{nn,mm}+g^{(2)}_{mn,mn}\biggr]
\end{array}\right.
\end{eqnarray}

By assuming the harmonic gauge \footnote{The gauge transformation
is $\tilde{h}_{\mu\nu}=h_{\mu\nu}-\zeta_{\mu,\nu}-\zeta_{\nu,\mu}$
when we perform a coordinate transformation as
$x'^\mu=x^\mu+\zeta^\mu$ with O($\zeta^2$)$\ll 1$. To obtain our
gauge and the validity of field equation for both perturbation
$h_{\mu\nu}$ and $\tilde{h}_{\mu\nu}$ the $\zeta_\mu$ have satisfy
the harmonic condition $\Box\zeta^\mu=0$.}

\begin{eqnarray}\label{gauge-harmonic}
g^{\rho\sigma}\Gamma^\mu_{\rho\sigma}\,=\,0
\end{eqnarray}
it is possible to simplify the components of Ricci tensor
(\ref{PPN-ricci-tensor}). In fact for $\mu=0$ one has

\begin{eqnarray}\label{gau1}
2g^{\sigma\tau}\Gamma^{t}_{\sigma\tau}\approx g^{(2)}_{tt,t}
-2g^{(3)}_{tm,m}+g^{(2)}_{mm,t}=0\,,
\end{eqnarray} and for
$\mu=i$

\begin{eqnarray}\label{gau2}
2g^{\sigma\tau}\Gamma^{i}_{\sigma\tau}\approx
g^{(2)}_{tt,i}+2g^{(2)}_{mi,m}-g^{(2)}_{mm,i}=0\,.
\end{eqnarray}
Differentiating Eq.(\ref{gau1}) with respect to $t$, $x^j$ and
(\ref{gau2}) and with respect to $t$, one obtains

\begin{eqnarray}\label{gau3}
g^{(2)}_{tt,tt}-2g^{(3)}_{tm,mt}+g^{(2)}_{mm,tt}=0\,,
\end{eqnarray}

\begin{eqnarray}\label{gau4}
g^{(2)}_{tt,tj}-2g^{(3)}_{mt,jm}+g^{(2)}_{mm,tj}=0\,,
\end{eqnarray}

\begin{eqnarray}\label{gau5}
g^{(2)}_{tt,ti}+2g^{(2)}_{mi,tm}-g^{(2)}_{mm,ti}=0\,.
\end{eqnarray}
On the other side, combining Eq.(\ref{gau4}) and Eq.(\ref{gau5}),
we get

\begin{eqnarray}\label{gau6}
g^{(2)}_{mm,ti}-g^{(2)}_{mi,tm}-g^{(3)}_{mt,mi}=0\,.
\end{eqnarray}
Finally, differentiating Eq.(\ref{gau2}) with respect to $x^j$,
one has\,:

\begin{eqnarray}\label{gau7}
g^{(2)}_{tt,ij}+2g^{(2)}_{mi,jm}-g^{(2)}_{mm,ij}=0
\end{eqnarray}
and redefining indexes as $j\rightarrow i$, $i\rightarrow j$ since
these are mute indexes, we get

\begin{eqnarray}\label{gau8}
g^{(2)}_{tt,ij}+2g^{(2)}_{mj,im}-g^{(2)}_{mm,ij}=0\,.
\end{eqnarray}
Combining  Eq.(\ref{gau7}) and  Eq.(\ref{gau8}), we obtain

\begin{eqnarray}\label{gau9}
g^{(2)}_{tt,ij}+g^{(2)}_{mi,jm}+g^{(2)}_{mj,im}-g^{(2)}_{mm,ij}=0\,.
\end{eqnarray}
Relations (\ref{gau3}), (\ref{gau6}), (\ref{gau9}) guarantee us to
rewrite Eqs. (\ref{PPN-ricci-tensor}) as

\begin{eqnarray}\label{PPN-ricci-tensor-HG}
\left\{\begin{array}{ll}R^{(2)}_{tt}|_{HG}=\frac{1}{2}\triangle
g^{(2)}_{tt}\\\\R^{(4)}_{tt}|_{HG}=\frac{1}{2}\triangle
g^{(4)}_{tt}+\frac{1}{2}g^{(2)}_{mn}g^{(2)}_{tt,mn}-\frac{1}{2}g^{(2)}_{tt,tt}-\frac{1}{2}|\bigtriangledown
g^{(2)}_{tt}|^2\\\\R^{(3)}_{ti}|_{HG}=\frac{1}{2} \triangle
g^{(3)}_{ti}\\\\R^{(2)}_{ij}|_{HG}=\frac{1}{2}\triangle
g^{(2)}_{ij}\end{array}\right.
\end{eqnarray}
and Eqs. (\ref{PPN-ricci-scalar}) becomes

\begin{eqnarray}\label{PPN-ricci-scalar-HG}
\left\{\begin{array}{ll}R^{(2)}|_{HG}=\frac{1}{2}\triangle
g^{(2)}_{tt}-\frac{1}{2}\triangle
g^{(2)}_{mm}\\\\R^{(4)}|_{HG}=\frac{1}{2}\triangle
g^{(4)}_{tt}+\frac{1}{2}g^{(2)}_{mn}g^{(2)}_{tt,mn}-\frac{1}{2}g^{(2)}_{tt,tt}-\frac{1}{2}|\bigtriangledown
g^{(2)}_{tt}|^2-\frac{1}{2}g^{(2)}_{tt}\triangle g^{(2)}_{tt}-\frac{1}{2}g^{(2)}_{mn}\triangle g^{(2)}_{mn}\end{array}\right.
\end{eqnarray}
where $\nabla$ and $\triangle$ are, respectively, the gradient and the Laplacian in flat space. The Einstein tensor components (\ref{einstein-tensor}) in the
harmonic gauge are

\begin{eqnarray}\label{PPN-einstein-tensor-HG}
\left\{\begin{array}{ll}G^{(2)}_{tt}|_{HG}=\frac{1}{4}\triangle
g^{(2)}_{tt}+\frac{1}{4}\triangle g^{(2)}_{mm}\\\\G^{(4)}_{tt}|_{HG}=...\\\\
G^{(3)}_{ti}|_{HG}=\frac{1}{2}\triangle
g^{(3)}_{ti}\\\\G^{(2)}_{ij}|_{HG}= \frac{1}{2}\triangle
g^{(2)}_{ij}+\frac{\delta_{ij}}{4}\biggl[\triangle
g^{(2)}_{tt}-\triangle g^{(2)}_{mm}\biggr]
\end{array}\right.
\end{eqnarray}

On the matter side, i.e.\ right-hand side of the field equations
(\ref{fieldequationGR}), we start with the general definition of
the energy-momentum tensor of a perfect fluid

\begin{eqnarray}
T_{\alpha \beta }=\left( \rho+\Pi\rho+p\right)u_\alpha
u_\beta-pg_{\alpha\beta }\,.
\end{eqnarray}
Following the procedure outlined in \cite{puetzfeld1}, we
derive the explicit form of the energy-momentum as follows

\begin{eqnarray}\label{PPN-tensor-matter}
\left\{\begin{array}{ll}T_{tt}=\rho+\rho(v^2-2U+\Pi)+\rho\biggl[v^2\biggl(\frac{p}{\rho}+v^2+2V+\Pi\biggr)+\sigma-2\Pi
U\biggr]\\\\
T_{ti}=-\rho v^i+\rho\biggl[-v^i\biggl(\frac{p}{\rho}+2V+v^2+\Pi\biggl)+h_{ti}\biggr]\\\\
T_{ij}=\rho
v^iv^j+p\delta_{ij}+\rho\biggl[v^iv^j\biggl(\Pi+\frac{p}{\rho}+4V+v^2+2U\biggr)
-2v^c\delta_{c(i}h_{0|j)}+2\frac{p}{\rho}V\delta_{ij}\biggr]
\end{array}\right.
\end{eqnarray}
We are now ready to make use of Einstein field equations
(\ref{fieldequationGR}), which we assume in the form

\begin{eqnarray}\label{fieldequationGR-2}R_{\mu\nu}=\mathcal{X}\biggl[T_{\mu\nu}-\frac{T}{2}g_{\mu\nu}\biggr]
\,.\end{eqnarray} From their interpretation as the energy density,
momentum density and momentum flux, then, we have $T_{tt}$,
$T_{ti}$ and $T_{ij}$ at various order

\begin{eqnarray}\label{PPN-matter-density}
\left\{\begin{array}{ll}T_{tt}\,=\,T^{(0)}_{tt}+T^{(2)}_{tt}+\text{O}(4)\\\\
T_{ti}\,=\,T^{(1)}_{ti}+\text{O}(3)\\\\T_{ij}=T^{(2)}_{ij}+\text{O}(4)
\end{array}\right.
\end{eqnarray}
where $T^{(N)}_{\mu\nu}$ denotes the term in $T_{\mu\nu}$ of order
$\bar{M}/\bar{r}^3\,\,\bar{v}^{N}$. In particular $T^{(0)}_{tt}$
is the density of rest-mass, while $T^{(2)}_{tt}$ is the
non-relativistic part of the energy density. What we need is

\begin{eqnarray}S_{\mu\nu}\,=\,T_{\mu\nu}-\frac{T}{2}g_{\mu\nu}\,.\end{eqnarray}
But $G\bar{M}/\bar{r}$ is of order $\bar{v}^2$, so
(\ref{PPN-metric}) and (\ref{PPN-matter-density}) give

\begin{eqnarray}\label{PPN-matter-density-2}
\left\{\begin{array}{ll}S_{tt}\,=\,S^{(0)}_{tt}+S^{(2)}_{tt}+\text{O(6)}\\\\
S_{ti}\,=\,S^{(1)}_{ti}+\text{O(3)}\\\\S_{ij}=S^{(0)}_{ij}+\text{O(2)}
\end{array}\right.
\end{eqnarray}
where $S^{(N)}_{\mu\nu}$ denotes the term in $S_{\mu\nu}$ of order
$\bar{M}/\bar{r}^3\,\,\bar{v}^{N}$. In particular

\begin{eqnarray}\label{PPN-matter-density-3}
\left\{\begin{array}{ll}S^{(0)}_{tt}\,=\,\frac{1}{2}T^{(0)}_{tt}\\\\
S^{(2)}_{tt}\,=\,\frac{1}{2}T^{(2)}_{tt}+\frac{1}{2}T^{(2)}_{mm}\\\\
S^{(1)}_{ti}=T^{(1)}_{ti}\\\\
S^{(0)}_{ij}=\frac{1}{2}\delta_{ij}T^{(0)}_{tt}
\end{array}\right.
\end{eqnarray}
Using the (\ref{PPN-ricci-tensor-HG}) and
(\ref{PPN-matter-density-2}) in the field equation
(\ref{fieldequationGR-2}) we find that the field equations in
harmonic coordinates are indeed consistent with the expansions we
are using, and give

\begin{eqnarray}
\left\{\begin{array}{ll}R^{(2)}_{tt}\,=\,\mathcal{X}S^{(0)}_{tt}\\\\
R^{(4)}_{tt}\,=\,\mathcal{X}S^{(2)}_{tt}\\\\
R^{(3)}_{ti}\,=\,\mathcal{X}S^{(0)}_{ti}\\\\
R^{(2)}_{ij}\,=\,\mathcal{X}S^{(0)}_{ij}
\end{array}\right.
\end{eqnarray}
and in particular

\begin{eqnarray}\label{PPN-fieldequationGR}
\left\{\begin{array}{ll}\triangle
g^{(2)}_{tt}\,=\,\mathcal{X}\,T^{(0)}_{tt}\\\\
\triangle
g^{(4)}_{tt}\,=\,\mathcal{X}\,\biggl[T^{(2)}_{tt}+T^{(2)}_{mm}\biggr]-g^{(2)}_{mn}g^{(2)}_{tt,mn}+g^{(2)}_{tt,tt}+
|\bigtriangledown g^{(2)}_{tt}|^2\\\\
\triangle g^{(3)}_{ti}\,=\,2\,\mathcal{X}\,T^{(1)}_{ti}\\\\
\triangle
g^{(2)}_{ij}\,=\,\mathcal{X}\,\delta_{ij}\,T^{(0)}_{tt}\end{array}\right.
\end{eqnarray}
From the first one of (\ref{PPN-fieldequationGR}), we find, as
expected, the Newtonian mechanics:

\begin{eqnarray}\label{gravitational-potential-metric}
g^{(2)}_{tt}\,=\,-\frac{\mathcal{X}}{4\pi}\int d^3\mathbf{x}'\frac{T^{(0)}_{tt}(\mathbf{x}')}{|\mathbf{x}
-\mathbf{x}'|}\,=\,-2G\int
d^3\mathbf{x}'\frac{T^{(0)}_{tt}(\mathbf{x}')}{|\mathbf{x}
-\mathbf{x}'|}\,\doteq\,2\Phi(\mathbf{x})\end{eqnarray} where
$\Phi(\mathbf{x})$ is the gravitational potential which, in the
case of point-like source with mass $M$, is

\begin{eqnarray}\Phi(\mathbf{x})\,=\,-\frac{GM}{|\mathbf{x}|}\,.\end{eqnarray}
From the third and fourth equations of (\ref{PPN-fieldequationGR})
we find that

\begin{eqnarray}
\left\{\begin{array}{ll}
g^{(3)}_{ti}\,=\,-\frac{\mathcal{X}}{2\pi}\,\int
d^3\mathbf{x}'\frac{T^{(1)}_{ti}(\mathbf{x}')}{|\mathbf{x}
-\mathbf{x}'|}\,\doteq\,Z_i(\mathbf{x})\\\\
g^{(2)}_{ij}\,=\,-\frac{\mathcal{X}}{4\pi}\,\delta_{ij}\,\int
d^3\mathbf{x}'\frac{T^{(0)}_{tt}(\mathbf{x}')}{|\mathbf{x}
-\mathbf{x}'|}\,=\,2\delta_{ij}\Phi(\mathbf{x})\end{array}\right.
\end{eqnarray}
The second equation of (\ref{PPN-fieldequationGR}) can be
rewritten as follows

\begin{eqnarray}\triangle\biggl[g^{(4)}_{tt}-2\Phi^2\biggr]\,=\,\mathcal{X}\,\biggl[T^{(2)}_{tt}+T^{(2)}_{mm}\biggr]
-8\Phi\triangle\Phi+2\Phi_{,tt}\end{eqnarray} and the solution for
$g^{(4)}_{tt}$ is

\begin{eqnarray}g^{(4)}_{tt}\,=&&2\Phi^2-\frac{\mathcal{X}}{4\pi}\int
d^3\textbf{x}'\frac{T^{(2)}_{tt}(\mathbf{x}')+T^{(2)}_{mm}(\mathbf{x}')}{|\textbf{x}-\textbf{x}'|}+\frac{2}{\pi}\int
d^3\textbf{x}'\frac{\Phi(\mathbf{x}')\triangle_{\mathbf{x}'}\Phi(\mathbf{x}')}{|\textbf{x}-\textbf{x}'|}\nonumber\\&&-
\frac{1}{2\pi}\int
d^3\textbf{x}'\frac{\Phi_{,tt}(\mathbf{x}')}{|\textbf{x}-\textbf{x}'|}\doteq
2\,\Theta(\mathbf{x})\,.
\end{eqnarray}
By using the equations at second order we obtain the final
expression for the correction at fourth order in the time-time
component of the metric:

\begin{eqnarray}\Theta(\textbf{x})\,=&&\Phi(\textbf{x})^2-\frac{\mathcal{X}}{8\pi}\int
d^3\textbf{x}'\frac{T^{(2)}_{tt}(\mathbf{x}')+T^{(2)}_{mm}(\mathbf{x}')}{|\textbf{x}-\textbf{x}'|}+\frac{\mathcal{X}}{\pi}
\int
d^3\textbf{x}'\frac{\Phi(\mathbf{x}')\,\,T^{(0)}_{tt}(\textbf{x}')}{|\textbf{x}-\textbf{x}'|}\nonumber\\&&-
\frac{1}{4\pi}\partial^2_{tt}\int d^3\textbf{x}'\frac{\Phi
(\mathbf{x}')}{|\textbf{x}-\textbf{x}'|},.
\end{eqnarray}
We can rewrite the metric expression (\ref{PPN-metric}) as follows

\begin{eqnarray}\label{PPN-metric-potential-GR}
  g_{\mu\nu}\sim \begin{pmatrix}
  1+2\Phi+2\,\Theta & \vec{Z}^T \\
  \vec{Z} & -\delta_{ij}(1-2\Phi)
\end{pmatrix}
\end{eqnarray}

Finally the Lagrangian of a particle in presence of a
gravitational field can be expressed as proportional to the
invariant distance $ds^{1/2}$, thus we have\,:

\begin{eqnarray}
L=\biggl(g_{\rho\sigma}\frac{dx^\rho}{dt}\frac{dx^\sigma}{dt}\biggr)^{1/2}&=&\biggl(g_{tt}+2g_{tm}v^m+g_{mn}v^mv^n\biggr)^{1/2}
=\nonumber\\&=&\biggl(1+g^{(2)}_{tt}+g^{(4)}_{tt}+2g^{(3)}_{tm}v^m-\textbf{v}^2+g^{(2)}_{mn}v^mv^n\biggr)^{1/2}\,,
\end{eqnarray}
which, to the O(2) order, reduces to the classic Newtonian
Lagrangian of a test particle
$L_{\text{New}}=\biggl(1+2\Phi-\textbf{v}^2\biggr)^{1/2}$, where
$v^m=\frac{dx^m}{dt}$ and $|\mathbf{v}|^2=v^mv^m$. As matter of
fact, post-Newtonian physics has to involve higher than O(2) order
terms in the Lagrangian. In fact we obtain

\begin{eqnarray}L\,&\sim&\,1+\biggl[\Phi-\frac{1}{2}\mathbf{v}^2\biggr]+\frac{3}{4}\biggl[\Theta+Z_m
v^m+\Phi\,\textbf{v}^2\biggr]\,.\end{eqnarray}

An important remark concerns the odd-order perturbation terms O(1)
or O(3). Since, these terms contain  odd powers of velocity
$\textbf{v}$ or of time derivatives, they are related to the
energy dissipation or absorption by the system. Nevertheless, the
mass-energy conservation prevents the energy and mass losses and,
as a consequence, prevents, in the Newtonian limit, terms of O(1)
and O(3) orders in the  Lagrangian. If one takes into account
contributions  higher than O(4) order, different theories give
different predictions. GR, for example, due to the conservation of
post-Newtonian energy, forbids terms of O(5) order; on the other
hand, terms of O(7) order can appear and are related to the energy
lost by means of the gravitational radiation.

\section{General remarks on the post $-$ Minkowskian approximation of Einstein
equation}\label{PM-limit-GR}

We suppose the metric to be close to the Minkowski metric
$\eta_{\mu\nu}$:

\begin{eqnarray}\label{PM-metric}
g_{\mu\nu}\,=\,\eta_{\mu\nu}+h_{\mu\nu}
\end{eqnarray}
with $h_{\mu\nu}$ small quantities ($O(h)^2\ll 1$). To first order
in $h$, the Christoffel symbols (\ref{christoffel}) are

\begin{eqnarray}\label{PM-christoffel}
\Gamma^\alpha_{\mu\nu}=\frac{1}{2}\eta^{\alpha\sigma}(h_{\mu\sigma,\nu}+h_{\nu\sigma,\mu}-h_{\mu\nu,\sigma})\,.
\end{eqnarray}
As long as we restrict ourselves to first order in $h$, we must
raise and lower all indices using $\eta_{\mu\nu}$, not
$g_{\mu\nu}$; that is

\begin{eqnarray}\eta^{\sigma\tau}h_{\sigma\tau}=h^\sigma_{\,\,\,\,\sigma}=h,\,\,\,\,\,\,\,\,\,\,\,\,\,\,\,\,\eta^{\sigma\tau}
\frac{\partial}{\partial x^\sigma}=\frac{\partial}{\partial
x_\tau},\,\,\,\,\,\,\,\,\,\,\,\,\,\,\,\,etc.\end{eqnarray} With
this assumptions, the Ricci tensor and scalar (\ref{riccitensor})
- (\ref{ricciscalar}) are then

\begin{eqnarray}\label{PM-ricci-tensor}
\left\{\begin{array}{ll}R^{(1)}_{\mu\nu}\,=\,h^\sigma_{(\mu,\nu)\sigma}-\frac{1}{2}\Box_\eta
h_{\mu\nu}-\frac{1}{2}h_{,\mu\nu}\\\\
R^{(1)}\,=\,{h_{\sigma\tau}}^{,\sigma\tau}-\Box_\eta h
\end{array}\right.
\end{eqnarray}
where
$\nabla^\alpha\nabla_\alpha\,\sim\,{{}_{,\sigma}}^{,\sigma}\,=\Box_\eta$
is the d'Alembertian operator in the flat space. The field
equation (\ref{fieldequationGR}) becomes

\begin{eqnarray}\label{PM-fieldequationGR}
G^{(1)}_{\mu\nu}=R^{(1)}_{\mu\nu}-\frac{1}{2}R^{(1)}\eta_{\mu\nu}=\mathcal{X}\,T^{(0)}_{\mu\nu}
\end{eqnarray}
where $T_{\mu\nu}$ is fixed at zero-order in
(\ref{PM-fieldequationGR}) since in this perturbation scheme the
first order on Minkowski space has to be connected with the zero
order of the standard matter energy momentum tensor\footnote{In
this perturbation scheme the first order on Minkowski space has to
be connected with the zero order of the standard matter energy
momentum tensor. This formalism descends from the theoretical
setting of Newtonian mechanics which requires the appropriate
scheme of approximation and coincides with a gravity theory
analyzed at the first order of perturbations in the curved
spacetime metric.}. Eqs. (\ref{PM-fieldequationGR}) in terms of
$h_{\mu\nu}$ are

\begin{eqnarray}\label{PM-fieldequationGR-2}
h^\sigma_{(\mu,\nu)\sigma}-\frac{1}{2}\Box_\eta
h_{\mu\nu}-\frac{1}{2}h_{,\mu\nu}-\frac{1}{2}[{h_{\sigma\tau}}^{,\sigma\tau}-\Box_\eta
h]\eta_{\mu\nu}=\mathcal{X}\,T^{(0)}_{\mu\nu}\,.
\end{eqnarray}
Since $T_{\mu\nu}$ is taken to the lowest order in $h_{\mu\nu}$,
so it is independent of $h_{\mu\nu}$, it has to satisfies the
ordinary conservation conditions:

\begin{eqnarray}\label{PM-conserv-law-energy}T^{\sigma\mu}_{\,\,\,\,\,\,\,\,,\sigma}\,=\,0\,.\end{eqnarray}
Note that it is this form of the conservation law that is needed
for the consistency of (\ref{PM-fieldequationGR-2}), because
(\ref{PM-conserv-law-energy}) implies

\begin{eqnarray}
{G^{(1)}}^{\mu\sigma}_{\,\,\,\,\,\,\,\,,\sigma}=0
\end{eqnarray}
whereas the linearized Ricci tensor satisfies Bianchi identities
(\ref{bianchi-identity}) of the form

\begin{eqnarray}
{R^{(1)}}^{\sigma\mu}_{\,\,\,\,\,\,\,\,\,,\sigma}=\frac{1}{2}\biggl[h^{\alpha\beta}_{\,\,\,\,\,\,\,\,,\alpha\beta}-\Box_\eta
h\biggr]^{,\mu}=\frac{1}{2}{R^{(1)}}^{,\mu}\,.
\end{eqnarray}
By choosing the transformation
$\tilde{h}_{\mu\nu}=h_{\mu\nu}-\frac{h}{2}\eta_{\mu\nu}$ and the
gauge condition $\tilde{h}^{\mu\nu}_{\,\,\,\,\,\,\,,\mu}=0$
(harmonic gauge (\ref{gauge-harmonic})) one obtains that field
equations read

\begin{eqnarray}\label{PM-fieldequationGR-3}
\Box\tilde{h}_{\mu\nu}=-2\mathcal{X}\,T^{(0)}_{\mu\nu}\,.
\end{eqnarray}
One solution is the \emph{retarded potential}

\begin{eqnarray}
\tilde{h}_{\mu\nu}(t,\textbf{x})=4G\int
d^3\textbf{x}'\,\frac{T^{(0)}_{\mu\nu}(\textbf{x}',t-|\textbf{x}-\textbf{x}'|)}{|\textbf{x}-\textbf{x}'|}
\end{eqnarray}
or in terms of perturbation $h_{\mu\nu}$

\begin{eqnarray}
h_{\mu\nu}(t,\textbf{x})=4G\int
d^3\textbf{x}'\,\frac{S^{(0)}_{\mu\nu}(\textbf{x}',t-|\textbf{x}-\textbf{x}'|)}{|\textbf{x}-\textbf{x}'|}\,.
\end{eqnarray}
The propagation of $h_{\mu\nu}$ is possible with a particle
massless.

\clearpage{\pagestyle{empty}\cleardoublepage}

\chapter{Spherical symmetry in  $f$ $-$ gravity}\label{sperical-symmetry-f-gravity}

Spherical symmetry in $f$-gravity is discussed in details
considering also the relations with the weak field limit
[\textbf{D}]. Exact solutions are obtained for constant Ricci
curvature scalar and for Ricci scalar depending on the radial
coordinate. In particular, we discuss how to obtain results which
can be consistently compared with General Relativity giving the
well known post-Newtonian and post-Minkowskian limits.
Furthermore, we implement a perturbation approach to obtain
solutions up to the first order starting from spherically
symmetric backgrounds. Exact solutions are given for several
classes of $f$-theories in both $R\,=$ constant and $R\,=\,R(r)$.

\section{The Ricci curvature scalar in spherical symmetry}

Starting by the definition of Ricci scalar (\ref{ricciscalar}) and
imposing the spherical symmetry (\ref{me5}), the Ricci scalar in
terms of the gravitational potentials ($g_{tt}$ and $g_{rr}$)
reads\,:

\begin{eqnarray}\label{riccispher}
R(t,\,r)\,&=&\,\frac{1}{2r^2g_{tt}^2g_{rr}^2}\biggl\{g_{rr}\biggl[\dot{g}_{tt}\dot{g}_{rr}-g_{tt}'^2\biggr]r^2+g_{tt}\biggl[
r(\dot{g}_{tt}^2-g_{tt}'g_{rr}')+2g_{rr}(2g_{tt}'+rg_{tt}''-r\ddot{g}_{rr})\biggr]\nonumber\\&&-4g_{tt}^2\biggl[g_{rr}^2-
g_{rr}+rg_{rr}'\biggr]\biggr\}
\end{eqnarray}
where, for the sake of brevity, we have discarded the explicit
dependence in $g_{tt}(t,\,r)$ and $g_{rr}(t,\,r)$ and the prime
indicates the derivative with respect to $r$ while the dot with
respect to $t$. If the metric (\ref{me5}) is time-independent,
i.e. $g_{tt}(t,r)\,=\,a(r)$, $g_{rr}(t,r)\,=\,b(r)$, the
(\ref{riccispher}) assumes the simpler form\,:

\begin{eqnarray}\label{ricscalin}R(r)&=&\frac{1}{2r^2a(r)^2b(r)^2}\biggl\{a(r)\biggl[2b(r)\biggl(2a'(r)+ra''(r)\biggr)
-ra'(r)b'(r)\biggr]-b(r)a'(r)^2r^2\nonumber\\&&-4a(r)^2\biggl(b(r)^2-b(r)+rb'(r)\biggr)\biggr\}\end{eqnarray}
where the radial dependence of the gravitational potentials is now
explicitly shown. This expression can be seen as a constraint for
the functions $a(r)$ and $b(r)$ once a specific form of Ricci
scalar is given. In particular, it reduces to a Bernoulli equation
of index two with respect to the metric potential $b(r)$\,:

\begin{eqnarray}\label{eqric}
&&b'(r)+\biggl\{\frac{r^2a'(r)^2-4a(r)^2-2ra(r)[2a(r)'+ra(r)'']}{ra(r)[4a(r)+ra'(r)]}\biggr\}b(r)\nonumber\\\nonumber\\&&
+\biggl\{\frac{2a(r)}{r}\biggl[\frac{2+r^2R(r)}{4a(r)+ra'(r)}\biggr]\biggr\}b(r)^2\,\doteq\,b'(r)+h(r)b(r)+l(r)b(r)^2=\,0\,.
\end{eqnarray}
A general solution of (\ref{eqric}) is:

\begin{eqnarray}\label{gensol}
b(r)\,=\,\frac{\exp[-\int dr\,h(r)]}{K+\int dr\,l(r)\,\exp[-\int
dr\,h(r)]}\,,
\end{eqnarray}
where $K$ is an integration constant while $h(r)$ and $l(r)$ are
the two functions which, respectively, define the coefficients of
the quadratic and the linear term with respect to $b(r)$, as in
the standard definition of the Bernoulli equation \cite{ince}.
Looking at the equation, we can notice that it is possible to have
$l(r)\,=\,0$ which implies to find out solutions with a Ricci
scalar scaling as ${\displaystyle -\frac{2}{r^2}}$ in term of the
radial coordinate. On the other side, it is not possible to have
$h(r)\,=\,0$ since otherwise we will get imaginary solutions. A
particular consideration deserves the limit $r\rightarrow\infty$.
In order to achieve a gravitational potential $b(r)$ with the
correct Minkowski limit, both $h(r)$ and $l(r)$ have to go to zero
provided that the quantity $r^2R(r)$ turns out to be constant:
this result implies $b'(r)=0$, and, finally, also the metric
potential $b(r)$ has  a correct Minkowski limit. In general, if we
ask for the asymptotic flatness of the metric as a feature of the
theory, the Ricci scalar has to evolve at infinity as $r^{-n}$
with $n\geqslant 2$. Formally it has to be:

\begin{eqnarray}\label{condricc}
\lim_{r\rightarrow\infty}r^2R(r)\,=\,r^{-n}
\end{eqnarray}
with $n\in\mathbb{N}$. Any other behavior of the Ricci scalar
could affect the requirement of the correct asymptotic flatness.
This result can be easily deduced from (\ref{eqric}). In fact, let
us consider  the simplest spherically symmetric case:

\begin{eqnarray}
ds^2\,=\,a(r)dt^2-\frac{dr^2}{a(r)}-r^2d\Omega\,.
\end{eqnarray}
The Bernoulli equation (\ref{eqric}) is easy to integrate and the
most general metric potential $a(r)$, compatible with the Ricci
scalar constraint (\ref{ricscalin}), is\,:

\begin{eqnarray}
a(r)\,=\,1+\frac{k_1}{r}+\frac{k_2}{r^2}+\frac{1}{r^2}\int
\biggl[\int r^2 R(r)dr\biggr]dr
\end{eqnarray}
where $k_1$ and $k_2$ are integration constants. Actually one gets
the standard result $a(r)=1$ (Minkowski) for $r\rightarrow \infty$
only if the condition (\ref{condricc}) is satisfied, otherwise we
get a diverging gravitational potential.

\section{Solutions with constant curvature scalar}

Let us assume a scalar curvature constant ($R\,=\,R_0$). The field
equations (\ref{fe}) and (\ref{fetr}) reduce to:

\begin{eqnarray}\label{fe-curvature-constant}
\left\{\begin{array}{ll}
f'(R_0)R_{\mu\nu}-\frac{1}{2}f(R_0)g_{\mu\nu}\,=\,\mathcal{X}\,T_{\mu\nu}\\\\
f'(R_0)R_0-2f(R_0)\,=\,\mathcal{X}\,T\end{array}\right.
\end{eqnarray}
Such equations can be arranged as:

\begin{eqnarray}\label{fe-curvature-constant-2}
\left\{\begin{array}{ll}
R_{\mu\nu}+\lambda g_{\mu\nu}\,=\,q\,\mathcal{X}\,T_{\mu\nu}\\\\
R_0\,=\,q\,\mathcal{X}\,T-4\lambda\end{array}\right.
\end{eqnarray}
where $\lambda\,=\,-\frac{f(R_0)}{2f'(R_0)}$ and
$q^{-1}\,=\,f'(R_0)$. Since we are analyzing the weak field limit
of HOG, it is reasonable  to consider Lagrangians which work as
the Hilbert\,-\,Einstein one when $R\rightarrow{0}$ (this even
means that we can suitably put the cosmological constant to zero),
that is:

\begin{eqnarray}
\lim_{R\rightarrow 0} f\sim R\,.
\end{eqnarray}
In such a case, the trace equation of
(\ref{fe-curvature-constant-2}) indicates that in the vacuum case
($T_{\mu\nu}=0$), one  obtains a solutions with constant curvature
$R\,=\,R_0$.

Let us now suppose that the above theory, for small curvature
values, evolves to a constant as $\lim_{R\rightarrow 0}
f\,=\,f_0$. Even in this case, considering only the trace equation
of (\ref{fe-curvature-constant}), some interesting features
emerge. In fact if we consider the expression for $f$\,:

\begin{eqnarray}\label{f}
f\,=\,f_0+f_1R+\mathcal{F}(R)\,.
\end{eqnarray}
where $f_0$ and $f_1$ are a coupling constants, while
$\mathcal{F}(R)$ is a generic analytic function of $R$ satisfying
the condition

\begin{eqnarray}\label{f1}
\lim_{R\rightarrow 0}R^{-2}\mathcal{F}(R)\,=\,\text{constant}\,,
\end{eqnarray}
it is evident that no zero\,-\,curvature solutions are obtainable
since\,:

\begin{eqnarray}
\mathcal{F}'(R_0)R_0-2\mathcal{F}(R_0)-f_1
R_0-2f_0\,=\,\mathcal{X}\,T\,.
\end{eqnarray}
Furthermore, in this case, even in absence of matter, there are no
Ricci flat  solution of the field equations since the higher order
derivative terms  give  curvature constant solutions corresponding
to a sort of effective cosmological constant.  Of course, this is
not the case for GR. In fact in the standard Einstein theory,
constant curvature solutions different from zero are in order only
in presence of matter because of the proportionality of the Ricci
scalar to  the trace of energy\,-\,momentum tensor  of matter.
Besides, one can get a similar situation in presence of a
cosmological constant put by hands into the dynamics.

In other words, the difference between GR and HOG is that the
Schwarzschild\,-\,de Sitter solution is not necessarily given by a
$\Lambda$\,-\,term while the effect of an ``effective"
cosmological constant can be achieved by the higher order
derivative contributions. This result has been extensively
investigated in several recent papers as, for example,
\cite{mul-vil2}. For a discussion, see also \cite{bar-ott}.

Let us consider now the search for a general solution of
(\ref{fe}) and (\ref{fetr}) considering a spherically symmetric
metric as (\ref{me5}). By substituting the metric into the field
equations. One obtains that the $\{t,r\}$ - component of
(\ref{fe}) gives $\frac{\dot{g_{rr}}(t,r)}{rg_{rr}(t,r)}\,=\,0$.
This means that $g_{rr}(t,\,r)$ has to be time independent and we
can write $g_{rr}(t,\,r)\,=\,b(r)$. On the other hand, by the
$\{\theta,\theta\}$ - component of (\ref{fe}), one gets the
relation $\frac{g_{tt}'(t,r)}{g_{tt}(t,r)}\,=\,\zeta(r)$ with
$\zeta(r)$ a given time independent function\,:

\begin{eqnarray}
g_{tt}(t,r)\,=\,\tilde{a}(t)\exp\biggl[\int\zeta(r)dr\biggr]=\tilde{a}(t)\frac{b(r)}{r^2}\exp\biggl[2\int
dr\frac{[1- r^2(\lambda+q\mathcal{X}p)]b(r)}{r}\biggr]\,,
\end{eqnarray}
where $\lambda$ and $q$ are defined as above and $p$ is  the
pressure of a  perfect fluid being the stress-energy tensor of
matter

\begin{eqnarray}\label{enmomten}
T_{\mu\nu}\,=\,(\rho+p)u_\mu u_\nu-pg_{\mu\nu}\,;
\end{eqnarray}
$\rho$ is the energy density and $u^\mu=dx^{\mu}/ds$ is the
4-velocity. Therefore, the function $g_{tt}(t,r)$ has to be a
separable function of time and radial coordinates, i.e.
$g_{tt}(t,r)\,=\,\tilde{a}(t)a(r)$. As a matter of facts, the
metric (\ref{me5}) becomes

\begin{eqnarray}\label{me2y}
ds^2\,=\,\tilde{a}(t)a(r)dt^2-b(r)dr^2-r^2d\Omega
\end{eqnarray}
which can be recast as

\begin{eqnarray}\label{me4x}
ds^2\,=\,a(r)d\tilde{t}^2-b(r)dr^2-r^2d\Omega\,,
\end{eqnarray}
by a suitable time\,-\,redefinition
$d\tilde{t}\,=\,\sqrt{\tilde{a}(t)}dt$. Nevertheless we redefine
the time $\tilde{t}$ as $t$.

This exact result states that whenever one has a constant scalar
curvature spacetime, any spherically symmetric background has to
be necessarily static. In other words, this means that the
Birkhoff theorem holds for the $f$ - theories with constant
curvature as it has to be (see \cite{haw-ell}). It has to be
noticed that such a result is in striking contrast with what has
been argued elsewhere about the fact that Birkhoff theorem does
not hold, in general, for HOG theories.

A remark is in order at this point. We have obtained  this result
considering a constant Ricci scalar spacetime and deducing some
conditions on the form of  gravitational potentials. Nevertheless
one can even reverse the problem. It is possible to argue that
\emph{whenever the gravitational potential $g_{tt}(t,\,r)$ is
described by a separable functions and $g_{rr}(t,\,r)$ is
time\,-\,independent, by the definition of the Ricci scalar, one
gets that $R\,=\,R_0$ and at the same time the final solutions of
the field equations will be static if the spherical symmetry is
invoked}.

Actually, for a complete analysis of the problem, one should take
into account  the remaining field equations descending from
(\ref{fe}) and (\ref{fetr}) which have to be solved by taking even
into account the definition of the Ricci scalar (\ref{ricscalin}).
In summary, we have to solve the system\,:

\begin{eqnarray}
\left\{\begin{array}{ll} R_{tt}+\lambda\,
a(r)-q\mathcal{X}[\rho+p(1-a(r))]\,=\,0\,
\\\\
R_{rr}-\lambda b(r)-q\mathcal{X}p\,b(r)\,=\,0\,\\\\
R_0-q\mathcal{X}(\rho-3p)+4\lambda\,=\,0\\\\
R(a(r),b(r))\,=\,R_0
\end{array} \right.
\end{eqnarray}
A general solution of the above set of equations is achieved for
$p\,=\,-\rho$ and reads\,:

\begin{eqnarray}\label{schwarz-desitt-solution-constant-curvature}
ds^2\,=\,\biggl(1+\frac{k_1}{r}+\frac{q\mathcal{X}\rho-\lambda}{3}r^2\biggr)dt^2-\frac{dr^2}{1+\frac{k_1}{r}+
\frac{q\mathcal{X}\rho-\lambda}{3}r^2}-r^2d\Omega\,.
\end{eqnarray}
In other words, any $f$\,-\,theory in the case of constant
curvature scalar ($R\,=\,R_0$) exhibits solutions with
cosmological constant as the solution Schwarzschild - de Sitter
(\ref{schwarz-desitt-solution}) if we consider the relation
$\Lambda\,=\,q\mathcal{X}\rho-\lambda\,=\,\frac{2\mathcal{X}\rho+f(R_0)}{2f'(R_0)}$.
This is one of the reason why the dark energy issue can be
addressed using these theory \cite{all-bor-fra, cap, cap-car-tro,
cap-car-car-tro, car-duv-tro-tur, noj-odi1, noj-odi3,
cap-card-tro1, car-dun-cap-tro}. This fact is well known using the
FRW metric \cite{bar-ott}.

If we neglect the cosmological constant $f_0$ and $f_1$ is set to
zero in the (\ref{f}), we obtain a new class of theories which, in
the limit $R\rightarrow{0}$, does not reproduce GR (from
(\ref{f1}), we have $\lim_{R\rightarrow 0}f\sim R^2$). In such a
case analyzing the whole set of (\ref{fe}) and (\ref{fetr}), one
can observe that both zero and constant $\neq 0$ curvature
solutions are possible. In particular,  if $R\,=\,R_0\,=\,0$ the
field equations are solved for every form of gravitational
potentials entering the spherically symmetric background
(\ref{me4x}), provided that the Bernoulli equation (\ref{eqric}),
relating these functions, is fulfilled for the particular case
$R(r)=0$. The solutions are thus defined by the relation

\begin{eqnarray}\label{gensol0}
b(r)\,=\,\frac{\exp[-\int
dr\,h(r)]}{K+4\int\frac{dr\,a(r)\,\exp[-\int
dr\,h(r)]}{r[a(r)+ra'(r)]}}\,.
\end{eqnarray}
In table \ref{solutions-constant-curvature}, we give some examples
of $f$ - theories admitting solutions with constant\,$\neq 0$ or
null scalar curvature. Each model admits Schwarzschild,
Schwarzschild\,-\,de Sitter, and the class of solutions given by
(\ref{gensol0}).

\begin{table}\label{solutions-constant-curvature}
\begin{center}
\begin{tabular}{ccccc}
  \hline\hline\hline
  & & & & \\
  & $f$\,-\,theory: & & Field equations: & \\
  & & & & \\
  & $R$ & $\longrightarrow$ & $R_{\mu\nu}=0$, $\text{with}$ $R=0$ & \\
  & & & & \\
  & $\xi_1 R+\xi_2 R^n$ & $\longrightarrow$ & $\begin{cases}
                R_{\mu\nu}=0 & \text{with}\,\,R=0,\,\,\,\xi_1\neq 0\\
                R_{\mu\nu}+\lambda g_{\mu\nu}=0 &
                \text{with}\,\,R=\biggl[\frac{\xi_1}{(n-2)\xi_2}\biggr]^{\frac{1}{n-1}}
                ,\,\,\,\xi_1\neq 0,\,\,\,n\neq2\\
                0=0 & \text{with}\,\,R=0,\,\,\,\xi_1=0\\
                R_{\mu\nu}+\lambda g_{\mu\nu}=0 & \text{with}\ R=R_0,\ \xi_1=0,\ n=2
                \end{cases}$& \\
  & & & & \\
  & $\xi_1R+\xi_2R^{-m}$ & $\longrightarrow$ & $R_{\mu\nu}+\lambda g_{\mu\nu}=0$ with $R=\biggl[-\frac{(m+2)\xi_2}{\xi_1}
  \biggr]^{\frac{1}{m+1}}$ & \\
  & & & & \\
  & $\xi_1 R+\xi_2 R^n+\xi_3 R^{-m}$ & $\longrightarrow$ & $R_{\mu\nu}+\lambda g_{\mu\nu}=0$, \text{with}\ $R=R_0$ so that
  & \\
  & & & $\xi_1R_0^{m+1}+(2-n)\xi_2R_0^{n+m}+(m+2)\xi_3=0$ & \\
  & & & & \\
  & $\frac{R}{\xi_1+R}$ & $\longrightarrow$ & $\begin{cases}
                                         R_{\mu\nu}=0 & \text{with}\ R=0\\
                                         R_{\mu\nu}+\lambda g_{\mu\nu}=0 & \text{with}\ R=-\frac{\xi_1}{2}
                                         \end{cases}$ & \\
  & & & & \\
  & $\frac{1}{\xi_1+R}$ & $\longrightarrow$ & $R_{\mu\nu}+\lambda g_{\mu\nu}=0$, \text{with}\ $R=-\frac{2\xi_1}{3}$ & \\
  & & & & \\
  \hline\hline\hline
\end{tabular}
\end{center}
\caption{Examples of $f$ - models admitting constant and zero
scalar curvature solutions. In the right hand side, the field
equations are given for each model. The power $n$, $m$ are natural
numbers while $\xi_i$ are generic real constants.}
\end{table}

\section{Solutions with curvature scalar function of  $r$}

Up to now we have discussed the behavior of $f$-gravity seeking
for spherically symmetric solutions (\ref{me5}) with  constant
scalar curvature. This situation is well known in GR and give rise
to the Schwarzschild solution $(R\,=\,0)$ and the
Schwarzschild\,-\,de Sitter solution $(R\,=\,R_0\,\neq\,0)$. The
problem can be generalized in $f$ - gravity investigating
considering the Ricci scalar as an arbitrary function of the
radial coordinate $r$.

This approach is interesting since, in general, HOG theories are
supposed to admit such kind of solutions and several examples have
been found in  literature \cite{cap-card-tro2, stel, cap-card-tro,
mul-vil1, mul-vil2}. Here we want to face the problem from general
point of view.

If we choose the Ricci scalar $R$ as a generic function of the
radial coordinate ($R\,=\,R(r)$), it is possible to show that also
in this case the  solution of the field equations (\ref{fe}) and
(\ref{fetr}) is time independent (if $T_{\mu\nu}=0$). In other
words, the Birkhoff theorem has to hold. The crucial point of the
approach is to study the off\,-\,diagonal $\{t,r\}$\,-\,component
of (\ref{fe}) as well as in the case of GR. This equation, for a
generic $f$ reads\,:

\begin{eqnarray}\label{field-equation-off-diagonal}
\frac{d}{dr}\biggl(r^2f'\biggr)\dot{g}_{rr}(t,r)\,=\,0\,,
\end{eqnarray}
and two possibilities are in order. Firstly, we can choose
$\dot{g}_{tt}(t,r)\neq 0$. This choice implies that
$f'\,\sim\,r^{-2}$. If this is the case, the remaining field
equation turn out to be not fulfilled and it can be easily
recognized that the dynamical system encounters a mathematical
incompatibility.

The only possible solution is given by $\dot{g}_{rr}(t,\,r)\,=\,0$
and then the gravitational potential has to be
$g_{rr}(t,r)\,=\,b(r)$. Considering also the
$\{\theta,\theta\}$\,-\,equation of (\ref{fe}) one can determine
that the gravitational potential $g_{tt}(t,r)$ can be factorized
with respect to the time, so that we get solutions of the type
(\ref{me2y}) which can be recast in the stationary spherically
symmetric form (\ref{me4x}) after a suitable coordinate
transformation.

As a matter of fact, even the more general radial dependent case
admit  time\,-\,independent solutions. From the trace equation and
the $\{\theta,\theta\}$\,-\,component, we deduce a relation which
links $a(r)$ and $b(r)$\,:

\begin{eqnarray}\label{birk}
a(r)\,=\,\frac{b(r)e^{\frac{2}{3}\int\frac{(Rf'-2f)b(r)}{R'f''}dr}}{r^4R'^2f''^2}\,,
\end{eqnarray}
(with $f''\neq 0$) and one which relates $b(r)$ and $f$ (see also
\cite{mul-vil1, mul-vil2} for a similar result)\,:

\begin{eqnarray}\label{rebf}
b(r)\,=\,\frac{6[f'(rR'f'')'-rR'^2f''^2]}{rf(rR'f''-4f')+2f'(rR(f'-rR'f'')-3R'f'')}\,.
\end{eqnarray}

As above, three further equations has to be satisfied to
completely solve the system (respectively the $\{t,t\}$ and
$\{r,r\}$ components of the field equations and the Ricci scalar
constraint) while the only unknown functions are $f$ and the Ricci
scalar $R(r)$.

If we now consider a fourth order model of the form
$f\,=\,R+\mathcal{F}(R)$, with $\mathcal{F}(R)\ll R$ we are
capable of satisfying the whole set of equations up to third order
in $\mathcal{F}$. In particular, we can solve the whole set of
equations: the relations (\ref{birk}) and (\ref{rebf}) will give
the general solution  depending only on the forms of $\mathcal{F}$
and $R\,=\,R(r)$, that is:

\begin{eqnarray}\label{solRr}
\left\{\begin{array}{ll}
a(r)\,=\,\frac{b(r)\,e^{-\frac{2}{3}\int\frac{[R+(2\mathcal{F}-R\mathcal{F}')]b(r)}{R'\mathcal{F}''}dr}}{r^4R'^2
\mathcal{F}''^2}\\\\
b(r)\,=\,-\frac{3(rR'\mathcal{F}'')_{,r}}{rR}
\end{array} \right.
\end{eqnarray}
Once the radial dependence of the scalar curvature is obtained,
(\ref{solRr}) allow to write down the solution of the field
equations and the gravitational potential, related to the function
$a(r)$, can be deduced. Furthermore one can check the physical
relevance of such a potential by means of astrophysical data, see
for example the analysis in \cite{kai-pii-rei-sun}.

\section{Perturbing the spherically symmetric solutions}\label{development-pertirbative}

The search for solutions in $f$ - gravity, in the case of Ricci
scalar dependent on the radial coordinate, can be faced by means
of a perturbation approach. There are several perturbation
techniques by which HOG can be investigated in the weak field
limit. A general approach is starting from analytical $f$ -
theories assuming that the background model slightly deviates from
the Einstein GR (this means to consider $f\,=\,R+\mathcal{F}(R)$
where $\mathcal{F}(R)\ll R$ as above). Another approach can be
developed starting from the background metric considered as the
0th\,-\,order solution. Both these approaches assume the weak
field limit of a given HOG theory as a correction to GR, supposing
that zero order approximation should yield the standard lore.

Both these methods can provide interesting results on the
astrophysical scales where spherically symmetric solutions
characterized by small values of the scalar curvature, can be
taken into account.

In the following, we will consider the first approach assuming
that the background metric matches, at zero order, the GR
solutions.

In general, searching for solutions by a perturbation technique
means to perturb the metric

\begin{eqnarray}\label{approx-metric}g_{\mu\nu}\,=\,g^{(0)}_{\mu\nu}+g^{(1)}_{\mu\nu}\,.\end{eqnarray}
This implies that the field equations (\ref{fe}) and (\ref{fetr})
split, up to first order, in two levels. Besides, a perturbation
on the metric acts also on the Ricci scalar $R$
(\ref{ricciscalar}):

\begin{eqnarray}R\,\sim\,R^{(0)}+R^{(1)}\,,\end{eqnarray}
and then we can Taylor expand the analytic $f$ about the
background value of $R$, i.e.:

\begin{eqnarray}\label{approx}
\left\{\begin{array}{ll}
f\,=\,\sum_{n}\frac{f^n(R^{(0)})}{n!}\biggl[R-R^{(0)}\biggr]^n=\sum_{n}\frac{{f^n}^{(0)}}{n!}{R^{(1)}}^n=f^{(0)}+f'^{(0)}
R^{(1)}+\frac{f''^{(0)}}{2}{R^{(1)}}^2\\\\\,\,\,\,\,\,\,\,\,\,\,\,\,\,\,\,\,\,\,\,\,\,\,\,\,\,\,\,\,\,\,\,\,\,\,\,\,\,\,\,
\,\,\,\,\,\,\,\,\,\,\,\,\,\,\,\,\,\,\,\,\,\,\,\,\,\,\,\,\,\,\,\,\,\,\,\,\,\,\,\,\,\,+\frac{f'''^{(0)}}{6}{R^{(1)}}^3+
\frac{{f^{IV}}^{(0)}}{24}{R^{(1)}}^4+\dots\\\\
f'\,=\,\sum_{n}\frac{f^{n+1}(R^{(0)})}{n!}\biggl[R-R^{(0)}\biggr]^n=f'^{(0)}
+f''^{(0)}R^{(1)}+\frac{f'''^{(0)}}{2}{R^{(1)}}^2+
\frac{{f^{IV}}^{(0)}}{6}{R^{(1)}}^3\\\\\,\,\,\,\,\,\,\,\,\,\,\,\,\,\,\,\,\,\,\,\,\,\,\,\,\,\,\,\,\,\,\,\,\,\,\,\,\,\,\,
\,\,\,\,\,\,\,\,\,\,\,\,\,\,\,\,\,\,\,\,\,\,\,\,\,\,\,\,\,\,\,\,\,\,\,\,\,\,\,\,\,=\,\frac{df}{dR^{(1)}}\\\\
f''\,=\,\sum_{n}\frac{f^{n+2}(R^{(0)})}{n!}\biggl[R-R^{(0)}\biggr]^n=
+f''^{(0)}+f'''^{(0)}R^{(1)}+
\frac{{f^{IV}}^{(0)}}{2}{R^{(1)}}^2\,=\,\frac{df'}{dR^{(1)}}\\\\
f'''\,=\,\sum_{n}\frac{f^{n+3}(R^{(0)})}{n!}\biggl[R-R^{(0)}\biggr]^n=f'''^{(0)}+
{f^{IV}}^{(0)}R^{(1)}\,=\,\frac{df''}{dR^{(1)}}\\\\
\end{array} \right.
\end{eqnarray}
However the above condition $\mathcal{F}(R)\ll R$ has to imply the
validity of the linear approximation
$\frac{f''(R^{(0)})R^{(1)}}{f'(R^{(0)})}\ll 1$. This is
demonstrated by assuming $f'=1+\mathcal{F}'$ and
$f''=\mathcal{F}''$. Immediately we obtain that the condition is
fulfilled for

\begin{eqnarray}\label{strong-condition}
\frac{\mathcal{F}''(R^{(0)})R^{(1)}}{1+\mathcal{F}'(R^{(0)})}\ll1\,.
\end{eqnarray}
For example, given a Lagrangian of the form
$f\,=\,R+\frac{R_0}{R}$, the (\ref{strong-condition}) means

\begin{eqnarray}
\frac{2R_0R^{(1)}}{R^{(0)}({R^{(0)}}^2-R_0)}\ll1\,,
\end{eqnarray}
while, for $f\,=\,R+\alpha R^2$, the (\ref{strong-condition})
means

\begin{eqnarray}
\frac{2\alpha R^{(1)}}{1+2\alpha R^{(0)}}\ll1\,.
\end{eqnarray}
This means that the validity of the approximation strictly depends
on the form of the models and the value of the parameters, in the
previous case $R_0$ and $\alpha$. For the considerations below, we
will assume that it holds. A detailed discussion for the Palatini
formalism is in \cite{sot, bus-bar}.

The zero order field equations read\,:

\begin{equation}\label{eqp0}
{f'}^{(0)}R^{(0)}_{\mu\nu}-\frac{1}{2}g^{(0)}_{\mu\nu}f^{(0)}+\mathcal{H}^{(0)}_{\mu\nu}\,=\,\mathcal{X}\,T^{(0)}_{\mu\nu}
\end{equation}
where

\begin{eqnarray}
\mathcal{H}^{(0)}_{\mu\nu}\,&=&\,-{f''}^{(0)}\biggl\{R^{(0)}_{,\mu\nu}-{\Gamma^{(0)}}^{\rho}_{\mu\nu}R^{(0)}_{,\rho}-g^{(0)}_
{\mu\nu}\biggl({g^{(0)\rho\sigma}}_{,\rho}R^{(0)}_{,\sigma}+g^{(0)\rho\sigma}R^{(0)}_{,\rho\sigma}+\nonumber\\\nonumber\\&&
+g^{(0)\rho\sigma}\ln\sqrt{-g}_{,\rho}R^{(0)}_{,\sigma}\biggl)\biggl\}-{f'''}^{(0)}\biggl\{R^{(0)}_{,\mu}R^{(0)}_{,\nu}-g^{(0)}
_{\mu\nu}g^{(0)\rho\sigma}R^{(0)}_{,\rho}R^{(0)}_{,\sigma}\biggl\}\,.
\end{eqnarray}
At first order one has\,:

\begin{eqnarray}\label{eqp1}
{f'}^{(0)}\biggl\{R^{(1)}_{\mu\nu}-\frac{1}{2}g^{(0)}_{\mu\nu}R^{(1)}\biggr\}+{f''}^{(0)}R^{(1)}R^{(0)}_{\mu\nu}
-\frac{1}{2}f^{(0)}g^{(1)}_{\mu\nu}+\mathcal{H}^{(1)}_{\mu\nu}\,=\,\mathcal{X}\,T^{(1)}_{\mu\nu}
\end{eqnarray}
with

\begin{eqnarray}
\mathcal{H}^{(1)}_{\mu\nu}\,&=&\,-{f''}^{(0)}\biggl\{R^{(1)}_{,\mu\nu}-{\Gamma^{(0)}}^{\rho}_{\mu\nu}R^{(1)}_{,\rho}-
{\Gamma^{(1)}}^{\rho}_{\mu\nu}R^{(0)}_{,\rho}-g^{(0)}_{\mu\nu}\biggl[{g^{(0)\rho\sigma}}_{,\rho}R^{(1)}_{,\sigma}+{g^{(1)
\rho\sigma}}_{,\rho}R^{(0)}_{,\sigma}+\nonumber\\\nonumber\\&&+g^{(0)\rho\sigma}R^{(1)}_{,\rho\sigma}+g^{(1)\rho\sigma}
R^{(0)}_{,\rho\sigma}+g^{(0)\rho\sigma}\biggl(\ln\sqrt{-g}^{(0)}_{,\rho}R^{(1)}_{,\sigma}+\ln\sqrt{-g}^{(1)}_{,\rho}R^{(0)}
_{,\sigma}\biggl)+\nonumber\\\nonumber\\&&+g^{(1)\rho\sigma}\ln\sqrt{-g}^{(0)}_{,\rho}R^{(0)}_{,\sigma}\biggr]-g^{(1)}_
{\mu\nu}\biggl({g^{(0)\rho\sigma}}_{,\rho}R^{(0)}_{,\sigma}+g^{(0)\rho\sigma}R^{(0)}_{,\rho\sigma}+\nonumber
\end{eqnarray}
\begin{eqnarray}
&&+g^{(0)\rho\sigma}\ln\sqrt{-g}^{(0)}_{,\rho}R^{(0)}_{,\sigma}\biggr)\biggr\}-{f'''}^{(0)}\biggl\{R^{(0)}_{,\mu}R^{(1)}
_{,\nu}+R^{(1)}_{,\mu}R^{(0)}_{,\nu}-g^{(0)}_{\mu\nu}g^{(0)\rho\sigma}\biggl(R^{(0)}_{,\rho}R^{(1)}_{,\sigma}+\nonumber\\
\nonumber\\&&+R^{(1)}_{,\rho}R^{(0)}_{,\sigma}\biggr)-g^{(0)}_{\mu\nu}g^{(1)\rho\sigma}R^{(0)}_{,\rho}R^{(0)}_{,\sigma}-g^{(1)}
_{\mu\nu}g^{(0)\rho\sigma}R^{(0)}_{,\rho}R^{(0)}_{,\sigma}\biggr\}-{f'''}^{(0)}R^{(1)}\biggl\{R^{(0)}_{,\mu\nu}+\nonumber\\
\nonumber\\&&-{\Gamma^{(0)}}^{\rho}_{\mu\nu}R^{(0)}_{,\rho}-g^{(0)}_{\mu\nu}\biggl({g^{(0)\rho\sigma}}_{,\rho}R^{(0)}_
{,\sigma}+g^{(0)\rho\sigma}R^{(0)}_{,\rho\sigma}+g^{(0)\rho\sigma}\ln\sqrt{-g}^{(0)}_{,\rho}R^{(0)}_{,\sigma}\biggl)\biggl\}
+\nonumber\\\nonumber\\&&-{f^{IV}}^{(0)}R^{(1)}\biggl\{R^{(0)}_{,\mu}R^{(0)}_{,\nu}-g^{(0)}_{\mu\nu}g^{(0)\rho\sigma}R^{(0)}_
{,\rho}R^{(0)}_{,\sigma}\biggl\}\,.
\end{eqnarray}
A part the analyticity, no hypothesis has been invoked on the form
of $f$. As a matter of fact, $f$ can be completely general. At
this level, to solve the problem, it is required the zero order
solution of (\ref{eqp0}) which, in general, could not be a GR
solution. This problem can be overcome assuming the same order of
perturbation on the $f$, that is:

\begin{eqnarray}\label{theapprox}
f\,=\,R+\mathcal{F}(R)\,,
\end{eqnarray}
where $\mathcal{F}$ is a generical function of the Ricci scalar as
above. Then we have

\begin{eqnarray}\label{approx2}
f\,=\,R^{(0)}+R^{(1)}+\mathcal{F}^{(0)}\,,\ \ \
f'\,=\,1+\mathcal{F}'^{(0)}\,,\ \ \
f''\,=\,\mathcal{F}''^{(0)}\,,\ \ \ \ \ \
f'''\,=\,\mathcal{F}'''^{(0)}\,,
\end{eqnarray}
and the (\ref{eqp0}) reduce to the form

\begin{eqnarray}
R^{(0)}_{\mu\nu}-\frac{1}{2}R^{(0)}g^{(0)}_{\mu\nu}\,=\,G^{(0)}_{\mu\nu}\,=\,\mathcal{X}\,T^{(0)}_{\mu\nu}\,.
\end{eqnarray}
On the other hand, the (\ref{eqp1}) reduce to

\begin{eqnarray}
R^{(1)}_{\mu\nu}-\frac{1}{2}g^{(0)}_{\mu\nu}R^{(1)}-\frac{1}{2}g^{(1)}_{\mu\nu}R^{(0)}-\frac{1}{2}g^{(0)}_{\mu\nu}
\mathcal{F}^{(0)}+\mathcal{F}'^{(0)}R^{(0)}_{\mu\nu}+\mathcal{H}^{(1)}_{\mu\nu}\,=\,\mathcal{X}\,T^{(1)}_{\mu\nu}
\end{eqnarray}
where

\begin{eqnarray}
\mathcal{H}^{(1)}_{\mu\nu}\,&=&\,-\mathcal{F}'''^{(0)}\biggl\{R^{(0)}_{,\mu}R^{(0)}_{,\nu}-g^{(0)}_{\mu\nu}g^{(0)\sigma\tau}
R^{(0)}_{,\sigma}R^{(0)}_{,\tau}\biggr\}-\mathcal{F}''^{(0)}\biggl\{R^{(0)}_{,\mu\nu}-{\Gamma^{(0)}}^\sigma_{\mu\nu}R^{(0)}
_{,\sigma}\nonumber\\\nonumber\\&&-g^{(0)}_{\mu\nu}\biggl({g^{(0)}}^{\sigma\tau}_{\,\,\,\,\,\,,\sigma}R^{(0)}_{,\tau}+
{g^{(0)}}^{\sigma\tau}R^{(0)}_{,\sigma\tau}+{g^{(0)}}^{\sigma\tau}\ln\sqrt{-g}^{(0)}_{,\sigma}R^{(0)}_{,\tau}\biggl)\biggr\}
\,.\end{eqnarray} The new system of field equations is evidently
simpler than the starting one and once  the zero order solution is
obtained, the solutions at the first order correction can be
easily achieved. In Tables \ref{table-solutions-1} and
\ref{table-solutions-2}, a list of solutions, obtained with this
perturbation method, is given considering different classes of $f$
- models.

Some remarks on these solutions are in order at this point. In the
case of $f$ models which  are evidently corrections to the Hilbert
- Einstein Lagrangian as $\Lambda+R+\epsilon R \ln R$ and
$R+\epsilon R^n$, with $\epsilon\ll 1$, one obtains  exact
solutions for the gravitational potentials $a(r)$ and $b(r)$
related by $a(r)\,=\,b(r)^{-1}$. The first order expansion is
straightforward as in GR. If the functions $a(r)$ and $b(r)$ are
not related, for $f\,=\,\Lambda+R+\epsilon R \ln R$, the first
order system is directly solved  without any prescription on the
perturbation functions $x(r)$ and $y(r)$. This is not the case for
$f\,=\,R+\epsilon R^n$ since, for this model, one can obtain an
explicit constraint on the perturbation function  implying the
possibility to deduce the form of the gravitational potential
$\Phi(\mathbf{x})$ from $a(r)\,=\,1+2\,\Phi(\mathbf{x})$. In such
a case, no corrections are found with respect to the standard
Newtonian potential. The theories $f\,=\,R^n$ and
$f\,=\,\frac{R}{(R_0+R)}$ show similar behaviors. The case
$f\,=\,R^2$ is peculiar and it has to be dealt independently.

\begin{table}
\begin{center}
\begin{tabular}{cccc}
  \hline\hline\hline
  & & & \\
  & $f$ - theory: & $\Lambda+R+\epsilon R \ln R$ & \\
  & & & \\
  & spherical potentials: & $a(r)=b(r)^{-1}=1+\frac{k_1}{r}-\frac{\Lambda r^2}{6}+\delta x(r)$ & \\
  & & & \\
  & solutions: & $x(r)=\frac{k_2}{r}+\frac{\epsilon\Lambda[\ln(-2\Lambda)-1]r^2}{6\delta}$ & \\
  & & & \\
  & first order metric: & $a(r)=1-\frac{\Lambda r^2}{6}+\delta x(r)$,\,\,\,\,\,$b(r)=\frac{1}{1-\frac{\Lambda r^2}{6}}
  +\delta y(r)$ & \\
  & & & \\
  & solutions: & $\begin{cases} x(r)=(\Lambda r^2-6)\biggl\{k_1+\int\frac{dr}{36r\delta(\Lambda r^2-6)}\biggl[4\delta(2
  \Lambda^2r^4-15\Lambda r^2\\\,\,\,\,\,\,\,\,\,\,\,\,\,\,\,\,\,\,\,+18)y(r)+r\{36r\epsilon\Lambda[\log(-2\Lambda)-1]
  \\\,\,\,\,\,\,\,\,\,\,\,\,\,\,\,\,\,\,\,+\delta(\Lambda r^2-6)^2y'(r)\}\biggr]\biggr\}\\\\
  y(r)=\frac{k_2\delta-6r^3\epsilon\Lambda[\ln(-2\Lambda)-1]}{r\delta(r^2\Lambda-6)^2}  \end{cases} $& \\
  & & & \\
  & & & \\
  & $f$ - theory: & $R+\epsilon R^n$ & \\
  & & & \\
  & spherical potentials: & $a(r)=b(r)^{-1}=1+\frac{k_1}{r}+\delta x(r)$ & \\
  & & & \\
  & solutions: & $x(r)=\frac{k_2}{r}$ & \\
  & & & \\
  & first order metric: & $a(r)=1+\delta\frac{x(r)}{r}$,\,\,\,\,\,$b(r)=1+\delta\frac{y(r)}{r}$ & \\
  & & & \\
  & solutions: & $x(r)=k_1+k_2r$,\,\,\,\,\,$y(r)=k_3$ & \\
  & & & \\
  & & & \\
  & $f$ - theory: & $R/(R_0+R)$ & \\
  & & & \\
  & first order metric: & $a(r)=1+\delta\frac{x(r)}{r}$,\,\,\,\,\,$b(r)=1+\delta\frac{y(r)}{r}$ & \\
  & & & \\
  & solutions: & $\begin{cases} x(r)=-\frac{4e^{-\frac{R_0^{1/2}r}{\sqrt{6}}}}{R_0}k_1-\frac{2\sqrt{6}e^{\frac{R_0^{1/2}r}
  {\sqrt{6}}}}{R_0^{3/2}}k_2+k_3r \\ y(r)=-\frac{2e^{-\frac{R_0^{1/2}r}{\sqrt{6}}}(6R_0^{1/2}+\sqrt{6}R_0\,r)}{3b^{3/2}}
  k_1-\frac{2e^{\frac{R_0^{1/2}r}{\sqrt{6}}}(\sqrt{6}-R_0^{1/2}r)}{R_0^{3/2}}k_2
  \end{cases}$ & \\
  & & & \\
  \hline\hline\hline
\end{tabular}
\end{center}
\caption{A list of exact solutions obtained \emph{via} the
perturbation approach for several classes of $f$ - theories; $k_i$
are integration constants; the potentials $a(r)$ and $b(r)$ are
defined by the metric (\ref{me4x}).\label{table-solutions-1}}
\end{table}

\begin{table}
\begin{center}
\begin{tabular}{cccc}
  \hline\hline\hline
  & & & \\
  & $f$ - theory: & $R^n$ & \\
  & & & \\
  & spherical potentials: & $a(r)=b(r)^{-1}=1+\frac{k_1}{r}+\frac{R_0r^2}{12}+\delta x(r)$ & \\
  & & & \\
  & solutions: & $\begin{cases} n=2,\,\,\,\,\,\,R_0\neq 0 \,\,\text{and}\,\,x(r)=\frac{3k_2-k_3}{3r}+\frac{k_3r^2}{12} \\
  \,\,\,\,\,\,\,\,\,\,\,\,\,\,\,\,\,\,\,\,\,\,\,\,\,\,\,\,\,\,\,\,\,\,\,\,\,\,\,\,\,\,\,\,\,\,\,\,\,\,\,\,\,\,\,
  +\frac{k_4}{r}\int dr\,r^2\biggl\{\int dr\frac{\exp\biggl[\frac{R_0r_0^2\ln(r-r_0)}{8+3R_0r_0^2}
  \biggr]}{r^5}\biggr\} \\
  \,\,\,\,\,\,\,\,\,\,\,\,\,\,\,\,\,\,\,\,\,\,\,\, \text{with}\,\,r_0\,\,$satisfying the condition$ \\
  \,\,\,\,\,\,\,\,\,\,\,\,\,\,\,\,\,\,\,\,\,\,\,\,\,6k_1+8r_0+R_0r_0^3=0
  \\\\ n\geq 2,\,\,\,\,\,\,
  \text{System solved only whit $R_0=0$} \\ \,\,\,\,\,\,\,\,\,\,\,\,\,\,\,\,\,\,\,\,\,\,\,\text{and no prescriptions on
  $x(r)$} \end{cases}$ & \\
  & & & \\
  & first order metric: & $a(r)=1+\delta\frac{x(r)}{r}$,\,\,\,\,\,$b(r)=1+\delta\frac{y(r)}{r}$ & \\
  & & & \\
  & solutions: & $\begin{cases} n=2 & y(r)=-\frac{R_0r^3}{6}-\frac{x(r)}{2}+\frac{1}{2}rx'(r)+k_1, \\
  & R(r)=\delta R_0 \\\\ n\neq 2 &
  y(r)=-\frac{1}{2}\int dr\,r^2R(r)-\frac{x(r)}{2}+\frac{1}{2}rx'(r)+k_1 \\ & \text{with $R(r)$ whatever}
  \end{cases}$ & \\
  & & & \\
  & first order metric: & $a(r)=1-\frac{r_g}{r}+\delta x(r)$,\,\,\,\,\,$b(r)=\frac{1}{1-\frac{r_g}{r}}+\delta y(r)$ & \\
  & & & \\
  & solutions: & $\begin{cases} n=2 & y(r)=\frac{rk_1}{3r_g^2-7r_gr+4r^2}+\frac{r^2k_2}{3(3r_g^2-7r_gr+4r^2)} \\
  & \,\,\,\,\,\,\,\,\,\,\,\,\,\,\,\,\,\,+\frac{r_gr^2x(r)+2(r_gr^3-r^4)x'(r)}{(3r_g-4r)(r_g-r)^2} \\\\ n\neq 2 &
  \text{whatever functions $x(r)$\,, $y(r)$ and $R(r)$} \end{cases}$ & \\
  & & & \\
  \hline\hline\hline
\end{tabular}
\end{center}
\caption{A list of exact solutions obtained \emph{via} the
perturbation approach for several classes of $f$ - theories; $k_i$
are integration constants; the potentials $a(r)$ and $b(r)$ are
defined by the metric (\ref{me4x}).\label{table-solutions-2}}
\end{table}

\clearpage{\pagestyle{empty}\cleardoublepage}

\chapter{The Noether Symmetries of $f$ $-$ gravity}

We search for spherically symmetric solutions of $f$-theories of
gravity via the Noether Symmetry Approach [\textbf{B}]. A general
formalism in the metric framework is developed considering a
point-like $f$\,-\,Lagrangian where spherical symmetry is
required. New exact solutions are given.

\section{The point-like $f$ Lagrangian in spherical symmetry}

As hinted in the plan of thesis, the aim of this chapter is to
work out an approach  to obtain spherically symmetric solutions in
HOG by means of Noether Symmetries. In order to develop this
approach, we need to deduce a point-like Lagrangian from the
action (\ref{actionfR}). Such a Lagrangian can be obtained by
imposing the spherical symmetry in the field action
(\ref{actionfR}). As a consequence, the infinite number of degrees
of freedom of the original field theory will be reduced to a
finite number. The technique is based on the choice of a suitable
Lagrange multiplier defined by assuming the Ricci scalar, argument
of the function $f$ in spherical symmetry. Elsewhere, this
approach has been successfully used  for the FRW metric with the
purpose to find out cosmological solutions \cite{cap-der-rub-scu,
cap-der, cap-der1, cap-lam}.

In general, a spherically symmetric spacetime  can be described
assuming that the metric (\ref{me3}) is time independent\,:

\begin{eqnarray}\label{me2x}
ds^2=A(r)dt^2-B(r)dr^2-M(r)d\Omega\,,
\end{eqnarray}
where $g_{tt}(t',r)\,\doteq\,A(r)$, $g_{rr}(t',r)\,\doteq\,B(r)$
and $g_{\Omega\Omega}(t',r)\,\doteq\,M(r)$. Obviously the
conditions $M(r)=r^2$ and $B(r)=A^{-1}(r)$ are requested to obtain
the standard Schwarzschild case of GR. Our goal is to reduce the
field action (\ref{actionfR}) to a form with a finite degrees of
freedom, that is the canonical action of the form

\begin{eqnarray}
\mathcal{A}=\int dr\mathcal{L}(A, A', B, B', M, M', R, R')
\end{eqnarray}
where the Ricci scalar $R$ and the potentials $A$, $B$, $M$ are
the set of independent variables defining the configuration space.
Prime indicates now the derivative with respect to the radial
coordinate $r$. In order to achieve the point-like Lagrangian in
this set of coordinates, we write, in the vacuum, the
(\ref{actionfR}) as

\begin{eqnarray}\label{lm}
\mathcal{A}=\int
d^4x\sqrt{-g}\biggl[f(R)-\lambda(R-\bar{R})\biggr]\,,
\end{eqnarray}
where $\lambda$ is the Lagrangian multiplier and $\bar{R}$ is the
Ricci scalar (\ref{ricciscalar}) expressed in terms of the metric
(\ref{me2x}):

\begin{eqnarray}\label{rs}
\bar{R}=\frac{A''}{AB}+2\frac{M''}{BM}+\frac{A'M'}{ABM}-\frac{A'^2}{2A^2B}-\frac{M'^2}{2BM^2}
-\frac{A'B'}{2AB^2}-\frac{B'M'}{B^2M}-\frac{2}{M}\,,
\end{eqnarray}
which can be recast in the more compact form

\begin{eqnarray}
\bar{R}=R^*+\frac{A''}{AB}+2\frac{M''}{BM}\,,
\end{eqnarray}
where $R^*$ collects first order derivative terms. The Lagrange
multiplier $\lambda$ is obtained by varying  the action (\ref{lm})
with respect to $R$. One gets $\lambda\,=\,f_R$\footnote{In this
chapter the derivatives of $f$ will be indicated like
$d^nf/dR^n\,=\,f_{R....R}$.}. By expressing the determinant $g$
and $\bar{R}$ in terms of $A$, $B$ and $M$, we have, from
(\ref{lm}),

\begin{eqnarray}\label{ac1}\mathcal{A}=\int
drA^{1/2}B^{1/2}M\biggl[f-f_{R}\biggl(R-R^*-\frac{A''}{AB}-2\frac{M''}{BM}\biggr)\biggr]=\nonumber\\
=\int dr
\biggl\{A^{1/2}B^{1/2}M\biggl[f-f_{R}(R-R^*)\biggr]-\biggl(\frac{f_{R}M}{A^{1/2}B^{1/2}}\biggr)'A'-
2\biggl(\frac{A^{1/2}}{B^{1/2}}f_{R}\biggr)
'M'\biggr\}\,.\end{eqnarray} The two lines differs for a
divergence term which we discard  integrating by parts. Therefore,
the point-like Lagrangian becomes\,:

\begin{eqnarray}\label{lag}
\mathcal{L}=-\frac{A^{1/2}f_{R}}{2MB^{1/2}}{M'}^2-\frac{f_{R}}{A^{1/2}B^{1/2}}A'M'-\frac{Mf_{RR
}}{A^{1/2}B^{1/2}}A'R'+\nonumber\\\nonumber\\-\frac{2A^{1/2}f_{RR}}{B^{1/2}}R'M'-A^{1/2}B^{1/2}[(2+MR)f_{R}-Mf]\,,
\end{eqnarray}
which is canonical since only the configuration variables and
their first order derivatives with respect to $r$ are present. Eq.
(\ref{lag}) can be recast in a more compact form introducing the
matrix formalism\,:

\begin{eqnarray}\label{la}
\mathcal{L}={\underline{q}'}^t\hat{T}\underline{q}'+V
\end{eqnarray}
where $\underline{q}=(A,B,M,R)$ and $\underline{q}'=(A',B',M',R')$
are the generalized positions and velocities associated to
$\mathcal{L}$. The index ``\emph{t}" indicates the transposed
column vector. The kinetic tensor is given by ${\displaystyle
\hat{T}_{ij}\,=\,\frac{\partial^2\mathcal{L}}{\partial
q'_i\partial q'_j}}$.  $V=V(q)$ is the potential depending only on
the configuration variables. The Euler\,-\,Lagrange equations read

\begin{eqnarray}\label{eul-lag-equations}\nonumber
\frac{d}{dr}\nabla_{q'}\mathcal{L}-\nabla_{q}\mathcal{L}=2\frac{d}{dr}\biggl(\hat{T}\underline{q}'\biggr)-\nabla_{q}V-
{\underline{q}'}^t\biggl(\nabla_{q}\hat{T}\biggr)\underline{q}'\,=\nonumber\\\nonumber\\=\,2\hat{T}\underline{q}''+2\biggl(
\underline{q}'\cdot\nabla_{q}\hat{T}\biggr)\underline{q}'-\nabla_{q}V-\underline{q}'^t\biggl(\nabla_{q}\hat{T}\biggr)
\underline{q}'=0\end{eqnarray} which furnish the  equations of
motion in term of  $A$, $B$, $M$ and $R$, respectively. The field
equation for $R$ corresponds to the constraint among the
configuration coordinates. It is worth noting that the Hessian
determinant of (\ref{lag}), ${\displaystyle
\left|\left|\frac{\partial^2\mathcal{L}}{\partial q'_i\partial
q'_j}\right|\right|}$, is zero. This result clearly depends on the
absence of the generalized velocity $B'$ into the point\,-\,like
Lagrangian. As matter of fact, using a point-like Lagrangian
approach implies that the metric variable $B$ does not contributes
to dynamics, but the  equation of motion for $B$ has to be
considered as a further constraint equation.

Beside the Euler\,-\,Lagrange equations (\ref{eul-lag-equations}),
one has to take into account the energy $E_\mathcal{L}$\,:

\begin{eqnarray}\label{ene}
E_\mathcal{L}={\underline
{q}'}\cdot\nabla_{q'}\mathcal{L}-\mathcal{L}
\end{eqnarray}
which can be easily recognized to be coincident with the
Euler-Lagrangian equation for the component $B$ of the generalized
position $\underline{q}$. Then  the Lagrangian (\ref{lag}) has
three degrees of freedom and not four, as we would expect "a
priori".

Now, since the motion equation describing the evolution of the
metric potential $B$ does not depends on its derivative, it can be
explicitly solved in term of $B$ as a function of other
coordinates\,:

\begin{eqnarray}\label{eqb}
B=\frac{2M^2f_{RR}A'R'+2Mf_{R}A'M'+4AMf_{RR}M'R'+Af_{R}M'^2}{2AM[(2+MR)f_{R}-Mf]}\,.
\end{eqnarray}
By inserting the (\ref{eqb}) into the Lagrangian (\ref{lag}), we
obtain a non-vanishing Hessian matrix removing the singular
dynamics. The new Lagrangian reads\footnote{Lowering  the
dimension of configuration space through the substitution
(\ref{eqb}) does not affect the  dynamics, since $B$ is a
non-evolving quantity. In fact, introducing the (\ref{eqb})
directly into the dynamical equations given by (\ref{lag}),  they
coincide with those derived by (\ref{lag2}).}

\begin{eqnarray}\mathcal{L}^*= {\bf L}^{1/2}\end{eqnarray}
with

\begin{eqnarray}\label{lag2}\nonumber
\mathbf{L}=\underline{q'}^t\,\,\hat{\mathbf{L}}\,\,\underline{q'}&=&\frac{[(2+MR)f_{R}-fM]}{M}\nonumber\\\nonumber\\&&
\times[2M^2f_{RR}A'R'+2MM'(f_{R}A'+2Af_{RR}R') +Af_{R}M'^2]\,.
\end{eqnarray}
Since ${\displaystyle \frac{\partial{\bf L}}{\partial r}=0}$,
${\bf L}$ is canonical (${\bf L}$ is the quadratic form of
generalized velocities, $A'$, $M'$ and $R'$ and then coincides
with the Hamiltonian), so that we can consider ${\bf L}$ as the
new Lagrangian with three  degrees of freedom. The crucial point
of such a replacement is that the Hessian determinant is now
non\,-\,vanishing, being\,:

\begin{eqnarray}
\left|\left|\frac{\partial^2 {\bf L}}{\partial q'_i\partial
q'_j}\right|\right|=3AM[(2+MR)f_{R}-Mf]^3f_{R}{f_{RR}}^2\,.
\end{eqnarray}
Obviously, we are supposing that $(2+MR)f_{R}-Mf\neq0$, otherwise
the above definitions of $B$, (\ref{eqb}), and ${\bf L}$,
(\ref{lag2}), lose of significance, besides we assume $f_{RR}\neq
0$ to admit a wide class of HOG models. The case $f\,=\,R$
requires a different investigation. In fact, considering the GR
point\,-\,like Lagrangian needs a further lowering of  degrees of
freedom of the system and the previous results cannot be
straightforwardly  considered. From (\ref{lag}), we get\,:

\begin{eqnarray}\label{lagr}
\mathcal{L}_{GR}=-\frac{A^{1/2}}{2MB^{1/2}}{M'}^2-\frac{1}{A^{1/2}B^{1/2}}A'M'-2A^{1/2}B^{1/2}\,,
\end{eqnarray}
whose Euler-Lagrange equations provide the standard equations of
GR for Schwarzschild metric. It is easy to see the absence of the
generalized velocity $B'$ in (\ref{lagr}). Again, the Hessian
determinant is zero. Nevertheless, considering, as above, the
constraint (\ref{eqb}) for $B$, it is possible to obtain a
Lagrangian  with a non-vanishing Hessian. In particular one has\,:

\begin{eqnarray}\label{eqbgr}B_{GR}=\frac{M'^2}{4M}+\frac{A'M'}{2A}\,,\end{eqnarray}

\begin{eqnarray}\label{lag2gr}\mathcal{L}_{GR}^*= {\bf L}_{GR}^{1/2}=\sqrt{\frac{M'(2MA'+AM')}{M}}\,,\end{eqnarray}
and then the Hessian determinant is

\begin{eqnarray}\left|\left|\frac{\partial^2 {\bf L}_{GR}}{\partial q'_i\partial q'_j}\right|\right|=-1\,,\end{eqnarray}
which is nothing else but a non-vanishing sub-matrix of the $f$
Hessian matrix.

Considering the Euler\,-\,Lagrange equations coming from
(\ref{eqbgr}) and (\ref{lag2gr}), one obtains the vacuum solutions
of GR (\ref{schwarz-solution-stand-coord}), that is\,:

\begin{eqnarray}\label{schsol}
A=k_{4}-\frac{k_{3}}{r+k_{1}}\,,\ \ \ \ \ \ \
B=\frac{k_{2}k_{4}}{A}\,,\ \ \ \ \ \ M=k_{2}(r+k_{1})^2\,.
\end{eqnarray}
In particular, the standard form of Schwarzschild solution is
obtained for $k_1=0$, $k_2=1$, $k_3=r_g$ and $k_4=1$.

A formal summary of the field equations descending from the
point\,-\,like Lagrangian and their relation with respect to the
ones of the standard approach is given in Tab.
(\ref{table-point-like}).

\begin{table}[h]\label{table-point-like}
\begin{center}
\begin{tabular}{ccc}
  \hline\hline\hline
  Field equations approach &  & Point-like Lagrangian approach \\
  $\downarrow$ &  &  $\downarrow$ \\
  $\delta\int d^4x\sqrt{-g}f=0$  & $\leftrightarrows$ &  $\delta\int dr\mathcal{L}=0$  \\
  $\downarrow$ &  &  $\downarrow$ \\
  $H_{\mu\nu}=\partial_\rho\frac{\partial(\sqrt{-g}f)}{\partial_\rho g^{\mu\nu}}-\frac{\partial (\sqrt{-g}f)}{\partial
  g^{\mu\nu}}=0$ &  & $\frac{d}{dr}\nabla_{q'}\mathcal{L}-\nabla_{q}\mathcal{L}=0$ \\  & $\leftrightarrows$ &   \\
  $H=g^{\alpha\beta}H_{\alpha\beta}=0$ &
  & $E_\mathcal{L}={\underline
  {q}'}\cdot\nabla_{q'}\mathcal{L}-\mathcal{L}$ \\
  $\downarrow$ &  & $\downarrow$  \\
  $H_{tt}=0$ & $\leftrightarrows$ & $\frac{d}{dr}\frac{\partial \mathcal{L}}{\partial A'}-\frac{\partial \mathcal{L}}
  {\partial A}=0$ \\
  $H_{rr}=0$ & $\leftrightarrows$ & $\frac{d}{dr}\frac{\partial \mathcal{L}}{\partial B'}-\frac{\partial \mathcal{L}}
  {\partial B}\propto E_\mathcal{L}=0$ \\
  $H_{\theta\theta}={\csc^{2}\theta}H_{\phi\phi}=0$ & $\leftrightarrows$ & $\frac{d}{dr}\frac{\partial \mathcal{L}}
  {\partial M'}-\frac{\partial \mathcal{L}}{\partial M}=0$ \\
  $H=A^{-1}H_{tt}-B^{-1}H_{rr}-2M^{-1}{\csc^{2}\theta}H_{\phi\phi}=0$ & $\leftrightarrows$ & A combination of the above
  equations \\
  \hline\hline\hline
\end{tabular}
\end{center}
\caption{ The field-equations approach and the point-like
Lagrangian approach differ since the symmetry, in our case the
spherical one, can be imposed whether in the field equations,
after standard variation with respect to the metric, or directly
into the Lagrangian, which becomes point-like. The energy
$E_\mathcal{L}$ corresponds to the $tt$\,-\,component of
$H_{\mu\nu}$. The absence of $B'$ in the Lagrangian implies the
proportionality between the constraint equation for $B$ and the
energy function $E_\mathcal{L}$. As a consequence, the number of
independent equations is three (as the number of unknown
functions). Finally it is obvious the correspondence between
$\theta\theta$ component and field equation for $M$.}\end{table}

The explicit form of field equations (\ref{fe}) - (\ref{fetr}) in
the vacuum with the metric (\ref{me2x}) are

\begin{eqnarray}H_{tt}&=&2A^2B^2Mf+\{BMA'^2-A[2BA'M'+M(2BA''-A'B')]\}f_{R}+\nonumber\\&&+(-2A^2MB'R'+4A^2BM'R'+4A^2BMR'')
f_{RR}+\nonumber\\&&+4A^2BMR'^2f_{RRR}=0\,,\end{eqnarray}

\begin{eqnarray}H_{rr}&=&2A^2B^2M^2f+(BM^2A'^2+AM^2A'B'+2A^2MB'M'+2A^2BM'^2+\nonumber\\&&-2ABM^2A''-4A^2BMM'') f_{R}+(2ABM^2
A'R'+\nonumber\\&&+4A^2BMM'R')f_{RR}=0\,,\end{eqnarray}

\begin{eqnarray}H_{\theta\theta}&=&2AB^2Mf+(4AB^2-BA'M'+AB'M'-2ABM'')f_{R}+\nonumber\\&&+(2BMA'R'-2AMB'R'+2ABM'R'+4ABMR'')
f_{RR}+\nonumber\\&&+4ABMR'^2f_{RRR}=0\,,\end{eqnarray}

\begin{eqnarray}H_{\phi\phi}=\sin^2\theta H_{\theta\theta}=0\,.\end{eqnarray}

\begin{eqnarray}H=&&4AB^2Mf-2AB^2MRf_{R}+3(BMA'R'-AMB'R'+\nonumber\\&&+2ABM'R'
+2ABMR'')f_{RR}+6ABMR'^2f_{RRR}=0
\end{eqnarray}

\section{The Noether Symmetry Approach}\label{noether-symmetry-approach}

In order to find out solutions for the  Lagrangian (\ref{lag2}),
we can search for symmetries related to cyclic variables and then
reduce dynamics. This approach allows, in principle,  to select
$f$-gravity models compatible with spherical symmetry. As a
general remark, the \emph{Noether Theorem} states that conserved
quantities are related to the existence of cyclic variables into
dynamics \cite{arnold, mar-sal-sim-vit, mor-fer-lov-mar-rub}. Let
us give a summary of the approach for finite dimensional dynamical
systems.

Let $\mathcal{L}(q^i,\dot{q}^i)$ be a canonical, non-degenerate
point-like Lagrangian where

\begin{eqnarray}\label{01}\frac{\partial\mathcal{L}}{\partial\lambda}=0\,\,\,\,\,\,\,\,\,\mbox{det}H_{ij}\doteq\mbox{det}
\left|\left|\frac{\partial^2\mathcal{L}}{\partial\dot{q}^i\partial\dot{q}^j}\right|\right|\neq
0\,,\end{eqnarray} with $H_{ij}$ as above, the Hessian matrix
related to $\mathcal{L}$. The dot indicates derivatives with
respect to the affine parameter $\lambda$ which, ordinarily,
corresponds to time $t$. In our case, it is the radial coordinate
$r$. In standard problems of analytical mechanics, $\mathcal{L}$
is in the form

\begin{eqnarray}\label{02}\mathcal{L}=T(\textbf{q},\dot{\textbf{q}})-V(\textbf{q})\,,\end{eqnarray}
where $T$ and $V$ are the "kinetic" and "potential energy"
respectively. $T$ is a positive definite quadratic form in
$\dot{\textbf{q}}$. The energy function associated with
$\mathcal{L}$ is

\begin{eqnarray}\label{03}E_\mathcal{L}\equiv\frac{\partial\mathcal{L}}{\partial\dot{q}^i}\dot{q}^i-\mathcal{L}\,,
\end{eqnarray}
which is  the total energy $T+V$. It has to be noted that
$E_\mathcal{L}$ is, in any case, a constant of motion. In this
formalism, we are going to consider only transformations which are
point-transformations. Any invertible and smooth transformation of
the "positions" $Q^{i}=Q^{i}(\textbf{q})$ induces a transformation
of the "velocities" such that

\begin{eqnarray}\label{04}\dot{Q}^i(\textbf{q})=\frac{\partial Q^i}{\partial q^j}\dot{q}^j\,;\end{eqnarray}
the matrix $\mathcal{J}=||\partial Q^i/\partial q^j ||$ is the
Jacobian of the transformation on the positions, and it is assumed
to be nonzero. The Jacobian $\widetilde{\mathcal{J}}$ of the
"induced" transformation is easily derived and $\mathcal{J}\neq 0
\rightarrow\widetilde{\mathcal{J}}\neq 0$. Usually, this condition
is not satisfied in the whole space but only in the neighbor of a
point. It is \emph{local transformation}. If one extends the
transformation to the maximal submanifold  such that
$\mathcal{J}\neq 0$, it is possible to get troubles for the whole
manifold due to possible different topologies
\cite{mor-fer-lov-mar-rub}.

A point transformation $Q^{i}=Q^{i}(\textbf{q})$ can depend on one
(or more than one) parameter. Let us assume that a point
transformation depends on a parameter $\varepsilon$, i.e.
$Q^{i}=Q^{i}(\textbf{q},\varepsilon)$, and that it gives rise to a
one-parameter Lie group. For infinitesimal values of
$\varepsilon$, the transformation is then generated by a vector
field: for instance, as well known, $\partial/\partial x$
represents a translation along the $x$ axis, $x(\partial/\partial
y)-y(\partial/\partial x)$ is a rotation around the $z$ axis and
so on. In general, an infinitesimal point transformation is
represented by a generic vector field on $Q$

\begin{eqnarray}\label{04b}\textbf{X}=\alpha^i(\textbf{q})\frac{\partial}{\partial
q^i}\,.\end{eqnarray} The induced transformation (\ref{04}) is
then represented by

\begin{eqnarray}\label{05}\textbf{X}^c=\alpha^i(\textbf{q})\frac{\partial}{\partial q^i}+
\biggl[\frac{d}{d\lambda}\alpha^{i}(\textbf{q})\biggr]\frac{\partial}{\partial
q^i}\,.\end{eqnarray} $\textbf{X}^c$ is called the "complete lift"
of $\textbf{X}$ \cite{mor-fer-lov-mar-rub}. A function
$f(\textbf{q}, \dot{\textbf{q}})$ is invariant under the
transformation $\textbf{X}^{c}$ if

\begin{eqnarray}\label{06}L_{\textbf{X}^c}f\doteq\alpha^i(\textbf{q})\frac{\partial f}{\partial q^{i}}+
\biggl[\frac{d}{d\lambda}\alpha^i(\textbf{q})\biggr]\frac{\partial
f}{\partial q^i}\,=\,0\;,\end{eqnarray} where $L_{\textbf{X}^c}f$
is the Lie derivative of $f$. In particular, if
$L_{\textbf{X}^c}{\mathcal{L}}=0$, $\textbf{X}^c$ is said to be a
\emph{symmetry} for the dynamics derived by $\mathcal{L}$.

In order to see how Noether's theorem and cyclic variables are
related, let us consider a Lagrangian $\mathcal{L}$ and its
Euler-Lagrange equations

\begin{eqnarray}\label{07}\frac{d}{d\lambda}\frac{\partial
\mathcal{L}}{\partial\dot{q}^i}-\frac{\partial\mathcal{L}}{\partial
q^i}=0\,.\end{eqnarray} Let us consider also the vector field
(\ref{05}). Contracting (\ref{07}) with the $\alpha^i$'s gives

\begin{eqnarray}\label{06a}\alpha^i\biggl(\frac{d}{d\lambda}\frac{\partial
\mathcal{L}}{\partial\dot{q}^i}-\frac{\partial\mathcal{L}}{\partial
q^i}\biggr)=0\,.\end{eqnarray} Being

\begin{eqnarray}\label{06b}\alpha^i\frac{d}{d\lambda}\frac{\partial
\mathcal{L}}{\partial\dot{q}^i}=\frac{d}{d\lambda}\biggl(\alpha^i\frac{\partial\mathcal{L}}{\partial\dot{q}^i}\biggr)-
\biggl(\frac{d\alpha^i}{d\lambda}\biggr)\frac{\partial\mathcal{L}}{\partial\dot{q}^i}\,,\end{eqnarray}
from (\ref{06a}), we obtain

\begin{eqnarray}\label{08}\frac{d}{d\lambda}\biggl(\alpha^i\frac{\partial\mathcal{L}}{\partial\dot{q}^i}\biggr)=
L_{\textbf{X}}\mathcal{L}\,.\end{eqnarray} The immediate
consequence is the \emph{Noether Theorem}\footnote{In the
following, with abuse of notation, we shall write $\textbf{X}$
instead of $\textbf{X}^c$, whenever no confusion is possible.}:

\vspace{3. mm}

\noindent If $L_{\textbf{X}}\mathcal{L}=0$, then the function

\begin{equation}\label{09}\Sigma_{0}=\alpha^i\frac{\partial\mathcal{L}}{\partial\dot{q}^i}\,,\end{equation}
is a constant of motion.

\vspace{3. mm}

\noindent \textbf{Remark.} Eq.(\ref{09}) can be expressed
independently of coordinates as a contraction of $\textbf{X}$ by a
Cartan one-form

\begin{eqnarray}\label{09a}\theta_\mathcal{L}\doteq\frac{\partial\mathcal{L}}{\partial\dot{q}^i}dq^i\,.\end{eqnarray}
For a generic vector field $\textbf{Y}=y^i\partial/\partial x^i$,
and one-form $\beta=\beta_idx^i$, we have, by definition,
$i_{\textbf{Y}}\beta=y^i\beta_i$. Thus Eq.(\ref{09}) can be
written as

\begin{eqnarray}\label{09b}i_\textbf{X}\theta_\mathcal{L}=\Sigma_{0}\,.\end{eqnarray}
By a point-transformation, the vector field $\textbf{X}$ becomes

\begin{eqnarray}\label{09c}\widetilde{\textbf{X}}=i_\textbf{X}dQ^k\frac{\partial}{\partial Q^k}+\biggl[\frac{d}{d\lambda}(i_x
dQ^k)\biggr]\frac{\partial}{\partial\dot{Q}^k}\,.\end{eqnarray} We
see that $\widetilde{\textbf{X}}'$ is still the lift of a vector
field defined on the "space of positions". If $\textbf{X}$ is a
symmetry and we choose a point transformation  such  that

\begin{eqnarray}\label{010}i_\textbf{X}dQ^1=1\, ; \,\,\, i_\textbf{X}dQ^i=0
\,\,\, i\neq1\,,\end{eqnarray} we get

\begin{eqnarray}\label{010a} \widetilde{\textbf{X}}=\frac{\partial}{\partial Q^1}
\,;\,\,\,\,\frac{\partial\mathcal{L}}{\partial
Q^1}=0\,.\end{eqnarray} Thus $Q^1$ is a cyclic coordinate and the
dynamics can be reduced \cite{arnold, mar-sal-sim-vit}.

\noindent\textbf{Remarks}:
\begin{enumerate}
\item The change of coordinates defined by (\ref{010}) is not unique. Usually
a clever choice is very important.
\item In general, the solution of (\ref{010}) is not well defined
on the whole space. It is \emph{local} in the sense explained
above.
\item It is possible that more than one $\textbf{X}$ is found,
say for instance $\textbf{X}_1$, $\textbf{X}_2$. If they commute,
i.e. $[\textbf{X}_1, \textbf{X}_2]=0$, then it is possible to
obtain two cyclic coordinates by solving the system

\begin{eqnarray}i_{\textbf{X}_1} dQ^1=1; \,\,\, i_{\textbf{X}_2} dQ^2 = 1;
\,\,\, i_{\textbf{X}_1} dQ^i = 0; \,\,\, i \neq 1;\,
i_{\textbf{X}_2} dQ^i = 0; \,\,\, i \neq 2\,. \end{eqnarray} The
transformed fields will be $\partial/\partial Q^{1}$,
$\partial/\partial Q^{2}$. If they do not commute, this procedure
is clearly not applicable, since commutation relations are
preserved by diffeomorphisms. Let us note that
$\textbf{X}_3=[\textbf{X}_1, \textbf{X}_2]$ is also a symmetry,
indeed, being
$L_{\textbf{X}_3}\mathcal{L}=L_{\textbf{X}_1}L_{\textbf{X}_2}\mathcal{L}-
L_{\textbf{X}_2}L_{\textbf{X}_1}\mathcal{L}=0$. If $\textbf{X}_3$
is independent of $\textbf{X}_1$, $\textbf{X}_2$, we can go on
until the vector fields close the Lie algebra. The usual way to
treat this situation is to make a Legendre transformation, going
to the Hamiltonian formalism and to a Lie algebra of Poisson
brackets. If we look for a reduction with cyclic coordinates, this
procedure is possible in the following way:
\begin{itemize}
\item we arbitrarily choose  one of the symmetries,
or a linear combination of them, and get new coordinates as above.
After the reduction, we get a new Lagrangian
$\widetilde{\mathcal{L}}(\textbf{Q})$;
\item we search again for symmetries in this new space, make a new reduction
and so on until possible;
\item if the search fails, we try again with another of the existing
symmetries.
\end{itemize}
\end{enumerate}
Let us now assume that $\mathcal{L}$ is of the form (\ref{02}). As
$\textbf{X}$ is of the form (\ref{05}), $L_\textbf{X}\mathcal{L}$
will be a homogeneous polynomial of second degree in the
velocities plus a inhomogeneous term in the $q^i$. Since such a
polynomial has to be identically zero, each coefficient must be
independently zero. If $n$ is the dimension of the configuration
space, we get $1+n(n+1)/2$ partial differential equations (PDE).
The system is overdetermined, therefore, if any solution exists,
it will be expressed in terms of integration constants instead of
boundary conditions. It is also obvious that an overall constant
factor in the Lie vector $\textbf{X}$ is irrelevant. In other
words, the Noether Symmetry Approach can be used to select
functions which assign the models and, as we shall see below, such
functions (and then the models) can be physically relevant. This
fact justifies the method at least \emph{a posteriori}.

\section{The Noether Approach for $f$ $-$ gravity in spherical symmetry}

Since the above considerations, if one assumes the spherical
symmetry, the role of the {\it affine parameter} is  played by the
coordinate radius $r$.  In this case, the configuration space is
given by $\mathcal{Q}=\{A, M, R\}$ and the tangent space by
$\mathcal{TQ}=\{A, A', M, M', R, R'\}$. On the other hand,
according to the Noether theorem, the existence of a symmetry for
dynamics described by the Lagrangian (\ref{lag2})  implies a
constant of motion. Let us apply the Lie derivative to the
(\ref{lag2}), we have\footnote{From now on, $\underline{q}$
indicates the vector $(A,M,R)$.}\,:

\begin{eqnarray}
L_{\mathbf{X}}{\bf L}\,=\,\underline{\alpha}\cdot\nabla_{q}{\bf L
}+\underline{\alpha}'\cdot\nabla_{q'}{\bf L}
=\underline{q}'^t\biggl[\underline{\alpha}\cdot\nabla_{q}\hat{{\bf
L }}+ 2\biggl(\nabla_{q}\alpha\biggr)^t\hat{{\bf L
}}\biggr]\underline{q}'\,,
\end{eqnarray}
which vanish if the functions ${\underline{\alpha}}$ satisfy the
following system

\begin{eqnarray}\label{sys}
\underline{\alpha}\cdot\nabla_{q}\hat{{\bf L}}
+2(\nabla_{q}{\underline{\alpha}})^t\hat{{\bf L
}}\,=\,0\,\longrightarrow\ \ \ \ \alpha_{i}\frac{\partial
\hat{{\bf L}}_{km}}{\partial
q_{i}}+2\frac{\partial\alpha_{i}}{\partial q_{k}}\hat{{\bf L
}}_{im}=0\,.
\end{eqnarray}
The system (\ref{sys}) assumes the following explicit form

\begin{eqnarray}\label{sys-complete}
\left\{\begin{array}{ll}
\Upsilon\biggl(\frac{\partial\alpha_2}{\partial
A}f_{R}+M\frac{\partial\alpha_3}{\partial A}f_{RR}\biggr)=0
\\\\
\frac{A}{M}\biggl[(2+MR)\alpha_3
f_{RR}-\frac{2\alpha_2}{M}f_{R}\biggr]f_{R}+\\
\,\,\,\,\,\,\,\,\,\,\,\,\,\,\,\,\,\,\,\,\,\,\,\,\,\,+\Upsilon\biggl[\biggl(\frac{\alpha_1}{M}+2\frac{\partial\alpha_1}
{\partial M}+\frac{2A}{M}\frac{\partial\alpha_2}{\partial
M}\biggr)f_{R}+A\biggl(\frac{\alpha_3}{M}+4\frac{\partial \alpha_3
}{\partial M}\biggr)f_{RR}\biggr]=0
\\\\
\Upsilon\biggl(M\frac{\partial \alpha_1}{\partial
R}+2A\frac{\partial \alpha_2}{\partial R}\biggr)f_{RR}=0
\\\\
\alpha_2(f-Rf_{R})f_{R}-\Upsilon\biggl[\biggl(\alpha_3+M\frac{\partial\alpha_3}
{\partial M}+2A\frac{\partial\alpha_3}{\partial
A}\biggr)f_{RR}+\\
\,\,\,\,\,\,\,\,\,\,\,\,\,\,\,\,\,\,\,\,\,\,\,\,\,\,+\biggl(\frac{\partial\alpha_2}{\partial
M}+\frac{\partial\alpha_1}{\partial A}+\frac{A}{M}\frac{\partial
\alpha_2 }{\partial A}\biggr)f_{R}\biggr]=0
\\\\
\biggl[M(2+MR)\alpha_3 f_{RR}-2\alpha_2 f_R\biggr]f_{RR}+\Upsilon\biggl[f_R\frac{\partial\alpha_2}{\partial R}+\\
\,\,\,\,\,\,\,\,\,\,\,\,\,\,\,\,\,\,\,\,\,\,\,\,\,\,+\biggl(2\alpha_2+M\frac{\partial\alpha_1}{\partial
A}+2 A\frac{\partial\alpha_2}{\partial A}+M\frac{\partial
\alpha_3}{\partial R}\biggr)\alpha_3 f_{RR}+Mf_{RRR}\biggr]=0
\\\\
2A[(2+MR)\alpha_3
f_{RR}-(f-Rf_{R})\alpha_2]f_{RR}+\Upsilon\biggl[\biggl(\frac{\partial\alpha_1}{\partial
R}+\frac{A}{M}\frac{\partial\alpha_2}{\partial
R}\biggr)f_{R}+\\
\,\,\,\,\,\,\,\,\,\,\,\,\,\,\,\,\,\,\,\,\,\,\,\,\,\,+\biggl(2\alpha_1
+2A\frac{\partial\alpha_3}{\partial R}+M\frac{\partial\alpha_1}
{\partial M}+2A\frac{\partial\alpha_2}{\partial
M}\biggr)f_{RR}+2A\alpha_3 f_{RRR}\biggr]=0
\end{array} \right.
\end{eqnarray}
where $\Upsilon\,=\,(2+MR)f_R-Mf$. Solving the system
(\ref{sys-complete}) means to find out the functions $\alpha_{i}$
which assign the Noether vector. However the system (\ref{sys})
implicitly depends on the form of $f$ and then, by solving it, we
get also $f$ theories compatible with spherical symmetry. On the
other hand, by choosing the $f$ form, we can explicitly solve
(\ref{sys}). As an example, one finds that the system (\ref{sys})
is satisfied if we chose

\begin{eqnarray}\label{solsy}f\,=\,f_0 R^s\ \ \ \ \  \underline{\alpha}=(\alpha_1,\alpha_2,\alpha_3)=
\biggl((3-2s)kA,\ -kM,\ kR\biggr)\,
\end{eqnarray}
with $s$ a real number, $k$ an integration constant and $f_0$ a
dimensional coupling constant\footnote{The dimensions are given by
$R^{1-s}$ in term of the Ricci scalar. For the sake of simplicity
we will put $f_0=1$ in the forthcoming discussion.}. This means
that for any $f=R^s$ exists, at least, a Noether symmetry and a
related constant of motion $\Sigma_{0}$\,:

\begin{eqnarray}\label{cm}\nonumber
\Sigma_{0}\,&=&\,\underline{\alpha}\cdot\nabla_{q'}{\bf
L}=\nonumber\\&&=2
skMR^{2s-3}[2s+(s-1)MR][(s-2)RA'-(2s^2-3s+1)AR']\,.\end{eqnarray}
A physical interpretation of $\Sigma_{0}$ is possible if one gives
an interpretation of this quantity in GR. In such a case, with
$s=1$, the above procedure has to be applied to the Lagrangian
(\ref{lag2gr}). We obtain the solution

\begin{eqnarray}\label{solsygr}\underline{\alpha}_{GR}=(-kA,\
kM)\,.\end{eqnarray} The functions $A$ and $M$ give the
Schwarzschild solution (\ref{schsol}), and then the constant of
motion acquires the form

\begin{eqnarray}\label{cmgr}\Sigma_{0}=r_g\,.\end{eqnarray}
In other words, in the case of Einstein gravity, the Noether
symmetry gives as a conserved quantity the Schwarzschild radius or
the mass of the gravitating system. Another solution can be find
out for $R=R_0$ where $R_0$ is a constant. In this case, the field
equations (\ref{fe}) reduce to

\begin{eqnarray}\label{fe1}R_{\mu\nu}+k_0 g_{\mu\nu}=0\,,\end{eqnarray}
where $k_0=-\frac{1}{2}f(R_0)/f_R(R_0)$. The general solution is

\begin{eqnarray}\label{scwdes}A(r)=\frac{1}{B(r)}=1+\frac{k_0}{r}+\frac{R_0}{12}r^2\,,\qquad M=r^2\end{eqnarray}
with the special case

\begin{eqnarray}A(r)=\frac{1}{B(r)}=1+\frac{k_0}{r}\,,\qquad M=r^2\,,\qquad R=0\,.\end{eqnarray}
The solution (\ref{scwdes}) is the well known Schwarzschild-de
Sitter one which is a solution in most of modified gravity
theories. It evades the Solar System constraints due to the
smallness of the effective cosmological constant. However, other
spherically symmetric solutions, different from this, are more
significant for Solar System tests. In the general case
$f\,=\,R^s$, the Lagrangian (\ref{lag2}) becomes

\begin{eqnarray}\textbf{L}&=&\frac{sR^{2s-3}[2s+(s-1)MR]}{M}\nonumber\\&&
\times[2(s-1)M^2A'R'+2MRM'A'+4(s-1)AMM'R'+ARM'^2]\,,\end{eqnarray}
and the expression (\ref{eqb}) for $B$ is

\begin{eqnarray}B=\frac{s[2(s-1)M^2A'R'+2MRM'A'+4(s-1)AMM'R'+ARM'^2]}{2AMR[2s+(s-1)MR]}\end{eqnarray}
As it can be easily checked, GR is recovered when $s=1$. Using the
constant of motion (\ref{cm}),  we solve in term of $A$ and obtain

\begin{eqnarray}A=R^{\frac{2s^2-3s+1}{s-2}}\biggl\{k_1+\Sigma_{0}\int\frac{R^{\frac{4s^2-9s+5}{2-s}}dr}{2ks(s-2)
M[2s+(s-1)MR]}\biggr\}\end{eqnarray} for $s\neq2$, with $k_1$ an
integration constant. For $s\,=\,2$, one finds

\begin{eqnarray}A=-\frac{\Sigma_{0}}{12kr^2(4+r^2R)RR'}\,.\end{eqnarray}
These relations allow to find out general solutions for the field
equations giving the dependence of the Ricci scalar on the radial
coordinate $r$. For example, a solution is found for
\begin{eqnarray}
s=5/4\,,\ \ \ \ M=r^2\,,\ \ \ \ R= 5 r^{-2}\,,
\end{eqnarray}
obtaining  the spherically symmetric metric
\begin{eqnarray}ds^2=\frac{1}{\sqrt{5}}(k_2+k_1 r)dt^2-
\frac{1}{2}\biggl(\frac{1}{1+\frac{k_2}{k_1r}}\biggr)dr^2-r^2d\Omega\,,\end{eqnarray}
with $k_2=\frac{32\Sigma_0}{225 k}$. It is worth noting that such
exact solution is in the range of $s$ values ruled out by Solar
System observations, as pointed out in \cite{cli-bar, cli-bar1,
cli-bar2}.

\section{Perspectives of Noether symmetries approach}

In this chapter, we have discussed a general method to find out
exact solutions in Extended Theories of Gravity when a spherically
symmetric background is taken into account. In particular, we have
searched for exact spherically symmetric solutions in $f$-gravity
by asking for the existence of Noether symmetries. We have
developed a general formalism and  given some examples of exact
solutions. The procedure consists in: $i)$ considering the
point-like $f$ Lagrangian where spherical symmetry has been
imposed; $ii)$ deriving the Euler-Lagrange equations; $iii)$
searching for a Noether vector field; $iv)$ reducing dynamics and
then integrating the equations of motion using conserved
quantities. Viceversa, the approach allows also to select families
of $f$ models where a particular symmetry (in this case the
spherical one) is present. As examples, we discussed power law
models and models with constant Ricci curvature scalar. However,
the above method can be further generalized. If a symmetry exists,
the Noether Approach allows, as discussed in \S\,
\ref{noether-symmetry-approach}, transformations of variables
where the cyclic ones are evident. This fact allows to reduce
dynamics and then  to get more easily exact solutions. For
example, since we know that $f\,=\,R^s$\,-\,gravity admit a
conserved quantity, a coordinate transformation can be induced by
the Noether symmetry. We  ask for the coordinate transformation\,:

\begin{eqnarray}
\textbf{L}=\textbf{L}(\underline{q}, \underline{q}')=\textbf{L}(A,
M, R, A', M',
R')\rightarrow\widetilde{\textbf{L}}=\widetilde{\textbf{L}}(\widetilde{M},
\widetilde{R}, \widetilde{A}', \widetilde{M}', \bar{R}')\,,
\end{eqnarray}
for the Lagrangian (\ref{lag2}), where the Noether symmetry, and
then the conserved quantity, corresponds to the cyclic variable
$\widetilde{A}$. If more than one symmetry exists, one can find
more than one cyclic variables. In our case, if three Noether
symmetries exist, we can transform the Lagrangian $\textbf{L}$ in
a Lagrangian with three cyclic coordinates, that is
$\widetilde{A}=\widetilde{A}({\underline{q}})$,
$\widetilde{M}=\widetilde{M}({\underline{q}})$ and
$\widetilde{R}=\widetilde{R}({\underline{q}})$ which are function
of the old ones. These new functions have to  satisfy the
following system

\begin{eqnarray}\label{sys2}
\left\{\begin{array}{ll}
(3-2s)A\frac{\partial\widetilde{A}}{\partial
A}-M\frac{\partial\widetilde{A}}{\partial
M}+R\frac{\partial\widetilde{A}}{\partial R}=1\,,
\\\\
(3-2s)A\frac{\partial\widetilde{q}_i}{\partial
A}-M\frac{\partial\widetilde{q}_i}{\partial
M}+R\frac{\partial\widetilde{q}_i}{\partial R}\,=\,0\,,
\end{array} \right.
\end{eqnarray}
with $i=2,3$ (we have put $k=1$). A solution of (\ref{sys2}) is
given by the set (for $s\neq 3/2$)

\begin{eqnarray}\label{so1}
\left\{ \begin{array}{ll} \widetilde{A}=\frac{\ln
A}{(3-2s)}+F_{A}(A^{\frac{\eta_A}{3-2s}}M^{\eta_A},A^{\frac{\xi_A}{2s-3}}M^{\xi_A})
\\\\
\widetilde{q}_i=F_{i}(A^{\frac{\eta_i}{3-2s}}M^{\eta_i},A^{\frac{\xi_i}{2s-3}}M^{\xi_i})
\end{array} \right.
\end{eqnarray}
and if $s=3/2$

\begin{eqnarray}\label{so2}
\left\{\begin{array}{ll} \widetilde{A}=-\ln M+F_{A}(A)G_A(MR)
\\\\
\widetilde{q}_i=F_{i}(A)G_i(MR)
\end{array} \right.
\end{eqnarray}
where $F_A$, $F_i$, $G_A$ and $G_i$ are arbitrary functions  and
$\eta_A$, $\eta_i$, $\xi_A$ and $\xi_i$ integration constants.

These considerations show that the Noether Symmetries Approach can
be applied to large classes of gravity theories.  Up to now the
Noether symmetries Approach has been  worked out in the case of
FRW\,-\,metric. In this chapter, we have concentrated our
attention to the development of the general formalism in the case
of spherically symmetric spacetimes. Therefore the fact that, even
in the case of a spherical symmetry, it is possible to achieve
exact solutions seems to suggest that this technique can represent
a paradigmatic approach to work out exact solutions in any theory
of gravity. At this stage, the systematic search for exact
solution is well beyond the aim of this thesis. A more
comprehensive analysis in this sense will be the argument of
forthcoming studies. The results presented in this chapter point
out that it does not hold in general for the specific $f$ theories
considered. However, the above technique could be a good approach
to select suitable classes of theories where such a theorem holds.

\clearpage{\pagestyle{empty}\cleardoublepage}

\chapter{$f$ $-$ gravity and scalar-tensor gravity: affinities and differences}\label{PPN-TS-fR-theory}

In the last years a very strong debate has been pursued about the
Newtonian limit of HOG models. According to some authors the
Newtonian limit of $f$ - gravity is equivalent to the one of
Brans-Dicke gravity with $\omega_{BD}\,=\,0$, so that the PPN
parameters of these models turn out to be ill defined. In this
chapter we show that this is indeed not true. We discuss that HOG
models are dynamically equivalent to a O'Hanlon Lagrangian which
is a special case of Scalar-tensor theory characterized by a
self-interaction potential and that, in the low energy and small
velocity limit, this will imply a non-standard behaviour
[\textbf{H}]. This result turns out to be completely different
from the one of a pure Brans-Dicke model and in particular
suggests that it is completely misleading to consider the PPN
parameters of this theory with $\omega_{BD}\,=\,0$ in order to
characterize the homologous quantities of $f$-gravity.

By using the definition of the PPN-parameters $\gamma$ and $\beta$
(\ref{schwarz-isotropic-PPN}) in term of $f$-theories, we show
that a family of third-order polynomial theories, in the Ricci
scalar $R$, turns out to be compatible with the PPN - limit and
the deviation from GR, theoretically predicted, can agree with
experimental data [\textbf{A}].

\section{PPN $-$ parameters in Scalar $-$ Tensor and Fourth Order Gravity}

If one takes into account a more general theory of gravity, the
calculation of the PPN - limit can be performed following a well
defined pipeline shown in the \S\,\ref{post-new-new-formalism}
which straightforwardly generalizes the standard GR. A significant
development in this sense has been pursued by Damour and Esposito
- Farese \cite{dam-esp, dam-esp1, dam-esp2, dam-esp3} which have
approached to the calculation of the PPN-limit of scalar - tensor
gravity by means of a conformal transformation (see
\S\,\ref{conformal-transformation-general-approach}) to the
standard Einstein frame. This scheme provides several interesting
results up to obtain an intrinsic definition of $\gamma,\,\beta$
in term of the non - minimal coupling function $F(\phi)$. The
analogy between scalar - tensor gravity and higher order theories
of gravity has been widely investigated \cite{tey-tou, schmidt1,
wands}.

Starting from this analogy, the PPN results for scalar - tensor
gravity can be extended to HOG \cite{cap-tro}. In fact,
identifying $\phi\,\rightarrow\,R$ \cite{wands}, it is possible to
extend the definition of the scalar-tensor PPN - parameters
\cite{dam-esp, sch-uza-ria} to the case of HOG\,:

\begin{eqnarray}\label{ppn-R1}
\gamma-1=-\frac{{f''}^2}{f'+2f''^2}\,, \qquad
\beta-1=\frac{1}{4}\left(\frac{f'\cdot
f''}{2f'+3f''^2}\right)\frac{d\gamma}{dR}.
\end{eqnarray}
In  \cite{cap-tro}, these definitions have been confronted with
the observational upper limits on $\gamma$ and $\beta$ coming from
Mercury Perihelion Shift \cite{shapiro} and Very Long Baseline
Interferometry \cite{shapiro1}. Actually, it is possible to show
that data and theoretical predictions from (\ref{ppn-R1}) agree in
the limits of experimental measures for several classes of fourth
order theories. Such a result tells us that extended theories of
gravity are not ruled out from Solar System experiments but a more
careful analysis of theories against experimental limits has to be
performed. A possible procedure could be to link the analytic form
of a generic fourth order theory with experimental data. In fact,
the matching between data and theoretical predictions, found in
\cite{cap-tro}, holds provided some restrictions for the model
parameters but gives no general constraints on the theory. In
general, the function $f$ could contain an infinite number of
parameters (\emph{i.e.} it can be conceived as an infinite  power
series \cite{schmidt1}) while, on the contrary, the number of
useful relations  is finite (in our case we have only two
relations). An attempt to deduce the form of the gravity
Lagrangian can be to consider the relations (\ref{ppn-R1}) as
differential equations for $f$, so that, taking into account the
experimental results, one could constrain, in principle, the model
parameters by the measured values of $\gamma$ and $\beta$. This
hypothesis is reasonable if the derivatives of $f$ function are
smoothly evolving with the Ricci scalar. Formally, one can
consider the r.h.s. of the definitions (\ref{ppn-R1}) as
differential relations which have to be matched with values of PPN
- parameters. In other words, one has to solve the equations
(\ref{ppn-R1}) where $\gamma$ and $\beta$ are two parameters.
Based on such an assumption, on can try to derive the largest
class of $f$ - theories compatible with experimental data. In
fact, by the integration of (\ref{ppn-R1}), one obtains a solution
parameterized  by $\beta$ and $\gamma$ which have to be confronted
with the experimental quantities $\beta_{exp}$ and $\gamma_{exp}$.

Assuming  $f'+2f''^2\neq 0$ and defining
$A\,=\,\Bigl|\frac{1-\gamma}{2\gamma-1}\Bigl|$ we obtain from
(\ref{ppn-R1}) a differential equation for $f$:

\begin{eqnarray}
f''^2\,=\,A\,f'\,.
\end{eqnarray}
The general solution of such an equation is a third order
polynomial $f\,=\,a\,R^3+b\,R^2+c\,R+d$ whose coefficients have to
satisfy the conditions\,: $a\,=\,b\,=\,c\,=\,0$ and $d \neq 0$
(trivial solution) or $a\,=\,\frac{A}{12}$ and
$b\,=\,\pm\,\frac{\sqrt{A\, c}}{2}$, with $c$, $d$ $\neq 0$. Thus,
the general solution for the non-trivial case, in natural units,
reads

\begin{eqnarray}\label{ger-sol1}
f\,=\,\frac{1}{12}\Bigl|\frac{1-\gamma}{2\gamma-1}\Bigl|R^3\pm\frac{\sqrt{c}}{2}
\sqrt{\Bigl|\frac{1-\gamma}{2\gamma-1}\Bigl|}R^2+cR+d\,.
\end{eqnarray}
It is evident that the integration constants $c$ and $d$ have to
be compatible with GR prescriptions and, eventually, with the
presence of a cosmological constant. Indeed, when $\gamma
\rightarrow 1$, which implies $f \rightarrow c\,R+d$, the
GR\,-\,limit is recovered. As a consequence the values of these
constants remain fixed ($c$\,=\,1 and $d$\,=\,$\Lambda$, where
$\Lambda$ is the cosmological constant). Therefore, the fourth
order theory provided by (\ref{ger-sol1}) becomes

\begin{eqnarray}\label{ger-sol2}
f_{\pm}\,=\,\frac{1}{12}\Bigl|\frac{1-\gamma}{2\gamma-1}\Bigl|R^3\pm
\frac{1}{2}\sqrt{\Bigl|\frac{1-\gamma}{2\gamma-1}\Bigl|}R^2+R+\Lambda\,,
\end{eqnarray}
where we have formally displayed the two branch form of the
solution depending on the sign of the coefficient entering the
second order term. Since the constants $a$, $b$, $c$, $d$ of the
general solution satisfy the relation $3\,a\,c-b^2\,=\,0$, one can
easily verify that it gives\,:

\begin{eqnarray}
\frac{d\gamma}{dR}\Bigl|_{f_{\pm}}=-\frac{d}{dR}\frac{f''^2}{f'+2f''^2}\Bigl|_{f_{\pm}}\,=\,0\,,
\end{eqnarray}
where the subscript $_{f_{\pm}}$ refers the calculation to the
solution (\ref{ger-sol2}). This result, compared with the second
differential equation (\ref{ppn-R1}), implies $4(\beta-1)=0\,,$
which means the compatibility of the solution even with this
second relation.

\section{Comparing with experimental measurements}

Up to now we have discussed a family of fourth order theories
(\ref{ger-sol2}) parameterized by the PPN - quantity $\gamma$; on
the other hand, for this class of Lagrangians, the parameter
$\beta$ is compatible with GR value being unity.

\begin{table}[ht]
\centering
\begin{tabular}{l|c}
\hline\hline\hline
  Mercury Perihelion Shift&
  $|2\gamma-\beta-1|<3\times10^{-3}$ \\\hline
  Lunar Laser Ranging &  $4\beta-\gamma-3\,=\,-(0.7\pm 1)\times{10^{-3}}$ \\\hline
  Very Long Baseline Interf. &  $|\gamma -1|\,=\,4\times10^{-4}$ \\\hline
  Cassini Spacecraft &  $\gamma-1\,=\,(2.1\pm 2.3)\times10^{-5}$ \\
\hline\hline\hline
\end{tabular}
\caption{\label{ppn} A schematic resume of recent experimental
constraints on the PPN - parameters. They are the perihelion shift
of Mercury \cite{shapiro}, the Lunar Laser Ranging
\cite{williams}, the upper limit coming from the Very Long
Baseline Interferometry \cite{shapiro1} and the results obtained
by the estimate of the Cassini spacecraft delay into the radio
waves transmission near the Solar conjunction \cite{ber-ies-tor}.}
\end{table}

\begin{figure}[htbp]
\centering
  \includegraphics[width=8cm]{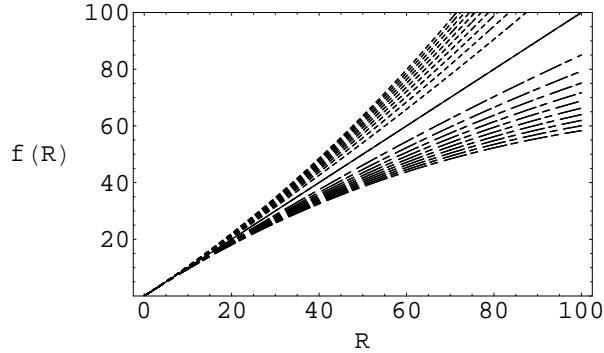}\\
  \caption{Plot of the two branch solution provided in (\ref{ger-sol2}). The $f_+$ (dotted line) branch
  family is up to GR solution (straight line), while the one indicated with $f_-$ (dotted-dashed line) remains below
  this line. The different plots for each family refer to different values of
  $\gamma$ fulfilling the condition $|\gamma-1|\leq 10^{-4}$
  and increased by step of $10^{-5}$.}
\label{fig1}
\end{figure}

Now, the further step directly characterizes such a class of
theories by means of the experimental estimates of $\gamma$. In
particular, by fixing $\gamma$ to its observational estimate
$\gamma_{exp}$, we will obtain the weight of the coefficients
relative to each of the non-linear terms in the Ricci scalar of
the Lagrangian (\ref{ger-sol2}). In such a way, since GR
predictions require exactly $\gamma_{exp}\,=\,\beta_{exp}\,=\,1$,
in the case of HOG, one could to take into account small
deviations from this values as inferred from experiments. Some
plots can contribute to the discussion of this argument. In figure
\ref{fig1}, the Lagrangian (\ref{ger-sol2}) is plotted. It is
parameterized for several values of $\gamma$ compatible with the
experimental bounds coming from the Mercury perihelion shift (see
Table 1 and \cite{shapiro}). The function is plotted in the range
$R\geq 0$. Since the property $f_+(R)\,=\,-f_-(-R)$ holds for the
function (\ref{ger-sol2}), one can easily recover the shape of the
plot in the negative region. As it is reasonable, the deviation
from GR becomes remarkable when scalar curvature is large.

\begin{figure}[htbp]
\centering
  \includegraphics[width=8cm]{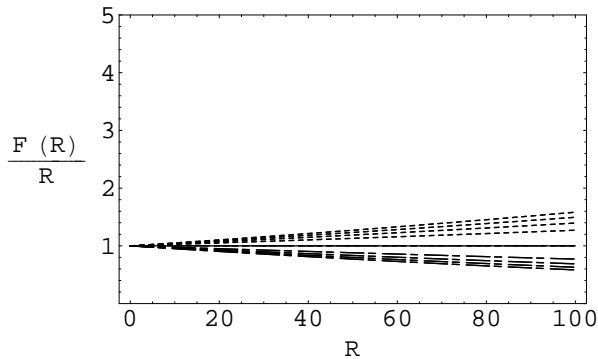}\\
  \caption{The ratio $f/R$. It is shown the deviation of the HOG from GR considering the PPN - limit.
  Dotted and dotted-dashed lines refer to the $f_{+}$ and $f_{-}$ branches plotted with respect to several values
  of $\gamma$ (the step in this case is $2.5\times{10^{-5}}$). }
  \label{fig2}
\end{figure}

\begin{figure}[htbp]
\centering
  \includegraphics[scale=1]{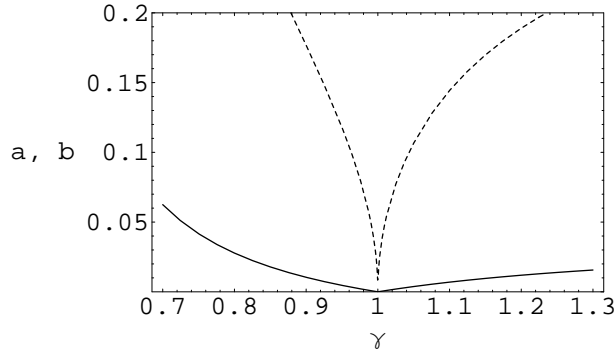}\\
  \caption{Plot of the modulus of coefficients: $a$ ($R^3$) (line) and $b$ ($R^2$) (dashed line). The choice
  of plotting the modulus of coefficients is the consequence of two solution for $f$.}
  \label{fig3}
\end{figure}

In order to display the differences between the theory
(\ref{ger-sol2}) and  Hilbert-Einstein one, the ratio $f/R$ is
plotted in figure \ref{fig2}. Again it is evident that the two
Lagrangians differ significantly for great values of the curvature
scalar. It is worth noting that the formal difference between the
PPN - inspired Lagrangian and the GR expression can be related to
the physical meaning of the parameter $\gamma$ which is the
deviation from the  Schwarzschild - like solution. It measures the
spatial curvature of the region which one is investigating, then
the deviation  from the local flatness can be due to the influence
of higher order contributions in Ricci scalar. On the other hand,
one can reverse the argument and notice that if such a deviation
is measured, it can be recast in the framework of HOG, and in
particular its ``amount" indicates the deviation from GR.
Furthermore, it is worth considering that, in the expression
(\ref{ger-sol2}), the modulus  of the coefficients in $\gamma$
(i.e. the strength of the term) decreases by increasing the degree
of $R$. In particular, the highest values of cubic and squared
terms in $R$ are, respectively, of order $10^{-4}$ and $10^{-2}$
(see figure \ref{fig3}) then GR remains a viable theory at short
distances (i.e. Solar System) and low curvature regimes.

A remark is in order at this point. The class of theories which we
have discussed is a third order function  of the Ricci scalar $R$
parameterized by the experimental values of the PPN parameter
$\gamma$. In principle, any  analytic $f$ can be compared with the
Lagrangian (\ref{ger-sol2}) provided suitable values of the
coefficients. However, more general results can be achieved
relaxing the condition $\beta\,=\,1$ which is an intrinsic feature
for (\ref{ger-sol2}) (see for example \cite{cap-tro}). These
considerations suggest to take into account, as physical theories,
functions of the Ricci scalar which slightly deviates from GR,
i.e. $f\,=\,f_0R^{(1+\epsilon)}$ with $\epsilon$ a small parameter
which indicates how much the theory deviates from GR
\cite{cli-bar}. In fact, supposing $\epsilon$ sufficiently small,
it is possible to approximate this expression

\begin{eqnarray}\label{fReps}
|R|^{(1+\epsilon)}\simeq
|R|\biggl(1+\epsilon\ln|R|+\frac{\epsilon^2\ln^2|R|}{2}+\dots\biggr)\,.
\end{eqnarray}
This  relation can be easily confronted with the solution
(\ref{ger-sol2}) since, also in this case, the corrections have
very small ``strength".

We can conclude this paragraph having shown how a polynomial
Lagrangian in the Ricci scalar $R$, compatible with the PPN -
limit, can be recovered in the framework of HOG. The  approach is
based on the formulation of the PPN - limit of such gravity models
developed in analogy with scalar-tensor gravity \cite{cap-tro}. In
particular, considering the local relations defining the PPN
fourth order parameters as differential expressions, one obtains a
third-order polynomial in the Ricci scalar which is parameterized
by the PPN - quantity $\gamma$ and compatible with the limit
$\beta\,=\,1$. The order of deviation from the linearity in $R$ is
induced by the deviations of $\gamma$ from the GR expectation
value $\gamma\,=\,1$. Actually, the PPN parameter $\gamma$ may
represent the  key parameter to discriminate among relativistic
theories of gravity. In particular, this quantity should be
significatively tested at Solar System scales by forthcoming
experiments like LATOR \cite{tur-sha-nor}. From a physical point
of view, any analytic function of $R$, by means of its Taylor
expansion, can be compared with (\ref{ger-sol2}). Therefore, a
theory like $f\,=\,f_0R^{(1+\epsilon)}$, indicating small
deviations from standard GR, is in agreement with the proposed
approach, so, in principle, the experimental $\gamma$ could
indicate the value of the parameter $\epsilon$. In conclusion, one
can reasonably state that generic fourth-order gravity models
could be viable candidate theories even in the PPN - limit. In
other words, due to the presented results, they cannot be  a
priori excluded  at Solar System scales.

\section{Newtonian limit of $f$ $-$ gravity by O'Hanlon theroy analogy}

Recently several authors claimed that HOGs and among these, in
particular, HOG models are characterized by an ill defined
behaviour in the Newtonian regime. In particular, in a series of
papers \cite{olm, chi, eri-smi-kam, jin-liu-li, sou-woo,
chi-smi-eri} it is addressed that Post-Newtonian corrections of
the gravitational potential violate experimental constraints since
 these quantities can be recovered by a direct analogy with
Brans-Dicke gravity \cite{bra-dic} simply supposing the Brans -
Dicke characteristic parameter $\omega_{BD}$ vanishing. Actually
despite the calculation of the Newtonian and the post-Newtonian
limit of $f$ - theory, performed in a rigorous manner, have showed
that this is not the case \cite{cap-tro, faraoni4, dick,
nav-van1}, it remains to clarify why the analogy with Brans -
Dicke gravity seems to fail its predictions. The issue is easily
overcame once the correct analogy between $f$-gravity and the
Brans-Dicke model is taken into account.

In literature, it is  suggested that HOG models can be rewritten
in term of a scalar-field Lagrangian non minimally coupled with
gravity but without any kinetic term by fact implying
$\omega_{BD}\,=\,0$ (as shown in the chapter 1). Actually, the
simplest case of scalar - tensor gravity models has been
introduced some decades ago by Brans and Dicke in order to give a
general mechanism capable of explaining the inertial forces by
means of a background gravitational interaction. The explicit
expression of such gravitational action is (\ref{BD-action}),
while the general action of $f$-gravity is (\ref{actionfR}). As
said above, $f$-gravity can be recast as a scalar-tensor theory by
introducing a suitable scalar field $\phi$ which nonminimally
couples with the gravity sector. It is important to remark that
such an analogy holds in a formalism in which the scalar field
displays no kinetic term but is characterized by means of a
self-interaction potential which determines the whole dynamics
(\emph{O'Hanlon Lagrangian}) \cite{tey-tou}. We can resume the
actions as follow

\begin{eqnarray}\label{assdf}
\left\{\begin{array}{ll} \mathcal{A}_{JF}^{f}=\int
d^4x\sqrt{-g}\biggl[f+\mathcal{X}\mathcal{L}_m\biggr]\\\\
\mathcal{A}_{JF}^{BD}=\int d^4x\sqrt{-g}\biggl[\phi
R-\omega_{BD}\frac{\phi_{;\alpha}\phi^{;\alpha}}{\phi}+\mathcal{X}\mathcal{L}_m\biggr]\\\\
\mathcal{A}_{JF}^{OH}=\int d^4x\sqrt{-g}\biggl[\phi
R+V(\phi)+\mathcal{X}\mathcal{L}_m\biggr]
\end{array}\right.
\end{eqnarray}

This consideration, therefore, implies that the scalar field
Lagrangian equivalent to the purely geometrical $f$-gravity turns
out to be quite different with respect to the ordinary Brans -
Dicke definition (\ref{BD-action}). This point represents a
crucial aspect of our analysis. In fact, as we afterwards will
see, such a difference will imply completely different results in
the Newtonian limit of the two models and, consequently, the
impossibility of extend predictions from the PPN approximation of
Brans-Dicke models to $f$-gravity. Considering natural units, the
O'Hanlon Lagrangian \cite{ohanlon} is the third of (\ref{assdf})
or (\ref{ohanlon}). The field equations are obtained by varying
Eq. (\ref{ohanlon}) with respect to both $g_{\mu\nu}$ and $\phi$
which now represent the dynamical variables (the same field
equations are given setting $\omega(\phi)\,=\,0$ and
$F(\phi)\,=\,\phi$ in the (\ref{TSfieldequation})). Thus, one
obtains

\begin{eqnarray}\label{ohanlon-field-equation}
\left\{\begin{array}{ll} \phi
G_{\mu\nu}-\frac{1}{2}V(\phi)g_{\mu\nu}-\phi_{;\mu\nu}+g_{\mu\nu}\Box\phi\,=\,\mathcal{X}\,T_{\mu\nu}\\\\
R+\frac{dV(\phi)}{d\phi}=0\\\\
\Box\phi+\frac{1}{3}\phi\frac{dV(\phi)}{d\phi}-\frac{2}{3}V(\phi)=\frac{\mathcal{X}}{3}T
\end{array}\right.
\end{eqnarray}
where the second line of (\ref{ohanlon-field-equation}) the field
equation for $\phi$. While the third equation is a combination of
the trace of the first one and of the second one. The two schemes
can be mapped one into the other considering the following
equivalences

\begin{eqnarray}\label{TS-fR-relation}
\left\{\begin{array}{ll}
\phi\,=\,f'\\\\
V(\phi)\,=\,f-f'R\\\\
\phi\frac{d V(\phi)}{d\phi}-2V(\phi)\,=\,f'R-2f
\end{array}\right.
\end{eqnarray}
where we are the Jacobian matrix of the transformation
$\phi\,\Longleftrightarrow\,f'$ is non-vanishing. Henceforth we
can consider instead of (\ref{fe}) - (\ref{fetr}) a new set of
field equations determined by the equivalence of $f$-gravity with
the O'Hanlon approach:

\begin{eqnarray}\label{fets}
\left\{\begin{array}{ll} \phi
R_{\mu\nu}+\frac{1}{6}\biggl(V(\phi)+\phi\frac{d V(\phi)}{d
\phi}\biggr)g_{\mu\nu}-\phi_{;\mu\nu}\,=\,\mathcal
{X}\Sigma_{\mu\nu}\\\\
\Box\phi+\frac{1}{3}\biggl(\phi\frac{d V(\phi)}{d
\phi}-2V(\phi)\biggr)=\frac{\mathcal{X}}{3}T
\end{array}\right.
\end{eqnarray}
where $\Sigma_{\mu\nu}=T_{\mu\nu}-\frac{1}{3}Tg_{\mu\nu}$.

Let us, now, calculate the Newtonian limit of Eqs. (\ref{fets}).
Considering the perturbations of the metric tensor $g_{\mu\nu}$
and of the scalar field $\phi$ with respect to a background value,
we search for solution at the O(2) - order in term of the metric
entries and of the scalar field itself (see \S\,
\ref{post-new-new-formalism})\,:

\begin{eqnarray}
  g_{\mu\nu}\sim \begin{pmatrix}
  1+g^{(2)}_{tt} & \vec{0}^T \\
  & \\
  \vec{0} & -\delta_{ij}+g^{(2)}_{ij}
\end{pmatrix}\end{eqnarray}

\begin{eqnarray}\phi\sim\phi^{(0)}+\phi^{(2)}\end{eqnarray}
where we neglected the vectorial component in the metric. The
differential operators turn out to be approximated as

\begin{eqnarray}
\Box\approx\partial^2_t-\Delta\,\,\,\,\,\,\,\,\text{and}\,\,\,\,\,\,\,\,\,\nabla_\mu\nabla_\nu\approx\partial^2_{\mu\nu}\,.
\end{eqnarray}
Actually in order to simplify calculations we can exploit the
intrinsic gauge freedom intrinsic in the metric definition. In
particular, we choose the harmonic gauge (\ref{gauge-harmonic})
and the expressions of Ricci tensor components given by Eqs.
(\ref{PPN-ricci-tensor-HG}). In relation with the adopted
approximation we coherently develop the self-interaction potential
at second order. In particular, the quantities in (\ref{fets})
read\,:

\begin{eqnarray}
\left\{\begin{array}{ll} \phi V(\phi)+\phi\frac{d V(\phi)}{d
\phi}\simeq V(\phi^{(0)})+\phi^{(0)}\frac{d V(\phi^{(0)})}{d
\phi}+\biggr[\phi^{(0)} \frac{d^2V(\phi^{(0)})}{d\phi^2}+2\frac{d
V(\phi^{(0)})}{d\phi}\biggl]\phi^{(2)}\\\\
\phi\frac{d V(\phi)}{d \phi}-2V(\phi)\simeq\phi^{(0)}\frac{d
V(\phi^{(0)})}{d \phi}-2V(\phi^{(0)})+\biggr[\phi^{(0)}
\frac{d^2V(\phi^{(0)})}{d\phi^2}-\frac{d
V(\phi^{(0)})}{d\phi}\biggl]\phi^{(2)}
\end{array}\right.
\end{eqnarray}
The field equations (\ref{fets}), solved at 0-th order of
approximation, provide the two solutions

\begin{eqnarray}
V(\phi^{(0)})=0\,\,\,\,\,\,\,\,\text{and}\,\,\,\,\,\,\,\,\,\frac{dV(\phi^{(0)})}{d\phi}\,=\,0
\end{eqnarray}
which fix the 0-th order terms in the development of the
self-interaction potential; therefore we have

\begin{eqnarray}
\left\{\begin{array}{ll} V(\phi)+\phi\frac{d V(\phi)}{d
\phi}\simeq\phi^{(0)}
\frac{d^2V(\phi^{(0)})}{d\phi^2}\phi^{(2)}\doteq3\lambda^2\phi^{(2)}\\\\
\phi\frac{d V(\phi)}{\delta \phi}-2V(\phi)\simeq\phi^{(0)}
\frac{\delta^2V(\phi^{(0)})}{d\phi^2}\phi^{(2)}\doteq
3\lambda^2\phi^{(2)}
\end{array}\right.
\end{eqnarray}
where constant factors $\phi^{(0)}\frac{d^2 V(\phi^{(0)})}{d
\phi^2}$ have been condensed within the quantity
$3\lambda^2$\footnote{The factor 3 is introduced to simplify an
analogous factor present in the field equations (\ref{fets}).}.
Such a constant can be easily interpreted as a mass term as will
become clearer in the following. Now, taking into account the
above simplifications, we can rewrite field equations (\ref{fets})
at the at O(2) - order in the form\,:

\begin{eqnarray}\label{fets1.2t}
\triangle
g^{(2)}_{tt}=\frac{2\mathcal{X}}{\phi^{(0)}}\Sigma_{tt}^{(0)}-\lambda^2\frac{\phi^{(2)}}{
\phi^{(0)}}\,,
\end{eqnarray}
\begin{eqnarray}\label{fets1.2r}
\triangle
g^{(2)}_{ij}=\frac{2\mathcal{X}}{\phi^{(0)}}\Sigma_{ij}^{(0)}+\lambda^2\frac{\phi^{(2)}}{
\phi^{(0)}}\delta_{ij}+2\frac{\phi^{(2)}_{,ij}}{\phi^{(0)}}
\end{eqnarray}
\begin{eqnarray}\label{fetstr1.2}
\triangle\phi^{(2)}-\lambda^2\phi^{(2)}=-\frac{\mathcal{X}}{3}T^{(0)}\,.
\end{eqnarray}
The scalar field solution can be easily obtained from the
(\ref{fetstr1.2}) as\,:

\begin{eqnarray}
\phi(\mathbf{x})=\phi^{(0)}+\frac{\mathcal{X}}{3}\int\frac{d^3\mathbf{k}}{(2\pi)^{3/2}}\frac{\tilde{T}^{(0)}(\mathbf{k})e^
{i\mathbf{k}\cdot \mathbf{x}}}{\mathbf{k}^2+\lambda^2}
\end{eqnarray}
while for $g^{(2)}_{tt}$ and $g^{(2)}_{ij}$ we have

\begin{eqnarray}
g^{(2)}_{tt}(\mathbf{x})=-\frac{\mathcal{X}}{2\pi\phi^{(0)}}\int
d^3\mathbf{x}'\frac{\Sigma^{(0)}_{tt}(\mathbf{x}')}{|\mathbf{x}-\mathbf{x}'
|}+\frac{\lambda^2}{4\pi\phi^{(0)}}\int
d^3\mathbf{x}'\frac{\phi^{(2)}(\mathbf{x}')}{|\mathbf{x}-\mathbf{x}'|}\,,
\end{eqnarray}
\begin{eqnarray}
g^{(2)}_{ij}(\mathbf{x})=&-&\frac{\mathcal{X}}{2\pi\phi^{(0)}}\int
d^3\mathbf{x}'\frac{\Sigma^{(0)}_{ij}(\mathbf{x}')}{|\mathbf{x}-\mathbf
{x}'|}-\frac{\lambda^2\delta_{ij}}{4\pi\phi^{(0)}}\int
d^3\mathbf{x}'\frac{\phi^{(2)}(\mathbf{x}')}{|\mathbf{x}-\mathbf{x}'|}\nonumber\\\nonumber\\&+&\frac{2}{\phi^{(0)}}\biggl
[\frac{x_ix_j}{\mathbf{x}^2}\phi^{(2)}(\mathbf{x})+\biggl(\delta_{ij}-\frac{3x_ix_j}{\mathbf{x}^2}\biggr)\frac{1}{|\mathbf{x}|
^3}\int_0^{|\mathbf{x}|}d|\mathbf{x}'||\mathbf{x}|'^2\phi^{(2)}(\mathbf{x}')\biggr]\,.
\end{eqnarray}
The above three solutions represent a completely general result.
In particular adopting the transformation rules
(\ref{TS-fR-relation}), one can straightforwardly obtain the
solutions in the pure $f$ - scheme.

Let us analyze the above results with a simple example. We can
consider a HOG Lagrangian of the form $f\,=\,a_1R+a_2R^2$ so that
the ``dummy" scalar field reads $\phi\,=\,a_1+2a_2R$. The
self-interaction potential turns out the be
$V(\phi)\,=\,-\frac{(\phi-a_1)^2}{4a_2}$ satisfying the conditions
$V(a_1)\,=\,0$ and $V'(a_1)\,=\,0$. In relation with the
definition of the scalar field, we can opportunely identify $a_1$
with a constant value $\phi^{(0)}\,=\,a_1$. Furthermore, the
scalar field "mass" can be expressed in term of the Lagrangian
parameters as $\lambda^2\,=\,\frac{1}{3}\phi^{(0)}
\frac{\delta^2V(\phi^{(0)})}{\delta\phi^2}\,=\,-\frac{a_1}{6a_2}$.
Since the Ricci scalar at the second order reads

\begin{eqnarray}
R\simeq
R^{(2)}=\frac{\phi^{(2)}}{2a_2}=\frac{\mathcal{X}}{6a_2}\int\frac{d^3\mathbf{k}}{(2\pi)^{3/2}}\frac{\tilde{T}^{(0)}
(\mathbf{k})e^{i\mathbf{k}\cdot
\mathbf{x}}}{\mathbf{k}^2+\lambda^2}\,,
\end{eqnarray}
if we consider a point-like mass $M$, the energy-momentum tensor
components become respectively $T_{tt}=\rho$, $T\sim\rho$ while
$\rho=M\delta(\mathbf{x})$, therefore we obtain

\begin{eqnarray}\label{ricci-scalar-solution-ohanlon}
R^{(2)}=\frac{GM}{3\pi^2a_2}\int
d^3\mathbf{k}\frac{e^{i\mathbf{k}\cdot
\mathbf{x}}}{\mathbf{k}^2+\lambda^2}=-\sqrt{\frac{\pi}{2}}\frac{r_g\lambda^2}{a_1}\frac{e^{-\lambda|\mathbf{x}|}}{|\mathbf{x}
|}\,.
\end{eqnarray}
The immediate consequence is that the solution for the scalar
field $\phi$ at second order is

\begin{eqnarray}
\phi^{(2)}=2a_2R^{(2)}=\sqrt{\frac{\pi}{2}}\frac{r_g}{3}\frac{e^{-\lambda|\mathbf{x}|}}{|\mathbf{x}|}
\end{eqnarray}
while the complete scalar field solution up to the second order of
perturbation is given by

\begin{eqnarray}
\phi=a+\sqrt{\frac{\pi}{2}}\frac{r_g}{3}\frac{e^{-\lambda|\mathbf{x}|}}{|\mathbf{x}|}\,.
\end{eqnarray}
Once the behavior of the scalar field has been obtained up to the
second order of perturbation, in the same way, one can deduce the
expressions for $g^{(2)}_{tt}$ and $g^{(2)}_{ij}$, where
$\Sigma^{(0)}_{tt}\,=\,\frac{2}{3}\rho$ and
$\Sigma^{(0)}_{ij}\,=\,\frac{1}{3}\rho\delta_{ij}\,=\,\frac{1}{2}\Sigma^{(0)}_{tt}\delta_{ij}$.
As matter of fact the metric solutions at the second order of
perturbation are

\begin{eqnarray}\label{gsol1}
\left\{\begin{array}{ll}
g_{tt}=1-\frac{2}{3a}\frac{r_g}{|\mathbf{x}|}-\sqrt{\frac{\pi}{2}}\frac{1}{3a}\frac{r_ge^{-\lambda|\mathbf{x}|}}{|\mathbf{x}|}
\\\\
g_{ij}=-\biggl\{1+\frac{1}{3a}\frac{r_g}{|\mathbf{x}|}-\sqrt{\frac{\pi}{2}}\frac{r_g}{3a}\biggl[\biggl(\frac{1}{|\mathbf{x}
|}-\frac{2}{\lambda^2|\textbf{x}|^3}-\frac{2}{\lambda|\textbf{x}|^2}\biggr)e^{-\lambda|\mathbf{x}|}-\frac{2}{\lambda^2
|\textbf{x}|^3}\biggr]\biggr\}\delta_{ij}\\\\\,\,\,\,\,\,\,\,\,\,\,\,\,\,\,\,+\frac{(2\pi)^{1/2}r_g}{3a}\biggl[\biggl(\frac{1}
{|\textbf{x}|}+\frac{3}{\lambda|\textbf{x}|^2}+\frac{3}{\lambda^2|\textbf{x}|^3}\biggr)e^{-\lambda|\textbf{x}|}-\frac{3}
{\lambda^2|\textbf{x}|^3}\biggr]\frac{x_ix_j}{|\textbf{x}|^2}
\end{array}\right.
\end{eqnarray}
This quantity, which is directly related to the gravitational
potential, shows that the gravitational potential of the O'Hanlon
Lagrangian is non-Newtonian like. Such a behavior prevents from
obtaining a natural definition of the PPN parameters as
corrections to the Newtonian potential. As matter of fact since it
is indeed not true that a generic $f$-gravity model corresponds to
a Brans-Dicke model with $\omega_{BD}\,=\,0$ coherently to its
Post-Newtonian approximation. In particular it turns out to be
wrong considering the PPN parameter
$\gamma\,=\,\frac{1+\omega_{BD}}{2+\omega_{BD}}$ (see, for
example, \cite{will}) of Brans - Dicke gravity and evaluating this
at $\omega_{BD}\,=\,0$ so that one has $\gamma\,=\,1/2$ as
suggested in \cite{olm, chi, jin-liu-li, sou-woo}.

Differently, because of the presence of the self-interaction
potential $V(\phi)$, in the O'Hanlon Lagrangian, a Yukawa like
correction appears in the Newtonian limit appears. Such a
correction in a completely different way even at the
post-Newtonian limit. As matter of fact, one obtains a completely
different gravitational potential with respect to the ordinary
Newtonian one and as matter of fact the fourth order corrections
in term of the $v/c$ ratio (Newtonian level), have to be evaluated
in a complete new general way. In other words, considering a
Brans-Dicke Lagrangian and an O'Hanlon one, despite their similar
structure, will imply completely different predictions in the weak
field and small velocity limits. Such a result represents a
significant argument against the claim that HOG models can be
ruled out only on the bases of the analogy with Brans-Dicke PPN
parameters.

an important consideration is in order now. The definition of
PPN-parameters $\gamma$ and $\beta$, in the GR realm, is intended
as a correction to the Newtonian-like behaviour of the
gravitational potentials (\ref{schwarz-isotropic-PPN}). In
particular, the PPN parameter $\gamma$ is related to the second
order correction of to the gravitational potential while $\beta$
is linked with the fourth order level of perturbation. Actually,
if we consider the limit $f\rightarrow R$, from Eqs.
(\ref{gsol1}), we have

\begin{eqnarray}
\left\{\begin{array}{ll}
g_{tt}=1-\frac{2}{3a}\frac{r_g}{|\mathbf{x}|}\\\\
g_{ij}=-\biggl(1+\frac{1}{3a}\frac{r_g}{|\mathbf{x}|}\biggr)\delta_{ij}
\end{array}\right.
\end{eqnarray}
Since $a$ is an arbitrary constant, in order to match the
Newtonian gravitational potential of GR, we should fix $a=2/3$.
This assumption implies

\begin{eqnarray}
\left\{\begin{array}{ll}
g_{tt}=1-\frac{r_g}{|\mathbf{x}|}\\\\
g_{ij}=-\biggl(1+\frac{1}{2}\frac{r_g}{|\mathbf{x}|}\biggr)\delta_{ij}
\end{array}\right.
\end{eqnarray}
which suggest that the PPN parameter $\gamma$, in this limit,
results $1/2$ which is in striking contrast with GR predictions
($\gamma\sim1$). Such a result is in reality not surprising. In
fact, the GR limit of the O'Hanlon Lagrangian requires $\phi\sim$
const and $V(\phi)\rightarrow 0$ but such approximations induce
mathematical inconsistencies in the field equations of $f$ -
gravity once these have been obtained by a general O'Hanlon
Lagrangian. In reality this is a general issue of O'Hanlon
Lagrangian. In fact it can be demonstrated that the field
equations (\ref{fets}) do not reduce to the standard GR ones (for
$V(\phi)\longrightarrow0$ and $\phi\sim\text{const}$) since we
have\,:

\begin{eqnarray}
\left\{\begin{array}{ll}
R_{\mu\nu}=\frac{\mathcal{X}}{a_1}\Sigma_{\mu\nu}\\\\
0=\frac{\mathcal{X}}{3}T
\end{array}\right.
\end{eqnarray}
But $\Sigma_{\mu\nu}$ components read
$\Sigma_{tt}=\frac{2}{3}\rho$ and
$\Sigma_{ij}=\frac{1}{3}\rho\delta_{ij}=\frac{1}{2}\Sigma_{tt}\delta_{ij}$
in place of $S_{tt}=\frac{1}{2}\rho$ and
$S_{ij}=\frac{1}{2}\rho\delta_{ij}=S_{tt}\delta_{ij}$ as usual,
while the GR field equations are the (\ref{fieldequationGR-2}).
Such a pathology is in order even when the GR limit is performed
from a pure Brans - Dicke Lagrangian. In such a case, in order to
match the Hilbert - Einstein Lagrangian, one needs $\phi\sim
const$ and $\omega_{BD}=0$, the immediate consequence is that the
PPN parameter $\gamma$ turns out to be $1/2$, while it is well
known that Brans - Dicke model fulfils low energy limit
prescriptions in the limit $\omega\rightarrow \infty$. Even in
this case, the problem, with respect to the GR prediction, is that
the GR limit of the model introduces inconsistencies in the field
equations. In other words, it is not possible to impose the same
transformation which leads the Brans-Dicke theory into GR at the
Lagrangian level on the solutions obtained by solving the field
equations descending from the general Lagrangian. The relevant
aspect of this discussion is that considering a $f$ model, in
analogy with the O'Hanlon Lagrangian and supposing that the
self-interaction potential is negligible, introduces a
pathological behaviour of the dynamical solutions and induces to
obtain a PPN parameter $\gamma\,=\,1/2$. This is what happens when
an effective approximation scheme is introduced in the field
equations in order to calculate the weak field limit of HOG by
means of Brans-Dicke model. Such a result seems, from another
point of view, to enforce the claim that HOG models have to be
carefully investigated in this limit and their analogy with
scalar-tensor gravity should be opportunely considered.

\section{Differences of a generic scalar $-$ tensor theory in the Jordan and Einstein frames}

up to now Since along the chapter we have discussed the weak field
and small velocity limit of HOG models in term of Brans-Dicke like
Lagrangian remaining in the Jordan frame. There we show what are
the predictions of the weak field and small velocity limit when a
conformal transformation (\ref{transconf}) is applied on the
O'Hanlon Lagrangian. In other words, we discuss HOG models
conformally transformed in the Einstein frame. The generic
scalar-tensor action $\mathcal{A}_{JF}^{ST}$ in the Jordan frame
(\ref{TSaction}) is linked to the generic action
$\mathcal{A}_{EF}^{ST}$ in the Einstein frame (\ref{TS-EF-action})
via the transformations (\ref{transconfTS}) between the quantities
in the two frames. In the case of the O'Hanlon theory in the
Jordan frame, (\ref{ohanlon}), \emph{i.e.} $F(\phi)\,=\,\phi$ and
$\omega{(\phi)}\,=\,0$, the action (\ref{TS-EF-action}) in the
Einstein frame is simplified and the transformation between the
two scalar fields reads

\begin{equation}\label{phivarphi}
\Omega(\varphi){d\varphi}^2\,=\,-\frac{3\Lambda}{2}\frac{{d\phi}^2}{\phi^2}\,.
\end{equation}
If, now, we suppose $\Omega(\varphi)\,=\,-\Omega_{0}<0$ we have

\begin{equation}\label{phivarphisol}
\phi\,=\,k\,e^{\pm Y\varphi}\,,
\end{equation}
where $Y=\sqrt{\frac{2\Omega_0}{3\Lambda}}$ and $k$ is an
integration constant. We obtain the transformed of (\ref{ohanlon})
in the Einstein frame is

\begin{eqnarray}\label{ohanlon-EF}
\mathcal{A}_{EF}^{OH}=\int
d^4x\sqrt{-\tilde{g}}\biggl[\Lambda\tilde{R}-\Omega_0\varphi_
{;\alpha}\varphi^{;\alpha}+\frac{\Lambda^2}{k^2}e^{\mp
2Y\varphi}V(k\,e^{\pm
Y\varphi})\nonumber\\+\frac{\mathcal{X}\Lambda^2}{k^2}e^{\mp
2Y\varphi}\mathcal{L}_m\biggl(\frac{\Lambda}{k}e^{\mp
Y\varphi}\tilde{g}_{\rho\sigma}\biggr)\biggr]\,.
\end{eqnarray}
The field equations on the other side are

\begin{eqnarray}
\left\{\begin{array}{ll}
\Lambda\tilde{G}_{\mu\nu}-\frac{1}{2}\frac{\Lambda^2}{k^2}e^{\mp
2Y\varphi}V(k\,e^{\pm
Y\varphi})\tilde{g}_{\mu\nu}-\Omega_0\biggl(\varphi_{;\mu}\varphi_{;\nu}-\frac{1}{2}
\varphi_{;\alpha}\varphi^{;\alpha}\tilde{g}_{\mu\nu}\biggr)=\mathcal{X}\,\tilde{T}^\varphi_{\mu\nu}
\\\\
2\Omega_0\tilde{\Box}\varphi+\frac{\Lambda^2}{k^2}e^{\mp
2Y\varphi}[\frac{\delta V}{\delta\phi}(k\,e^{\pm Y\varphi})\mp 2 Y
V(k\,e^{\pm
Y\varphi})]+\mathcal{X}\tilde{\mathcal{L}}_{m,\varphi}=0
\\\\
\tilde{R}=-\frac{\mathcal{X}}{2\Lambda}\tilde{T}^\varphi+\frac{\Omega_0}{\Lambda}\varphi_{;\alpha}\varphi^{;\alpha}-\frac{2
\Lambda}{k^2}e^{\mp 2Y\varphi}V(ke^{\pm Y\varphi})
\end{array} \right.
\end{eqnarray}
where the matter tensor, which now coupled with the scalar field
$\varphi$, in the Einstein frame reads

\begin{eqnarray}
\tilde{T}^\varphi_{\mu\nu}\,=\,\frac{-1}{\sqrt{-\tilde{g}}}\frac{\delta(\sqrt{-\tilde{g}}\tilde{\mathcal{L}}
_m)}{\delta\tilde{g}^{\mu\nu}}=\frac{\Lambda^2}{2k^2}e^{\mp
2Y\varphi}\biggl[\mathcal{L}_m\biggl(\frac{\Lambda}{k}e^{\mp Y
\varphi}\tilde{g}_{\rho\sigma}\biggr)\tilde{g}_{\mu\nu}\nonumber\\-2\frac{\delta}{\delta\tilde{g}^{\mu\nu}}\mathcal{L}_m
\biggl(\frac{\Lambda}{k}e^{\mp
Y\varphi}\tilde{g}_{\rho\sigma}\biggr)\biggr]\,,
\end{eqnarray}
and
\begin{eqnarray}
\tilde{\mathcal{L}}_{m,\varphi}=\mp\frac{\Lambda^2Y}{k^2}e^{\mp
2Y\varphi}\biggl[2\mathcal{L}_m\biggl(\frac{ \Lambda}{k}e^{\mp
Y\varphi}\tilde{g}_{\rho\sigma}\biggr)+\frac{\Lambda}{k}e^{\mp
Y\varphi}\tilde{g}_{\rho\sigma}\frac{\delta\mathcal{L}_m}{\delta
g_{\rho\sigma}}\biggl(\frac{\Lambda}{k}e^{\mp Y
\varphi}\tilde{g}_{\rho\sigma}\biggr)\biggr]\,.
\end{eqnarray}
Actually, in order to calculate the weak field and small velocity
limit of the model in the Einstein frame, we can develop the two
scalar fields at the second order $\phi\sim \phi^{(0)}+\phi^{(2)}$
and $\varphi\sim \varphi^{(0)}+\varphi^{(2)}$ with respect to a
background value. This choice gives the relations\,:

\begin{eqnarray}
\left\{\begin{array}{ll}\varphi^{(0)}\,=\,\pm Y^{-1}\ln{\frac{\phi^{(0)}}{k}}\\\\
\varphi^{(2)}\,=\,\pm
Y^{-1}\frac{\phi^{(2)}}{\phi^{(0)}}\end{array}\right.
\end{eqnarray}
Let us consider the conformal transformation
$\tilde{g}_{\mu\nu}\,=\,\frac{\phi}{\Lambda}g_{\mu\nu}$
(\ref{transconf}). From this relation and considering the
(\ref{phivarphisol}) one obtains, if $\phi^{(0)}=\Lambda$, that

\begin{eqnarray}\label{confg}
\left\{\begin{array}{ll}\tilde{g}^{(2)}_{tt}\,=\,g^{(2)}_{tt}+\frac{\phi^{(2)}}{\phi^{(0)}}\\\\
\tilde{g}^{(2)}_{ij}\,=\,g^{(2)}_{ij}-\frac{\phi^{(2)}}{\phi^{(0)}}\delta_{ij}\end{array}\right.
\end{eqnarray}
As matter of fact, since $g_{tt}^{(2)}\,=\,2 \Phi^{JF}$,
$g_{ij}^{(2)}\,=\,2 \Psi^{JF}\,\delta_{ij}$ and
$\tilde{g}_{tt}^{(2)}\,=\,2 \Phi^{EF}$,
$\tilde{g}_{ij}^{(2)}\,=\,2 \Psi^{EF}\,\delta_{ij}$ from
(\ref{confg}) it descends a relevant relation which links the
gravitational potentials of Jordan and Einstein frame\,:

\begin{eqnarray}\label{gravpotconf1}
\left\{\begin{array}{ll}\Phi^{EF}\,=\,\Phi^{JF}+\frac{\phi^{(2)}}{2\phi^{(0)}}\,=\,\Phi^{JF}\pm\frac{Y}{2}\varphi^{(2)}\\\\
\Psi^{EF}\,=\,\Psi^{JF}-\frac{\phi^{(2)}}{2\phi^{(0)}}\,=\,\Psi^{JF}\mp\frac{Y}{2}\varphi^{(2)}\end{array}\right.
\end{eqnarray}
If we introduce the variations of two potentials:
$\Delta\Phi\,=\,\Phi^{JF}-\Phi^{EF}$ and
$\Delta\Psi\,=\,\Psi^{JF}-\Psi^{EF}$ we obtain the most relevant
result of this paragraph:

\begin{eqnarray}
\Delta\Phi\,=\,-\,\Delta\Psi\,=\,-\,\frac{\phi^{(2)}}{2\phi^{(0)}}\,=\,\mp\,\frac{Y}{2}\varphi^{(2)}\,
\propto\,a_2\,\propto\,f''(0)\,.
\end{eqnarray}

From the above expressions, one can notice that there is an
evident difference between the behaviour of the two gravitational
potentials in the two frames. Such a result suggests that, at the
Newtonian level, it is possible to discriminate between the two
mathematical frame thus one can deduce what is the true physical
one. In particular, once, the gravitational potential is
calculated in the Jordan frame and the dynamical evolution of
$\phi$ is taken into account at the suitable perturbation level,
these can be substituted in the first of (\ref{gravpotconf1}) so
that to obtain its Einstein frame evolution. The final step is
that the two potentials have to be matched with experimental data
in order to investigate what is the true physical solution. A
similar result has been provided in a recent paper \cite{cap-tsu}.

\clearpage{\pagestyle{empty}\cleardoublepage}

\chapter{The Newtonian limit of Fourth Order Gravity
theory}\label{newtonian-limit-fourth-order-gravity}

The Newtonian limit of HOG is worked out discussing its viability
with respect to the standard results of GR. We exclusively
investigate the limit in the metric approach, refraining from
exploiting the formal equivalence of higher - order theories by
considering the analogy with specific scalar - tensor theories,
\emph{i.e.} we work in the Jordan frame in order to avoid possible
misleading interpretations of the results. Considering the Taylor
expansion of a generic $f$-gravity, it is possible to obtain
general solutions in term of the metric coefficients up to the
third order of approximation. Furthermore, we show that the
Birkhoff theorem is not a general result for $f$-gravity since
time-dependent evolution for spherically symmetric solutions can
be achieved depending on the order of perturbations [\textbf{C}],
[\textbf{G}] . Furthermore we provide explicit solutions for
several different types of Lagrangians containing powers of the
Ricci scalar as well as combinations of the other curvature
invariants [\textbf{E}]. In particular, we develop the Green
function method for fourth - order theories in order to find out
solutions. Finally, the consistency of the results with respect to
GR is discussed. In particular, the solution relative to the
$g_{tt}$ component gives a gravitational potential always
corrected with respect to the Newtonian one of the linear theory
$f\,=\,R$.

\section{The Newtonian limit of $f$ $-$ gravity in spherically symmetric
background}\label{newtonian-lmit-standard-coordinates}

Exploiting the formalism of Newtonian and post-Newtonian
approximation described in paragraph
(\ref{post-new-new-formalism}), we can develop a systematic
analysis in the limits of weak field and small velocities for the
$f$-gravity. We are going to assume, as background, a spherically
symmetric spacetime and we are going to investigate the vacuum
case. Considering the metric (\ref{me5}), we have, for a given
$g_{\mu\nu}$\,:

\begin{eqnarray}\label{definexpans}
\left\{\begin{array}{ll}g_{tt}(t,
r)\,\simeq\,1+g^{(2)}_{tt}(t,r)+g^{(4)}_{tt}(t,r)
\\\\
g_{rr}(t,r)\,\simeq\,-1+g^{(2)}_{rr}(t,r)\\\\
g_{\theta\theta}(t,r)\,=\,-r^2\\\\
g_{\phi\phi}(t,r)\,=\,-r^2\sin^2\theta
\end{array}\right.
\end{eqnarray}
while considering Eqs. (\ref{PPN-metric-contro}):

\begin{eqnarray}
\left\{\begin{array}{ll}g^{tt}\simeq
1-g^{(2)}_{tt}+[{g^{(2)}_{tt}}^2-g^{(4)}_{tt}]
\\\\
g^{rr}\simeq-1-g^{(2)}_{rr}
\end{array} \right.
\end{eqnarray}
The determinant reads

\begin{eqnarray}
g\simeq
r^4\sin^2\theta\{-1+[g^{(2)}_{rr}-g^{(2)}_{tt}]+[g^{(2)}_{tt}g^{(2)}_{rr}-g^{(4)}_{tt}]\}\,.
\end{eqnarray}
The Christoffel symbols (\ref{PPN-christoffel}) are

\begin{eqnarray}
\left\{\begin{array}{ll} \begin{array}{ccc}
  {\Gamma^{(3)}}^{t}_{tt}=\frac{g^{(2)}_{tt,t}}{2}\, & \,\,\, & {\Gamma^{(2)}}^{r}_{tt}+{\Gamma^{(4)}}^{r}_{tt}=
  \frac{g^{(2)}_{tt,r}}{2}+\frac{g^{(2)}_{rr}g^{(2)}_{tt,r}+g^{(4)}_{tt,r}}{2} \\
  & & \\
  {\Gamma^{(3)}}^{r}_{tr}=-\frac{g^{(2)}_{rr,t}}{2}\, & \,\,\, & {\Gamma^{(2)}}^{t}_{tr}+{\Gamma^{(4)}}^{t}_{tr}
  =\frac{g^{(2)}_{tt,r}}{2}+\frac{g^{(4)}_{tt,r}-g^{(2)}_{tt}g^{(2)}_{tt,r}}{2} \\
  & & \\
  {\Gamma^{(3)}}^{t}_{rr}=-\frac{g^{(2)}_{rr,t}}{2}\, & \,\,\, & {\Gamma^{(2)}}^{r}_{rr}+{\Gamma^{(4)}}^{r}_{rr}
  =-\frac{g^{(2)}_{rr,r}}{2}-\frac{g^{(2)}_{rr}g^{(2)}_{rr,r}}{2} \\
  & & \\
  \Gamma^{r}_{\phi\phi}=\sin^2\theta \Gamma^{r}_{\theta\theta}\, & \,\,\, & {\Gamma^{(0)}}^{r}_{\theta\theta}+
  {\Gamma^{(2)}}^{r}_{\theta\theta}+{\Gamma^{(4)}}^{r}_{\theta\theta}=-r-rg^{(2)}_{rr}-r{g^{(2)}_{rr}}^2 \\
\end{array}
\end{array} \right.
\end{eqnarray}
while the Ricci tensor component, (\ref{PPN-ricci-tensor}), are

\begin{eqnarray}
\left\{\begin{array}{ll}
R^{(2)}_{tt}=\frac{rg^{(2)}_{tt,rr}+2g^{(2)}_{tt,r}}{2r}\\\\
R^{(4)}_{tt}=\frac{-r{g^{(2)}_{tt,r}}^2+4g^{(4)}_{tt,r}+rg^{(2)}_{tt,r}g^{(2)}_{rr,r}+2g^{(2)}_{rr}[2g^{(2)}_{tt,r}+
rg^{(2)}_{tt,rr}]+2rg^{(4)}_{tt,rr}+2rg^{(2)}_{rr,tt}}{4r}\\\\
R^{(3)}_{tr}=-\frac{g^{(2)}_{rr,t}}{r}\\\\
R^{(2)}_{rr}=-\frac{rg^{(2)}_{tt,rr}+2g^{(2)}_{rr,r}}{2r}\\\\
R^{(2)}_{\theta\theta}=-\frac{2g^{(2)}_{rr}+r[g^{(2)}_{tt,r}+g^{(2)}_{rr,r}]}{2}\\\\
R^{(2)}_{\phi\phi}=\sin^2\theta R^{(2)}_{\theta\theta}
\end{array} \right.
\end{eqnarray}
and, finally, the Ricci scalar expression is

\begin{eqnarray}
\left\{\begin{array}{ll}
R^{(2)}=\frac{2g^{(2)}_{rr}+r[2g^{(2)}_{tt,r}+2g^{(2)}_{rr,r}+rg^{(2)}_{tt,rr}]}{r^2}\\\\
R^{(4)}=\frac{1}{2r^2}\biggr[4{g^{(2)}_{rr}}^2+2rg^{(2)}_{rr}[2g^{(2)}_{tt,r}+4g^{(2)}_{rr,r}+rg^{(2)}_{tt,rr}]+r\{-r{g^{(2)}
_{tt,r}}^2+4g^{(4)}_{tt,r}+\\\,\,\,\,\,\,\,\,\,\,\,\,\,\,\,\,\,\,\,\,\,\,\,\,\,\,\,\,\,\,\,\,\,\,\,\,\,\,\,\,\,+rg^{(2)}_
{tt,r}g^{(2)}_{rr,r}-2g^{(2)}_{tt}[2g^{(2)}_{tt,r}+rg^{(2)}_{tt,rr}]+2rg^{(4)}_{tt,rr}+2rg^{(2)}_{rr,tt}\}\biggr]\end{array}
\right.\end{eqnarray}

In order to derive the Newtonian and post-Newtonian approximation
for a generic function $f$, one should specify the
$f$\,-\,Lagrangian into the field equations (\ref{fe}). This is a
crucial point because once a certain Lagrangian is chosen, one
will obtain a particular approximation referred to such a choice.
This means to lose any general prescription and to obtain
corrections to the Newtonian potential, $\Phi(\mathbf{x})$, which
refer "univocally" to the considered $f$ - function.
Alternatively, one can restrict to analytic $f$ - functions
expandable with respect to a certain value $R\,=\,R_0\,=$ constant
or, at least, its non-analytic part, if exists at all, goes to
zero faster than $R^n$, with $n\geq 2$ at $R\rightarrow 0$. In
general, such theories are physically interesting and allow to
recover the GR results and the correct boundary and asymptotic
conditions. Then we assume

\begin{eqnarray}\label{sertay}
f\,=\,\sum_{n}\frac{f^n(R_0)}{n!}(R-R_0)^n\simeq
f_0+f_1R+f_2R^2+f_3R^3+\dots\,,
\end{eqnarray}
One has to note that the development (\ref{sertay}), also if
similar to (\ref{approx}), is very different from the one in the
Chapter \ref{sperical-symmetry-f-gravity}: in fact $R^{(0)}$ is a
general space-time function linked to the background metric
$g^{(0)}_{\mu\nu}$ (\ref{approx-metric}), while here $R_0$ is a
constant value of scalar curvature. Besides the coefficients
$f_0$, $f_1$, $f_2$, $f_3$ are not proportional, respectively, to
zero-th, first, second, third coefficient of Taylor development of
$f$. In fact, we have

\begin{eqnarray}\label{TS-function-gravity}
\left\{\begin{array}{ll}
f_0\,=\,f(R_0)-R_0f'(R_0)+\frac{1}{2}R_0^2f''(R_0)-\frac{1}{6}R_0^3f'''(R_0)
\\\\
f_1\,=\,f'(R_0)-R_0f''(R_0)+\frac{1}{2}R_0^2f'''(R_0)\\\\
f_2\,=\,\frac{1}{2}f''(R_0)-\frac{1}{2}R_0f'''(R_0)\\\\
f_3\,=\,\frac{1}{6}f'''(R_0)
\end{array} \right.
\end{eqnarray}
If we consider a flat background, then $R_0\,=\,0$ and the
coefficients $f_0$, $f_1$, $f_2$, $f_3$ are the terms of Taylor
series. But if we are finding the solutions at Newtonian and
(possibility) post-newtonian level we have to consider a vanishing
background scalar curvature. It is possible to obtain the
Newtonian and post-Newtonian approximation of $f$ - gravity
considering such an expansion (\ref{sertay}) into the field
equations (\ref{fe}) and expanding the system up to the orders
O(0), O(2), O(3) and O(4). This approach provides general results
and specific (analytic) Lagrangians are selected by  the
coefficients $f_i$ in (\ref{sertay}). Developing the equations in
the case of vanishing matter, i.e. $T_{\mu\nu}\,=\,0$, we have

\begin{eqnarray}\label{sys1}
\left\{\begin{array}{ll} H^{(0)}_{\mu\nu}=0,\,\,&\,\,H^{(0)}=0
\\\\
H^{(2)}_{\mu\nu}=0,\,\,&\,\,H^{(2)}=0\\\\
H^{(3)}_{\mu\nu}=0,\,\,&\,\,H^{(3)}=0\\\\
H^{(4)}_{\mu\nu}=0,\,\,&\,\,H^{(4)}=0
\end{array} \right.
\end{eqnarray}
and, in particular, from the O(0) order approximation, one obtains

\begin{eqnarray}\label{eq0}
f_0=0\,,
\end{eqnarray}
which trivially follows from the above assumption that the
space-time is asymptotically Minkowskian (asymptotically flat
background). This result suggests a first  consideration. \emph{If
the Lagrangian is developable around a vanishing value of the
Ricci scalar ($R_0=0$) the relation (\ref{eq0}) will imply that
the cosmological constant contribution has to be zero whatever is
the $f$ - gravity theory}.

If we now consider the O(2) - order approximation, the equations
system (\ref{sys1}), in the vacuum case, results to be

\begin{eqnarray}\label{eq23}
\left\{\begin{array}{ll}
f_1rR^{(2)}-2f_1g^{(2)}_{tt,r}+8f_2R^{(2)}_{,r}-f_1rg^{(2)}_{tt,rr}+4f_2rR^{(2)}=0
\\\\
f_1rR^{(2)}-2f_1g^{(2)}_{rr,r}+8f_2R^{(2)}_{,r}-f_1rg^{(2)}_{tt,rr}=0
\\\\
2f_1g^{(2)}_{rr}-r[f_1rR^{(2)}-f_1g^{(2)}_{tt,r}-f_1g^{(2)}_{rr,r}+4f_2R^{(2)}_{,r}+4f_2rR^{(2)}_{,rr}]=0
\\\\
f_1rR^{(2)}+6f_2[2R^{(2)}_{,r}+rR^{(2)}_{,rr}]=0
\\\\
2g^{(2)}_{rr}+r[2g^{(2)}_{tt,r}-rR^{(2)}+2g^{(2)}_{rr,r}+rg^{(2)}_{tt,rr}]=0
\end{array} \right.
\end{eqnarray}
The last equation of the system (\ref{eq23}) is the definition of
Ricci scalar (\ref{ricciscalar}) at O(2) - order. The trace
equation (the fourth line in the (\ref{eq23})), in particular,
provides a differential equation with respect to the Ricci scalar
which allows to solve, if $\text{sign}[f_1]=-\text{sign}[f_2]$,
the system (\ref{eq23}) at O(2) - order\,:

\begin{eqnarray}\label{yukawa-solution-O(2)-order}
\left\{\begin{array}{ll}
g^{(2)}_{tt}=\delta_0-\frac{\delta_1}{f_1r}+\frac{\delta_2(t)}{3\lambda}\frac{e^{-\lambda
r}}{\lambda r}+\frac{\delta_3(t)}{6\lambda^2}\frac{e^{\lambda
r}}{\lambda r}
\\\\
g^{(2)}_{rr}=-\frac{\delta_1}{f_1r}-\frac{\delta_2(t)}{3\lambda}\frac{\lambda
r+1}{\lambda r}e^{-\lambda
r}+\frac{\delta_3(t)}{6\lambda^2}\frac{\lambda r-1}{\lambda
r}e^{\lambda r}
\\\\
R^{(2)}=\delta_2(t)\frac{e^{-\lambda
r}}{r}+\frac{\delta_3(t)}{2\lambda}\frac{e^{\lambda r}}{r}
\end{array}
\right.
\end{eqnarray}
where

\begin{eqnarray}\label{yukawa-length}\lambda\,\doteq\,\sqrt{-\frac{f_1}{6f_2}}\,,\end{eqnarray}
with the dimension of \emph{length}$^{-1}$. Let us notice that the
integration constant $\delta_0$ has to be dimensionless,
$\delta_1$ has the dimension of \emph{length}, while the time -
dependent functions $\delta_2$ and $\delta_3$, respectively, have
the dimensions of \emph{length}$^{-1}$ and \emph{length}$^{-2}$.
The functions $\delta_i(t)$ ($i\,=\,2,3$) are completely arbitrary
since the differential equation system (\ref{eq2}) contains only
spatial derivatives. Besides, the integration constant $\delta_0$
can be set to zero, as in the theory of the potential,  since it
represents an unessential additive quantity. When we consider the
limit $f\rightarrow R$, in the case of a point-like source of mass
$M$, we recover the perturbed version of standard Schwarzschild
solution (\ref{schwarz-solution-stand-coord}) at O(2) - order with
$\delta_1\,=\,r_g$. In order to match at infinity the Minkowskian
prescription of the metric, we discard the Yukawa growing mode
present in (\ref{yukawa-solution-O(2)-order}), we have\,:

\begin{eqnarray}\label{yukawa-solution-O(2)-order-1}
\left\{\begin{array}{ll}ds^2\,=\,\biggl[1-\frac{r_g}{f_1r}+\frac{\delta_2(t)}{3\lambda}\frac{e^{-\lambda
r}}{\lambda
r}\biggr]dt^2-\biggl[1+\frac{r_g}{f_1r}+\frac{\delta_2(t)}{3\lambda}\frac{\lambda
r+1}{\lambda r}e^{-\lambda
r}\biggr]dr^2-r^2d\Omega\\\\R\,=\,\frac{\delta_2(t)e^{-\lambda
r}}{r}\end{array}\right.
\end{eqnarray}
At this point one can provide the solution in term of the
gravitational potential. In such a case, we have an explicit
Newtonian-like term into the definition, according to previous
results obtained with less rigorous methods \cite{stel, qua-sch}.
The first of (\ref{yukawa-solution-O(2)-order}) provides the
second order solution in term of the metric expansion (see the
definition (\ref{definexpans})), but, this term coincides with the
gravitational potential at the Newtonian order
(\ref{gravitational-potential-metric}). In particular, since we
have $g_{tt}\,=\,1+2\Phi$, the gravitational potential of a HOG
theory, analytic in the Ricci scalar $R$, is

\begin{eqnarray}\label{gravpot}
\Phi\,=\,-\frac{GM}{f_1r}+\frac{\delta_2(t)}{6\lambda}\frac{e^{-\lambda
r}}{\lambda r}\,.
\end{eqnarray}

As first remark, one has to notice that the structure of the
potential (\ref{gravpot}), for a given $f$ - theory, is determined
by the parameter $\lambda$, (\ref{yukawa-length}), which depends
on the first and the second derivative of $f$, once developed
around a vanishing value of Ricci scalar. The potential
(\ref{gravpot}) is coherent with respect to the results in
\cite{qua-sch, schmidt2}, obtained for higher order Lagrangians as
$\digamma\,=\,R+\sum_{k=0}^p\,a_kR\,\Box^kR$. In this last case,
it is demonstrated that the number of Yukawa corrections to the
gravitational potential was strictly related to the order of the
theory.

From (\ref{yukawa-solution-O(2)-order-1}) one can notice that the
Newtonian limit of any analytic $f$ - theory is related only to
the first and second term of the Taylor expansion (\ref{sertay})
of the given theory. \emph{In other words, the gravitational
potential is always characterized by two Yukawa corrections and
only the first two terms of the Taylor expansion of a generic $f$
Lagrangian turn out to be relevant. This is indeed a general
result.}

Let us now consider the  system (\ref{sys1}) at third order
contributions. The first important issue is that, at this order,
one has to consider even the off-diagonal equation:

\begin{eqnarray}\label{off-d}
f_1g^{(2)}_{rr,t}+2f_2rR^{(2)}_{,tr}=0\,,
\end{eqnarray}
which relates the time derivative of the Ricci scalar to the time
derivative of $g^{(2)}_{rr}$. From this relation, it is possible
to draw a  relevant consideration. One can deduce that, if the
Ricci scalar depends on time, so it is the same for the metric
components and even the gravitational potential turns out to be
time-dependent. This result agrees with the analysis provided in
the Chapter \ref{sperical-symmetry-f-gravity} where a complete
description of the weak field limit of HOG has been provided in
term of the dynamical evolution of the Ricci scalar. Moreover it
was been demonstrated that if one supposes the time independent
Ricci scalar (\ref{field-equation-off-diagonal}), static
spherically symmetric solutions result of the form (\ref{solRr}).

Eq. (\ref{off-d}) confirms this result and provides the formal
theoretical explanation of such a behavior. In particular,
together with Eqs. (\ref{yukawa-solution-O(2)-order-1}), it
suggests that if one considers the problem at a lower level of
approximation (i.e. the second order) the background spacetime
metric can have static solutions according to the Birkhoff
theorem; this is no more verified when the problem is faced with
approximations of higher order. In other words, the debated issue
to prove the validity of the Birkhoff theorem in higher order
theories of gravity, finds here its physical answer. In the
Chapter \ref{sperical-symmetry-f-gravity} and here, the validity
of this theorem is demonstrated for $f$ - theories only when the
Ricci scalar is time independent or, in addition, when the
solutions of Eqs. (\ref{fe}) are investigated up to the second
order of approximation in the metric coefficients
(\ref{definexpans}). \emph{Therefore, the Birkhoff theorem is not
a general result for HOG but, on the other hand, in the limit of
small velocities and weak fields (which is enough to deal with the
Solar System gravitational experiments), one can assume that the
gravitational potential is effectively time independent according
to (\ref{yukawa-solution-O(2)-order-1}) and (\ref{gravpot})}.

The above results fix a fundamental difference between GR and HOG
theories. While, in GR, a spherically symmetric solution
represents a stationary and static configuration difficult to be
related to a cosmological background evolution, this is no more
true in the case of HOG. In the latter case, a spherically
symmetric background can have time-dependent evolution together
with the radial dependence. In this sense, a relation between a
spherical solution and the cosmological Hubble flow can be easily
achieved.

From the system (\ref{sys1}), one can notice that the general
solution is characterized only by the first three orders of the
$f$ expansion (\ref{sertay}). Such a result is in agreement with
the $f$ reconstruction which can be induced by the post-Newtonian
parameters adopting a scalar-tensor analogy as discussed in
Chapter \ref{PPN-TS-fR-theory}.

\subsection{Newtonian and post $-$ Newtonian limit in the harmonic gauge}

Up to now, the discussion has been developed without any gauge
choice. In order to overcome the difficulties related to the
nonlinearities of calculations, we can work considering some gauge
choice obtaining less general solutions for the metric entries. If
we consider the gauge (\ref{gauge-harmonic}) we can use the Ricci
tensor components (\ref{PPN-ricci-tensor-HG}) and the Ricci scalar
expression (\ref{PPN-ricci-scalar-HG}). The gauge choice does not
affect the Christoffel symbols. Thus, by solving the field
equations (\ref{fe}) one obtains

\begin{eqnarray}\label{armonic}
\left\{\begin{array}{ll}g_{tt_{|_{HG}}}(t,r)=1+k^{(2)}_2+k^{(4)}_6-\frac{k^{(2)}_1+2k^{(2)}_1k^{(2)}_4+k^{(4)}_5}{r}+
\frac{{k^{(2)}_1}^2-2k^{(2)}_1k^{(2)}_3}{2r^2}-2k^{(2)}_1k^{(2)}_4\frac{\log r}{r}\\\\
g_{rr_{|_{HG}}}(t,r)=-1+k^{(2)}_4-\frac{k^{(2)}_3}{r}\end{array}
\right.
\end{eqnarray}
where the constants $k^{(2)}_i$ are relative to approximation
level O(2), while $k^{(4)}_i$ to O(4). The Ricci scalar is zero
both at $O(2)$ and at $O(4)$ approximation orders.

Eqs. (\ref{armonic}) suggest some interesting remarks. It is easy
to check that the GR prescriptions are immediately recovered for a
particularly choice of integration constants. The $g_{rr}$
component displays only the second order term, as required by a GR
- like behavior, while the $g_{tt}$ component shows also the
fourth order corrections which determine the second post-Newtonian
parameter $\beta$ \cite{will}. It has to be stressed here that a
full post-Newtonian formalism requires to take into account matter
in the system ({\ref{sys1}}): the presence of matter links the
second and fourth order contributions in the metric coefficients
\cite{will}.

\section{The Newtonian limit of quadratic gravity}\label{newtonian-limit-quadratic-gravity}

Since terms resulting from $R^n$ with $n \geq 3$ \emph{do not
contribute} in the Newtonian limit, as seen previously, the most
general choice for the Lagrangian is

\begin{eqnarray}\label{quadratic-theory}f\,=\,a_1R+a_2R^2+a_3R_{\alpha\beta}R^{\alpha\beta}\end{eqnarray}
where $a_1$, $a_2$, $a_3$\footnote{Note that
$[a_2]=[a_3]\,=\,\text{length}^{2}$ and
$[a_1]\,=\,\text{length}^{0}$} are constants. We have to note that
the field equations (\ref{fe}) (with $f\,=\,R^2$),
(\ref{fericcisquare}) and (\ref{ferienmansquare}) satisfy, in four
dimensions, the condition

\begin{eqnarray}H^{R^2}_{\mu\nu}-4H^{Ric}_{\mu\nu}+H^{Rie}_{\mu\nu}=0\,,\end{eqnarray}
then only two of the three expressions are independent
\cite{lanczos2}. Such a quantity is related to the Gauss-Bonnet
topological invariant \cite{bir-dav}. We can consider Eq.
(\ref{quadratic-theory}) as the \emph{most general quadratic
theory of gravity}. The field equations of
(\ref{quadratic-theory}) are a linear combination of
(\ref{fieldequationGR}), (\ref{fe}) and (\ref{fericcisquare}),
that is

\begin{eqnarray}\label{fe-quadratic-theory}a_1G_{\mu\nu}+a_2H^{R^2}_{\mu\nu}+a_3H^{Ric}_{\mu\nu}\,=\,\mathcal{X}\,T_{\mu\nu}
\,.\end{eqnarray} If we introduce the generalization of the
gravitational potentials in the isotropic metric (\ref{me4}) by
the quantities $\Phi$ and $\Psi$ linked to $g^{(2)}_{tt}$ and
$g^{(2)}_{ij}$, we can investigate the solution of field equations
(\ref{fe-quadratic-theory}) in the Newtonian limit:

\begin{eqnarray}\label{metric-newtonian-order}ds^2\,=\,\biggl[1+2\Phi\biggr]dt^2-\biggl[1-2\Psi\biggr]\delta_{ij}dx^idx^j\,.
\end{eqnarray}
Up to the Newtonian order the left-hand side of the field
equations (\ref{fieldequationGR}), (\ref{fe}),
(\ref{fericcisquare}) and (\ref{ferienmansquare}) are

\begin{enumerate}

\item for the GR - theory:
\begin{eqnarray}\label{eq1}\left\{\begin{array}{ll}G_{tt}\sim G^{(2)}_{tt}\,=\,2\triangle\Psi\\\\
G_{ij}\sim
G^{(2)}_{ij}\,=\,\triangle(\Phi-\Psi)\delta_{ij}-(\Phi-\Psi)_{,ij}\end{array}
\right.\end{eqnarray}

\item for $R^2$ - theory:

\begin{eqnarray}\label{eq2}\left\{\begin{array}{ll}H^{R^2}_{tt}\sim H^{R^2(2)}_{tt}\,=\,4\triangle^2(\Phi-2\Psi)\\\\
H^{R^2}_{ij}\sim
H^{R^2(2)}_{ij}\,=\,4\biggl[\triangle^2(2\Psi-\Phi)\delta_{ij}+(\triangle\Phi-2\triangle\Psi)_{,ij}\biggr]\end{array}
\right.\end{eqnarray}

\item for $R_{\alpha\beta}R^{\alpha\beta}$ - theory:

\begin{eqnarray}\label{eq3}\left\{\begin{array}{ll}H^{Ric}_{tt}\sim H^{Ric(2)}_{tt}\,=\,2\triangle^2(\Phi-\Psi)\\\\
H^{Ric}_{ij}\sim
H^{Ric(2)}_{ij}\,=\,\triangle^2(3\Psi-\Phi)\delta_{ij}+(\triangle\Phi-3\triangle\Psi)_{,ij}\end{array}
\right.\end{eqnarray}

\item for $R_{\alpha\beta\gamma\delta}R^{\alpha\beta\gamma\delta}$ - theory:

\begin{eqnarray}\left\{\begin{array}{ll}H^{Rie}_{tt}\sim H^{Rie(2)}_{tt}\,=\,4\triangle^2\Phi\\\\
H^{Rie}_{ij}\sim
H^{Rie(2)}_{ij}\,=\,4\biggl[\triangle^2\Psi\delta_{ij}-(\triangle\Psi)_{,ij}\biggr]\end{array}
\right.\end{eqnarray}

\end{enumerate}

If we take into account the results (\ref{eq1}), (\ref{eq2}) and
(\ref{eq3}) for the geometric side and the results
(\ref{PPN-tensor-matter}) for the matter side, the explicit form
of the field equations (\ref{fe-quadratic-theory}) up to the
Newtonian order is

\begin{eqnarray}\label{PPN-fe-quadratic-theory}
\left\{\begin{array}{ll}2a_1\triangle\Psi-2(4a_2+a_3)\triangle^2\Psi
+2(2a_2+a_3)\triangle^2\Phi\,=\,\mathcal{X}\rho\,,\\\\
\triangle\biggl[a_1(\Psi-\Phi)+(4a_2+a_3)\triangle\Phi-(8a_2+3a_3)\triangle\Psi\biggr]\delta_{ij}\\
\,\,\,\,\,\,\,\,\,\,-\biggl[a_1(\Psi-\Phi)+(4a_2+a_3)\triangle\Phi-(8a_2+3a_3)\triangle\Psi\biggr]_{,ij}\,=\,0
\end{array} \right.
\end{eqnarray}
By introducing two new auxiliary functions ($\tilde{\Phi}$ and
$\tilde{\Psi}$), the equations (\ref{PPN-fe-quadratic-theory})
become

\begin{eqnarray}\label{PPN-fe-quadratic-theory-1}
\left\{\begin{array}{ll}\frac{2a_2}{2a_2+a_3}\triangle^2\tilde{\Psi}-\frac{2a_3(3a_2+a_3)}{a_1(2a_2+a_3)}\triangle^2\tilde
{\Phi}-\frac{4a_2+a_3}{2a_2+a_3}\triangle\tilde{\Phi}-\frac{a_1}{2a_2+a_3}\triangle\tilde{\Psi}\,=\,\mathcal{X}
\rho\,\\\\
\triangle\biggl[\tilde{\Phi}+\triangle\tilde{\Psi}\biggr]\delta_{ij}-\biggl[\tilde{\Phi}+\triangle\tilde{\Psi}\biggr]_{,ij}
\,=\,0
\end{array} \right.
\end{eqnarray}
where $\tilde{\Phi}$ and $\tilde{\Psi}$ are linked to $\Phi$ and
$\Psi$ via

\begin{eqnarray}\label{trans}
\left\{\begin{array}{ll}
\Phi\,=\,-\frac{(8a_2+3a_3)\tilde{\Phi}+a_1\tilde{\Psi}}{2a_1(2a_2+a_3)}\,\\\\
\Psi\,=\,-\frac{(4a_2+a_3)\tilde{\Phi}+a_1\tilde{\Psi}}{2a_1(2a_2+a_3)}
\end{array} \right.
\end{eqnarray}
Obviously we must require $a_1(2a_2+a_3)\neq 0$, which is the
determinant of the transformations (\ref{trans}). Let us introduce
the new function $\Xi$ defined as follows:

\begin{eqnarray}\label{definition}\Xi:=\tilde{\Phi}+\triangle\tilde{\Psi}.\end{eqnarray}
A this point, we can use the new function $\Xi$ to uncouple the
system (\ref{PPN-fe-quadratic-theory}). With the choice
$\tilde{\Phi}\,=\,\Xi-\triangle\tilde{\Psi}$, it is possible to
rewrite equations (\ref{PPN-fe-quadratic-theory}) as follows

\begin{eqnarray}
\left\{\begin{array}{ll} \frac{2a_3(3a_2+a_3)}
{a_1(2a_2+a_3)}\triangle^3\tilde{\Psi}+\frac{6a_2+a_3}{2a_2+a_3}\triangle^2\tilde{\Psi}-\frac{a_1}{2a_2+a_3}\triangle
\tilde{\Psi}\,=\,\mathcal{X}\rho+\tau\,,\\\\
\triangle\,\Xi\,\delta_{ij}-\Xi_{,ij}\,=\,0
\end{array} \right.
\end{eqnarray}
where
$\tau\,\doteq\,\frac{4a_2+a_3}{2a_2+a_3}\triangle\Xi+\frac{2a_3(3a_2+a_3)}
{a_1(2a_2+a_3)}\triangle^2\Xi$. We are interested in the solution
of (\ref{PPN-fe-quadratic-theory-1}) in terms of the Green
function $\mathcal{G}(\mathbf{x},\mathbf{x}')$ defined by

\begin{eqnarray}\label{gr1}\tilde{\Psi}(\mathbf{x})=Y\int d^3\mathbf{x}'
\mathcal{G}(\mathbf{x},\mathbf{x}')\sigma(\mathbf{x}')\,,\end{eqnarray}
where

\begin{eqnarray}\label{den1}\sigma(\mathbf{x})\,\doteq\,\mathcal{X}\rho(\mathbf{x})+\tau(\mathbf{x})\,,\end{eqnarray} and
$Y$ being a constant, which we introduce for dimensional reasons.
Then Eqs. (\ref{PPN-fe-quadratic-theory}) are equivalent to

\begin{eqnarray}\label{green-equation}
\left\{\begin{array}{ll}\frac{2a_3(3a_2+a_3)}
{a_1(2a_2+a_3)}\triangle^3\mathcal{G}(\mathbf{x},\mathbf{x}')+\frac{6a_2+a_3}{2a_2+a_3}\triangle^2\mathcal{G}(\mathbf{x},
\mathbf{x}')-\frac{a_1}{2a_2+a_3}\triangle
\mathcal{G}(\mathbf{x},\mathbf{x}')\,=\,Y^{-1}\delta(\mathbf{x}-\mathbf{x}')\\\\
\triangle\,\Xi(\textbf{x})\,\delta_{ij}-\Xi(\textbf{x})_{,ij}\,=\,0
\end{array} \right.
\end{eqnarray}
where $\delta(\mathbf{x}-\mathbf{x}')$ is the 3-dimensional Dirac
$\delta$-function. The general solution of
(\ref{PPN-fe-quadratic-theory-1}) for $\Phi(\mathbf{x})$ and
$\Psi(\mathbf{x}$), in terms of the Green function
$\mathcal{G}(\mathbf{x}, \mathbf{x}')$ and the function
$\Xi(\mathbf{x})$, are

\begin{eqnarray}\label{general-solution-newtonian-limit}
\left\{\begin{array}{ll}
\Phi(\mathbf{x})\,=\,Y\,\frac{(8a_2+3a_3)\triangle_\mathbf{x}-a_1}{2a_1(2a_2+a_3)}\int
d^3\mathbf{x}'\mathcal{G}(\mathbf{x},\mathbf{x}')\biggl[\mathcal{X}\rho(\mathbf{x}')\\\,\,\,\,\,\,\,\,\,\,\,\,\,\,\,\,\,\,\,\,
+\frac{4a_2+a_3}{2a_2+a_3}\triangle_{\mathbf{x}'}\Xi(\mathbf{x}')+\frac{2a_3(3a_2+a_3)}{a_1(2a_2+a_3)}\triangle^2_
{\mathbf{x}'}\Xi(\mathbf{x}')\biggr]\\\,\,\,\,\,\,\,\,\,\,\,\,\,\,\,\,\,\,
\,\,-\frac{8a_2+3a_3}{2a_1(2a_2+a_3)}\Xi(\mathbf{x})\\\\
\Psi(\mathbf{x})\,=\,Y\,\frac{(4a_2+a_3)\triangle_\mathbf{x}-a_1}{2a_1(2a_2+a_3)}\int
d^3\mathbf{x}'\mathcal{G}(\mathbf{x},\mathbf{x}')\biggl[\mathcal{X}\rho(\mathbf{x}')\\\,\,\,\,\,\,\,\,\,\,\,\,\,\,\,\,\,\,
\,\,+\frac{4a_2+a_3}{2a_2+a_3}\triangle_{\mathbf{x}'}\Xi(\mathbf{x}')+\frac{2a_3
(3a_2+a_3)}{a_1(2a_2+a_3)}\triangle^2_{\mathbf{x}'}\Xi(\mathbf{x}')\biggr]\\\,\,\,\,\,\,\,\,\,\,\,\,\,\,\,\,\,\,
\,\,-\frac{4a_2+a_3}{2a_1(2a_2+a_3)}\Xi(\mathbf{x})
\end{array} \right.
\end{eqnarray}

Eqs. (\ref{PPN-fe-quadratic-theory-1}) represent a coupled set of
fourth order differential equations. The total number of
integration constant is eight. With the substitution
(\ref{definition}), it has been possible to decouple the set of
equations, but now the differential order is changed. The total
differential order is the same, indeed we have one equation of
sixth order and another equation of second order, while previously
we had two equations of fourth order. The number of integration
constants is conserved. We can conclude that, with our approach,
also introducing the new quantities $\tilde{\Phi}$, $\tilde{\Psi}$
does not contradict the paradigm of a metric theory of HOG. The
price is that now the r.h.s. of $tt$\,-\,component of field
equation has been modified: there is an additional matter term
$\tau$ coming from the $ij$\,-\,component. (see the redefinition
of the matter density (\ref{den1})). In Table
\ref{tablefieldequation}, we show particular cases of Eqs.
(\ref{PPN-fe-quadratic-theory-1}) for different choices of
coupling constants of the theory with vanishing the determinant of
transformations (\ref{trans}).

\begin{table}[htbp]
\centering
\begin{tabular}{c|c|c}
\hline\hline\hline
 Case & Choices of constants & Corresponding field equations \\
 \hline
 & & \\
 A & $\begin{array}{ll}a_2=0\\a_3=0\end{array}$ & $\begin{array}{ll}
 \triangle\Psi=\frac{\mathcal{X}}{2a_1}\rho\,,\\\triangle\biggl[\Phi(\mathbf{x})+\frac{G}{a_1}\int d^3\mathbf{x}'\frac{\rho
 (\mathbf{x}')}{|\mathbf{x}-\mathbf{x}'|} \biggr]\delta_{ij}\\\,\,\,\,\,\,\,-\biggl[\Phi(\mathbf{x})+\frac{G}{a_1}\int d^3x'
 \frac{\rho(x')}{|\mathbf{x}-\mathbf{x}'|}\biggr]_{,ij}=0\end{array}$ \\
 \hline
 & & \\
 B & $\begin{array}{ll}a_1=0\\a_3=0\end{array}$ & $\begin{array}{ll}\triangle^2(2\Psi-\Phi)=-\frac{\mathcal{X}}{4a_2}\rho\,,
 \\\triangle\biggl[\triangle(2\Psi-\Phi)\biggr]\delta_{ij}-\biggl[\triangle(2\Psi-\Phi)\biggr]_{,ij}=0
 \end{array}$ \\
 \hline
 & & \\
 C & $\begin{array}{ll}a_1=0\\a_2=0\end{array}$ &
 $\begin{array}{ll}\triangle^2(\Phi-\Psi)=\frac{\mathcal{X}}{2b_1}\rho\,,\\\triangle\biggl[\triangle
 (\Phi-3\Psi)\biggr]\delta_{ij}-\biggl[\triangle(\Phi-3\Psi)\biggr]_{,ij}=0\end{array}$ \\
 \hline
 & & \\
 D & $a_3=-2a_2$ &
 $\begin{array}{ll}2a_2\triangle^2\Psi-a_1\triangle\Psi=-\frac{\mathcal{X}}{2}\rho\,,\\\nabla^2\biggl[a_1\Phi(\mathbf{x})-2a_2
 \triangle\Phi(\mathbf{x})+G\int d^3\mathbf{x}'\frac{\rho(\mathbf{x}')}{|\mathbf{x}-\mathbf{x}'|}\biggr]\delta_{ij}\\\,\,\,\,
 \,\,\,-\biggl[
 a_1\Phi(\mathbf{x})-2a_2\triangle\Phi(\mathbf{x})+G\int d^3\mathbf{x}'\frac{\rho(\mathbf{x}')}{|\mathbf{x}-\mathbf{x}'|}
 \biggr]_{,ij}=0\end{array}$ \\
 \hline
 & & \\
 E & $\begin{array}{ll}a_1= 0\\a_3=-4a_2\end{array}$ & $\begin{array}{ll}\triangle^2\Phi=-\frac{\mathcal{X}}{4a_2}\rho\,,
 \\\triangle\biggl[\triangle\Psi\biggr]\delta_{ij}- \biggl[\triangle\Psi\biggr]_{,ij}=0\end{array}$ \\
 \hline
 & & \\
 F & $\begin{array}{ll}a_1=0\\a_3=-2a_2\end{array}$ & $\begin{array}{ll}\triangle^2\Psi=-\frac{\mathcal{X}}{4a_2}\rho\\\
 \triangle\biggl[\triangle(\Psi-\Phi)\biggr]\delta_{ij}-\biggl[\triangle(\Psi-\Phi)\biggr]_{,ij}=0\end{array}$ \\
 \hline
 & & \\
 G & $\begin{array}{ll}a_1=0\\a_3=-\frac{8a_2}{3}\end{array}$ & $\begin{array}{ll}\triangle^2(2\Psi+\Phi)=-\frac{3\mathcal{X}}
 {4a_2} \rho\\\triangle\biggl[\triangle\Phi\biggr]\delta_{ij}-\biggl[\triangle\Phi\biggr]_{,ij}=0\end{array}$ \\
 \hline\hline\hline
 \end{tabular}
\caption{\label{tablefieldequation}Explicit form of the field
equations for different choices of the coupling constants for
which the determinant of the transformations (\ref{trans})
vanishes.}
\end{table}

\section{Green functions for systems with spherical
symmetry}\label{grenn-function-paragraph}

We are interested in the solutions of field Eqs.
(\ref{PPN-fe-quadratic-theory-1}) at Newtonian order by using the
method of Green functions and assuming a system with spherical
symmetry:
$\mathcal{G}(\mathbf{x},\mathbf{x}')=\mathcal{G}(|\mathbf{x}-\mathbf{x}'|)$.
Let us introduce the radial coordinate
$r\doteq|\mathbf{x}-\mathbf{x}'|$; with this choice, the first
equation of  (\ref{green-equation}) for $r\neq 0$ becomes

\begin{eqnarray}\label{eq16}
2a_3(3a_2+a_3)\triangle_r^3\mathcal{G}(\mathbf{x},\mathbf{x}')+(6a_2+a_3)\triangle_r^2\mathcal{G}(\mathbf{x},
\mathbf{x}')-a^2_1\triangle_r
\mathcal{G}(\mathbf{x},\mathbf{x}')\,=\,0\,,\end{eqnarray} where
$\triangle_r\,=\,r^{-2}\partial_r(r^{-2}\partial_r)$ is the radial
component of the Laplacian in polar coordinates. The solution of
(\ref{eq16}) is:

\begin{eqnarray}\label{sol9}\mathcal{G}(r)&=&K_1-\frac{1}{r}\biggr[K_2+\frac{a_3}{a_1}\biggl(K_3e^{-\sqrt{-\frac{a_1}{a_3}}r}
+K_4e^{\sqrt{-\frac{a_1}{a_3}}r}\biggr)\nonumber\\&&-\frac{2(3a_2+a_3)}{a_1}\biggl(K_5e^{-\sqrt{\frac{a_1}{2(3a_2+a_3)}}r}
+K_6e^{\sqrt{\frac{a_1}{2(3a_2+a_3)}}r}\biggr)\biggr]
\end{eqnarray}
where $K_i$ are constants. We note that, if $a_2\,=\,a_3\,=\,0$,
the Green function of the Newtonian mechanics is found. It is the
same of the Electromagnetism. The integration constants $K_i$ have
to be fixed by imposing the boundary conditions at infinity and in
the origin. In fact Eqs. (\ref{sol9}) is a solution of
(\ref{eq16}) and not of the first equation in
(\ref{green-equation}). A physically acceptable solution has to
satisfy the condition
$\mathcal{G}(\mathbf{x},\mathbf{x}')\rightarrow 0$ if
$|\mathbf{x}-\mathbf{x}'|\rightarrow \infty$, then the constants
$K_1$, $K_4$, $K_6$ in (\ref{sol9}) have to vanish. To obtain the
conditions on the constants $K_2$, $K_3$, $K_5$ we consider the
Fourier transformation of $\mathcal{G}(\mathbf{x},\mathbf{x}')$,
that is

\begin{eqnarray}\mathcal{G}(\mathbf{x},\mathbf{x}')\,=\,\mathcal{G}(\mathbf{x}-\mathbf{x}')\,=\,\int\frac{d^3\mathbf{k}}
{(2\pi)^{3/2}}\,\,\tilde{\mathcal{G}}(\mathbf{k})\,\,e^{i\mathbf{k}\cdot(\mathbf{x}-\mathbf{x}')}\,.\end{eqnarray}
Eq. (\ref{green-equation}), in terms of Fourier transform, becomes

\begin{eqnarray}\int\frac{d^3\mathbf{k}}{(2\pi)^{3/2}}\,\,e^{i\mathbf{k}\cdot(\mathbf{x}-\mathbf{x}')}
\biggl\{\tilde{\mathcal{G}}(\mathbf{k})\biggl[-\frac{2a_3(3a_2+a_3)}{a_1(2a_2+a_3)}\mathbf{k}^6+\frac{6a_2+a_3}{2a_2+a_3}
\mathbf{k}^4+\nonumber\\+\frac{a_1}{2a_2+a_3}\mathbf{k}^2\biggr]-Y^{-1}_I\biggr\}\,=\,0\,.\end{eqnarray}
The Green function can be expressed as follows:

\begin{eqnarray}\label{intfou}\mathcal{G}(\mathbf{x}-\mathbf{x}')\,=\,-Y^{-1}\int\frac{d^3\mathbf{k}}{(2\pi)^{3/2}}\frac{
e^{i\mathbf{k}\cdot(\mathbf{x}-\mathbf{x}')}}
{\frac{2a_3(3a_2+a_3)}{a_1(2a_2+a_3)}\mathbf{k}^6-\frac{6a_2+a_3}{2a_2+a_3}\mathbf{k}^4-\frac{a_1}{2a_2+a_3}
\mathbf{k}^2}.\end{eqnarray} Since we are investigating systems
with spherical symmetry, it is better to introduce polar
coordinates in the $\mathbf{k}$-space. Eq. (\ref{intfou}) becomes

\begin{eqnarray}\mathcal{G}(\mathbf{x}-\mathbf{x}')&=&-\frac{Y^{-1}}{(2\pi)^{3/2}}\int_0^\infty\frac{d|\mathbf{k}|}
{\frac{2a_3(3a_2+a_3)}{a_1(2a_2+a_3)}\mathbf{k}^4-\frac{6a_2+a_3}{2a_2+a_3}\mathbf{k}^2-\frac{a_1}{2a_2+a_3}}\int_\Omega
d\Omega\,\,\,e^{i\mathbf{k}\cdot(\mathbf{x}-\mathbf{x}')}=\nonumber\\\\\nonumber
&=&-\frac{Y^{-1}}{\sqrt{2\pi}}\frac{a_1(2a_2+a_3)}{a_3(3a_2+a_3)}\frac{1}{|\mathbf{x}-\mathbf{x}'|}\int_0^\infty\frac{d
|\mathbf{k}|\sin|\mathbf{k}||\mathbf{x}-\mathbf{x}'|}{|\mathbf{k}|\biggl[\mathbf{k}^2-\frac{a_1}{a_3}\biggr]\biggl
[\mathbf{k}^2+\frac{a_1}{2(3a_2+a_3)}\biggr]}\,.
\end{eqnarray} The analytic expression of
$\mathcal{G}(\mathbf{x}-\mathbf{x}')$ depends on the nature of the
poles of $|\mathbf{k}|$ and on the values of the arbitrary
constants $a_1$, $a_2$, $a_3$. If we define two new quantities
$\lambda_1$, $\lambda_2 \in \,\mathcal{R}$:

\begin{eqnarray}\label{scale2}
\lambda_1^2\doteq-\frac{a_1}{a_3}\,,\,\,\,\,\,\,\,\,\,\,\,\,\,\,\,\,\,\lambda_2^2\doteq\frac{a_1}{2(3a_2+a_3)}\,,
\end{eqnarray} we obtain a particular expression of
(\ref{intfou}):

\begin{eqnarray}
\mathcal{G}(\mathbf{x}-\mathbf{x}')\,=\,-\sqrt{\frac{\pi}{18}}\frac{Y^{-1}}{|\mathbf{x}-\mathbf{x}'|}\biggl[\frac{\lambda_2^2
-\lambda_1^2}{\lambda_1^2\lambda_2^2}-\frac{e^{-\lambda_1|\mathbf{x}-\mathbf{x}'|}}{\lambda_1^2}+\frac{e^{-\lambda_2
|\mathbf{x}-\mathbf{x}'|}}{\lambda_2^2}\biggr]\,.
\end{eqnarray}
This Green function corresponds to the one in (\ref{sol9}) but now
we have also the conditions in the origin. Obviously, we have
three possibilities  to introduce $\lambda_1$ and $\lambda_2$. In
Table \ref{tablegrennfunction}, we provide the complete set of
Green functions $\mathcal{G}(\mathbf{x}-\mathbf{x}')$, depending
on the choices of the coefficients $a_2$ and $a_3$ (with a fixed
sign of $a_1$).

\begin{table}[htbp]
 \centering
 \begin{tabular}{c|c|c}
 \hline\hline\hline
 Case & Choices of constants & Green function \\
 \hline
 & & \\
 A & $\begin{array}{ll}a_3<0\\\\3a_2+a_3>0\end{array}$ & $
 \mathcal{G}^A(\mathbf{x}-\mathbf{x}')=\sqrt{\frac{\pi}{18}}\frac{Y^{-1}}{|\mathbf{x}-\mathbf{x}'|}\biggl[\frac{\lambda_1
 ^2-\lambda_2^2}{\lambda_1^2\lambda_2^2}+\frac{e^{-\lambda_1|\mathbf{x}+\mathbf{x}'|}}{\lambda_1^2}-\frac{e^{-\lambda_2|
 \mathbf{x}-\mathbf{x}'|}}{\lambda_2^2}\biggr]$ \\
 \hline
 & & \\
 B & $\begin{array}{ll}a_3>0\\\\3a_2+a_3<0\end{array}$ & $ \mathcal{G}^B(\mathbf{x}-\mathbf{x}')=\sqrt{\frac{\pi}{18}}
 \frac{Y^{-1}}{|\mathbf{x}-\mathbf {x}'|}\biggl[\frac{\lambda_2^2-\lambda_1^2}{\lambda_1^2\lambda_2^2}-\frac{\cos(
 \lambda_1|\mathbf{x}-\mathbf{x}'|)}{\lambda_1^2}+\frac{\cos(\lambda_2|\mathbf{x}-\mathbf{x}'|)}{\lambda_2^2}\biggr]$ \\
 \hline
 & & \\
 C & $\begin{array}{ll}a_3<0\\\\3a_2+a_3<0\end{array}$ & $ \mathcal{G}^C(\mathbf{x}-\mathbf{x}')=\sqrt{\frac{\pi}{18}}\frac
 {Y^{-1}}{|\mathbf{x}-\mathbf{x}'|}\biggl[-\frac{\lambda_1^2+\lambda_2^2}{\lambda_1^2\lambda_2^2}+\frac{e^{-\lambda_1|
 \mathbf{x}-\mathbf{x}'|}}{\lambda_1^2}+\frac{\cos(\lambda_2|\mathbf{x}-\mathbf{x}'|)}{\lambda_2^2}\biggr]$ \\
 \hline\hline\hline
 \end{tabular}
 \caption{\label{tablegrennfunction}The complete set of Green
 functions for equations (\ref{intfou}). It is possible to have a further choice for the scale lengths which turns out to be
 dependent on the two knows length scales. In fact, if we perform the substitution
 $\lambda_1\rightleftharpoons\lambda_2$, we obtain a fourth choice. In addition, for a correct
 Newtonian component, we assumed $a_1\,>\,0$. In fact when $a_2\,=\,a_3=0$
 the field equations (\ref{PPN-fe-quadratic-theory}) give us the Newtonian theory if $a_1\,=\,1$.}
\end{table}

When one considers a point-like source,
$\rho\propto\delta(\mathbf{x})$, and by setting
$\Xi(\mathbf{x})\,=\,0$, the potentials
(\ref{general-solution-newtonian-limit}) are proportional to
$\mathcal{G}(\mathbf{x}-\mathbf{x}')$. Without losing of
generality, we have:

\begin{eqnarray}\label{potpointsource}\Phi(\mathbf{x})\sim\frac{1}{\mathbf{|x|}}+\frac{e^{-\lambda_1\mathbf{|x|}}}
{\mathbf{|x|}}+\frac{e^{-\lambda_2\mathbf{|x|}}}{\mathbf{|x|}}\,,\end{eqnarray}
an analogous behavior is obtained for the potential
$\Psi(\mathbf{x})$. We note that, in the vacuum case, we found
Yukawa-like corrections to Newtonian mechanics  but with two scale
lengths related to the quadratic corrections in the Lagrangian
(\ref{quadratic-theory}). See also the above expressions
(\ref{scale2}). This behavior is strictly linked to the sixth
order of (\ref{green-equation}), which depends on the coupled form
of the system of equations (\ref{PPN-fe-quadratic-theory}). In
fact if we consider the Fourier transform of the potentials $\Phi$
and $\Psi$:

\begin{eqnarray}\Phi(\mathbf{x})\,=\,\int\frac{d^3\mathbf{k}}{(2\pi)^{3/2}}\,\,\hat{\Phi}(\mathbf{k})\,\,e^{i\mathbf{k}\cdot
\mathbf{x}}\,,\,\,\,\,\,\,\,\,\Psi(\mathbf{x})\,=\,\int\frac{d^3\mathbf{k}}{(2\pi)^{3/2}}\,\,\hat{\Psi}(\mathbf{k})\,\,
e^{i\mathbf{k}\cdot\mathbf{x}}\,,\end{eqnarray}
the solutions are

\begin{eqnarray}\label{general-solution-newtonian-limit-point-like}
\left\{\begin{array}{ll}
\Phi(\mathbf{x})\,=\,-\frac{\mathcal{X}}{2}\int\frac{d^3\mathbf{k}}{(2\pi)^{3/2}}\frac{[a_1+
(8a_2+3a_3)\mathbf{k}^2]\tilde{\rho}(\mathbf{k})e^{i\mathbf{k}\cdot\mathbf{x}}}{\mathbf{k}^2(a_1-a_3\mathbf{k}^2)[a_1+2(3a_2+
a_3)\mathbf{k}^2]}\\\\
\Psi(\mathbf{x})\,=\,-\frac{\mathcal{X}}{2}\int\frac{d^3\mathbf{k}}{(2\pi)^{3/2}}\frac{[a_1+
(4a_2+a_3)\mathbf{k}^2]\tilde{\rho}(\mathbf{k})e^{i\mathbf{k}\cdot\mathbf{x}}}{\mathbf{k}^2(a_1-a_3\mathbf{k}^2)[a_1+2(3a_2+
a_3)\mathbf{k}^2]}
\end{array} \right.
\end{eqnarray}
where $\tilde{\rho}(\mathbf{k})$ is the Fourier transform of the
matter density. We can see that the solutions have the same poles
as (\ref{intfou}). Finally, if ${\displaystyle
\tilde{\rho}(\mathbf{k})=\frac{M}{(2\pi)^{3/2}}}$ (the Fourier
transform of a point-like source) the solutions
(\ref{general-solution-newtonian-limit-point-like}) are similar to
(\ref{potpointsource}). In fact, if we suppose that $a_3\,\neq\,0$
and $3a_2+a_3\,\neq\,0$, the solutions
(\ref{general-solution-newtonian-limit-point-like}) are

\begin{eqnarray}
\left\{\begin{array}{ll}
\Phi(\mathbf{x})\,=\,-\frac{GM}{a_1|\mathbf{x}|}\biggl(1-\frac{4}{3}e^{-\lambda_1|\mathbf{x}|}+\frac{1}{3}
e^{-\lambda_2|\mathbf{x}|}\biggr)\\\\
\Psi(\mathbf{x})\,=\,-\frac{GM}{a_1|\mathbf{x}|}\biggl(1-\frac{2}{3}e^{-\lambda_1|\mathbf{x}|}-\frac{1}{3}
e^{-\lambda_2|\mathbf{x}|}\biggr)
\end{array} \right.
\end{eqnarray}
Then the metric (\ref{metric-newtonian-order}) becomes

\begin{eqnarray}
ds^2\,=&&\biggl[1-\frac{r_g}{a_1|\mathbf{x}|}\biggl(1-\frac{4}{3}e^{-\lambda_1|\mathbf{x}|}+\frac{1}{3}
e^{-\lambda_2|\mathbf{x}|}\biggr)\biggr]dt^2+\nonumber\\\nonumber\\&&\,\,\,\,\,\,\,\,\,\,\,\,-\biggl[1+\frac{r_g}{a_1|
\mathbf{x}|}\biggl(1-\frac{2}{3}e^{-\lambda_1|\mathbf{x}|}-\frac{1}{3}e^{-\lambda_2|\mathbf{x}|}\biggr)\biggr]\delta_{ij}dx^i
dx^j\end{eqnarray} It is interesting to note that, if $a_3\,=\,0$
($\lambda_1\rightarrow\infty$), we have the missing of a scale
length (a pole is missed) with only a Yukawa-like term as for the
Electrodynamics. The Green function, in this case, is:

\begin{eqnarray}\label{greenfunctionfR}\tilde{\mathcal{G}}(\mathbf{k})_{a_3=0}\,
=\,\frac{2a_2Y^{-1}}{6a_2\mathbf{k}^4+a_1\mathbf{k}^2}\,,\end{eqnarray}
and the Lagrangian becomes: $f\,=\,a_1R+a_2R^2$. Since at the
level of the Newtonian limit, as discussed, the powers of Ricci
scalar higher then two do not contribute, we can conclude that
(\ref{greenfunctionfR}) is the Green function for any $f$-gravity
at Newtonian order, if $f$ is an analytic function of the Ricci
scalar. We found the same situation in the newtonian limit of $f$
- theory in standard coordinates (\S\,
\ref{newtonian-lmit-standard-coordinates}). In fact if we consider
the analogy

\begin{eqnarray}a_1\,=\,f_1,\,\,\,\,\,\,\,\,\,\,\,\,a_2\,=\,f_2\end{eqnarray}
and if $\text{sign}[a_2]\,=\,-\text{sign}[a_1]$ we found the same
characteristic scale - length: $\lambda\,=\,\lambda_2$. Finally
the presence of the pole is achieved considering a particular
choice of the constants in the theory, e.g. $a_3\,=\,-2a_2$. In
Table \ref{tablefieldequation} (Case D), we provide the field
equations for this choice and the relative Green function is:

\begin{eqnarray}\label{greenfunctionD}\tilde{\mathcal{G}}_{(2a_2\nabla^4-a_1\nabla^2)}(\mathbf{k})\propto\frac{1}{2a_2
\mathbf{k}^4+a_1\mathbf{k}^2}\,.\end{eqnarray} The spatial
behavior of (\ref{greenfunctionfR}) - (\ref{greenfunctionD}) is
the same but the coefficients are different since the theories are
different. In conclusion we need the Green function for the
differential operator $\triangle^2$. The only possible physical
choice for the squared Laplacian  is:

\begin{eqnarray}\label{greenfunctionlaplsqu}\tilde{\mathcal{G}}_{(\triangle^2)}(\mathbf{x}-\mathbf{x}')\propto\frac{1}
{|\mathbf{x}-\mathbf{x}'|}\,,\end{eqnarray} since the other choice
is proportional to $|\mathbf{x}-\mathbf{x}'|$ and cannot to be
accepted. Considering the last possibility, we will end up with a
force law increasing with distance \cite{havas}. Summary, we have
shown the general approach to find out solutions of the field
equations by using the Green functions. In particular, the vacuum
solutions with point-like source have been used to find out
directly the potentials, however it remains the most important
issue to find out  solutions  when we consider systems with
extended matter distribution.

\section{Solutions using the Green function}\label{sec:solgreen}

Unlike the \S\,\ref{newtonian-lmit-standard-coordinates} we are
going to find solutions with Green functions method. Before to
investigate the general solution of Eqs.
(\ref{PPN-fe-quadratic-theory}) we want discuss, in the first
subparagraph, all cases shown in the Table
\ref{tablefieldequation}. While in the next subsections we will
analyze the solution in presence of matter using the Green
functions shown in Table \ref{tablegrennfunction}.

\subsection{Particular solutions}

In Table \ref{tablefieldequationsolution} we provide solutions, in
terms of the Green function of the corresponding differential
operator, of the field equations shown in Table
\ref{tablefieldequation}. Case A corresponds to the Newtonian
theory and the arbitrary constant $a_1$ can be absorbed in the
definition of matter Lagrangian. The solutions are:

\begin{eqnarray}_A\Phi(\mathbf{x})\,=\,_A\Psi(\mathbf{x})\,=\,-G\int d^3\mathbf{x}'\frac{\rho(\mathbf{x}')}{|\mathbf{x}-
\mathbf{x}'|}\,.\end{eqnarray} For Case D, instead, we have the
field equations  of a sort of modified electrodynamic-like
representation. The solution can be expressed as follows:

\begin{eqnarray}\label{potsolD}_D\Phi(\mathbf{x})\,=\,_D\Psi(\mathbf{x})\,=\,-G\int d^3\mathbf{x}'\biggl[\frac{1-e^{-\sqrt
{\frac{a_1}{2a_2}}|\mathbf{x}-\mathbf{x}'|}}{|\mathbf{x}-\mathbf{x}'|}\biggr]\rho(\mathbf{x}')\,.\end{eqnarray}
The solutions make sense only if $a_1/a_2\,>\,0$, then we can
introduce a new scale-length. A particular expression of
(\ref{potsolD}), for a fixed matter density $\rho(\mathbf{x})$,
will be found in a more general context in the next section.
Nevertheless these two cases are the only ones which exhibit the
Newtonian limit (obviously the first one!), while for the
remaining cases there are serious problems with divergences and
incompatibilities. In fact, Case B presents an incompatibility
between the solution obtained from the $tt$ - component and the
one from the $ij$\,-\,component. The incompatibility can be
removed if we consider, as the Green function for the differential
operator $\nabla^4$, the trivial solution:
$\mathcal{G}_{(\triangle^2)}|_B\,=\,\text{const.}$ With this
choice, the arbitrary integration constant $\Phi_0$ can be
interpreted as $-GM$. However another problem remains: namely the
divergence at the origin. The interpretation of the constant
$\Phi_0$ as a total mass and not as a generic integral $\int
d^3\mathbf{x}'\rho(\mathbf{x}')$ does not avoid the singularity.
We can conclude, then, the solution

\begin{eqnarray}2_B\Psi(\mathbf{x})-_B\Phi(\mathbf{x})\,=\,-\frac{GM}{|\mathbf{x}-\mathbf{x}'|}\end{eqnarray}
holds only in vacuum. Before continuing our  analysis of the
various cases, the term $\int
d^3\mathbf{x}'\mathcal{G}_{(\triangle^2)}(\mathbf{x}-\mathbf{x}')\rho(\mathbf{x}')$
has to be discussed for the  choice (\ref{greenfunctionlaplsqu}).
A generic field equation with $\triangle^2$ (from Table
\ref{tablefieldequation}) is

\begin{eqnarray}\triangle_\mathbf{x}^2\Phi(\mathbf{x})\propto\triangle_\mathbf{x}^2\int d^3\mathbf{x}'\frac{\rho(\mathbf{x}')}
{|\mathbf{x}-\mathbf{x}'|}\,\propto\,\triangle_\mathbf{x}\rho(\mathbf{x})\,\neq\,4\pi\rho(\mathbf{x});\end{eqnarray}
we  conclude that the only consistent possibility is to set
$\rho(\mathbf{x})=0$. In the remaining cases, we can only consider
 vacuum solutions.

\begin{table}[htbp]
 \centering
 \begin{tabular}{c|c|c}
 \hline\hline\hline
 Case & Solutions & Newtonian behavior \\
 \hline
 & & \\
 A & $\begin{array}{ll}
 \Phi(\mathbf{x})=\Psi(\mathbf{x})=-\frac{G}{a_1}\int d^3\mathbf{x}'\frac{\rho(\mathbf{x}')}{|\mathbf{x}-\mathbf{x}'|}
 \end{array}$ & yes \\
 \hline
 & & \\
 B & $\begin{array}{ll}
 2\Psi(\mathbf{x})-\Phi(\mathbf{x})=\frac{\Phi_0}{|\mathbf{x}|}\\\\ 2\Psi(\mathbf{x})-\Phi(\mathbf{x})=-\frac{2\pi G}{a_2}\int
 d^3\mathbf{x}'\mathcal{G}_{(\nabla^4)}(\mathbf{x}-\mathbf{x}')\rho
 (\mathbf{x}')\end{array}$ & no \\
 \hline
 & & \\
 C & $\begin{array}{ll}
 \Phi(\mathbf{x})=\frac{\Phi_0}{|\mathbf{x}|}+\frac{6\pi G}{a_3}\int d^3\mathbf{x}'\mathcal{G}_{(\nabla^4)}(\mathbf{x}-
 \mathbf{x}')\rho(\mathbf{x}')\\\\\Psi(\mathbf{x})=\frac{\Phi_0}{|\mathbf{x}|}+\frac{2\pi G}{a_3}\int d^3\mathbf{x}'\mathcal
 {G}_{(\nabla^4)}(\mathbf{x}-\mathbf{x}')\rho(\mathbf{x}')\end{array}$ & no \\
 \hline
 & & \\
 D & $\begin{array}{ll}
 \Phi(\mathbf{x})=-4\pi G\int d^3\mathbf{x}'\mathcal{G}_{(2a_2\nabla^4-a_1\nabla^2)}(\mathbf{x}-\mathbf{x}')\rho
 (\mathbf{x}') \\\\\Psi(\mathbf{x})=-4\pi G\int d^3\mathbf{x}'\mathcal{G}_{(2a_2\nabla^4-a_1\nabla^2)}(\mathbf{x}-\mathbf{x}')
 \rho(\mathbf{x}')\end{array}$ & yes \\
 \hline
 & & \\
 E & $\begin{array}{ll}
 \Phi(\mathbf{x})=-\frac{2\pi G}{a_2}\int d^3\mathbf{x}'\mathcal{G}_{(\nabla^4)}(\mathbf{x}-\mathbf{x}')\rho(\mathbf{x}')\\\\
 \Psi(\mathbf{x})=\frac{\Phi_0}{|\mathbf{x}|}\end{array}$ & no\\
 \hline
 & & \\
 F & $\begin{array}{ll}
 \Phi(\mathbf{x})=\frac{\Phi_0}{|\mathbf{x}|}-\frac{2\pi G}{a_2}\int d^3\mathbf{x}'\mathcal{G}_{(\nabla^4)}(\mathbf{x}-
 \mathbf{x}')\rho(\mathbf{x}')\\\\\Psi(\mathbf{x})=-\frac{2\pi G}{a_2}\int
 d^3\mathbf{x}'\mathcal{G}_{(\nabla^4)}(\mathbf{x}-\mathbf{x}')\rho
 (\mathbf{x}')\end{array}$ & no\\
 \hline
 & & \\
 G & $\begin{array}{ll}
 \Phi(\mathbf{x})=\frac{\Phi_0}{|\mathbf{x}|}\\\\
 \Psi(\mathbf{x})=-\frac{1}{2}\frac{\Phi_0}{|\mathbf{x}|}-\frac{3\pi G}{a_2}\int d^3\mathbf{x}'\mathcal{G}_{(\nabla^4)}
 (\mathbf{x}-\mathbf{x}')\rho(\mathbf{x}')\end{array}$ & no\\
 \hline\hline\hline
 \end{tabular}
 \caption{\label{tablefieldequationsolution}Here we provide the
 solutions of the field equations in Table
 \ref{tablefieldequation}. The solutions are found by setting
 $\Xi=0$ in the $ij$\,-\,component of the field equation (\ref{green-equation}). The solutions are displayed in terms
 of the Green functions. $\Phi_0$ is a generic integration constant.}
\end{table}

\subsection{The general solution by Green function $\mathcal{G}^A(\mathbf{x}-\mathbf{x}')$}

In this section, we explicitly determine the gravitational
potential in the inner and in the outer region of a spherically
symmetric matter distribution. The first consequence of the
extended gravity theories which we are considering is the
non-validity of the Gauss theorem. In fact, in the Newtonian limit
of GR, the equation for the gravitational potential, generated by
a point-like source

\begin{eqnarray}\triangle_\mathbf{x}\mathcal{G}_{New. mech.}(\mathbf{x}-\mathbf{x}')=-4\pi\delta(\mathbf{x}-\mathbf{x}')
\end{eqnarray}
is not satisfied by the new Green functions developed above. If we
consider the flux of gravitational field $\mathbf{g}_{New. mech.}$
defined as

\begin{eqnarray}\mathbf{g}_{New. mech.}\,=\,-\frac{GM(\mathbf{x}-\mathbf{x}')}{|\mathbf{x}-\mathbf{x}'|^3}\,=\,-GM
\mathbf{\nabla}_{\mathbf{x}}\mathcal{G}_{New.
mech.}(\mathbf{x}-\mathbf{x}')\,,\end{eqnarray} we obtain, as
standard, the Gauss theorem:

\begin{eqnarray}\int_\Sigma d\Sigma\,\,\,\,\mathbf{g}_{New. mech.}\cdot\hat{n}\propto M\,,\end{eqnarray}
where $\Sigma$ is a generic two-dimensional surface and $\hat{n}$
its surface normal vector. The flux of field $\mathbf{g}_{New.
mech.}$ on the surface $\Sigma$ is proportional to the matter
content $M$, inside the surface independently of the particular
shape of surface (Gauss theorem, or Newton theorem for the
gravitational field \cite{bin-tre}). On the other hand, if we
consider the flux defined by the new Green function, its value is
not proportional to the enclosed mass but depends on the
particular choice of the surface:

\begin{eqnarray}\int_\Sigma d\Sigma\,\,\,\,\mathbf{g}_{New. mech.}\cdot\hat{n}\propto M_\Sigma\,.\end{eqnarray}
Hence $M_\Sigma$ is a mass-function depending on the surface
$\Sigma$. Then we have to find the solution inside/outside the
matter distribution by evaluating the quantity

\begin{eqnarray}\int d^3\mathbf{x}'\mathcal{G}^A(\mathbf{x}-\mathbf{x}')\rho(\mathbf{x}')\,,\end{eqnarray}
and by imposing the boundary condition on the separation surface.
By considering solutions (\ref{general-solution-newtonian-limit})
with the Green function $\mathcal{G}^A(\mathbf{x}-\mathbf{x}')$
from Table \ref{tablegrennfunction} and by assuming
$\Xi(\mathbf{x})=0$, we have

\begin{eqnarray}
\left\{\begin{array}{ll}
_A\Phi(\mathbf{x})\,=\,Y\,\mathcal{X}\,\frac{(8a_2+3a_3)\triangle_\mathbf{x}-a_1}{2a_1(2a_2+a_3)}\int
d^3\mathbf{x}'\mathcal{G}^A(\mathbf{x},\mathbf{x}')\rho(\mathbf{x}')\\\\\,\,\,\,\,\,\,\,\,\,\,\,\,\,\,\,\,\,\,=
\,(\mu_1+\mu_2\triangle_\mathbf{x})\,\,G\int
d^3\mathbf{x}'\mathcal{G}^A(\mathbf{x}-\mathbf{x}')\rho(\mathbf{x}')\\\\
_A\Psi(\mathbf{x})\,=\,Y\,\mathcal{X}\,\frac{(4a_2+a_3)\triangle_\mathbf{x}-a_1}{2a_1(2a_2+a_3)}\int
d^3\mathbf{x}'\mathcal{G}^A(\mathbf{x},\mathbf{x}')\rho(\mathbf{x}')\\\\\,\,\,\,\,\,\,\,\,\,\,\,\,\,\,\,\,\,\,=
\,(\mu_1+\mu_3\triangle_\mathbf{x})\,\,G\int
d^3\mathbf{x}'\mathcal{G}^A(\mathbf{x}-\mathbf{x}')\rho(\mathbf{x}')
\end{array} \right.
\end{eqnarray}
where $\mu_1:=-\frac{4\pi Y}{2a_2+a_3}=-\frac{12\pi
Y}{a_1}\frac{\lambda_1^2\lambda_2^2}{\lambda_1^2-\lambda_2^2}$,
$\mu_2:=\frac{4\pi Y(8a_2+3a_3)}{a_1(2a_2+a_3)}=\frac{4\pi
Y}{a_1}\frac{4\lambda_1^2-\lambda_2^2}{\lambda_1^2-\lambda_2^2}$,
$\mu_3:=\frac{4\pi Y(4a_2+a_3)}{a_1(2a_2+a_3)}=\frac{4\pi
Y}{a_1}\frac{2\lambda_1^2+\lambda_2^2}{\lambda_1^2-\lambda_2^2}$.
We have to note that our working hypothesis, $\Xi(\mathbf{x})=0$,
is not particular, since when we considered the Hilbert-Einstein
Lagrangian in \S\,\ref{newtonian-limit-quadratic-gravity} to give
the Newtonian solution, we  imposed an analogous condition. For
the potential $\Phi(\mathbf{x})$, one obtains:

\begin{eqnarray}\label{poteAgeneral}_A\Phi(\mathbf{x})&=&(\mu_1+\mu_2\triangle_\mathbf{x})\,\,G\int
d^3\mathbf{x}'\rho(\mathbf{x}')\biggr[\sum_{i=0}^2\mathcal{G}^A
_i\frac{e^{-\lambda_i|\mathbf{x}-\mathbf{x}'|}}{|\mathbf{x}-\mathbf{x}'|}\biggr]\nonumber\\&=&\sum_{i=0}^2\mathcal{G}^A_i
(\mu_1+\mu_2\triangle_\mathbf{x})\mathcal{T}^{A,\,\lambda_i}(\mathbf{x})=\sum_{i=0}^2\mathcal{G}^A_i\Phi^{A,\,
\lambda_i}(\mathbf{x})\,,\end{eqnarray} where

\begin{eqnarray}\mathcal{T}^{A,\,\lambda_i}(\mathbf{x})\,\doteq\,G\int d^3\mathbf{x}'\rho(\mathbf{x}')\frac{e^{-\lambda_i|
\mathbf{x}-\mathbf{x}'|}}{|\mathbf{x}-\mathbf{x}'|}\,,\end{eqnarray}

\begin{eqnarray}\Phi^{A,\,\lambda_i}(\mathbf{x})\,\doteq\,(\mu_1+\mu_2\triangle_\mathbf{x})\mathcal{T}^{A,\,\lambda_i}
(\mathbf{x})\,,\end{eqnarray} and $\mathcal{G}^A_i$ are the
coefficients. Here the numbers $\lambda_i$ assume the above values
$0$, $\lambda_1$, $\lambda_2$. Supposing a matter density
$\rho(\mathbf{x})=\rho(|\mathbf{x}|)$ and denoting the radius of
the sphere with total mass $M$ by $\xi$, we have

\begin{eqnarray}\label{int1}\mathcal{T}^{A,\,\lambda_i}(\mathbf{x})&=&G\int_0^\xi
d|\mathbf{x}'||\mathbf{x}'|^2\rho(|\mathbf{x}'|)\int_0^{2\pi}
d\phi'\nonumber\\&&\times\int_0^\pi
d\theta'\sin\theta'\frac{e^{-\lambda_i\sqrt{|\mathbf{x}|^2+|\mathbf{x}'|^2-2|\mathbf{x}||\mathbf{x}'|\cos\alpha}}}
{\sqrt{|\mathbf{x}|^2+|\mathbf{x}'|^2-2|\mathbf{x}||\mathbf{x}'|\cos\alpha}}\,,\end{eqnarray}
where
$\cos\alpha=\cos\theta\cos\theta'+\sin\theta\sin\theta'\cos(\phi-\phi')$
and $\alpha$ is the angle between two vectors $\mathbf{x}$,
$\mathbf{x}'$. In the spherically symmetric case, we can choose
$\theta=0$ without losing generality. The (\ref{int1}) becomes

\begin{eqnarray}\mathcal{T}^{A,\,\lambda_i}(\mathbf{x})&=&\frac{2\pi G}{\lambda_i|\mathbf{x}|}\int_0^\xi d|\mathbf{x}'|
|\mathbf{x}'|\rho(|\mathbf{x}'|)\biggl[e^{-\lambda_i||\mathbf{x}|-|\mathbf{x}'||}-e^{-\lambda_i(|\mathbf{x}|
+|\mathbf{x'}|)}\biggr]\,.\end{eqnarray} For a constant radial
profile of  density, ${\displaystyle
\rho(|\mathbf{x}|)=\frac{3M}{4\pi\xi^3}}$, we have:

\begin{eqnarray}\mathcal{T}^{A,\,\lambda_i}(\mathbf{x})=\left\{\begin{array}{ll}\frac{3GM}{\lambda_i^2\xi^3}\biggl[1-(1+
\lambda_i\xi)e^{-\lambda_i\xi}\frac{\sinh(\lambda_i|\mathbf{x}|)}{\lambda_i|\mathbf{x}|}\biggr]\,\,\,\,\,\,\,\,\,\,\,\,\,\,
|\mathbf{x}|<\xi\\\\\frac{3GM}{{\lambda_i}^2\xi^3}[\lambda_i\xi\cosh(\lambda_i\xi)-\sinh(\lambda_i\xi)]\frac{e^{-\lambda_i
|\mathbf{x}|}}{\lambda_i|\mathbf{x}|}\,\,\,\,\,\,\,\,\,\,|\mathbf{x}|>\xi\,,\end{array}\right.
\end{eqnarray} in the inner and in the outer region respectively. The limit

\begin{eqnarray}\label{newlimit}\lim_{\lambda_i\rightarrow
0}\mathcal{T}^{A,\,\lambda_i}(\mathbf{x})=\left\{\begin{array}{ll}\frac{3GM}{2\xi}-\frac{GM}{2\xi^3}|\mathbf{x}|^2\,\,\,\,\,
\,\,\,\,\,\,\,\,\,\,\,|\mathbf{x}|<\xi\\\\\frac{GM}{|\mathbf{x}|}\,\,\,\,\,\,\,\,\,\,\,\,\,\,\,\,\,\,\,\,\,\,\,\,\,\,\,\,\,
\,\,\,\,\,\,\,\,\,\,\,\,\,\,\,\,\,|\mathbf{x}|>\xi\,,\end{array}\right.\end{eqnarray}
gives us the internal and the external Newtonian behavior. The
internal and the external potential for a given $\lambda_i$ is

\begin{eqnarray}\left\{\begin{array}{ll}
\Phi_{in}^{A,\lambda_i}(\mathbf{x})=\frac{3GM}{\lambda_i^2\xi^3}\biggl[\mu_1-e^{-\lambda_i\xi}(1+
\lambda_i\xi)(\mu_1+\lambda_i^2\mu_2)\frac{\sinh(\lambda_i|\mathbf{x}|)}{\lambda_i|\mathbf{x}|}\biggr]\\\\
\Phi_{out}^{A,\lambda_i}(\mathbf{x})=\frac{3GM}{\lambda_i^2\xi^3}[\lambda_i\xi\cosh(\lambda_i\xi)-
\sinh(\lambda_i\xi)](\mu_1+
{\lambda_i}^2\mu_2)\frac{e^{-\lambda_i|\mathbf{x}|}}{\lambda_i|\mathbf{x}|}\end{array}\right.
\end{eqnarray}
The boundary condition on the surface $|\mathbf{x}|=\xi$ is

\begin{eqnarray}\label{boundcod}_A\Phi_{in}(\xi)-_A\Phi_{out}(\xi)\,=\,-\frac{3GM}{\xi^3}\mu_2\sum_{i=0}^3\mathcal{G}^A_i\,,
\end{eqnarray}
but the last relation is identically vanishing (see Table
\ref{tablegrennfunction}). The internal and external potential is
given by:

\begin{eqnarray}\label{potfinA}
\left\{\begin{array}{ll}
_A\Phi_{in}(\mathbf{x})\,=\,-\frac{\sqrt{2}\,\,\pi^{3/2}}{a_1}\frac{GM}{\xi^3}\biggl[\frac{\lambda_1^2(2+3\lambda_2^2\xi^2)
-8\lambda_2^2}{\lambda_1^2\lambda_2^2}-|\mathbf{x}|^2+8e^{-\lambda_1\xi}(1+\lambda_1\xi)\frac{\sinh(\lambda_1|\mathbf{x}|)}
{\lambda_1^3|\mathbf{x}|}\\\\\,\,\,\,\,\,\,\,\,\,\,\,\,\,\,\,\,\,\,\,\,\,\,\,\,\,\,\,\,\,
-2e^{-\lambda_2\xi}(1+\lambda_2\xi)\frac{\sinh(\lambda_2|\mathbf{x}|)}{\lambda_2^3|\mathbf{x}|}\biggr]\\\\\\
_A\Phi_{out}(\mathbf{x})=-\frac{2\sqrt{2}\,\,\pi^{3/2}}{a_1}\frac{GM}{|\mathbf{x}|}
+\frac{8\sqrt{2}\,\,\pi^{3/2}}{a_1}\frac{GM}{\lambda_1^3\xi^3}[\lambda_1\xi\cosh(\lambda_1\xi)-\sinh(\lambda_1
\xi)]\frac{e^{-\lambda_1|\mathbf{x}|}}{|\mathbf{x}|}\\\\\,\,\,\,\,\,\,\,\,\,\,\,\,\,\,\,\,\,\,\,\,\,\,\,\,\,\,\,\,\,
-\frac{2\sqrt{2}\,\,\pi^{3/2}}{a_1}\frac{GM}{\lambda_2^3\xi^3}[\lambda_2\xi\cosh(\lambda_2\xi)-\sinh(\lambda_2\xi)]\frac{e^
{-\lambda_2|\mathbf{x}|}}{|\mathbf{x}|}
\end{array}\right.
\end{eqnarray}
The relations (\ref{potfinA}) give the solutions for the
gravitational potential $\Phi$ inside and outside the constant
spherically symmetric matter distribution. A similar relation is
found for $\Psi$ by substituting $\mu_2\rightarrow\mu_3$. We note
that the corrections to the Newtonian terms are ruled by
$\mathcal{G}^A(\mathbf{x}-\mathbf{x}')$. If we perform a Taylor
expansion for $\lambda|\mathbf{x}|\ll 1$, we have:

\begin{eqnarray}\frac{\sinh(\lambda|\mathbf{x}|)}{\lambda|\mathbf{x}|}\simeq \text{constant}+|\mathbf{x}|^2+\dots\,.
\end{eqnarray}
For fixed value of the distance $|\mathbf{x}|$, the external
potential $\Phi^A_{out}(\mathbf{x})$ depends on the value of the
radius $\xi$, then we have that the  Gauss theorem does not work.
In this case, the potential depends on the total mass and on the
distribution of  matter in the space. In particular, if the matter
distribution takes a bigger volume, the potential
$|\Phi^A_{out}(\mathbf{x})|$ increases and viceversa. We can write

\begin{eqnarray}\lim_{\xi\rightarrow\infty}\frac{\lambda\xi\cosh(\lambda\xi)-\sinh(\lambda\xi)}{\lambda^3\xi^3}=\infty\,;
\end{eqnarray}
obviously the limit of $\xi$ has to be interpreted up to the
maximal value  where the generic position $|\mathbf{x}|$ in the
space is fixed. If we consider the limit $\xi\rightarrow 0$ (the
point-like source limit), we obtain

\begin{eqnarray}\lim_{\xi\rightarrow
0}3\frac{\lambda\xi\cosh(\lambda\xi)-\sinh(\lambda\xi)}{\lambda^3\xi^3}=1\,.
\end{eqnarray}
For $_A\Phi_{out}(\mathbf{x})$, we have

\begin{eqnarray}\lim_{\xi\rightarrow 0}\,\,_A\Phi_{out}(\mathbf{x})&=&-\frac{2\sqrt{2}\,\,\pi^{3/2}}{a_1}\frac{GM}
{|\mathbf{x}|}+\frac{8\sqrt{2}\,\,\pi^{3/2}}{3a_1}\frac{GMe^{-\lambda_1|\mathbf{x}|}}{|\mathbf{x}|}\nonumber\\&&-
\frac{2\sqrt{2}\,\,\pi^{3/2}}{3a_1}\frac{GMe^{-\lambda_1|\mathbf{x}|}}{|\mathbf
{x}|}\,.\end{eqnarray} The last expression is compatible with the
discussion in \S\,\ref{grenn-function-paragraph}.

\subsection{Further solutions by the Green functions $\mathcal{G}^B(\mathbf{x}-\mathbf{x}')$ and $\mathcal{G}^C
(\mathbf{x}-\mathbf{x}')$}

For the sake of completeness, let us derive  solutions for the
other  Green functions. Starting from  Case B in Table
\ref{tablegrennfunction}, we have:

\begin{eqnarray}\mathcal{T}^{B,\,\lambda_i}(\mathbf{x})=\left\{\begin{array}{ll}\frac{3GM}{\lambda_i^2\xi^3}\biggl\{-1+
[\cos(\lambda_i\xi)+\lambda_i\xi\sin(\lambda_i\xi)]\frac{\sin(\lambda_i|\mathbf{x}|)}{\lambda_i|\mathbf{x}|}\biggr\}\,\,\,
|\mathbf{x}|<\xi\\\\\frac{3GM}{{\lambda_i}^2\xi^3}[\sin(\lambda_i\xi)-\lambda_i\xi\cos(\lambda_i\xi)]\frac{\cos(\lambda_i
|\mathbf{x}|)}{\lambda_i|\mathbf{x}|}\,\,\,\,\,\,\,\,\,\,\,\,\,\,\,\,\,\,\,\,\,\,\,\,|\mathbf{x}|>\xi\,,\end{array}\right.
\end{eqnarray} in the inner and outer region. Also in this case, if
we consider the limit of $\lambda_i\rightarrow 0$, one obtains the
Newtonian limit (\ref{newlimit}). The internal and external
potential for given $\lambda_i$ is

\begin{eqnarray}\left\{\begin{array}{ll}
\Phi_{in}^{B,\lambda^i}(\mathbf{x})\,=\,\frac{3GM}{\lambda_i^2\xi^3}\biggl\{-\mu_1+(\mu_1-{\lambda_i}^2\mu_2)[\cos(\lambda_i
\xi)+\lambda_i\xi\sin(\lambda_i\xi)]\frac{\sin\lambda_i|\mathbf{x}|}{\lambda_i|\mathbf{x}|}\biggr\}\\\\\Phi_{out}^{B,\lambda
_i}(\mathbf{x})=\frac{3GM}{\lambda_i^2\xi^3}(\mu_1-{\lambda_i}^2\mu_2)[\sin(\lambda_i\xi)-\lambda_i\xi\cos(
\lambda_i\xi)]\frac{\cos(\lambda_i|\mathbf{x}|)}{\lambda_i|\mathbf{x}|}\end{array}\right.
\end{eqnarray}
The boundary condition on the surface $|\mathbf{x}|=\xi$ is

\begin{eqnarray}\label{boundcodB}_B\Phi_{in}(\xi)-_B\Phi_{out}(\xi)\,=\,-\frac{3GM}{\xi^3}\mu_2\sum_{i=0}^3\mathcal{G}^B_i\,
=\,0\,,\end{eqnarray} (see Table \ref{tablegrennfunction}). The
internal and external potential are given by

\begin{eqnarray}\label{potfinB}\left\{\begin{array}{ll}
_B\Phi_{in}(\mathbf{x})\,=\,-\frac{\sqrt{2}\,\,\pi^{3/2}}{a_1}\frac{GM}{\xi^3}\biggl\{\frac{\lambda_1^2(3\lambda_2^2
\xi^2-2)+8\lambda_2^2}{\lambda_1^2\lambda_2^2}-|\mathbf{x}|^2-\frac{8}{\lambda_1^2}[\cos(\lambda_1\xi)+\\\,\,\,\,\,\,\,\,\,\,
\,\,\,\,\,\,\,\,\,\,\,\,\,\,\,\,\,\,\,\,\,+\lambda_1\xi\sin(\lambda_1\xi)]\frac{\sin(\lambda_1|\mathbf{x}|)}{\lambda_1
|\mathbf{x}|}+\frac{2}{\lambda_2^2}[\cos(\lambda_2\xi)+\lambda_2\xi\sin(\lambda_2\xi)]\frac{\sin(\lambda_2|\mathbf{x}|)}
{\lambda_2|\mathbf{x}|}\biggr\}\\\\_B\Phi_{out}(\mathbf{x})\,=\,-\frac{2\sqrt{2}\pi^{3/2}}{a_1}\frac{GM}{|\mathbf{x}|}+
\frac{8\sqrt{2}\pi^{3/2}}{a_1}\frac{GM}{\lambda_1^3\xi^3}[\sin(\lambda_1\xi)-\lambda_1\xi\cos(\lambda_1\xi)]\frac{\cos(
\lambda_1|\mathbf{x}|)}{|\mathbf{x}|}+\\\,\,\,\,\,\,\,\,\,\,\,\,\,\,\,\,\,\,\,\,\,\,\,\,\,\,\,\,\,\,\,-\frac{2\sqrt{2}
\pi^{3/2}}{a_1}\frac{GM}{\lambda_2^3\xi^3}[\sin(\lambda_2\xi)-\lambda_2\xi\cos(\lambda_2\xi)]\frac{\cos(\lambda_2|\mathbf{x}|)}
{|\mathbf{x}|}\biggr]\end{array}\right.
\end{eqnarray}
The above considerations hold also for the first of
(\ref{potfinB}). The correction term to the Newtonian potential in
the external solution, second line of (\ref{potfinB}), can be
interpreted as the Fourier transform of the matter density. In
fact, we have:

\begin{eqnarray}\int\frac{d^3\mathbf{x}'}{(2\pi)^{3/2}}\rho(\mathbf{x}')e^{-i\mathbf{k}\cdot\mathbf{x}'}=\frac{3M}{(2\pi)^
{2/3}}\frac{\sin(|\mathbf{k}|\xi)-|\mathbf{k}|\xi\cos(|\mathbf{k}|\xi)}{|\mathbf{k}|^3\xi^3}\,,\end{eqnarray}
and in the limit

\begin{eqnarray}\lim_{\xi\rightarrow 0}\int\frac{d^3\mathbf{x}'}{(2\pi)^{3/2}}\rho(\mathbf{x}')e^{-i\mathbf{k}\cdot
\mathbf{x}'}=\frac{M}{(2\pi)^{2/3}}\,,
\end{eqnarray}
we obtain again the external solution for point-like source as
limit of external solution (\ref{potfinB}):

\begin{eqnarray}\lim_{\xi\rightarrow 0}\,\,_B\Phi_{out}(|\mathbf{x}|)&=&-\frac{2\sqrt{2}\,\,\pi^{3/2}}{a_1}\frac{GM}
{|\mathbf{x}|}+\frac{4\sqrt{2}\pi^{3/2}}{3a_1}\frac{GM\cos(\lambda_1|\mathbf{x}|)}{|\mathbf{x}|}\nonumber\\&&-\frac{\sqrt{2}
\pi^{3/2}}{3a_1}\frac{GM\cos(\lambda_2|\mathbf{x}|)}{|\mathbf{x}|}
\,.\end{eqnarray} Finally for Case C in Table
\ref{tablegrennfunction}, we have

\begin{eqnarray}\label{potfinC}\left\{\begin{array}{ll}
_C\Phi_{in}(\mathbf{x})\,=\,-\frac{\sqrt{2}\,\,\pi^{3/2}}{a_1}\frac{GM}{\xi^3}\biggl\{\frac{\lambda_1^2(3\lambda_2^2\xi^2-2)-
8\lambda_2}{\lambda_1^2\lambda_2^2}-|\mathbf{x}|^2+\frac{8}{\lambda_1^2}e^{-\lambda_1\xi}(1+\lambda_1\xi)\frac{\sinh(\lambda_1
|\mathbf{x}|)}{\lambda_1|\mathbf{x}|}\\\\\,\,\,\,\,\,\,\,\,\,\,\,\,\,\,\,\,\,\,\,\,\,\,\,\,\,\,\,\,\,\,+\frac{2}{\lambda_2^2}
[\cos(\lambda_2\xi)+\lambda_2\xi\sin(\lambda_2\xi)]\frac{\sin(\lambda_2|\mathbf{x}|)}{\lambda_2|\mathbf{x}|}\biggl\}\\\\
_C\Phi_{out}(\mathbf{x})\,=\,-\frac{2\sqrt{2}\,\,\pi^{3/2}}{a_1}\frac{GM}{|\mathbf{x}|}+\frac{8\sqrt{2}\,\,\pi^{3/2}}{a_1}
\frac{GM}{\lambda_1^3\xi^3}[\lambda_1\xi\cosh(\lambda_1\xi)-\sinh(\lambda_1\xi)]\frac{e^{-\lambda_1|\mathbf{x}|}}
{|\mathbf{x}|}\\\\\,\,\,\,\,\,\,\,\,\,\,\,\,\,\,\,\,\,\,\,\,\,\,\,\,\,\,\,\,\,\,-\frac{2\sqrt{2}\,\,\pi^{3/2}}{a_1}\frac{GM}
{\lambda_2^3\xi^3}[\sin(\lambda_2\xi)-\lambda_2\xi\cos(\lambda_2\xi)]\frac{\cos(\lambda_2|\mathbf{x}|)}{|\mathbf{x}|}
\end{array}\right.
\end{eqnarray}
The limit of point-like source is valid also in this last case:

\begin{eqnarray}\lim_{\xi\rightarrow 0}\,\,_C\Phi_{out}(\mathbf{x})\,&=&\,-\frac{2\sqrt{2}\,\,\pi^{3/2}}{a_1}\frac{GM}
{|\mathbf{x}|}+\frac{8\sqrt{2}\,\,\pi^{3/2}}{3a_1}\frac{GMe^{-\lambda_1|\mathbf{x}|}}{|\mathbf{x}|}\nonumber\\&&
-\frac{2\sqrt{2}\,\,\pi^{3/2}}{3a_1}\frac{GM\cos(\lambda_2|\mathbf{x}|)}{|\mathbf{x}|}\,.\end{eqnarray}
These results means that for suitable distance scales, the Gauss
theorem is recovered and the theory agrees with the standard
Newtonian limit of GR.

\section{Post $-$ Newtonian scheme of $f$ $-$ gravity}

In this last paragraph of sixth Chapter we want to trace a
methodological approach to "perturbed" Eqs. (\ref{fe}) when we
consider $f\sim f_1R+f_2R^2+f_3R^3+\dots$ (see the Taylor expanse
(\ref{sertay})) and the metric tensor completed at post-Newtonian
order (\ref{PPN-metric}). Obviously the matter tensor is Eq.
(\ref{PPN-tensor-matter}). Eqs. (\ref{fe}) - (\ref{fetr}) at O(2)
- order become

\begin{eqnarray}\label{PPN-field-equation-general-theory-fR-O2}
\left\{\begin{array}{ll}
H^{(2)}_{tt}=f_1R^{(2)}_{tt}-\frac{f_1}{2}R^{(2)}-2f_2\triangle
R^{(2)}=\mathcal{X}\,T^{(0)}_{tt}\\\\
H^{(2)}_{ij}=f_1R^{(2)}_{ij}+\biggl[\frac{f_1}{2}R^{(2)}+2f_2\triangle
R^{(2)}\biggr]\delta_{ij}-2f_2R^{(2)}_{,ij}=\mathcal{X}\,T^{(0)}_{ij}\\\\
H^{(2)}=-6f_2\triangle R^{(2)}-f_1R^{(2)}=\mathcal{X}\,T^{(0)}
\end{array}\right.
\end{eqnarray}
at O(3) - order

\begin{eqnarray}\label{PPN-field-equation-general-theory-fR-O3}
H^{(3)}_{ti}\,=\,f_1R^{(3)}_{ti}-2f_2{R^{(2)}}_{,ti}\,=\,\mathcal{X}\,T^{(1)}_{ti}
\end{eqnarray}
and by remember the expressions for the Christoffel symbols
(\ref{PPN-christoffel}) and
$\ln\sqrt{-g}\sim\frac{1}{2}[g^{(2)}_{tt}-g^{(2)}_{mm}]+\dots$,
finally, at O(4) - order,

\begin{eqnarray}\label{PPN-field-equation-general-theory-fR-O4}
\left\{\begin{array}{ll}
H^{(4)}_{tt}\,=\,f_1R^{(4)}_{tt}+2f_2R^{(2)}R^{(2)}_{tt}-\frac{f_1}{2}R^{(4)}-\frac{f_1}{2}g^{(2)}_{tt}R^{(2)}-\frac{f_2}{2}
{R^{(2)}}^2\\\\\,\,\,\,\,\,\,\,\,\,\,\,\,\,\,\,\,\,\,\,\,-2f_2\biggl[g^{(2)}_{mn,m}{R^{(2)}}_{,n}+\triangle
R^{(4)}+g^{(2)}_{mn}{R^{(2)}}_{,mn}-\frac{1}{2}\nabla
g^{(2)}_{mm}\cdot\nabla R^{(2)}\biggr]\\\\\,\,\,\,\,\,\,\,\,\,\,\,
\,\,\,\,\,\,\,\,\,-6f_3\biggl[|\nabla R^{(2)}|^2+R^{(2)}\triangle
R^{(2)}\biggr]=\mathcal{X}\,T^{(2)}_{tt}\\\\\\
H^{(4)}=-6f_2\triangle R^{(4)}-f_1R^{(4)}-18f_3\biggl[|\nabla
R^{(2)}|^2+R^{(2)}\triangle
R^{(2)}\biggr]\\\\\,\,\,\,\,\,\,\,\,\,\,\,\,\,\,\,\,\,\,\,\,+6f_2\biggl[R^{(2)}_{,tt}-g^{(2)}_{mn}R^{(2)}_{,mn}-\frac{1}{2}
\nabla(g^{(2)}_{tt}-g^{(2)}_{mm})\cdot\nabla
R^{(2)}-g^{(2)}_{mn,m}R^{(2)}_{,n}\biggr]=\mathcal{X}\,T^{(2)}
\end{array}\right.
\end{eqnarray}
The solution for the Ricci scalar $R^{(2)}$, in the last line of
(\ref{PPN-field-equation-general-theory-fR-O2}) is similar to
(\ref{ricci-scalar-solution-ohanlon}). In fact we have

\begin{eqnarray}
R^{(2)}(\textbf{x})=\mathcal{X}\int\frac{d^3\textbf{k}}{(2\pi)^{3/2}}\frac{\tilde{T}^{(0)}(\textbf{k})e^{i\textbf{k}\cdot
\textbf{x}}}{6f_2\textbf{k}^2-f_1}=&-&\frac{8\pi G
\lambda^2}{f_1}\int\frac{d^3\textbf{k}}
{(2\pi)^{3/2}}\frac{\tilde{T}^{(0)}(\textbf{k})e^{i\textbf{k}\cdot\textbf{x}}}{\textbf{k}^2+\lambda^2}=\nonumber\\\nonumber\\
&-&\frac{2G}{f_1}\int
d^3\mathbf{x}'T^{(0)}(\mathbf{x}')\frac{e^{-\lambda|\mathbf{x}-\mathbf{x}'|}}{|\mathbf{x}-\mathbf{x}'|}
\end{eqnarray}
where, again, $\lambda$ is one defined previously
(\ref{yukawa-length}) or (\ref{scale2}). Considering a
generalization of the metric (\ref{PPN-metric-potential-GR})

\begin{eqnarray}
g_{\mu\nu}= \begin{pmatrix}
  1+2\Phi+2\,\Theta & \vec{Z}^T \\
  \vec{Z} & -\delta_{ij}+g^{(2)}_{ij}
\end{pmatrix}
\end{eqnarray}
and requiring the harmonic gauge (\ref{gauge-harmonic}) the
solution for the metric tensor is found. Now, from the expression
of time-time component of Ricci tensor at Newtonian level
(\ref{PPN-ricci-tensor-HG}) we obtain the modified newtonian
potential as solution of the first line of
(\ref{PPN-field-equation-general-theory-fR-O2})

\begin{eqnarray}\Phi(\mathbf{x})=-\frac{2G}{f_1}\int d^3\textbf{x}'\frac{T^{(0)}_{tt}(\textbf{x}')}{|\textbf{x}-
\textbf{x}'|}-\frac{1}{8\pi}\int
d^3\textbf{x}'\frac{R^{(2)}(\textbf{x}')}{|\textbf{x}-
\textbf{x}'|}-\frac{1}{3\lambda^2}R^{(2)}(\textbf{x})\,.\end{eqnarray}
We can check immediately that when $f\rightarrow R$ we find
$\Phi(\textbf{x})\rightarrow-G\int
d^3\textbf{x}'\frac{\rho(\textbf{x}')}{|\textbf{x}-
\textbf{x}'|}$. This outcome is a generalization of solution
(\ref{gravpot}). From the
(\ref{PPN-field-equation-general-theory-fR-O3}) we find the
"vectorial" solution

\begin{eqnarray}
Z_i=-\frac{4G}{f_1}\int
d^3\textbf{x}'\frac{T^{(1)}_{ti}(\textbf{x}')}{|\textbf{x}-
\textbf{x}'|}+\frac{1}{6\pi\lambda^2}\int
d^3\textbf{x}'\frac{R^{(2)}_{,ti'}(\textbf{x}')}{|\textbf{x}-
\textbf{x}'|}\,,
\end{eqnarray}
and from second line of
(\ref{PPN-field-equation-general-theory-fR-O2}) the spatial
tensorial solution

\begin{eqnarray}
g^{(2)}_{ij}=&-&\frac{4G}{f_1}\int
d^3\textbf{x}'\frac{T^{(0)}_{ij}(\textbf{x}')}{|\textbf{x}-
\textbf{x}'|}+\frac{\delta_{ij}}{4\pi}\int
d^3\textbf{x}'\frac{R^{(2)}(\textbf{x}')}{|\textbf{x}-
\textbf{x}'|}+\frac{2\delta_{ij}}{3\lambda^2}R^{(2)}(\textbf{x})\nonumber\\\nonumber\\&-&\frac{2}{3\lambda^2}\biggl[\frac{x_i
x_j}{\mathbf{x}^2}R^{(2)}(\mathbf{x})+\biggl(\delta_{ij}-\frac{3x_ix_j}{\mathbf{x}^2}\biggr)\frac{1}{|\mathbf{x}|^3}\int_0^
{|\mathbf{x}|}d|\mathbf{x}'||\mathbf{x}|'^2R^{(2)}
(\mathbf{x}')\biggr]\,.
\end{eqnarray}
We can affirm that it is possible to have solution non-Ricci-flat
in vacuum: \emph{HOG mimics a matter source}.

From the fourth order of field equation, we note also the Ricci
scalar ($R^{(4)}$) propagates with the same $\lambda$ (the second
line of (\ref{PPN-field-equation-general-theory-fR-O4})) and the
solutions at second order originates a supplementary matter source
in r.h.s. of (\ref{fe}):

\begin{eqnarray}
\frac{1}{\lambda^2}\triangle
R^{(4)}-R^{(4)}\,=\,\frac{18f_3}{f_1}T_a+\frac{1}{\lambda^2}T_b+\frac{\mathcal{X}}{f_1}\,T^{(2)}
\end{eqnarray}
where the functions $T_a$, $T_b$ are known. The Ricci scalar at
fourth order is

\begin{eqnarray}
R^{(4)}(\textbf{x})=-\frac{\mathcal{X}}{4\pi f_1}\int
d^3\mathbf{x}'T^{(2)}(\mathbf{x}')\frac{e^{-\lambda|\mathbf{x}-\mathbf{x}'|}}{|\mathbf{x}-\mathbf{x}'|}-\frac{18f_3}{4\pi
f_1}\int
d^3\mathbf{x}'T^{(4)}_a(\mathbf{x}')\frac{e^{-\lambda|\mathbf{x}-\mathbf{x}'|}}{|\mathbf{x}-\mathbf{x}'|}\nonumber\\\nonumber\\
-\frac{1}{4\pi\lambda^2}\int
d^3\mathbf{x}'T^{(4)}_b(\mathbf{x}')\frac{e^{-\lambda|\mathbf{x}-\mathbf{x}'|}}{|\mathbf{x}-\mathbf{x}'|}
\end{eqnarray}
Also in this case we can have a non-vanishing curvature in absence
of matter. At last the $tt$ - component at fourth order can be
reformulated as follows

\begin{eqnarray}
\triangle
\Theta\,=\,\frac{\mathcal{X}}{f_1}\,T^{(2)}_{tt}+\text{contributions
from previously order}
\end{eqnarray}
then it is possible to find a generale solution for $tt$ -
component at fourth order of metric tensor.

With this last paragraph we wanted to resume a methodological
approach to Post-Newtonian limit of $f$-gravity, if $f$ is an
analytical function of Ricci scalar. The development of $f$ is
performed in $R\,=\,0$ and the Ricci tensor components are
expressed in the harmonic gauge.

\clearpage{\pagestyle{empty}\cleardoublepage}

\chapter{The post-Minkowskian approximation in $f$ $-$ gravity: Gravitational Waves in higher order gravity}

In this chapter, we develop the post-Minkowskian limit of HOG
theories [\textbf{F}]. It is well known that when dealing with GR
such an approach provides massless spin-two waves as propagating
degree of freedom of the gravitational field while ETGs imply
other additional propagating modes in the gravity spectra. We show
that a general analytic HOG model, together with a standard
massless graviton, is characterized by a massive scalar particle
with a finite-distance interaction. We briefly discuss how such
massive gravitational mode can have relevant consequences both on
cosmological and small scales distances affecting the stochastic
background of gravitational waves and representing a valid
alternative to Dark Matter on galactic scales. Furthermore we
develop an analytic definition of the energy-momentum tensor of
the gravitational field in such a scheme. Such a tensor represents
a basic quantity in order to calculate the gravitational time
delay in Pulsar timing.

\section{The Post $-$ Minkowskian approximation in spherically symmetric solution}

In Chapter \ref{newtonian-limit-fourth-order-gravity}, we have
found the spherically symmetric solution of $f$-gravity in the
Newtonian limit with the metric (\ref{me5}). Here we want to
discuss a different limit of these theories, pursued when the
small velocity assumption is relaxed and only the weak field
approximation is retained. This situation is related to the
Minkowski limit of the underlying gravity theory as well as the
discussion of the Chapter
\ref{newtonian-limit-fourth-order-gravity} was related to the
Newtonian one. In order to develop such an analysis, we can
reasonably resort to the metric (\ref{me5}), considering the
gravitational potentials $g_{tt}(t,r)$ and $g_{rr}(t,r)$ in the
suitable form

\begin{eqnarray}
\left\{\begin{array}{ll}g_{tt}(t,r)\,=\,1+g^{(1)}_{tt}(t,r)\\\\
g_{rr}(t,r)\,=\,1+g^{(1)}_{rr}(t,r)\end{array} \right.
\end{eqnarray}
with $g^{(1)}_{tt}, g^{(1)}_{rr}\ll 1$. Let us now perturb the
field equations (\ref{fe}), with respect to approach (\ref{eqp0})
- (\ref{eqp1}), considering, again, the Taylor expansion
(\ref{sertay}) for a generic $f$ - theory. For the vacuum case
($T_{\mu\nu}\,=\,0$), at the first order with respect to
$g^{(1)}_{tt}$ e $g^{(1)}_{rr}$, it is

\begin{eqnarray}\label{fe-GW-standard-coordinates}
\left\{\begin{array}{ll}f_0=0\\\\
f_1\biggl\{R^{(1)}_{\mu\nu}-\frac{1}{2}g^{(0)}_{\mu\nu}R^{(1)}\biggr\}+\mathcal{H}^{(1)}_{\mu\nu}=0\end{array}
\right.
\end{eqnarray}
where

\begin{eqnarray}
\mathcal{H}^{(1)}_{\mu\nu}=-f_2\biggl\{R^{(1)}_{,\mu\nu}-{\Gamma^{(0)}}^{\rho}_{\mu\nu}R^{(1)}_{,\rho}
-g^{(0)}_{\mu\nu}\biggl[{g^{(0)\rho\sigma}}_{,\rho}R^{(1)}_{,\sigma}+g^{(0)\rho\sigma}R^{(1)}_{,\rho\sigma}+\\+g^{(0)\rho
\sigma}\ln\sqrt{-g}^{(0)}_{,\rho}R^{(1)}_{,\sigma}\biggr]\biggr\}\,.
\end{eqnarray}
In this approximation, the Ricci scalar turns out to be  zero
while  the derivatives, in the previous relations, are calculated
at $R\,=\,0$. Again we find $f$-Lagrangians without the
cosmological contribution as in (\ref{eq0}).

Let us now consider the limit for large $r$, i.e. we study the
problem far from the source of the gravitational field. In such a
case the (\ref{fe-GW-standard-coordinates}) become

\begin{eqnarray}\label{fe-GW-standard-coordinates-1}
\left\{\begin{array}{ll}\frac{\partial^2g^{(1)}_{tt}}{\partial r^2}-\frac{\partial^2g^{(1)}_{rr}}{\partial t^2}=0\\\\
f_1\biggl[g^{(1)}_{tt}-g^{(1)}_{rr}\biggr]-8f_2\biggl[\frac{\partial^2g^{(1)}_{rr}}{\partial
r^2}+\frac{\partial^2g^{(1)}_{tt}}{\partial
t^2}-2\frac{\partial^2g^{(1)}_{rr}}{\partial
t^2}\biggr]\,=\,\text{any function of time $t$}\end{array} \right.
\end{eqnarray}
The (\ref{fe-GW-standard-coordinates-1}) are two coupled wave
equations in term of the two functions $g^{(1)}_{tt}$ and
$g^{(1)}_{rr}$. Therefore, we can ask for a wave-like solutions
for the gravitational potentials

\begin{eqnarray}
\left\{\begin{array}{ll}g^{(1)}_{tt}\,=\,\int\frac{d\omega
dk}{2\pi}\tilde{g}^{(1)}_{tt}(\omega,\,k)e^{i(\omega
t-kr)}\\\\
g^{(1)}_{rr}\,=\,\int\frac{d\omega
dk}{2\pi}\tilde{g}^{(1)}_{rr}(\omega,k)e^{i(\omega
t-kr)}\end{array} \right.
\end{eqnarray}
and substituting these into the
(\ref{fe-GW-standard-coordinates-1}). We find the condition

\begin{eqnarray}
  \left\{\begin{array}{ll}
  \tilde{g}^{(1)}_{tt}(\omega,\,k)\,=\,\tilde{g}^{(1)}_{rr}(\omega,\,k)\,\,\,\,\,\,\,\,\,\,\,\,\,\omega\,=\,\pm\,k
  \\\\
  \tilde{g}^{(1)}_{tt}(\omega,\,k)\,=\,\biggl[1+\frac{3}{4}\frac{\lambda^2}{k^2}\biggr]\tilde{g}^{(1)}_{rr}(\omega,k)\,\,\,
  \,\,\,\,\omega\,=\,\pm\,\sqrt{k^2+\frac{3}{4}\lambda^2}
  \end{array} \right.
\end{eqnarray}
where $\lambda$ is defined in the (\ref{yukawa-length}). In
particular, for $f_1=0$ or $f_2=0$ one obtains solutions with a
dispersion relation $\omega\,=\,\pm k$. In other words, for $f_i
\neq 0$ ($i\,=\,1,2$), that is in the case of non-linear $f$, the
above dispersion relation suggests that massive modes are in
order. In particular the mass of the graviton is
$m_{grav}\,=\,\frac{\sqrt{3}}{2}\lambda$ and, coherently, it is
obtained for a modified real gravitational potential. As matter of
fact, a gravitational potential deviating from the Newtonian
regime induces a massive degree of freedom into the particle
spectrum of the gravity sector with interesting perspective for
the detection and the production of gravitational waves
\cite{cap-cor-del}. It has to be remarked that the presence of
massive gravitons in the wave spectrum of HOG is a well known
result since the paper of \cite{stel}. Nevertheless it is our
opinion that this issue has been always considered under a
negative perspective and has been not sufficiently investigated.

In the post-Minkowskian approximation, as expected, the
gravitational field propagates by means of wave-like solutions.
This result suggests that investigating the gravitational waves
behavior of HOG can represent an interesting issue where a new
phenomenology (massive gravitons) has to be seriously taken  into
account. Besides, such massive degrees of freedom could be a
realistic and testable candidate for cold dark matter, as
discussed in \cite{dub-tiy-tka}.

\section{The post $-$ Minkowskian limit of $f$ $-$ gravity toward gravitational waves}

Let us formally develop, in this section, the post-Minkowskian
limit of HOG models. Such investigation completes the analysis of
the weak field regime of $f$-gravity and it has to be considered
in this present work thesis. \emph{The post-Minkowskian limit of
whatever gravity theory arises when the regime of small field is
considered without any prescription in term of the propagation
velocity of the field. This case has to be clearly distinguished
with respect to the Newtonian limit which, differently, requires
both the small velocity and the weak field approximations.} Often
in literature such a distinction is not clearly remarked and
several cases of pathological analysis can be accounted. The
post-Minkowskian limit of GR naturally furnishes massless waves as
the propagating behavior of gravity in this regime. We can now
develop an analogous study (see paragraph \ref{PM-limit-GR})
considering in place of the Hilbert-Einstein Lagrangian a general
function of the Ricci scalar. Actually, in order to perform a
post-Minkowskian development of field equations one has to
implement the field equations (\ref{fe}) with a small perturbation
on the Minkowski background $\eta_{\mu\nu}$ (\ref{PM-metric}). It
is reasonable to assume that the $f$ - Lagrangian is an analytic
expression in term of the Ricci scalar (\ref{sertay}) (\emph{i.e.}
Taylor expandable around the Ricci scalar value $R\,=\,R_0\,=0$).
In such a case field equations (\ref{fe}), at the first order of
approximation in term of the perturbation become\,:

\begin{eqnarray}\label{PM-field-equation}
f_1\biggl[R^{(1)}_{\mu\nu}-\frac{R^{(1)}}{2}\eta_{\mu\nu}\biggr]-2f_2\biggl[R^{(1)}_{,\mu\nu}-\eta_{\mu\nu}\Box_\eta
R^{(1)}\biggr]\,=\,\mathcal{X}\,T^{(0)}_{\mu\nu}\,.
\end{eqnarray}
From zero-order of (\ref{fe}) one gets again $f(0)\,=\,0$ while
$T_{\mu\nu}$ is fixed at zero-order in (\ref{PM-field-equation})
as in the paragraph \ref{PM-limit-GR}. The explicit expressions of
Ricci tensor and scalar are the same of (\ref{PM-ricci-tensor}).
The (\ref{PM-field-equation}) can be rewritten in a more suitable
form by introducing the constant $\lambda$ (\ref{yukawa-length}):

\begin{eqnarray}\label{PM-field-equation-1}
&&h^\sigma_{(\mu,\nu)\sigma}-\frac{1}{2}\Box_\eta
h_{\mu\nu}-\frac{1}{2}h_{,\mu\nu}-\frac{1}{2}({h_{\sigma\tau}}^{,\sigma\tau}-\Box_\eta
h)\eta_{\mu\nu}+\nonumber\\&&\,\,\,\,\,\,\,\,\,\,\,\,\,\,\,\,\,\,\,\,\,\,\,\,\,\,\,\,\,\,\,\,\,\,\,\,\,\,\,\,+\frac{1}
{3\lambda^2}(\partial^2_{\mu\nu}-\eta_{\mu\nu}\Box_\eta)({h_{\sigma\tau}}^{,\sigma\tau}-\Box_\eta
h)\,=\,\frac{\mathcal{X}}{f_1}T^{(0)}_{\mu\nu}
\end{eqnarray}
and by choosing the harmonic gauge (\ref{gauge-harmonic}):
$\tilde{h}_{\mu\nu}=h_{\mu\nu}-\frac{h}{2}\eta_{\mu\nu}$ with the
condition $\tilde{h}^{\mu\nu}_{\,\,\,\,\,\,\,,\mu}=0$, one obtains
that field equations and the trace equation respectively read

\begin{eqnarray}\label{PM-field-equation-2}
\left\{\begin{array}{ll}\Box_\eta\tilde{h}_{\mu\nu}+\frac{1}{3\lambda^2}(\eta_{\mu\nu}\Box_\eta-\partial^2_{\mu\nu})\Box_\eta
\tilde{h}\,=\,-\frac{2\mathcal{X}}{f_1}\,T^{(0)}_{\mu\nu}
\\\\
\Box_\eta\tilde{h}+\frac{1}{\lambda^2}\Box_\eta^2\tilde{h}\,=\,-\frac{2\mathcal{X}}{f_1}T^{(0)}\end{array}
\right.
\end{eqnarray}
In order to deduce the analytic solutions of
(\ref{PM-field-equation-2}), we can now adopt a dual space
(momentum space) description, this approach can simplify the
equations system and, above all, allows to directly observe what
are the physical properties of our problem. In such a scheme we
have\,:

\begin{eqnarray}\label{femomentum}
\left\{\begin{array}{ll}k^2\tilde{h}_{\mu\nu}(k)+\frac{1}{3\lambda^2}(k_\mu
k_\nu-k^2\eta_{\mu\nu})k^2\tilde{h}(k)\,=\,\frac{2\mathcal{X}}{f_1}T^{(0)}_{\mu\nu}(k)
\\\\ k^2\tilde{h}(k)(1-\frac{k^2}{\lambda^2})\,=\,\frac{2\mathcal{X}}{f_1}T^{(0)}(k)\end{array} \right.
\end{eqnarray}
where

\begin{eqnarray}\label{solgen}
\left\{\begin{array}{ll}\tilde{h}_{\mu\nu}(k)\,=\,\int\frac{d^4x}{(2\pi)^2}\tilde{h}_{\mu\nu}(x)\,\,e^{-ikx}
\\\\
T^{(0)}_{\mu\nu}(k)\,=\,\int\frac{d^4x}{(2\pi)^2}T^{(0)}_{\mu\nu}(x)\,\,e^{-ikx}\end{array}
\right. \end{eqnarray} are the Fourier transforms of the
perturbation $\tilde{h}_{\mu\nu}(x)$ and of the matter tensor
$T^{(0)}_{\mu\nu}$. We have defined, as usual, $k\,x=\omega
t-\textbf{k}\cdot\textbf{x}$ and $k^2=\omega^2-\textbf{k}^2$. On
the other side $\tilde{h}(k)$ and $T^{(0)}(k)$ are the traces of
$\tilde{h}_{\mu\nu}(k)$ and $T^{(0)}_{\mu\nu}(k)$. In the momentum
space one can easily recognize the solutions of
(\ref{femomentum}); the expression for $\tilde{h}_{\mu\nu}(k)$
turns out to be

\begin{eqnarray}
\tilde{h}_{\mu\nu}(k)\,=\,\frac{2\mathcal{X}}{f_1}\frac{T^{(0)}_{\mu\nu}(k)}{k^2}+\frac{2\mathcal{
X}}{3f_1}\frac{k_\mu
k_\nu-k^2\eta_{\mu\nu}}{k^2(k^2-\lambda^2)}T^{(0)}(k),
\end{eqnarray}
which fulfils the condition
$\tilde{h}^{\mu\nu}_{\,\,\,\,\,\,\,,\mu}=0$
($\tilde{h}^{\mu\nu}(k)\,\,k_\mu=0$). The true perturbation
variable $h_{\mu\nu}(k)$ can ne obtained inverting the relation
with the tilded variables, in particular inserting the matter
functions
$S^{(0)}_{\mu\nu}(k)\,=\,T^{(0)}_{\mu\nu}(k)-\frac{1}{2}\eta_{\mu\nu}T^{(0)}(k)$
and $S^{(0)}(k)\,=\,\eta^{\mu\nu}S^{(0)}_{\mu\nu}(k)$, one
obtains:

\begin{eqnarray}
h_{\mu\nu}(k)\,=\,\frac{2\mathcal{X}}{f_1}\frac{S^{(0)}_{\mu\nu}(k)}{k^2}-\frac{\mathcal{
X}}{3f_1}\frac{k^2\eta_{\mu\nu}+2k_\mu
k_\nu}{k^2(k^2-\lambda^2)}S^{(0)}(k)\,,
\end{eqnarray}
which represents a wavelike solution, in the momentum space, with
a massless and a massive contribute since the pole in the
denominator of the second term, whose mass is directly related
with the pole itself. The explicit wavelike solution can be
obtained returning the the configuration space inverting the
Fourier transform of $h_{\mu\nu}$.

Let us remark that field equations (\ref{fe}), for a generical $f$
- Lagrangian, can be rewritten isolating the Einstein tensor in
the l.h.s. as usual for \emph{Curvature Quintessence} \cite{cap,
cap-car-tro, cap-car-car-tro}. In such a case higher than second
order differential contributes, in term of the metric tensor, are
considered in the r.h.s. as a source component of the space-time
dynamics as well as the energy momentum tensor of ordinary matter
does\,:

\begin{eqnarray}\label{fe-curv-tensor}
G_{\mu\nu}\,=\,T^{(m)}_{\mu\nu}+T^{(curv)}_{\mu\nu}\,,
\end{eqnarray}
where

\begin{eqnarray}
\left\{\begin{array}{ll}
T^{(m)}_{\mu\nu}\,=\,\frac{\mathcal{X}\,T_{\mu\nu}}{f'}\\\\
T^{(curv)}_{\mu\nu}\,=\,\frac{1}{2}g_{\mu\nu}\frac{f-f'R}{f'}+\frac{f'_{;\mu\nu}-g_{\mu\nu}\Box
f'}{f'}\end{array} \right.
\end{eqnarray}
Actually if we consider the perturbed metric (\ref{PM-metric}) and
develop the Einstein tensor up to the first order of perturbation
we have

\begin{eqnarray}
G_{\mu\nu}\sim
G^{(1)}_{\mu\nu}=h^\sigma_{(\mu,\nu)\sigma}-\frac{1}{2}\Box_\eta
h_{\mu\nu}-\frac{1}{2}h_{,\mu\nu}-\frac{1}{2}({h_{\sigma\tau}}^{,\sigma\tau}-\Box_\eta
h)\eta_{\mu\nu}
\end{eqnarray}
while the curvature tensor will give the other contributes

\begin{eqnarray}
T^{(curv)}_{\mu\nu}\sim\frac{1}{3\lambda^2}(\eta_{\mu\nu}\Box_\eta-\partial^2_{\mu\nu})({h_{\sigma\tau}}^{,\sigma\tau}-
\Box_\eta h)
\end{eqnarray}
This expression easily  allow to recognize that, in the dual space
of momentum, such a quantity will be responsible of the pole-like
term which implies the introduction of a massive degree of freedom
into the particle spectrum of gravity. In fact, inserting these
two expressions into the the field equations
(\ref{fe-curv-tensor}) and considering the (\ref{PM-ricci-tensor})
we obtain the solution\,:

\begin{eqnarray}
\Box_\eta
h_{\mu\nu}(x)\,=\,-\frac{2\mathcal{X}}{f_1}\biggl[S^{(0)}_{\mu\nu}(x)+\Sigma^\lambda_{\mu\nu}(x)\biggr]
\end{eqnarray}
where $\Sigma^\lambda_{\mu\nu}(x)$ is related with the Curvature
tensor and is defined as

\begin{eqnarray}
\Sigma^\lambda_{\mu\nu}(x)\,=\,-\frac{1}{6}\int\frac{d^4k}{(2\pi)^2}\frac{k^2\eta_{\mu\nu}+2k_\mu
k_\nu}{k^2-\lambda^2}S^{(0)}(k)\,\,e^{ikx}\,.
\end{eqnarray}
The general solution for the metric perturbation $h_{\mu\nu}(x)$,
when equation are given as in (\ref{fe-curv-tensor}), can be
rewritten as

\begin{eqnarray}\label{gensolFT}
h_{\mu\nu}(x)=\frac{2\mathcal{X}}{f_1}\int\frac{d^4k}{(2\pi)^2}\frac{S^{(0)}_{\mu\nu}(k)}{k^2}\,\,e^{ikx}-\frac{\mathcal{
X}}{3f_1}\int\frac{d^4k}{(2\pi)^2}\frac{k^2\eta_{\mu\nu}+2k_\mu
k_\nu}{k^2(k^2-\lambda^2)}S^{(0)}(k)\,\,e^{ikx}\,,
\end{eqnarray}
which displays in the second term a pole whose properties can be
easily evaluated in vacuum.  In fact, in such a case (i.e.
$T_{\mu\nu}=0$), the (\ref{PM-field-equation-2}) becomes

\begin{eqnarray}
\left\{\begin{array}{ll}k^2[\tilde{h}_{\mu\nu}(k)+\frac{1}{3\lambda^2}(k_\mu
k_\nu-k^2\eta_{\mu\nu})\tilde{h}(k)]=0 \\\\
k^2\tilde{h}(k)(1-\frac{k^2}{\lambda^2})=0\end{array} \right.
\end{eqnarray}
showing that allowed solutions are of two types along the
relations\,:

\begin{eqnarray}
\left\{\begin{array}{ll} \omega=\pm |\textbf{k}| \\\\
h_{\mu\nu}(x)=\int\frac{d^4k}{(2\pi)^2}h_{\mu\nu}(k)\,\,e^{ikx}&\,\,\,\,\text{with}
\,\,\,\,\,\,h(k)=0\end{array} \right.
\end{eqnarray}
and

\begin{eqnarray}
\left\{\begin{array}{ll} \omega\,=\,\pm
\sqrt{\textbf{k}^2+\lambda^2} \\\\
h_{\mu\nu}(x)\,=\,-\int\frac{d^4k}{(2\pi)^2}\biggl[\frac{\lambda^2\eta_{\mu\nu}+2k_\mu
k_\nu}{6\lambda^2}\biggr]\,h(k)\,\,e^{ikx}
&\,\,\,\,\text{with}\,\,\,\,\,\,h(k)\neq 0\end{array} \right.
\end{eqnarray}
It is evident, that the first solution represents a massless
graviton according with standard prescriptions of GR while the
second one gives a massive degree of freedom with $m^2\doteq
\lambda^2$. In this sense we can furtherly rewrite the
(\ref{PM-field-equation-2}) introducing $\phi\doteq\Box\tilde{h}$
so that the general system can be rearranged in the following way

\begin{eqnarray}
\left\{\begin{array}{ll}\Box_\eta\tilde{h}_{\mu\nu}\,=\,-\frac{2\mathcal{X}}{f_1}T^{(0)}_{\mu\nu}+\biggl[\frac{\partial^2_
{\mu\nu}-\eta_{\mu\nu}\Box_\eta}{3m^2}\biggr]\phi\\\\
(\Box_\eta+m^2)\phi\,=\,-\frac{2\mathcal{X}}{f_1}m^2T^{(0)}\end{array}
\right.
\end{eqnarray}
which suggests that the higher order contributes act in the
post-Minkowskian limit as a massive scalar field whose mass
depends on the degree of deviation ($f',\ f''$), calculated on the
background unperturbed metric, of the initial $f$ - Lagrangian
with respect to the standard Hilbert-Einstein expression.

It is important to remark that the peculiarity of a massive
contribute in the wave spectrum of HOG is strictly related with
the peculiar behavior of the trace equation with respect to GR. In
fact in the case of HOG theories the trace equation establishes a
constraint for the Ricci scalar under the form of a dynamical
equation. This relation allows a more complex evolution of the
system since the Ricci scalar is not univocally fixed by the trace
equation as in GR. In fact, while in the framework of GR vacuum
solutions imply $R\,=\,0$ (\emph{i.e.} it holds the Poisson
equation in the Newtonian Limit), in the HOG models $R$ can assume
a generical dynamical evolution according with (\ref{fetr}), which
assumes the zero value only under certain hypotheses on the
nonlinear Lagrangian. This behaviour, as widely displayed in
chapter \ref{newtonian-limit-fourth-order-gravity}, implies, as
natural consequence, a modify of the Poisson equation in the
Newtonian limit to a form which allows a modified gravitational
potential in such a regime. This characteristic is directly
related with the massive degree of freedom obtained in the
post-Minkowskian limit of these theories. In other words, such
peculiarity is a different representation, at a different scale
(or energy range), of the same effect. One can easily notice that
the characteristic length of the modified gravitational potential
enters in the wave solution exactly as the mass parameter
$m^2\,\doteq\,-\frac{f_1}{6f_2}$ of the additive component in the
gravitational wave spectrum.

\section{Strong gravitational waves in a general $f$ $-$ gravity}

We are interesting to study the field equation for a small
perturbation $h_{\mu\nu}$ on the background metric
$g^{(0)}_{\mu\nu}$ ($O(h)^2\ll 1$). Where $g^{(0)}_{\mu\nu}$ is a
solution of GR with Ricci scalar $R=R^{(0)}=0$ (this solution is
in the vacuum). Then the relativistic invariant is described as

\begin{eqnarray}\label{me-GW}ds^2\,=\,g_{\sigma\tau}dx^\sigma dx^\tau=(g^{(0)}_{\sigma\tau}+h_{\sigma\tau})dx^\sigma dx^\tau
\end{eqnarray}
Obviously the lowering and rising of the index have been made with
the metric background $g^{(0)}_{\mu\nu}$. The field equation
(\ref{fe}) at zero order, if we consider the development shown in
the paragraph \ref{development-pertirbative}, is

\begin{eqnarray}\label{fe-hard-GW}f'^{(0)}R^{(0)}_{\mu\nu}-\frac{1}{2}g^{(0)}_{\mu\nu}f^{(0)}=\mathcal{X}\,T^{(0)}_{\mu\nu}
\end{eqnarray} from the which the trace equation states
$f^{(0)}=-\frac{\mathcal{X}}{2}T$. But in the vacuum the trace is
vanishing and we have to impose the condition $f^{(0)}\,=\,0$ (we
neglect the cosmological contribute). Then the field equation
(\ref{fe-hard-GW}) becomes

\begin{eqnarray}R^{(0)}_{\mu\nu}=0\end{eqnarray}
and the metric $g^{(0)}_{\mu\nu}$ is solution for the field
equation (\ref{fe}) at zero order in the vacuum (Schwarzschild
solution). At the first order we have

\begin{eqnarray}\label{fe-hard-GW-2}
f'^{(0)}\biggl\{R^{(1)}_{\mu\nu}-\frac{1}{2}g^{(0)}_{\mu\nu}R^{(1)}\biggr\}-f''^{(0)}
\biggl\{R^{(1)}_{;\mu\nu}-g^{(0)}_{\mu\nu}\Box_{g^{(0)}}R^{(1)}\biggr\}=\mathcal{X}\,T^{(1)}_{\mu\nu}
\end{eqnarray}
where the derivatives covariant have been calculated with respect
to metric $g^{(0)}_{\mu\nu}$. Now the expressions for Ricci tensor
and scalar are

\begin{eqnarray}
\left\{\begin{array}{ll}R^{(1)}_{\mu\nu}=h^\sigma_{(\mu;\nu)\sigma}-\frac{1}{2}\Box_{g^{(0)}}h_{\mu\nu}-\frac{1}{2}
h_{;\mu\nu}\\\\
R^{(1)}={h_{\sigma\tau}}^{;\sigma\tau}-\Box_{g^{(0)}}h
\end{array}\right.
\end{eqnarray}
and the general field equation perturbed (\ref{fe-hard-GW-2}) is

\begin{eqnarray}\label{fe-hard-GW-3}
&&2h^\sigma_{(\mu;\nu)\sigma}-\Box_{g^{(0)}}h_{\mu\nu}-h_{;\mu\nu}-g^{(0)}_{\mu\nu}\biggl({h_{\sigma\tau}}^
{;\sigma\tau}-\Box_{g^{(0)}}h\biggr)+\nonumber\\\nonumber\\&&\,\,\,\,\,\,\,\,\,\,\,\,\,\,\,\,\,\,\,\,+\frac{2}{3\tilde{
\lambda}^2}\biggl(\nabla_\mu\nabla_\nu-g^{(0)}_{\mu\nu}\Box_{g^{(0)}}\biggl)\biggl({h_{\sigma\tau}}^{;\sigma\tau}-
\Box_{g^{(0)}}h\biggr)=\frac{\mathcal{X}}{f'^{(0)}}\,T^{(1)}_{\mu\nu}
\end{eqnarray}
where $\tilde{\lambda}=-\frac{f'^{(0)}}{3f''^{(0)}}$. If
$f\rightarrow R$ one has $f''\rightarrow0$,
$\tilde{\lambda}\rightarrow\infty$ and the (\ref{fe-hard-GW-3})
becomes

\begin{eqnarray}
G^{(1)}_{\mu\nu}=R^{(1)}_{\mu\nu}-\frac{1}{2}g^{(0)}_{\mu\nu}R^{(1)}=\mathcal{X}\,T^{(1)}_{\mu\nu}
\end{eqnarray}
\emph{i.e.} the first order for Einstein equation in GR with the
same conditions for the metric $g^{(0)}_{\mu\nu}$. The trace of
(\ref{fe-hard-GW-3}) is

\begin{eqnarray}\label{fe-tr-hard-GW-3}
\Box_{g^{(0)}}\biggl({h_{\sigma\tau}}^{;\sigma\tau}-\Box_{g^{(0)}}h\biggr)+\tilde{\lambda}^2\biggl({h_{\sigma\tau}}^{;\sigma
\tau}-\Box_{g^{(0)}}h\biggr)=-\frac{\tilde{\lambda}^2\mathcal{X}}{f'^{(0)}}\,T^{(1)}\,.\end{eqnarray}
By using the harmonic gauge condition (\ref{gauge-harmonic}) from
the equations (\ref{fe-hard-GW-3}) and (\ref{fe-tr-hard-GW-3}) one
has

\begin{eqnarray}
\left\{\begin{array}{ll}
2\tilde{h}^\sigma_{(\mu;\nu)\sigma}-\Box_{g^{(0)}}\tilde{h}_{\mu\nu}+\frac{1}{\tilde{\lambda}^2}\biggl(\nabla_\mu\nabla_\nu-
g^{(0)}_{\mu\nu}\Box_{g^{(0)}}\biggl)\Box_{g^{(0)}}\tilde{h}=\frac{\mathcal{X}}{f'_0}T^{(1)}_{\mu\nu}
\\\\
(\Box_{g^{(0)}}+\tilde{\lambda}^2)\Box_{g^{(0)}}\tilde{h}=-\frac{\mathcal{X}}{f'^{(0)}}\tilde{\lambda}^2T^{(1)}
\end{array}\right.
\end{eqnarray}

\section{Energy-momentum tensor of $f$ $-$ gravity}

In order to detect gravitational waves the construction of a
number of sensitive detectors for gravitational waves (GWs) is
underway today. At the moment there are several laser
interferometers already built like the VIRGO detector (Italy), the
GEO 600 detector (Germany), the two LIGO detectors (United
States), the TAMA 300 detector (Japan) and many bar detectors are
currently in operation too. Since very soon there will be a huge
amount of experimental data the results of these detectors will
have a fundamental impact on astrophysics and gravitation physics.
GW detectors  will be of fundamental importance in order to probe
GR and, above all, to check every alternative theory of
gravitation \cite{cap-cor, cap-1, tob-suz-kur}. A possible target
of these experiments is the so called stochastic background of
gravitational waves \cite{allen, all-ott, maggiore, grishchuk,
grishchuk-1, corda} which can be related with the inflationary
scenario settled in the early universe evolution. Actually there
is another very challenging test dealing with gravitational waves
phenomenology: the gravitational time delay in Pulsar timing. This
experiment is one of the most important evidence of GR validity,
since allows to verify the correction to the orbital period of
pulsars as predicted by Einstein gravity theory \cite{hulse}.
Therefore, this experiment represents an unescapable test in order
to check a whatever viable gravity theory. An analytic calculation
of this problem has been performed in the case of the Brans -
Dicke theory with positive results since there are not significant
constraint on the Brans - Dicke parameter $\omega$ \cite{zaglauer,
wil-zag}. Actually, in order to calculate what is the physical
effect of HOG model on a pulsar system one has calculate the
energy-momentum tensor of the gravitational field. This quantity
will characterize the energy loss due to the gravitational
irradiation. Although the procedure to calculate the
energy-momentum tensor of the gravitational field in GR is often
debated, one can extend the formalism developed for a generical
field theory and obtain this quantity varying functionally on the
gravity Lagrangian in term of the Lagrange operator obtaining a so
called pseudo-tensor whose properties does not completely fulfils
dippheomorphisms invariance\footnote{This quantity is tipically
referred as the Landau-Lifshitz energy-momentum tensor,
nevertheless other kinds of energy-momentum tensor of the
gravitational field can be defined}. Such calculation need to be
extended when dealing with an HOG model since field equations are
of order higher than two.

In standard field theory, given a generical Lagrangian
$\mathcal{L}=\mathcal{L}(g_{\mu\nu},g_{\mu\nu,\rho},g_{\mu\nu,\rho\sigma})$
depending even on accelerations, field equations are obtained
considering a variational principle which considers all the
explicit functions. Thus, in the case of a HOG Lagrangian which
depends on the metric and its derivatives up to the second order
one has

\begin{eqnarray}
&&\delta\int d^4x\sqrt{-g}f\,=\,\delta\int
d^4x\mathcal{L}(g_{\mu\nu},g_{\mu\nu,\rho},g_{\mu\nu,\rho\sigma})\sim\nonumber\\\nonumber\\&&\int
d^4x\biggl(\frac{\partial\mathcal{L}}{\partial
g_{\rho\sigma}}-\partial_\lambda\frac{\partial\mathcal{L}}{\partial
g_{\rho\sigma,\lambda}}+\partial^2_{\lambda\xi}\frac{\partial\mathcal{L}}{\partial
g_{\rho\sigma,\lambda\xi}}\biggr)\delta g_{\rho\sigma}=\int
d^4x\sqrt{-g}H^{\rho\sigma}\delta_{\rho\sigma}=0
\end{eqnarray}
where $\sim$ means we neglected a pure divergence. Then we can set
\,:

\begin{eqnarray}
\int
d^4x\partial_\lambda\biggl[\biggl(\frac{\partial\mathcal{L}}{\partial
g_{\rho\sigma,\lambda}}-\partial_\xi\frac{\partial
\mathcal{L}}{\partial g_{\rho\sigma,\lambda\xi}}\biggr)\delta
g_{\rho\sigma}+\frac{\partial\mathcal{L}}{\partial
g_{\rho\sigma,\lambda\xi}}\delta
g_{\rho\sigma,\xi}\biggr]\rightarrow\,0\,.\end{eqnarray} As matter
of facts, one can write generalized Euler-Lagrange equations for
this framework\,:

\begin{eqnarray}
H^{\rho\sigma}=\frac{1}{\sqrt{-g}}\biggl[\frac{\partial
\mathcal{L}}{\partial
g_{\rho\sigma}}-\partial_\lambda\frac{\partial
\mathcal{L}}{\partial
g_{\rho\sigma,\lambda}}+\partial^2_{\lambda\xi}\frac{\partial\mathcal{L}}{\partial
g_{\rho\sigma,\lambda\xi}}\biggr]\,=\,0\,,\end{eqnarray} which
coincide with the field equations (\ref{fe}) in the vacuum.
Actually, even in the case of such general model it is possible to
define an energy momentum tensor of the field, in particular
$t^\lambda_{\,\,\,\,\,\alpha}$ turns out to be defined as
follows\,:

\begin{eqnarray}\label{ET-general-lagrangina-definition}
t^\lambda_{\,\,\,\,\,\alpha}\,=\,\frac{1}{\sqrt{-g}}\biggl[\biggl(\frac{\partial\mathcal{L}}{\partial
g_{\rho\sigma,\lambda}}-\partial_\xi\frac{\partial\mathcal{L}}{\partial
g_{\rho\sigma,\lambda\xi}}\biggr)g_{\rho\sigma,\alpha}+\frac{\partial\mathcal
{L}}{\partial
g_{\rho\sigma,\lambda\xi}}g_{\rho\sigma,\xi\alpha}-\delta^\lambda_\alpha\mathcal{L}\biggr]\,.
\end{eqnarray}
The (\ref{ET-general-lagrangina-definition}) quantity together the
energy-momentum tensor of matter $T_{\mu\nu}$ satisfies a
conservation according with standard requirements. In fact since
in presence of matter $H_{\mu\nu}\,=\,\mathcal{X}\,T_{\mu\nu}$,
one has

\begin{eqnarray}
(\sqrt{-g}t^\lambda_{\,\,\,\,\,\alpha})_{,\lambda}\,=\,-\sqrt{-g}H^{\rho\sigma}g_{\rho\sigma,\alpha}=-\mathcal{X}
\sqrt{-g}T^{\rho\sigma}g_{\rho\sigma,\alpha}=-2\mathcal{X}(\sqrt{-g}T^\lambda_{\,\,\,\,\,\alpha})_{,\lambda}
\end{eqnarray}
and as a consequence

\begin{eqnarray}
[\sqrt{-g}(t^\lambda_{\,\,\,\,\,\alpha}+2\mathcal{X}\,T^\lambda_{\,\,\,\,\,\alpha})]_{,\lambda}=0
\end{eqnarray}
which demonstrates the conservation law. We can now write down the
expression of the energy-momentum tensor
$t^\lambda_{\,\,\,\,\,\alpha}$ of the gravitational field in term
of the $f$ - gravity action and the respective derivatives, in
such a way to have a completely general expression\,:

\begin{eqnarray}\label{ET-f(R)-lagrangina-definition}
t^\lambda_{\,\,\,\,\,\alpha}=&&f'\biggl\{\biggl[\frac{\partial
R}{\partial
g_{\rho\sigma,\lambda}}-\frac{1}{\sqrt{-g}}\partial_\xi
\biggl(\sqrt{-g}\frac{\partial R}{\partial
g_{\rho\sigma,\lambda\xi}}\biggr)\biggl]g_{\rho\sigma,\alpha}+\frac{\partial
R }{\partial
g_{\rho\sigma,\lambda\xi}}g_{\rho\sigma,\xi\alpha}\biggr\}\nonumber\\\nonumber\\&&-f''R_{,\xi}\frac{\partial
R }{\partial
g_{\rho\sigma,\lambda\xi}}g_{\rho\sigma,\alpha}-\delta^\lambda_\alpha
f\,.\end{eqnarray}

Let us notice that while in GR $t^\lambda_{\,\,\,\,\,\alpha}$ is
non-covariant quantity, the relative generalization in HOG models
turns out to satisfy the covariance prescription behaving as an
ordinary tensor. One can easily verify that such an expression
reduces to the usual definition of the Landau-Lifshitz
energy-momentum tensor of GR when $f\,=\,R$.

\begin{eqnarray}
{t^\lambda_{\,\,\,\,\,\alpha}}_{|_{\text{GR}}}=\frac{1}{\sqrt{-g}}\biggl(\frac{\partial\mathcal{L}_{\text{GR}}}{\partial
g_{\rho\sigma,\lambda}}g_{
\rho\sigma,\alpha}-\delta^\lambda_\alpha\mathcal{L}_{\text{GR}}\biggr)\end{eqnarray}
where the GR Lagrangian has been considered in its effective form,
i.e. the symmetric part of the Ricci tensor, which effectively
characterizes the variation principle leading to the motion
equations
$\mathcal{L}_{\text{GR}}=\sqrt{-g}g^{\mu\nu}(\Gamma^\rho_{\mu\sigma}\Gamma^\sigma_{\rho\nu}-\Gamma^\sigma_{\mu\nu}\Gamma^
\rho_{\sigma\rho})$ \cite{dirac}.

It is important to stress that GR definition of the
energy-momentum tensor and HOG definition are quite different.
These discrepancies are due to the presence, in the second case,
of higher than second order differential term in the gravity
action, which cannot be discarded by means of a boundary
integration as it is done in GR. We have remarked above that the
effective Lagrangian of GR turns out to be the symmetric part of
the Ricci scalar since the second order terms present in the
definition of $R$ can be discarded by means of integration by
part.

A generic analytic $f$ - Lagrangian is characterized from the
dynamical point of view only by the first two terms of its Taylor
expansion once the perturbation is implemented at the linear
level, \emph{i.e.} $f\sim f'(0)\,R+\mathcal{F}(R)$, where the
function $\mathcal{F}$ satisfies the condition:
$\lim_{R\rightarrow 0}\mathcal{F}\rightarrow R^2$. As a
consequence one can rewrite the explicit expression of
(\ref{ET-f(R)-lagrangina-definition}) as\,:

\begin{eqnarray}\label{ET-f(R)-lagrangina-definition-1}
{t^\lambda_{\,\,\,\,\,\alpha}}=&&f'(0){t^\lambda_\alpha}_{|_{\text{GR}}}+\mathcal{F}'\biggl\{\biggl[\frac{\partial
R}{\partial g_{\rho
\sigma,\lambda}}-\frac{1}{\sqrt{-g}}\partial_\xi\biggl(\sqrt{-g}\frac{\partial
R}{\partial
g_{\rho\sigma,\lambda\xi}}\biggr)\biggl]g_{\rho\sigma,\alpha}+\frac{\partial
R }{\partial
g_{\rho\sigma,\lambda\xi}}g_{\rho\sigma,\xi\alpha}\biggr\}\nonumber\\\nonumber\\&&-\mathcal{F}''R_{,\xi}\frac{\partial
R }{\partial
g_{\rho\sigma,\lambda\xi}}g_{\rho\sigma,\alpha}-\delta^\lambda_\alpha
\mathcal{F}
\end{eqnarray}
Let us recall the general expression of the Ricci scalar
(\ref{ricciscalar}) splitting its linear ($R^*$) and quadratic
($\bar{R}$) dependence once a perturbed metric is considered

\begin{eqnarray}\label{defRicciscalar}R\,=\,R^*+\bar{R}\end{eqnarray}
(notice that $\mathcal{L}_{\text{GR}}=-\sqrt{-g}\bar{R}$).
Actually, in the case of GR ${t^\lambda_\alpha}_{|_{\text{GR}}}$
the Landau - Lifshitz tensor shows as a first non vanishing term a
$h^2$ contribute. A similar result can be obtained in the case of
HOG models. In fact considering the expression
(\ref{ET-f(R)-lagrangina-definition-1}) one obtains that at the
lower expansion order ${t^\lambda_{\,\,\,\,\,\alpha}}$ reads\,:

\begin{eqnarray}\label{ET-f(R)-lagrangina-definition-2}
{t^\lambda_{\,\,\,\,\,\alpha}}\sim{t^\lambda_{\,\,\,\,\,\alpha}}_{|h^2}&=&f'(0)\,{t^\lambda_{\,\,\,\,\,\alpha}}_{|_
{\text{GR}}}+f''(0)\,R^*\biggl[\biggl(-\partial_\xi\frac{\partial
R^*}{\partial
g_{\rho\sigma,\lambda\xi}}\biggr)g_{\rho\sigma,\alpha}+\frac{\partial
R^*}{\partial
g_{\rho\sigma,\lambda\xi}}g_{\rho\sigma,\xi\alpha}\biggr]\nonumber\\\nonumber\\
&&-f''(0)R^*_{,\xi}\frac{\partial R^*}{\partial
g_{\rho\sigma,\lambda\xi}}g_{\rho\sigma,\alpha}-\frac{1}{2}f''(0)\delta^\lambda_\alpha
{R^*}^2=\nonumber\\\nonumber\\&=&f'(0){t^\lambda_{\,\,\,\,\,\alpha}}_{|_{\text{GR}}}+f''(0)\biggl[R^*\biggl(\frac{\partial
R^*}{\partial g_{\rho\sigma,\lambda\xi}}g_{\rho\sigma,\xi\alpha}-\frac{1}{2}R^*\delta^\lambda_\alpha\biggr)\nonumber\\
\nonumber\\&&-\partial_\xi\biggl(R^*\frac{\partial R^*}{\partial
g_{\rho\sigma,\lambda\xi}}\biggr)g_{\rho\sigma,\alpha}\biggr]\,.\end{eqnarray}
Now, since for a perturbed metric (\ref{PM-metric}) $R^*\sim
R^{(1)}$, where $R^{(1)}$ is defined as in
(\ref{PM-ricci-tensor}), one has

\begin{eqnarray}
\left\{\begin{array}{ll}\frac{\partial R^*}{\partial
g_{\rho\sigma,\lambda\xi}}\sim\frac{\partial R^{(1)}}{\partial
h_{\rho\sigma,\lambda\xi}}=\eta^{\rho\lambda}\eta^{\sigma\xi}-\eta^{\lambda\xi}\eta^{\rho\sigma}\\\\\frac{\partial
R^*}{\partial
g_{\rho\sigma,\lambda\xi}}g_{\rho\sigma,\xi\alpha}\sim
h^{\lambda\xi}_{\,\,\,\,\,\,,\xi\alpha}-h^{,\lambda}_{\,\,\,\,\,\alpha}
\end{array}\right.
\end{eqnarray}
and the first significant term in
(\ref{ET-f(R)-lagrangina-definition-2}) is of second order in the
perturbation. We can now write down the expression of the energy -
momentum tensor explicitly in term of the perturbation $h$:

\begin{eqnarray}
{t^\lambda_{\,\,\,\,\,\alpha}}&\sim&f'(0){t^\lambda_\alpha}_{|_{\text{GR}}}+f''(0)\{(h^{\rho\sigma}_{\,\,\,\,\,\,\,,\rho
\sigma}-\Box
h)[h^{\lambda\xi}_{\,\,\,\,\,\,\,,\xi\alpha}-h^{,\lambda}_{\,\,\,\,\,\,\,\alpha}-\frac{1}{2}\delta^\lambda_\alpha(h^{\rho
\sigma}_{\,\,\,\,\,\,\,,\rho\sigma}-\Box
h)]\nonumber\\\nonumber\\&&-h^{\rho\sigma}_{\,\,\,\,\,\,\,,\rho\sigma\xi}h^{\lambda\xi}_{\,\,\,\,\,\,\,,\alpha}+h^{\rho\sigma
\,\,\,\,\,\,\,\,\,\,\lambda}_{\,\,\,\,\,\,\,,\rho\sigma}h_{,\alpha}+h^{\lambda\xi}_{\,\,\,\,\,\,\,,\alpha}\Box
h_{,\xi}-\Box h^{,\lambda}h_{,\alpha}\}\,,
\end{eqnarray}
in term of the tilded perturbation $\tilde{h}_{\mu\nu}$ the new
contribution reads\,:

\begin{eqnarray}
{t^\lambda_{\,\,\,\,\,\alpha}}_{|_f}=\frac{1}{2}\biggl[\frac{1}{2}\tilde{h}^{,\lambda}_{\,\,\,\,\alpha}\Box\tilde{h}-
\frac{1}{2}\tilde{h}_{,\alpha}\Box\tilde{h}^{,\lambda}-\tilde{h}^{\lambda}_{\,\,\,\,\,\sigma,\alpha}\Box\tilde{h}^{,\sigma}
-\frac{1}{4}(\Box\tilde{h})^2\delta^\lambda_\alpha\biggr]\,.
\end{eqnarray}
As matter of facts the energy-momentum tensor of the gravitational
field, which expresses the energy transport of this field during
its propagation, can have a natural generalization in the case of
HOG models. We have adopted in our case the Landau-Lifshitz
definition, however some other approaches are in order as outlined
in \cite{mul-put-vag-vil}. The general definition of
$t^\lambda_{\,\,\,\,\,\alpha}$ obtained above consists of a sum
considering the GR contribute plus a term which takes into account
corrections induced by the higher differential order of $f$ -
theories\,:

\begin{eqnarray}
t^\lambda_{\,\,\,\,\,\alpha}\,=\,f'(0)\,{t^\lambda_{\,\,\,\,\,\alpha}}_{|_{\text{GR}}}+f''(0)\,{t^\lambda_{\,\,\,\,\,\alpha}}
_{|_f}\,,
\end{eqnarray}
and again when $f\,=\,R$ we obtains
$t^\lambda_{\,\,\,\,\,\alpha}\,=\,{t^\lambda_\alpha}_{|_{\text{GR}}}$
as already discussed.

Quantities obtained along this section represent the basic
elements in order to develop an analytic calculation of the
gravitational time delay in the pulsar timing in the framework of
HOG models. Nevertheless such analysis is beyond the purposes of
the current study and will be argument of a forthcoming
investigation.

\clearpage{\pagestyle{empty}\cleardoublepage}

\chapter{Discussions and conclusions}

ETGs are good candidates to solve several shortcomings of modern
astrophysics and cosmology since they seem, in a natural way, to
address the problem of cosmological dynamics without introducing
unknown forms of dark matter and dark energy (see e.g.
\cite{cap-card-tro2, cap}). Nevertheless, a "final" alternative
theory solving all the issues has not been found out up to now and
the debate on modifying gravitational sector or adding new (dark)
ingredients is still open. Beside this general remark related to
the paradigm (extending gravity and/or adding new components),
there is the methodological issue to "recover" the standard and
well-tested results of GR in the framework of these enlarged
schemes. The recover of a self-consistent Newtonian limit (or a
weak field limit)is the test bed of any theory which pretends to
enlarge or correct the GR. In fact GR has been consistently tested
in physical situations implying, essentially, spherical symmetry
and weak field limit \cite{will}. One of the fundamental and
obvious issue that any theory of gravity should satisfy is the
fact that, in absence of gravitational field or very far from a
given distribution of sources, the spacetime has to be
asymptotically flat (Minkowski). Any alternative or modified
gravitational theory (beside the diffeomorphism invariance and the
general covariance) should address these physical requirements to
be consistently compared with GR. This is a crucial point which
several times is not considered when people is constructing the
weak field limit of alternative theories of gravity.

In our opinion, such a task has to be pursued in the natural frame
of the theory otherwise the results could be misleading.
Specifically, we have to develop the limit in the Jordan frame
without conformal transformations to the Einstein frame since such
transformations could alter the interpretation of the results.

In this PhD thesis, we have considered the Taylor expansion of a
generic $f$-theory, obtaining general solutions in term of the
metric coefficients up to the second order of approximation when
matter is neglected. In particular, the solution relative to the
$g_{tt}$ metric component gives the gravitational potential which
is corrected with respect to the Newtonian one of $f\,=\,R$. The
general gravitational potential is given by a Yukawa-like terms,
combined with the Newtonian potential, which is effectively
achieved at small distances. Besides also starting from the
standard corrections to the Hilbert - Einstein Action (the
well-known quadratic ones), but now the matter is present, we have
faced, in same systematic way, the problem to find out solutions.
The solutions are found using the Green function method and we
have derived several solutions where the Newton potential is
corrected by combinations Yukawa-like terms. We have classified
the results considering $i)$ the parameters in the Lagrangian,
$ii)$ the field equations and $iii)$ the resulting potential. In
relation to the sign of the characteristic coefficients entering
the $g_{tt}$ component, one can obtain real or complex solutions.
In both cases, the resulting gravitational potential has physical
meanings. A discussion on the non-validity of the Gauss theorem
has been given. Furthermore, for spherically symmetric
distributions of matter, we discussed the inner and the outer
solutions. Furthermore, it has been shown that the \emph{Birkhoff
theorem} is not a general result for $f$ - gravity. This is a
fundamental difference between GR and HOG. While in GR a
spherically symmetric solution is static, here time-dependent
evolution can be achieved depending on the order of perturbations
[\textbf{C}, \textbf{E}, \textbf{G}].

From other hand it is possible also to calculate Newtonian limit
of such theories with a redefinition of the degrees of freedom by
some scalar field leading to the so called O'Hanlon Lagrangian
\cite{ohanlon}. In fact, considering this latter approach, we get
a scalar-tensor theory with vanishing kinetic term and a potential
term linked to $f$-theory. Also in this case we found a
Yukawa-like correction to classic Newtonian potential.
Nevertheless when we turn off the modification of Hilbert-Einstein
Lagrangian we do not obtain the right Newtonian potential. In fact
only in this limit $f\rightarrow R$ it has sense speaking about
the Eddington parameter $\gamma$ and its value is $1/2$ and not
$\gamma\sim 1$ as observed. The origin of inconsistency is in the
not-well defined field equation when $f\rightarrow R$. In fact
this problem in present also in Brans-Dicke theory and only by
requiring $\omega_{BD}\rightarrow\infty$ we obtain the GR
[\textbf{H}].

We have discussed the differences between the post-Newtonian and
the post - Minkowskian limit in $f$ - gravity. The main result of
such an investigation is the presence of massive degrees of
freedom in the spectrum of gravitational waves which are strictly
related to the modifications occurring into the gravitational
potential. This occurrence could constitute an interesting
opportunity for the detection and investigation of gravitational
waves. To do this it needs to generalize the energy-momentum
tensor for a generic $f$-gravity. In the last chapter we tried to
find a new expression for a HOG [\textbf{F}].

Starting from Tensor-multi-scalar theory of gravity \cite{dam-esp}
we can show how a polynomial Lagrangian in the Ricci scalar $R$,
compatible with the PPN-limit, can be recovered in the framework
of HOG. The approach is based on the formulation of the PPN-limit
of such gravity models developed in analogy with scalar-tensor
gravity \cite{cap-tro}. In particular, considering the local
relations defining the PPN fourth order parameters as differential
expressions, one obtains a third-order polynomial in the Ricci
scalar which is parameterized by the PPN-quantity $\gamma$ and
compatible with the limit $\beta\,=\,1$. The order of deviation
from the linearity in $R$ is induced by the deviations of $\gamma$
from the GR expectation value $\gamma\,=\,1$. Actually, the PPN
parameter $\gamma$ may represent the key parameter to discriminate
among relativistic theories of gravity [\textbf{A}].

Besides we investigated also the viability to find spherical
solutions is in $f$-theories with an perturbation methodic
analysis with respect to standard results of GR when we consider
the limits $r\rightarrow\infty$ and $f\rightarrow R$. Essentially,
spherical solutions can be classified, with respect to the Ricci
curvature scalar $R$, as $R\,=\,0$, $R\,=\,R_0\neq 0$, and
$R\,=\,R(r)$, where $R_0$ is a constant and $R(r)$ is a function
of the radial coordinate $r$. In order to achieve exact spherical
solutions, a crucial role is played by the relations existing
between the metric potentials and between them and the Ricci
curvature scalar. In particular, the relations between the metric
potentials and the Ricci scalar can be used as a constraint: this
gives a Bernoulli equation. Solving it, in principle, spherically
symmetric solutions can be obtained for any analytic $f$ function,
both for constant curvature scalar and for curvature scalar
depending on $r$. Such spherically symmetric solutions can be used
as background to test how generic $f$ theories of gravity deviate
from GR. Particularly interesting are theories that imply
$f\rightarrow R$ in the weak field limit. In such cases, the
experimental comparison is straightforward and also experimental
results, evading GR constraints, can be framed in a
self-consistent picture \cite{ber-boh-lob}. Finally, we have
constructed a perturbation approach in which we search for
spherical solutions at the $0th$-order and then we search for
solutions at the first order. The scheme is iterative and could
be, in principle, extended to any order in perturbations. The
crucial request is to take into account $f$ - theories which are
Taylor expandable about some value $R\,=\,R_0$ of the curvature
scalar. A important remark is in order at this point. Considering
interior and exterior solutions, the junction conditions are
related to the integration constants of the problem and strictly
depend on the source (e.g. the form of $T_{\mu\nu}$). We have not
considered this aspect here since we have, essentially, searched
for vacuum solutions. However, such a problem has to be carefully
faced in order to deal with physically consistent solutions. For
example, the Schwarzschild solution $R\,=\,0$, which is one of the
exterior solutions which we have considered, always satisfies the
junction conditions with physically interesting interior metric.
This is not the case for several spherically symmetric solutions
which could give rise to unphysical junction conditions and not be
in agreement with Newton's law of gravitation, also
asymptotically. In these cases, such solutions have to be
discarded [\textbf{D}].

We have discussed a general method to find out exact solutions in
ETGs when a spherically symmetric background is taken into
account. In particular, we have searched for exact spherically
symmetric solutions in $f$-gravity by asking for the existence of
Noether symmetries. We have developed a general formalism and
given some examples of exact solutions. The procedure consists in:
$i)$ considering the point-like $f$ Lagrangian where spherical
symmetry has been imposed; $ii)$ deriving the Euler-Lagrange
equations; $iii)$ searching for a Noether vector field; $iv)$
reducing dynamics and then integrating the equations of motion
using conserved quantities. Viceversa, the approach allows also to
select families of $f$ models where a particular symmetry (in this
case the spherical one) is present. As examples, we discussed
power law models and models with constant Ricci curvature scalar.
However, the above method can be further generalized. If a
symmetry exists, the Noether Approach allows transformations of
variables where the cyclic ones are evident. This fact allows to
reduce dynamics and then  to get more easily exact solutions.
These considerations show that the Noether Symmetries Approach can
be applied to large classes of gravity theories.  Up to now the
Noether symmetries Approach has been worked out in the case of
FRW\,-\,metric. In this PhD work, we have concentrated our
attention to the development of the general formalism in the case
of spherically symmetric spacetimes. Therefore the fact that, even
in the case of a spherical symmetry, it is possible to achieve
exact solutions seems to suggest that this technique can represent
a paradigmatic approach to work out exact solutions in any theory
of gravity. A more comprehensive analysis in this sense  will be
the argument of forthcoming studies [\textbf{B}].

\clearpage{\pagestyle{empty}\cleardoublepage}

\backmatter

\chapter{List of papers}

\begin{description}

\item[A] \emph{Fourth-order gravity and experimental constraints on
Eddington parameters} - S. Capozziello, A. Stabile, A. Troisi,
Modern Physics Letters A \textbf{21}, 2291 (2006);

\item[B] \emph{Spherically symmetric solutions in $f(R)$-gravity via
Noether symmetry approach} - S. Capoz-ziello, A. Stabile, A. Troisi, Classical and Quantum Gravity \textbf{24}, 2153 (2007);

\item[C] \emph{Newtonian limit of $f(R)$-gravity} - S. Capozziello, A. Stabile, A. Troisi, Physical Review D \textbf{76}, 104019
(2007);

\item[D] \emph{Spherical symmetry in $f(R)$-gravity} - S. Capozziello, A. Stabile, A. Troisi, Classical and Quantum Gravity \textbf{25},
085004 (2008);

\item[E] \emph{The Newtonian limit of metric gravity theories with quadratic
Lagrangians} - S. Capozziello, A. Stabile, Submitted to Classical and Quantum Gravity;

\item[F]\emph{The post Minkowskian limit of $f(R)$-gravity} - S. Capozziello, A. Stabile, A. Troisi, (in preparation);

\item[G]\emph{A general solution in the Newtonian limit
of $f(R)$-gravity} - S. Capozziello, A. Stabile, A. Troisi, Submitted to Modern Physics Letters A;

\item[H] \emph{Comparing scalar-tensor gravity and $f(R)$-gravity in the Newtonian limit} -
Capozziello S. Capozziello, A. Stabile, A. Troisi, (in preparation).

\end{description}

\clearpage{\pagestyle{empty}\cleardoublepage}

\chapter{Acknowledges}

During the period of my PhD I have meet many people who
contributed to increase my consciousness in physics. It would be a
real huge enterprise to mention all of them, but I can not relieve
me to do so at least for these people who most had exert a
particular influence on me.

First of all I want to thank Giovanni Scelza, a friend for my
entire life with whom I shared the hope to succeed in our
porpoises one day, and I feel sad because now he is so far away.

I want to thank all the components of my group of research: the
\emph{GrAsCo} (\emph{Group of Astrophysics and Cosmology}) -
Vincenzo Fabrizio Cardone, Roberto Molinaro, Vincenzo Salzano,
Antonio Troisi - for all the times that I could interact with them
and derive benefit from their competencies.

I want to thank also with all my heart my company of the L1-19
room (ex AI-8 room): Mauro Palo, Luca Parisi, Francesco Romeo ...
and Lorenzo Farias Albano, with whom apart from the innumerable
occasions of comparison and discussions on different aspects of
physics it has increase during our PhD a sincere friendship.

About my permanence aboard, at the University of Oslo I want to
thank Dirk Puetzfeld for his kind disposability and the hole
\emph{Institute of Theoretical Astrophysics} personified in
professor \O ystein Elgaroy.

At last I want to thank the two persons who real believed in my
capacities. I mean the professors Gaetano Scarpetta and Salvatore
Capozziello. Prof. Scarpetta for his constant interest in my
progress, his suggestions on how to approach the current problems
in the camp of physics, but also for his personal interest toward
me. Prof. Capozziello for his constant attention on me during all
the period that starts from the preparation of my degree thesis
when he was my supervisor, till this moment that I conclude my PhD
and for his advices that helped me to overpass many difficult
moments.

\clearpage{\pagestyle{empty}\cleardoublepage}


\begin{thebibliography}{99}


\bibitem{weinberg} Weinberg S., \emph{Gravitation and Cosmology}, Wiley 1972, New York

\bibitem{weyl} Weyl H., Math. Zeit. \textbf{2}, 384 (1918)

\bibitem{weyl1} Weyl H., \emph{Raum-Zeit-Materie}, Springer 1921 Berlin

\bibitem{pauli} Pauli W., Phys. Zeit. \textbf{20}, 457 (1919)

\bibitem{bach} Bach R., Math. Zeit. \textbf{9}, 110 (1921)

\bibitem{eddington} Eddington A. S., \emph{The mathematical theory of relativity}, Cambridge University Press 1924 London

\bibitem{lanczos} Lanczos C., Z. Phys. \textbf{73}, 147 (1931)

\bibitem{schmidt} Schmidt H. J., ArXiv: 0407095 [gr-qc]

\bibitem{guth} Guth A., Phys. Rev. D \textbf{23}, 347 (1981)

\bibitem{cap-card-tro2} Capozziello S., Cardone V. F., Troisi A., JCAP \textbf{0608}, 001 (2006)


\bibitem{cap-card-tro3} Capozziello S., Cardone V. F., Troisi A., Mon.Not. Roy. Astron. Soc. \textbf{375}, 1423 (2007)

\bibitem{noj-odi6} Nojiri S., Odintsov S. D., Int. Jou. Geom. Meth. Mod. Phys. \textbf{4}, 115 (2007)

\bibitem{heh-von-ker-nes} Hehl F. W., von der Heyde P., Kerlick G. D., Nester J. M., Rev. Mod. Phys. \textbf{48}, 393 (1976)

\bibitem{heh-mcc-mie-nee} Hehl F. W., Mccrea J. D., Mielke E. W., Neeman Y., Phys. Rep. \textbf{4258}, 1 (1995)

\bibitem{trautman} Trautman A., Comp.rend. heb. sean. \textbf{257}, 617 (1963)

\bibitem{puetzfeld} Puetzfeld D., New Astron. Rev. \textbf{49}, 59 (2005)

\bibitem{buc-odi-sha} Buchbinder I. L., Odintsov S. D., Shapiro I. L., Effective Action in Quantum Gravity, IOP Publishing 1992 Bristol

\bibitem{ama-elg-mot-mul} Amarzguioui M., Elgaroy O., Mota D. F., Multamaki T., Astron. and Astrophys \textbf{454}, 707 (2006)

\bibitem{mag-fer-fra} Magnano G., Ferraris M., Francaviglia M., Gen. Rel. Grav. \textbf{19}, 465 (1987)

\bibitem{all-bor-fra} Allemandi G., Borowiec A., Francaviglia M., Phys. Rev. D \textbf{70}, 103503 (2004)


\bibitem{sot1} Sotiriou T. P., Class. Quant. Grav. \textbf{23}, 1253 (2006)

\bibitem{sot-lib} Sotiriou T. P., Liberati S., Ann. Phys. 322, 935 (2007)

\bibitem{dam-esp} Damour T., Esposito - Far\`{e}se G., Class. Quantum Grav. \textbf{9}, 2093 (1992)

\bibitem{bondi} Bondi H., \emph{Cosmology}, Cambridge Univ. Press 1952, Cambridge

\bibitem{bra-dic} Brans C., Dicke R. H., Phys. Rev. \textbf{124}, 925 (1961)

\bibitem{cap-der-rub-scu} Capozziello S., de Ritis R., Rubano C., Scudellaro P., La Rivista del Nuovo Cimento \textbf{4},1 (1996)

\bibitem{sciama} Sciama D. W., Mon. Not. R. Ast. Soc. \textbf{113}, 34 (1953)

\bibitem{buchdahl1} Buchdahl H. A., Nuovo Cim. \textbf{23}, 141 (1962)

\bibitem{dewitt} de Witt B. S., \emph{Dynamical theory of groups and fields}, Gordon and Breach (1965) New York

\bibitem{bicknell} Bicknell G. V., Journ. phys. A \textbf{7}, 1061 (1974)


\bibitem{havas} Havas P., Gen. Rel. Grav. \textbf{8}, 631 (1977)

\bibitem{stel} Stelle K., Gen. Rel. Grav. \textbf{9}, 353 (1978)

\bibitem{tey-tou} Teyssandier P., Tourrenc Ph., J. Math. Phys. \textbf{24}, 2793 (1983)

\bibitem{dur-ker} Duruisseau J. P., Kerner R., Gen. Rel. Grav. \textbf{15}, 797 (1983)

\bibitem{lanczos2} Lanczos C., Ann. Math. \textbf{39}, 842 (1938)

\bibitem{bir-dav} Birrell N. D., Davies P.C.W., \emph{Quantum Fields in Curved Space}, Cambridge Univ. Press 1982, Cambridge

\bibitem{vilkovisky} Vilkovisky G., Class. Quantum Grav. \textbf{9}, 895 (1992)

\bibitem{gas-ven} Gasperini M., Veneziano G., Phys. Lett. B \textbf{277}, 256 (1992)

\bibitem{ruz-ruz} Ruzmaikina T. V., Ruzmaikin A. A., JETP \textsc{30}, 372 (1970)

\bibitem{got-sch-sta} Gottl\"ober S., Schmidt H. J., Starobinsky A. A., Class. Quantum Grav. \textbf{7}, 893 (1990)


\bibitem{ame-bat-cap-got-mul-occ-sch} Amendola L., Battaglia - Mayer A., Capozziello S., Gottl\"ober S., M\"uller V., Occhionero F., Schmidt H. J., Class. Quantum Grav. \textbf{10}, L43 (1993)

\bibitem{bat-sch} Battaglia - Mayer A., Schmidt H. J., Class. Quantum Grav. \textbf{10}, 2441 (1993)

\bibitem{schmidt1} Schmidt H. J., Class. Quantum Grav. \textbf{7}, 1023 (1990)

\bibitem{bar-ott} Barrow J., Ottewill A. C., J. Phys. A: Math. Gen. \textbf{16}, 2757 (1983)

\bibitem{starobinsky} Starobinsky A. A., Phys. Lett. B \textbf{91}, 99 (1980)

\bibitem{la-ste} La D., Steinhardt P. J., Phys. Rev. Lett. \textbf{62}, 376 (1989)

\bibitem{maeda} Maeda K., Phys. Rev. D \textbf{39}, 3159 (1989)

\bibitem{wands} Wands D., Class. Quantum Grav. \textbf{11}, 269 (1994)

\bibitem{cap-der-mar} S. Capozziello, R. de Ritis, A.A. Marino., Gen. Rel. Grav. \textbf{30}, 1247 (1998)

\bibitem{cap-noj-odi} Capozziello S., Nojiri S., Odintsov S. D., Phys. Lett. B \textbf{634}, 93 (2006)


\bibitem{am-cap-lit-occ} Amendola L., Capozziello S., Litterio M., Occhionero F., Phys. Rev. D \textbf{45}, 417 (1992)

\bibitem{rie} Riess  A. G. et al., Astron. Journ. \textbf{116}, 1009 (1998)

\bibitem{rie1} Riess A. G. et al., Astroph. Journ. \textbf{607}, 665 (2004)

\bibitem{per} Perlmutter S. et al., Astroph. Journ. \textbf{517}, 565 (1999)

\bibitem{per1} Perlmutter S. et al. Astron. Astrophys. \textbf{447}, 31 (2006

\bibitem{cap} Capozziello S., Int. J. Mod. Phys. D \textbf{11}, 483 (2002)

\bibitem{cap-car-tro} Capozziello S., Carloni S., Troisi A., Rec. Res. Devel. Astronomy. \& Astrophys. \textbf{1}, 625 (2003)

\bibitem{cap-car-car-tro} Capozziello S., Cardone V. F., Carloni S., Troisi A., Int. J. Mod. Phys. D \textbf{12}, 1969 (2003)

\bibitem{car-duv-tro-tur} Carroll, S. M., Duvvuri, V., Trodden, M., Turner, M. S., Phys. Rev. D \textbf{70}, 043528 (2004)

\bibitem{noj-odi1} Nojiri S., Odintsov S. D., Phys. Rev. D \textbf{68}, 123512 (2003)


\bibitem{noj-odi3} Nojiri S., Odintsov S. D., Phys. Lett. B \textbf{576}, 5 (2003)

\bibitem{cap-card-tro1} Capozziello S., Cardone V. F., Troisi A., Phys. Rev. D \textbf{71}, 043503 (2005)

\bibitem{car-dun-cap-tro} Carloni S., Dunsby P. K. S., Capozziello S., Troisi A., Class. Quant. Grav. \textbf{22}, 4839 (2005)

\bibitem{vollick} Vollick D. N., Phys. Rev. D \textbf{68}, 063510 (2003)

\bibitem{men-wan2} Meng X. H., Wang P., Class. Quant. Grav. \textbf{20}, 4949 (2003)

\bibitem{men-wan3} Meng X. H., Wang P., Class. Quant. Grav. \textbf{21}, 951 (2004)

\bibitem{fla} Flanagan E. E., Phys. Rev. Lett. \textbf{92}, 071101 (2004)

\bibitem{fla1} Flanagan E. E., Class. Quant. Grav. \textbf{21}, 417 (2004)

\bibitem{kre-alv} Kremer G. M., Alves D. S. M., Phys. Rev. D \textbf{70}, 023503 (2004)

\bibitem{all-bor-fra1} Allemandi G., Borowiec A., Francaviglia M., Phys. Rev. D \textbf{70}, 043524 (2004)


\bibitem{cap-card-fra} Capozziello S., Cardone V. F., Francaviglia M., Gen. Rel. Grav. \textbf{38}, 711 (2006)

\bibitem{col} Cole S. et al., Mon. Not. Roy. Astron. Soc. \textbf{362}, 505 (2005)

\bibitem{spe} Spergel D. N. et al. Astroph Journ. Suppl. \textbf{148}, 175 (2003)

\bibitem{spe1} Spergel D. N. et al., ApJS \textbf{170}, 377 (2007)

\bibitem{cap-fra} Capozziello S., Francaviglia M., ArXiv: 0706.1146 [astro-ph]

\bibitem{noj-odi} Nojiri S., Odintsov S. D., Int. J. Meth. Mod. Phys. \textbf{4}, 115 (2007)

\bibitem {farhoudi} Farhoudi M., Gen. Relativ. Grav. \textbf{38}, 1261 (2006)

\bibitem{kno} Knop  R. A. et al, Astroph. Journ. \textbf{598}, 102 (2003)

\bibitem{ton} Tonry J. L. et al., Astroph. Journ. \textbf{594}, 1 (2003)

\bibitem{deb} de Bernardis P. et al., Nature \textbf{404}, 955 (2000)


\bibitem{sto} Stompor R. et al., Astroph. Journ. \textbf{561}, L7 (2001)

\bibitem{hin} Hinshaw G. et al. ApJS \textbf{148}, 135 (2003)

\bibitem{sah-sta} Sahni V., Starobinski A., Int. J. Mod. Phys. D \textbf{9}, 373 (2000)

\bibitem{pad} Padmanabhan T., Phys. Rev. D \textbf{66}, 021301 (2002)

\bibitem{pee-rat} Peebles P. J. E., Rathra B., Rev. Mod. Phys., \textbf{75}, 559 (2003)

\bibitem{cop-sam-tsu} Copeland E. J., Sami M., Tsujikawa S., Int. Journ. Mod. Phys. D \textbf{15}, 1753 (2006)

\bibitem{pee-rat1} Peebles P. J. E., Rathra B., Phys. Rev., \textbf{37}, 3406 (1998)

\bibitem{cal-dav-ste} Caldwell E. R., Steinhardt P. J., Phys. Rev. Lett., \textbf{80}, 1582 (1998)

\bibitem{kam-mos-pas} Kamenshchik A., Moschella U., Pasquier V., Phys. Lett. B \textbf{511}, 265 (2001)

\bibitem{bil-tup-vio} Bili\'c N., Tupper G., Viollier R. D., Phys. Lett. B \textbf{535}, 17 (2002)


\bibitem{ben-ber-sen} Bento M. C., Bertolami O., Sen A. A., Phys. Rev. D \textbf{67}, 063003 (2003)

\bibitem{car-tor-tro-cap} Cardone V. F., Tortora C., Troisi A., Capozziello S., Phys. Rev. D \textbf{73}, 043508 (2006)

\bibitem{cal} Caldwell R. R., Phys. Lett. B \textbf{545}, 23 (2002)

\bibitem{sin-sam-dad} Singh P., Sami M., Dadhich N., Phys. Rev. D \textbf{68}, 023522 (2003)

\bibitem{noj-odi-tsu} Nojiri S., Odintsov S. D, Tsujikawa S., Phys. Rev. D \textbf{71}, 063004 (2005)

\bibitem{faraoni2} Faraoni V., Class. Quant. Grav. \textbf{22}, 0607016 (2005)

\bibitem{dva-gab-por} Dvali G. R., Gabadadze G., Porrati M., Phys. Lett. B \textbf{485}, 208 (2000)

\bibitem{dva-gab-kol-nit} Dvali G. R., Gabadadze G., Kolanovic M., Nitti F., Phys. Rev. \textbf{64}, 084004 (2001)

\bibitem{dva-gab-kol-nit1} Dvali G. R., Gabadadze G., Kolanovic M., Nitti F., Phys. Rev. \textbf{64}, 024031 (2002)

\bibitem{maa} Maartens R., Living Rev. Rel. \textbf{7}, 7 (2004)


\bibitem{mul-vil} Multamaki T., Vilja I., Phys. Rev. D \textbf{73}, 024018, (2006)

\bibitem{cap-car-tro1} Capozziello S., Cardone V. F., Troisi A., Phys. Rev. D \textbf{71}, 043503 (2005)

\bibitem{cap-car-car-tro1} Capozziello S., Cardone V. F., Carloni S., Troisi A., Phys. Lett. A \textbf{326}, 292 (2004)

\bibitem{lue-sco-sta} Lue A., Scoccimarro R., Starkman G., Phys. Rev. D \textbf{69}, 044005 (2004

\bibitem{fre-lew} Freese K., Lewis M., Phys. Lett. B \textbf{540}, 1 (2002)

\bibitem{cap-car-pie-ser-tro} Capozziello S., Cardone V. F., Piedipalumbo E., Sereno M., Troisi A., Int. J. Mod. Phys. D \textbf{12}, 381 (2003)

\bibitem{li-bar} Li B., Barrow J. D., Phys. Rev. D \textbf{75}, 084010 (2007)

\bibitem{noj-odi2} Nojiri S., Odintsov S. D., hep-th/06012113 (2006)

\bibitem{li-bar-mot} Li B., Barrow J. D., Mota D. F., submitted to Phys. Rev. D, gr-qc/0705.3795

\bibitem{che-rat} Chen G., Rathra B., ApJ \textbf{582}, 586 (2003)


\bibitem{pod-dal-mor-rat} Podariu S., Daly R. A., Mory M. P., Rathra B., ApJ \textbf{584}, 577 (2003)

\bibitem{faraoni3} Faraoni V., Phys. Rev. D \textbf{72}, 124005 (2005)

\bibitem{cog-zer}  Cognola G., Zerbini S., J. Phys. A \textbf{39}, 6245 (2006)

\bibitem{cog-gas-zer} Cognola G., Gastaldi M., Zerbini S., ArXiv: 0701138 [gr-qc]

\bibitem{li-bar1} Li B., Barrow J., ArXiv: 0701111 [astrp-ph]

\bibitem{car-bun-tro} Carloni S., Dunsby P. K. S., Troisi A., ArXiv: 0707.0106 [gr-qc]

\bibitem{bea-ber-pog-sil-tro} Bean R., Bernat D., Pogosian L., Silvestri A., Trodden M., Phys. Rev. D \textbf{75}, 064020 (2007)

\bibitem{son-hu-saw} Song Y. S., Hu W., Sawicki I., Phys. Rev. D \textbf{75}, 04404 (2007)

\bibitem{cap-card-tro} Capozziello S., Cardone V. F., Troisi A., Mon. Not. R. Ast. Soc. \textbf{375}, 1423 (2007)

\bibitem{fri-sal} Frigerio Martins C., Salucci P., ArXiv: 0703243 [astro-ph]


\bibitem{sobouti} Sobouti Y.,  A\,\&\,A \textbf{464}, 921 (2007)

\bibitem{men-ros} Mendoza S., Rosas Guevara Y. M., A\,\&\,A \textbf{472}, 367 (2007)

\bibitem{sal-lap-ton-gen-yeg-kle} Salucci P., Lapi A., Tonini C., Gentile G., Yegorova I., Klein U., ArXiv: 0703115 [astro-ph]

\bibitem{ber-boe-har-lob} Bertolami O., Boehmer Ch. G., Harko T., Lobo F. S. N., Phys.Rev. D \textbf{75}, 104016 (2007)

\bibitem{noj-odi5} Nojiri S., Odintsov S. D., Gen. Rel. Grav. \textbf{36}, 1765 (2004)

\bibitem{anderson} Anderson J. D. et al., Phys. Rev. Lett. \textbf{81}, 2858 (1998)

\bibitem{anderson1} Anderson J. D. et al., Phys. Rev. D \textbf{65}, 082004 (2002)

\bibitem{ber-boh-lob} Bertolami O., B\"ohmer Ch. G., Lobo F. S. N., gr\,-\,qc/0704.1733

\bibitem{will} Will C. M., \emph{Theory and Experiments in Gravitational Physics} Cambridge Univ. Press 1993, Cambridge

\bibitem{san} Sanders R. H., Ann. Rev. Astr. Ap. \textbf{2}, 1 (1990)


\bibitem{man-kaz} Mannheim P. D., Kazanas D., Ap. J. \textbf{342}, 635 (1989)

\bibitem{qua-sch} Quant I., Schmidt H. J., Astron. Nachr. \textbf{312}, 97 (1991)

\bibitem{sch-ehl-fal} Schneider P., Ehlers J., Falco E. E., \emph{Gravitational Lenses} Springer-Verlag 1992 Berlin

\bibitem{kra-whi} Krauss L. M., White M., Ap. J. \textbf{397}, 357 (1992)

\bibitem{olm} Olmo G. J., Phys. Rev. Lett. \textbf{95}, 261102 (2005)

\bibitem{olm2} Olmo G. J., Phys. Rev. D \textbf{72}, 083505 (2005)

\bibitem{cap-tro} Capozziello S., Troisi A., Phys. Rev. D \textbf{72}, 044022 (2005)

\bibitem{all-fra-rug-tar} Allemandi G., Francaviglia M., Ruggiero M., Tartaglia A., Gen. Rel. Grav. \textbf{37}, 1891 (2005)

\bibitem{mul-vil1} Multam\"aki T., Vilja I. Phys. Rev. D \textbf{74}, 064022 (2006)

\bibitem{mul-vil2} Multam\"aki T., Vilja I. ArXiv: 0612775 [astro-ph]


\bibitem{sot-far-lib} Sotiriou T. P., Faraoni V., Liberati S., arXiv: 0707.2748 [gr-qc]

\bibitem{sot} Sotiriou T. P., Gen. Rel. Grav. \textbf{38}, 1407 (2006)

\bibitem{chi} Chiba T., Phys. Lett. B \textbf{575}, 1 (2005)

\bibitem{eri-smi-kam} Erickcek A. L., Smith T. L., Kamionkowski M., Phys.Rev. D \textbf{74}, 121501 (2006)

\bibitem{cli-bar} Clifton T., Barrow J. D., Phys. Rev. D \textbf{72}, 103005 (2005)

\bibitem{cli-bar1} Clifton T., Barrow J. D., Class. Quant. Grav. \textbf{23}, 2951 (2006)

\bibitem{cli-bar2} Clifton T., Barrow J. D., Class. Quant. Grav. \textbf{23}, L1 (2006)

\bibitem{faraoni1} Faraoni V., ArXiv: 0607016 [gr-qc]

\bibitem{jin-liu-li} Jin X. H., Liu D. J., Li X. Z., ArXiv: 0610854 [astro-ph]

\bibitem{chi-eri-smi} Chiba T., Erickeck A. L., smith T. L., ArXiv: 0611867 [astro-ph]


\bibitem{faraoni4} Faraoni V., Phys. Rev. D \textbf{74}, 023529 (2006)

\bibitem{kai-pii-rei-sun} Kainulainen K., Piilonen J., Reijonen V., Sunhede D., Phys. Rev. D \textbf{76}, 024020  (2007)

\bibitem{whitt} Whitt B., Phys. Rev. D \textbf{38}, 3000 (1988)

\bibitem{bar-ham1} Barraco D. E., Hamity V. H., Phys. Rev. D \textbf{62}, 044027 (2000)

\bibitem{puetzfeld1} Puetzfeld D., The Cosmological post-Newtonian equations of hydrodynamics in General Relativity, (submitted) (2006)

\bibitem{ince} Ince E. L., \emph{Ordinary Differential Equations}, Dover (1956) New York

\bibitem{haw-ell} Hawking S. W., Ellis G. F. R., 1973 T\emph{he large scale structure of space-time}, Cambridge Univ. Press 1973 Cambridge

\bibitem{buchdahl} Buchdahl H. A., J. Phys. A \textbf{12}, 1229 (1979)

\bibitem{einstein1} Einstein A., Ann. der Physik \textbf{49}, 769 (1916)

\bibitem{eisenhart} Eisenhart E., \emph{Riemannian Geometry}, Princeton Univ. Press 1955 Princeton


\bibitem{schroedinger} Schr\"odinger E., \emph{Space-Time Structure}, Cambridge Univ. Press 1960 Cambridge

\bibitem{landau} Landau Lev D., Lif\v{s}its E. M, \emph{Theoretical physics} vol. II,

\bibitem{levicivita} Levi Civita T., \emph{The Absolute Differential Calculus}, Blackie and Son 1929, London

\bibitem{kle} Klein O., New Theories in Physics \textbf{77}, Intern.Inst. of Intellectual Co-operation, League of Nations (1938)

\bibitem{cartan} Cartan E., Ann. Ec. Norm. \textbf{42}, 17 (1925)

\bibitem{palatini} Palatini A., Rend. Circ. Mat. Palermo \textbf{43}, 203 (1919)

\bibitem{arnold} Arnold V. I., \emph{Mathematical Methods of Classical Mechanics}, Springer Verlag 1978 Berlin

\bibitem{app-fre} Appelquist T., Chodos A., Freund P. G. O., \emph{Modern Kaluza-Klein Theories} 1978 Addison-Wesley Reading

\bibitem{kaku} Kaku M., Quantum Field Theory, 1933 Oxford Univ. Press, Oxford

\bibitem{gre-sch-wit} Green M. B., Schwarz J. H., Witten E., \emph{Superstring Theory}, Cambridge Univ. Press 1987 Cambridge


\bibitem{rub-scu} Rubano C., Scudellaro P., Gen. Rel. Grav. \textbf{37}, 521 (2005)

\bibitem{schmidt2} Schmidt H. J., ArXiv: 0602017 [gr-qc]

\bibitem{fer-fra} Ferraris M., Francaviglia M., Mechanics, Analysis and Geometry: 200 Years after Lagrange, Editor: M. Francaviglia, Elsevier Science Publishers 1991

\bibitem{cap-der} Capozziello S., de Ritis R., Class. Quant. Grav. \textbf{11}, 107 (1994)

\bibitem{mag-sok} Magnano G., Soko{\l}owski L. M., Phys. Rev. D \textbf{50}, 5039 (1994)

\bibitem{all-cap-cap-fra} Allemandi G., Capone M., Capozziello S., Francaviglia M., Gen. Rel. Grav. \textbf{38}, 33 (2006)

\bibitem{fer-fra-vol} Ferraris M, Francaviglia M., Volovich I., Class. Quantum Grav. \textbf{11}, 1505 (1994)

\bibitem{cap-lam-sto} Capozziello S., Lambiase G., Stornaiolo C., Ann. Phys. \textbf{10}, 713 (2001)

\bibitem{friedrichs} Friedrichs K.,  Math. Ann. \textbf{98}, 566 (1927)

\bibitem{kilmister} Kilmister C. W., J. Math. Phys. \textbf{12}, 1 (1963)


\bibitem{dautcourt} Dautcourt G., Acta Phys. Polon. \textbf{25}, 637 (1964)

\bibitem{kuenzle} Kuenzle H. P., Gen. Rel. Grav. \textbf{7}, 445 (1976)

\bibitem{ehlers} Ehlers J., Ann. N. Y. Acad. Scien. \textbf{336}, 279 (1980)

\bibitem{ehlers1} Ehlers J., Grundlagenprobleme der modernen Physik, Eds.\ J.\ Nitsch, J.\ Pfarr, E.W.\ Stachow, B. I. - Wissenschaftsverlag, Mannheim, 65 (1981)

\bibitem{dick} Dick R., Gen. Rel. Grav. \textbf{36}, 217 (2004

\bibitem{bus-bar} Bustelo A. J., Barraco D. E., Class. Quant.Grav. \textbf{24}, 2333 (2007)

\bibitem{cap-der1} Capozziello S., de Ritis R., Phys. Lett. \textbf{177}, 1 (1993)

\bibitem{cap-lam} Capozziello S., Lambiase G., Gen. Relativ. Grav. \textbf{32}, 295 (2000)

\bibitem{mar-sal-sim-vit} Marmo G., Saletan E. J., Simoni A., Vitale B., \emph{Dynamical Systems. A Differential Geometric Approach to Symmetry and Reduction} 1985 Wiley New York

\bibitem{mor-fer-lov-mar-rub} Morandi G., Ferrario C., Lo Vecchio G., Marmo G., Rubano C., Phys. Rep. \textbf{188}, 149 (1990)


\bibitem{dam-esp1} Damour T., Esposito - Far\`{e}se G., Phys. Rev. Lett. \textbf{70}, 2220 (1993)

\bibitem{dam-esp2} Damour T., Esposito - Far\`{e}se G. Phys. Rev. D \textbf{54}, 1474 (1996)

\bibitem{dam-esp3} Damour T., Esposito - Far\`{e}se G. Phys. Rev. D \textbf{58}, 042001 (1998)

\bibitem{sch-uza-ria} Schimd C., Uzan J. P., Riazuelo A.,  Phys. Rev. D \textbf{71}, 083512 (2005)

\bibitem{shapiro} Shapiro I. I., Gen. Rel. Grav. \textbf{12}, Ashby N.,, Eds. Cambridge University Press (1993)

\bibitem{shapiro1} Shapiro S. S. et al., Phys. Rev. Lett. D \textbf{92}, 121101 (2004)

\bibitem{williams} Williams J. G., et al., Phys. Rev. D \textbf{53}, 6730 (1996)

\bibitem{ber-ies-tor} Bertotti B., Iess L., Tortora P., Nature \textbf{425}, 374 (2003)

\bibitem{tur-sha-nor} Turyshev S. G., Shao M., Nordtvedt K.L., ArXiv: 0601035 [gr-qc]

\bibitem{sou-woo} Soussa M. E., Woodard R. P., Gen. Rel. Grav. \textbf{36}, 855 (2004)


\bibitem{chi-smi-eri} Chiba T.,, Smith T. L., Erickcek A. L., Phys. Rev. D \textbf{75}, 124014 (2007)

\bibitem{nav-van1} Navarro I., Van Acoleyen K., Phys. Lett. B \textbf{622}, 1 (2005)

\bibitem{ohanlon} O'Hanlon J., Phys. Rev.Lett. \textbf{29}, 137 (1972)

\bibitem{cap-tsu} Capozziello S., Tsujikawa S., arXiv: 0712.2268 [gr-qc]

\bibitem{bin-tre} Binney J., Tremaine S., \emph{Galactic dynamics}, Princeton University Books 1987

\bibitem{cap-cor-del} Capozziello S., Corda Ch., de Laurentis M., Mod. Phys. Lett. A \textbf{22}, 1097 (2007)

\bibitem{dub-tiy-tka} Dubovsky S. L., Tiyakov P. G., Tkachev I. I., Phys. Rev. Lett. \textbf{94}, 181102 (2005)

\bibitem{cap-cor} Capozziello S., Corda C., Int. J. Mod. Phys. D \textbf{15}, 1119 (2006)

\bibitem{cap-1} Capozziello S., \textit{Newtonian Limit of Extended Theories of Gravity} in \textit{Quantum Gravity Research Trends} Ed. A. Reimer, pp. 227 - 276 Nova Science Publishers Inc., NY (2005)

\bibitem{tob-suz-kur} Tobar M. E., Suzuki T., Kuroda K,. Phys. Rev. D \textbf{59}, 102002 (1999)


\bibitem{maggiore} Maggiore M., Phys. Rep. \textbf{331}, 283 (2000)

\bibitem{grishchuk} Grishchuk L. et al., Phys. Usp. \textbf{44} 1 (2001)

\bibitem{grishchuk-1} Grishchuk L. et al., Usp. Fiz. Nauk  \textbf{171} 3 (2001)

\bibitem{corda} Corda C., \textit{VIRGO Report:} \textit{{}``Relic gravitational waves: a {}``snapshot'' of the primordial universe''} VIRGO-NOTE-PIS 1390 (2003)

\bibitem{hulse} Hulse R. A., Taylor J. H., ApJ \textbf{195}, L51 (1975)

\bibitem{zaglauer} Davis W. F., Ph. D. thesis MIT 1979

\bibitem{wil-zag} Will C. M., Zaglauer M. W., Apj \textbf{346}, 366 (1989)

\bibitem{dirac} Dirac P. A. M., \emph{General Theory of Relativity}, John Wiley \& Sons

\bibitem{mul-put-vag-vil} Multamaki T., Putaja A., Vagenas E. C., Vilja I., ArXiv: 0712.0276 [gr-qc]

\bibitem{allen} Allen B. - Proceedings of the Les Houches School on Astrophysical Sources of Gravitational Waves, eds. Jean-Alain Marck and Jean-Pierre Lasota (Cambridge University Press, Cambridge, England (1998)


\bibitem{all-ott} Allen D., Ottewill A. C., Phys. Rev. D \textbf{56}, 545 (1997)

\end{thebibliography}
\end{document}